# UNIVERSITY OF NOTTINGHAM

## FACULTY OF SCIENCE

### School of Physics and Astronomy

**High-harmonic cosmic strings and gravitational waves**

by

**Despoina Pazouli**

Thesis for the degree of Doctor of Philosophy

October 2020

UNIVERSITY OF NOTTINGHAM

# <u>ABSTRACT</u>

FACULTY OF SCIENCE

School of Physics and Astronomy

<u>Thesis for the degree of Doctor of Philosophy</u>

**HIGH-HARMONIC COSMIC STRINGS AND GRAVITATIONAL WAVES**

by Despoina Pazouli

In this thesis we describe high-harmonic cosmic string loops in a general relativistic context, and study the implications of high-harmonic content for the predicted gravitational wave signal from cosmic string networks. Initially, we introduce the variational principle, spacetime concepts and other mathematical tools that we will need for the calculations in the following chapters. We use the variational principle to derive the Einstein field equations and study the solutions of the linearized Einstein equations in the homogeneous and the non-homogeneous case in the harmonic gauge, which are the gravitational waves. We also define basic spacetime concepts and the idea of the standard cosmological model. We introduce the FLRW universe and the $\Lambda CDM$ universe. We then describe the Nambu-Goto cosmic string in a curved spacetime, its equations of motion and its energy-momentum tensor. Fixing the spacetime to be flat, and fixing the gauge, we find the motion of the cosmic string and we present and discuss special solutions. Using the odd-harmonic family of cosmic string loops, we calculate the number of cusps per period and the values of the second derivatives of the left- and right-moving harmonic modes at the cusp, and study their dependence on the harmonic order. We then develop a toy model that calculates the stable daughter loops produced from a parent loop using a statistical approach based on a binary tree description of the loop chopping. We also use the toy model to calculate the average number of cusps produced from a system of loops that self intersect over their course of existence. Assuming a harmonic distribution of the loops that chop off the long string network, we can calculate the average number of cusps produced per loop period from an infinitesimal spacetime volume from a network of loops. We then implement our results in the calculation of the gravitational wave signal of cusps from cosmic string loops as received on Earth. We derive the gravitational waveform emitted from a cusp as observed away from the cusp, in any direction of observation. We then propagate this result in an FLRW spacetime to reach an expression of its amplitude on Earth. Assuming two different cosmic string network models, we implement our above mentioned high-harmonic results to find the amplitude of the signal and the rate at which these signals reach an observer on Earth.

# Contents

# List of Figures

# Declaration of Authorship

I, Despoina Pazouli, declare that the thesis entitled *High-harmonic cosmic strings and gravitational waves* and the work presented in the thesis are both my own, and have been generated by me as the result of my own original research. I confirm that:

- this work was done wholly or mainly while in candidature for a research degree at this University;

- where any part of this thesis has previously been submitted for a degree or any other qualification at this University or any other institution, this has been clearly stated;

- where I have consulted the published work of others, this is always clearly attributed;

- where I have quoted form the work of others, the source is always given. With the exception of such quotations, this thesis is entirely my own work;

- I have acknowledged all main sources of help;

- where the thesis is based on work done by myself jointly with others, I have made clear exactly what was done by others and what I have contributed myself;

- parts of this work have been published as:

    - Despoina Pazouli, Anastasios Avgoustidis, Edmund J. Copeland, "The cusp properties of high harmonic loops" (to appear)

    - Despoina Pazouli, Anastasios Avgoustidis, Edmund J. Copeland, Konstantinos Palapanidis, "GWBs from high harmonic cosmic string loops" (to appear)

Signed: ...............................................................................................................

Date: ...............................................................................................................

# Acknowledgements

I would like to thank my supervisors, Dr Anastasios Avgoustidis and Prof. Edmund Copeland, for their guidance and support during my doctoral studies. Their encouragement and help have been crucial for the completion of this thesis. I would like to thank the external examiner Prof. Christophe Ringeval and the internal examiner Dr. Adam Moss for reading my thesis and for their helpful comments and discussion. I would also like to thank Prof. Anne Green for her comments on and time spent reading my first report.

I am very thankful to Maria Palapanidou, Olympia Kostraki and Panagiotis Palapanidis, for their warm hospitality and company during the preparation of this thesis, I have always felt like being at home at their place.

I would like to express my gratitude to my family, my mother Elsa, my father Panagiotis and my siblings Prodromos and Dimitra, who have always supported me. Their presence has been crucial for the completion of this thesis. All our happy moments together have always given me the strength to continue.

I am endlessly indebted to my collaborator and friend, Konstantinos Palapanidis, for our very useful discussions during the course of my studies and for his continuous help during the development of the models present in the thesis. Without him the toy model for the self-intersection of loops presented in this thesis would not be as complete as it is now. I am very grateful to him for helping me develop it. His constant support and guidance during all the difficult moments I encountered over the past few years was of vital importance.

# Introduction

Cosmic strings are line-like topological defects produced by symmetry breaking phase transitions in a wide range of early universe models [5, 6, 7, 8, 9, 10]. They were envisioned as objects in the universe for the first time by Tom Kibble in 1976 [11]. Motivated by the theory of line vortices in superconductors, he showed that a similar stable structure, which he named a cosmic string, could emerge in the universe during the phase transitions occuring in the early stages of its evolution soon after the Big Bang. Cosmic strings could be produced during the symmetry-breaking phase transitions of many particle theory models, such as those that predict the breaking of a grand unification symmetry [6, 12].

Cosmic strings appear as vortex solutions of the Abelian Higgs equations, driven by a scalar potential. However, it was shown that an Abelian Higgs vortex can be approximated by the solutions of the Nambu-Goto action, which is the action of a two-dimensional relativistic spacetime surface, up to first order in the curvature [13]. This approximation is called the wire approximation or the zero-thickness approximation and it perceives the string as a geometric relativistic object inde-pendent of the structure of the Higgs potential. This description of the cosmic string is called the relativistic string or the Nambu-Goto string.

The mass and size of cosmic strings depends on the energy scale at which the phase transition that led to their formation occurred. If the phase transition occured at $10^{-37}$ to $10^{-35}$ seconds after the Big Bang, which corresponds to a grand unification energy scale of $10^{15}$GeV, then the thickness of a cosmic string should be comparable to the Compton wavelength of a grand unified theory particle, which is about $10^{-29}$cm. The mass per unit length of the cosmic string, which is called the string tension and is usually denoted by $\mu$, is also determined by the grand unified theory. A grand unified theory cosmic string can be as large as the size of the observable universe, stretching outside of it. Such a string would have a mass inside the horizon of $10^{16}$ solar masses, which is of the order of the mass of a cluster of galaxies [14]. Because the length of a cosmic string is significantly larger than its width, it can be approached as a zero width object, which justifies the zero-width approximation.

Superstrings are the basic constituents in fundamental string theory (also called M-theory), which is a theory of unification of all particle interactions. Although superstrings are of elementary size, it was shown that they can grow to macroscopic size in the universe and play the role of cosmic strings [15]. This object is called a cosmic superstring. A potential observation of cosmic superstrings could provide direct evidence for the fundamental theory [16, 12, 17].

Cosmic strings form tangles or networks that evolve dynamically and can produce a host of potentially observable signals. In particular, they are active, incoherent sources of cosmological perturbations and so their predicted effects on the cosmic microwave background (CMB) are very different to those of passive, coherent perturbations generated by cosmic inflation [18, 19, 20, 21]. This has allowed cosmic strings to be strongly constrained, having a maximum allowed contribution to the CMB anisotropy at the level of $\sim 1\%$ [22, 23].

While cosmic strings have been ruled out as the main source of the observed CMB anisotropy, they remain an important subject in modern cosmology. Indeed, the formation of string networks is a generic prediction in a wide range of models of the early universe [24, 25, 26, 27, 28, 29, 30], and so they are extremely interesting from a theoretical point of view. At the same time, they have a rich phenomenology with observational signals relevant to several areas of cosmology and astrophysics [8]. Thus, cosmic strings open an exciting observational window into the early universe. Their observation would be a major discovery in Physics, and as a bonus it would also provide important quantitative information about the physics of the early universe (e.g. the energy scale of the symmetry breaking phase transition that produced the string network). On the other hand, even failure to observe strings is of significant scientific value: as observational sensitivity improves and bounds on cosmic strings become tighter, we are excluding more of the parameter space of our models of the early universe.

The discovery of gravitational waves has reinforced interest in the physics of cosmic strings. In particular, the evolution of the string network leads to the production of closed string loops, which decay primarily via gravitational wave emission. At this point we should acknowledge this is not a uniformly accepted outcome for the string decay. Being modelled as Abelian-Higgs strings, there are also claims in the literature that the dominant form of decay is via the fields themselves and not gravitationally [31]. In this thesis, we will be considering the case where the primary decay route is considered to be through gravitational radiation, giving rise to a stochastic background of gravitational waves. Most of the emission comes from specific events on the string, known as cusps, arising when the local velocity of the string momentarily hits the speed of light, thus producing a burst of beamed gravitational radiation. Kinks, points on the string where the tangent plane to the worldsheet is discontinuous, are also known to contribute significantly to the gravitational wave signal. These signals are now being targeted by gravitational wave detectors including LIGO [32, 33, 3] and in the future LISA [34, 35]. Such targeting brings with it the need for a better quantitative understanding of gravitational radiation from string networks. Some of the most stringent observational bounds on cosmic strings come from their predicted stochastic gravitational wave background which can be constrained either indirectly through pulsar timing observations [36, 37, 38, 39, 40] or directly by gravitational wave detectors [33]. However, these constraints are also the most sensitive to largely unknown parameters, like for example the typical size of string loops in the network. Indeed, while the evolution of the long string component of the network is well understood, the quantitative modelling of the loop component remains uncertain. In particular, the typical size of loops depends on the loop production function [9], which has been the subject of debate over many years [41, 42, 43, 44, 45, 31, 46, 47, 48, 49].

At present, the models used for deriving constraints and forecasts on string networks based on their stochastic gravitational wave background are the small loop model of [4] and the large loop models, referred to as model 1/I, 2/II and 3/III in references [3, 48] respectively. They differ significantly in their assumed/derived loop distribution functions (the number density of loops as a function of loop size and time). As we are entering this exciting era of direct gravitational wave detection it is imperative that modelling be improved and the associated loop parameters become better quantified. In this thesis, we focus on two key parameters, the number of string cusps per oscillation period and the sharpness of cusps, both of which are important components of the overall gravitational signal from cusps. While there has been a considerable amount of work on the role of cusps on networks of cosmic string loops (see for example [50, 51, 52, 53, 54, 4, 55, 1, 46, 56, 57, 47, 49, 35, 58, 59]), the distribution of cusps on higher harmonic loops has not to date been studied in detail. An early attempt to address the issue can be found in Copi and Vachaspati [1] who used numerical simulations to characterise attractor non-self intersecting loop shapes, studying their length, velocity, kink, and cusp distributions.

In reality, we do not really know the harmonic distribution at formation of cosmic string loops, but we do know there could be loops formed off the long string network or as individual loops in the early universe that have many harmonics on them. The traditional picture of such loops is that as they evolve, the majority of them undergo a period where they self intersect. The initial loop then breaks into two daughter loops, with the accompanying formation of a pair of kinks on each daughter loop. These may well then self intersect, and a cascade process takes place whereby the initial high harmonic loop ends up into a class of much smaller non-self intersecting loops. The belief has been that the cusps associated with such non-self intersecting loops play an important role through the strong beams of high frequency gravitational waves they produce, which leads to both a stringy non-Gaussian distribution in the stochastic ensemble of gravitational waves generated by a cosmological network of oscillating loops, but also, and crucially, it can include occasional sharp bursts of gravitational waves from the cusp regions that stand out above the confusion of gravitational wave noise made by smaller overlapping bursts. The results of Damour and Vilenkin [4] suggest that if only 10% of all string loops have cusps, the gravitational wave bursts would be detectable by Advanced LIGO or LISA for string tensions down below $G\mu \sim 10^{-13}$.

In determining these constraints, there are two important parameters whose values need to be assumed. Given the spacetime position of a string $X_\mu(\sigma, \tau)$, where $\sigma$, $\tau$ are the worldsheet coordinates of the string, as we will shortly see the general solution for the string is given in terms of right and left moving modes travelling along it, $\vec{a}(u)$ and $\vec{b}(v)$ where $u = \sigma - \tau$ and $v = \sigma + \tau$. Now two key parameters involved in the gravitational wave calculation are the second derivative with time of the string position evaluated at the cusp $\partial_t^2 X$ (a measure of "cusp sharpness"), and the average number of cusps formed per loop period $T_\ell = \ell/2$ where $\ell$ is the invariant length of the loop. In Equation (3.21) of [4], the authors argue that the generic order of magnitude estimate for $|\vec{a}''| \sim 2\pi/\ell \sim |\vec{b}''|$, where $\vec{a}'' \equiv \frac{d^2\vec{a}}{du^2}$ and $\vec{b}'' \equiv \frac{d^2\vec{b}}{dv^2}$ which are evaluated at the cusp. In other words they expect the coefficient to be of order unity ($2\pi/\ell$ is the natural unit for the string to have). In terms of the number of cusps per loop oscillation, given by $c$ in Equation (5.14) of [4], the authors consider typical values of $c \sim 1$. This is meant to account for the possibility that all of the loops have of order one cusp per oscillation ($c \sim 1$) or, for example, only 10% of the loops have a cusp on them per oscillation ($c \sim 0.1$). This number, as shown in Figure 1 of [4], can have a significant impact on the strength of the gravitational wave amplitude of bursts emitted by cosmic string cusps. Similarly, knowing the true range of values of $|\vec{a}''|$ and $|\vec{b}''|$ is also important for a proper estimate of the strength of the signal emerging from cusp bursts. This is clear from Equations (3.11, 3.12) and (3.23) of [4] where the logarithmic Fourier transform of the gravitational wave burst asymptotic waveform for the cusp emission, hence the amplitude of the wave arriving on Earth, depends on terms of the form $1/(|\vec{a}''||\vec{b}''|)^{1/3}$. In particular if it turns out that a significant fraction of the cusps had associated values $|\vec{a}''| \ll 1$ and $|\vec{b}''| \ll 1$ (in units of $2\pi/\ell$) then it could have a significant impact on the strength of the signal produced.

It is apparent from the above discussion that to accurately quantify gravitational radiation from string networks one must understand: (a) how these two parameters (number of cusps per oscillation period and cusp sharpness) behave as functions of the loop harmonic number and (b) what is the expected loop distribution in terms of their harmonic content, or, at the very least, what is the harmonic content of a typical loop in the network. In this thesis, we aim to address (a), and explore different options for (b). A key goal is to analyse these two parameters, the number of cusps and the values of $|\vec{a}''|$ and $|\vec{b}''|$ at the cusp, for a wide range of high harmonic loops to establish whether there is a correlation between the number of cusps and the harmonic order of the loop, and what distribution of values we have for the magnitude of $|\vec{a}''|$ and $|\vec{b}''|$ evaluated at the cusps, as well as the effect of these values to the gravitational wave signal from a cosmic string network.

In the following we will present the dynamics and gravitational wave radiation emitted from cosmic strings in the zero-width approximation. In chapter 2, we review basic spacetime concepts, which are needed for a relativistic approach to cosmic strings. We also discuss Eulerian variations and derive the Euler-Lagrange equations. These will provide the foundation for deriving the equations of motion of general relativity and of the cosmic string. We also discuss mathematical tools that we will use, the Fourier series and Fourier transform, the delta function and the Heaviside theta, and basic concepts of probability and statistics. Finally, we introduce two numerical methods that we will use in the following to evaluate quantities that are unreachable with analytical means, the Simpson's rule and the Monte Carlo method.

In chapter 3, we introduce the Einstein Hilbert action, to which we apply Eulerian variations to derive the Einstein field equations, for a vacuum spacetime and for spacetime with matter. We introduce the concept of the energy-momentum tensor. We then calculate the linearized Einstein field equations and discuss the gauge issue. We show that the homogeneous solutions of the linearized Einstein field equations are plane waves, which we call gravitational waves, the radiative solutions of general relativity. We discuss their similarity with electromagnetic waves and the particle counterpart of the photon, the graviton. Then, we solve the non-homogeneous linearized Einstein field equations using Green's functions. We show that for a periodic energy-momentum tensor (which is the case for cosmic strings) the solutions away from the source of the gravitational waves is a plane wave.

In chapter 4, we introduce the cosmological concepts. We provide a short his-tory of the formulation of the FLRW universe and the definition of the standard cosmological model. We then introduce the FLRW metric based on the isotropy assumption for the universe, as well as the energy-momentum tensor of the FLRW universe. We discuss the properties of the different matter sources in the universe. We then discuss the dynamics of the FLRW universe, and we present the qualitative universe outcomes from the Friedmann equations. We then define redshift and derive Hubble's law. Finally, we discuss the $\Lambda$CDM model, the most accepted of the standard cosmological models, and solve the single-fluid Friedmann equations to describe the different universe eras.

In chapter 5, we describe the Nambu-Goto relativistic string. We derive the equations of motion and the energy-momentum tensor from the Nambu-Goto action. Assuming a flat background spacetime, we fix the gauge and we solve the Nambu-Goto equations in the conformal gauge for closed strings. We show that the solutions are superpositions of Fourier modes, we define the fundamental period of a cosmic string loop and introduce its left- and right-moving waves. We then introduce the concept of the Kibble-Turok sphere and discuss the presence of cusps and kinks and self-intersections on loops. We discuss the issue of constraining the loop parameters from the Virasoro conditions and how it can be solved. Then, we provide families of cosmic string loop solutions of particular interest. We focus in particular on the odd-harmonic family of loops, for which we calculate the cusps per period it supports, for a given harmonic order, and the value of its left- and right-movers at the point of the cusp. We discuss the cosmic string network, how strings evolve as they propagate in spacetime, and focus on the "one-scale" model and its implications for small and large loops. We discuss in particular the models of cosmic string networks introduced in [4] and the Model 1 from [3]. Since we would like to compare our estimations for the cusp implications on the gravitational wave signal from loops with [4] and [3], we will use for the rest of the parameters the same values the authors of the aforementioned publications used. We develop a toy model that uses some statistical assumptions to predict the self-intersection probability of a parent loop to daughter loops in a cascade of choppings until a loop that no further chops is produced. Assuming a distribution of harmonics in the parent loops we can use the results of the toy model to estimate the value of the cusps per loop period from a spacetime volume $dV(z)$ from the cosmic string network. Finally, we discuss possible issues and improvements of this toy model.

In chapter 6, we calculate the gravitational wave produced from a cusp on a cosmic string loop as seen by an observer away from the loop, following [4]. We also perform the calculation for the gravitational wave from a kink. We will call these gravitational waves, gravitational wave bursts, abbreviated GWBs. We will focus on the GWB contribution from cusps. We then display the gravitational wave amplitude formula for the gravitional wave propagated in an FLRW universe, as it appears to an observer on Earth. Using the above result, we can find the rate at which these GWBs reach the Earth, as well as their signal strength. We calculate the rate using the assumptions of the small loop model from [4], and also the assumptions of Model 1 from [3]. We compare our results for the values of $c$, $|\vec{a}''|$ and $|\vec{b}''|$ obtained in chapter 4, with the results of these two models.

Finally, in chapter 7, we summarize the important results of this thesis and their implications for the observation of GWBs from cusps on cosmic string loops. We also discuss possible next steps that could be implemented to improve our assumptions and calculations.

It has been our aim to write this thesis in a way that it is as self-explanatory as possible. The concept of this is to build a "ladder" that will lead as we move forward in the thesis to the final results that we obtain. The original results that we introduce are in section 5.5, where we calculate the number of cusps per period and the values of $|\vec{a}''|$ and $|\vec{b}''|$ for the odd harmonic string, in section 5.7, where we develop a toy model to predict the loop chopping, and in sections 6.5 and 6.6, where we calculate and discuss the gravitational wave signal implications of our cosmic string results.

CHAPTER 2

---

# Mathematical concepts

---

This chapter will offer a discussion of the mathematical concepts and tools that we will use in this thesis. We start with some concepts of geometry that are used for the description of gravity. We continue with the variational principle and the Euler-Lagrange equations, which give us the means of deriving the equations of motion for the physical systems that we will use. Then, we define the Fourier series and Fourier transform, and determine the convention we will adopt in this thesis, as well as define the Dirac delta function and the Heaviside theta function. We also discuss some probability concepts, define the probability distribution, and give examples of probability distributions that we will need for the following chapters, discuss properties of data sets and define the statistical hypothesis. Finally, we will describe two numerical methods, the Simpson's rule, used for evaluating integrals numerically, and the Monte Carlo method, a stochastic numerical method. We will use these methods in the upcoming chapters.

## 2.1 Spacetime concepts

In General Relativity we adopt the concept of spacetime, a 4-dimensional Lorentzian manifold with metric $g_{ab}$, which embodies the notion of distance on a manifold [60, 61, 62, 63]. The basic properties of the Lorentzian metric are that it is a real symmetric, non-degenerate tensor, with mixed signature. By symmetric tensor we mean that the metric satisfies the identity $g_{ab} = g_{ba}$. Also, this implies that the metric is diagonalizable, because any real symmetric matrix is diagonalizable. Choosing any basis in spacetime that diagonalizes the metric tensor, the non-degeneracy of the metric implies that all of its diagonal elements will be non-zero. The number of positive and negative elements is invariant of the basis we choose, which is called the signature of the metric. In this thesis, we assume the metric signature to be $(-, +, +, +)$. The non-degeneracy of the metric also implies that we can derive the inverse metric tensor through the relation

$$g^{ab}g_{bc} = \delta^a_c, \tag{2.1.1}$$

where $\delta^b_c$ is the Kronecker delta, with $\delta^b_c = 1$ for $b = c$ and $\delta^b_c = 0$ for $b \neq c$. Note that in the above expression we have introduced the Einstein summation convention to simplify the expressions with indexed terms in formulas. Hereafter, we will adopt this notation in any formula with indices. Since the metric tensor is not degenerate, its determinant will be non zero $g = \det g \neq 0$. Also, its sign will be the same in any coordinate basis. From our signature choice, the determinant of the metric will be negative, $g < 0$. We can use the metric and its inverse, defined in (2.1.1) to lower or raise indices of tensors, i.e. $V_a = g_{ab}V^b$ and $V^a = g^{ab}V_a$. We will use the Latin characters $a$, $b$, $c$ ... to denote the indices of tensorial quantities, i.e. quantities that are independent of the basis choice (also called covariant quantities), as we did in equation (2.1.1). We will use Greek letters $\mu$, $\nu$ ... to denote components of tensorial quantities in particular bases or non-tensorial quantities, such as the spacetime coordinates. Tensorial quantities have different components in different coordinate systems. Considering a vector $V^a$, its components $V^\mu$ expressed in the coordinate system $\{x^\mu\}$ transform to its components $V'^\mu$ in the coordinate system $\{x'^\mu\}$ through the relation

$$V'^\mu = \frac{\partial x'^\nu}{\partial x^\rho} V^\rho \tag{2.1.2}$$

where $x'^{\mu}$ are functions of $x^{\mu}$, $x'^{\mu} = f(x^{\mu})$. Equivalently, the components of a tensor $V_a$, called a covector, expressed in the coordinate system $\{x^{\mu}\}$ transform through the relation

$$V'_{\mu} = \frac{\partial x^{\nu}}{\partial x'^{\rho}} V_{\nu} \qquad (2.1.3)$$

to its components $V'_{\mu}$ in the coordinate system $\{x'^{\mu}\}$. For any tensor it holds that its upstairs indices transform under the rule (2.1.2) and its downstairs under the rule (2.1.3). A scalar field transforms under the rule $\phi'(x'^{\mu}) = \phi(x^{\mu})$, i.e. it remains unchanged in any coordinate system. We should keep in mind that an expression written in abstract index notation is always true in Greek indices. The inverse though is not always true. The usage of abstract index notation will help us to differentiate between covariant and component expressions in upcoming chapters where we will choose specific observers and fix the gauge freedom.

An infinitesimally small distance between two points on spacetime, $ds$, is defined as

$$ds^2 = g_{ab}dx^a dx^b, \qquad (2.1.4)$$

where $dx^a$ are infinitesimal coordinate differences, which are vectors on spacetime, and it is invariant under the choice of coordinate system $\{x^{\mu}\}$. Due to the mixed signature of the metric, the distance $ds^2$ can obtain positive, negative or zero values. The two points in spacetime are timelike separated if $ds^2 < 0$, spacelike separated if $ds^2 > 0$ and lightlike separated if $ds^2 = 0$. We will call a vector $V^a$ timelike if $g_{ab}V^aV^b < 0$, spacelike if $g_{ab}V^aV^b > 0$ and lightlike (or null) if $g_{ab}V^aV^b = 0$. We define the relativistic dot product to be $V \cdot W = g_{\mu\nu}V^{\mu}W^{\nu}$, and we will also use the notation $V^2 = V \cdot V$. We will use this notation for vectors only. For a tensor $T_{abc...}$ we can define its symmetric part over a pair of indices $a$ and $b$

$$T_{(ab)c...} = \frac{1}{2}\left(T_{abc...} + T_{bac...}\right) = T_{(ba)c...} \qquad (2.1.5)$$

and its antisymmetric part

$$T_{[ab]c...} = \frac{1}{2}\left(T_{abc...} - T_{bac...}\right) = -T_{[ba]c...}. \qquad (2.1.6)$$

The definitions hold for any pair of indices. The tensor is a sum of its symmetric and antisymmetric parts

$$T_{abc...} = T_{(ab)c...} + T_{[ab]c...}. \qquad (2.1.7)$$

A curve in spacetime is characterized spacelike if the tangent vector to the curve is everywhere spacelike. Equivalently, we define the timelike and lightlike (or null) curves. The curve a particle sweeps in spacetime is called the world-line of the particle. This curve is always timelike, because no particle can move faster than the speed of light. Along a particle's timelike curve, we define the proper time $d\tau^2 = -ds^2 = -g_{ab}dx^a dx^b$. If we use $\tau$ to parametrize the curve, then the tangent vector to the curve is

$$u^a = \frac{dx^a}{d\tau},\qquad(2.1.8)$$

called the particle's 4-velocity, which is always a timelike vector by definition, satisfying $u_a u^a = -1$.

The notion of partial differentiation is generalized in spacetime to the covariant derivative of a tensor defined as

$$\begin{aligned}\nabla_e T^{cd...}_{ab...} =& \partial_e T^{cd...}_{ab...} - \Gamma^f_{ae} T^{cd...}_{fb...} - \Gamma^f_{be} T^{cd...}_{af...} - ... \\ &+ \Gamma^c_{fe} T^{fd...}_{ab...} + \Gamma^d_{fe} T^{cf...}_{ab...} + ...\end{aligned}\qquad(2.1.9)$$

where $\Gamma^a_{bc}$ are the Christoffel symbols, which are defined as

$$\Gamma^a_{bc} = \frac{1}{2} g^{ad} \left( \partial_c g_{db} + \partial_b g_{cd} - \partial_d g_{bc} \right),\qquad(2.1.10)$$

and $\partial_\mu = \frac{\partial}{\partial x^\mu}$, is the partial derivative of a quantity in terms of coordinate $x^\mu$. A subtlety here is that the Christoffel symbols are in fact not tensorial quantities. However, we denote them using abstract index notation to avoid mixing Greek indices and Latin indices in the same expression, for ease of notation. What one needs to keep in mind is that although the definition (2.1.10) is not a covariant expression, the definition (2.1.9) is covariant.

The curvature of a Lorentzian manifold is expressed through the Riemann tensor which in index notation is

$$R^a_{bcd} = \partial_c \Gamma^a_{bd} - \partial_d \Gamma^a_{bc} + \Gamma^e_{bd} \Gamma^a_{ec} - \Gamma^e_{bc} \Gamma^a_{ed}.\qquad(2.1.11)$$

In Euclidean space and in Minkowski space, we can choose a coordinate basis for which the Christoffel symbols vanish at every point in spacetime. It can therefore be concluded that the Riemann tensor vanishes as well in these spaces.

We will also define the contraction of the Riemann tensor, the Ricci tensor

$$R_{ab} = R^c_{\ acb} \tag{2.1.12}$$

and the Ricci scalar,

$$R = g^{ab} R_{ab}, \tag{2.1.13}$$

which is the trace of the Ricci tensor. Both of the above quantities will appear in the Einstein field equations, which we will derive in section 3.1.

## 2.2 Euler-Lagrange equations

In this section we will present the variational principle, which leads to the Euler-Lagrange equations of motion for a scalar field, a fundamental result of the calculus of variations which involves problems where we seek the extremization of a quantity. We will show that the Euler-Lagrange equations are second-order partial differential equations, the solutions of which are functions that extremize a functional, i.e. find its stationary values.

A functional is a mapping from a function to a real number, which we will denote by $S[\phi]$ in the following. A differentiable functional is stationary at its local maxima or minima, which implies that finding the solutions of the Euler-Lagrange equations provide us with the solutions that maximize or minimize the functional. This is equivalent to Fermat's theorem of calculus, which proves that the value of a function is extremized when calculated at its local maxima and minima. The latter for a function $f(x)$ are found by solving the equation where the first derivative of $f$ in terms of $x$, $f'(x)$, vanishes.

In the terminology of physics, the above analysis is useful for finding the stationary points of the action, a functional that describes a physical problem. In this case, the solutions of the Euler-Lagrange equations provide us with the equations of motion of the physical problem. The Lagrangian function, $\mathcal{L}$, is also defined which is a function of the system coordinates, of their derivatives and of time, and it describes the dynamics of the physical system. The Euler-Lagrange equations will be useful in deriving both the Einstein field equations, which are tensor equa-tions that describe the geometry of spacetime with or without a matter source, and the Nambu-Goto equations of the cosmic string, which describe the motion of the cosmic string in spacetime, in forthcoming chapters.

Assuming a functional, i.e. a quantity whose arguments are functions and not just variables, $S[\phi]$, defined by the integral over all spacetime $\Omega$ [64, 65]

$$S[\phi] = \int_\Omega \mathcal{L}(\phi, \nabla_a \phi) \sqrt{-g} \, d^4x, \qquad (2.2.1)$$

our goal is to find the function $\phi(x^a)$ for which the functional $S[\phi]$ is extremized. Note the bracket notation that we use, which reminds us that the argument of $S$ is a function and not a variable. The function $\mathcal{L}(\phi, \nabla_a \phi)$ corresponds to the total Lagrangian of the system we want to study, and $\phi$ and $\nabla_a \phi$ are treated as independent variables of $\mathcal{L}$. Also, $g = \det g$ is the determinant of the spacetime metric, and $\sqrt{-g} \, d^4x$ is the invariant spacetime volume element at each point of $\Omega$.[1] Demanding that the action of the system is stationary with respect to the field $\phi$, i.e. $\delta S[\phi] = 0$, it follows that

$$\int_\Omega \delta \left( \mathcal{L}(\phi, \nabla_a \phi) \sqrt{-g} \right) \, d^4x = 0$$
$$\int_\Omega \left[ \frac{\partial \mathcal{L}}{\partial \phi} \delta \phi + \frac{\partial \mathcal{L}}{\partial \nabla_a \phi} \nabla_a \left( \delta \phi \right) \right] \sqrt{-g} \, d^4x = 0. \qquad (2.2.2)$$

We use the notation $\delta$, instead of $d$, to highlight that this is the variation of a function, and not the differential of a variable. In the above we have also used the fact that the spacetime metric is independent of the field $\phi$. Furthermore, we have taken into account that the variation $\delta$ commutes with the partial differentiation operator [65]

$$\delta \left( \partial_a \phi \right) = \partial_a \left( \delta \phi \right). \qquad (2.2.3)$$

Note that the chain rule of differentiation also holds for the variation of a composite function,

$$\delta \left( f(\phi) \right) = \frac{\partial f}{\partial \phi} \delta \phi, \qquad (2.2.4)$$

where $f(\phi)$ is a function of the field $\phi$.[2] Integrating by parts the second term of

---

[1] By the term invariant volume we mean that any piece of spacetime volume is independent of the parameters chosen to calculate it.

[2] We will see that in general relativity the square root of the metric tensor determinant

equation (2.2.2), we reach the expression

$$\int_{\Omega} \left[ \frac{\partial \mathcal{L}}{\partial \phi} \delta \phi + \nabla_a \left( \frac{\partial \mathcal{L}}{\partial \left( \nabla_a \phi \right)} \delta \phi \right) - \nabla_a \left( \frac{\partial \mathcal{L}}{\partial \left( \nabla_a \phi \right)} \right) \delta \phi \right] \sqrt{-g} \, d^4 x = 0. \quad (2.2.5)$$

We can now apply the divergence theorem (also known as Gauss's theorem) [66], which allows us to switch from integration on a space $\Omega$ to integration on its boundary $\partial \Omega$

$$\int_{\Omega} \left( \nabla_a \delta F^a \right) \sqrt{-g} \, d^4 x = \int_{\partial \Omega} \left( \delta F^a \right) n_a \sqrt{|h|} d^3 x. \quad (2.2.6)$$

In the above, $F^a$ is a 4-field, $n_a$ is the outward unit vector normal to $\partial \Omega$ and $\sqrt{|h|} \, d^3 x$ is the invariant volume element of the boundary $\partial \Omega$. According to the theorem, $\partial \Omega$ can be any of the boundaries of $\Omega$. We can apply the divergence theorem (2.2.6) to equation (2.2.5) with $\delta F^a = \left( \partial \mathcal{L} / \partial \left( \nabla_a \phi \right) \right) \delta \phi$. Assuming that the boundary $\partial \Omega$ extends to infinity and that the integrated quantities vanish at infinity (which in general holds for physical fields), we find the following expression

$$\int_{\Omega} \left[ \frac{\partial \mathcal{L}}{\partial \phi} - \nabla_a \left( \frac{\partial \mathcal{L}}{\partial \left( \nabla_a \phi \right)} \right) \right] \delta \phi \sqrt{-g} \, d^4 x = 0. \quad (2.2.7)$$

The above integral directly leads us to the Euler-Lagrange equations for a scalar field

$$\frac{\partial \mathcal{L}(\phi, \nabla_a \phi)}{\partial \phi} - \nabla_a \left( \frac{\partial \mathcal{L}(\phi, \nabla_a \phi)}{\partial \left( \nabla_a \phi \right)} \right) = 0. \quad (2.2.8)$$

The above expression provides us with a differential equation for the motion of the field. In upcoming chapters we will use the above result to study the evolution of systems of interest, such as deriving the Einstein equations and the motion of a cosmic string in spacetime.

## 2.3 Fourier series and Fourier transform

Fourier analysis is the field in Mathematics which involves the representation of a function as an expansion in terms of trigonometric functions. It is named

---

appears and we will use the above rule when calculating the variation of the Einstein-Hilbert action with the metric tensor.

after Joseph Fourier, who realized that the study of heat transfer is simplified by describing functions as sums of sinusoids, i.e. their Fourier Series. In this section we will focus on discussing the Fourier series of functions and the general Fourier transform, which will be useful tools for this thesis.

The Fourier series of a function is the expansion of the function as a series of sines and cosines

$$f(x) = \frac{a_0}{2} + \sum_{n=1}^{\infty} a_n \cos{(nx)} + \sum_{n=1}^{\infty} b_n \sin{(nx)} \tag{2.3.1}$$

where

$$a_n = \frac{1}{\pi} \int_0^{2\pi} f(x) \cos{(nx)}\, dx \tag{2.3.2}$$

for $n \in \{0, \mathbb{N}\}$ and

$$b_n = \frac{1}{\pi} \int_0^{2\pi} f(x) \sin{(nx)}\, dx \tag{2.3.3}$$

for $n \in \mathbb{N}$.[3] This result is based on the orthogonality of the sine and cosine functions, and the properties of complete orthogonal systems of functions (see chapter 5 of [64]). The conditions for a function to decompose into the series of equation (2.3.1) are that it has a finite number of finite discontinuities and a finite number of extrema in the interval $[0, 2\pi]$ [64, 67]. We can also use the complex exponential function to describe the Fourier series expansion of a function. In this case we have

$$f(x) = \sum_{n=-\infty}^{+\infty} A_n e^{inx} \tag{2.3.4}$$

with

$$A_n = \frac{1}{2\pi} \int_0^{2\pi} f(x) e^{-inx} dx, \tag{2.3.5}$$

---

[3]Note that we define the natural numbers $\mathbb{N}$ to start from 1, i.e. without including 0.

since

$$\int_0^{2\pi} f(x)e^{-imx}dx = \int_0^{2\pi}\left(\sum_{n=-\infty}^{\infty} A_n e^{-inx}\right)e^{imx}dx$$

$$= \sum_{n=-\infty}^{\infty} A_n \int_0^{2\pi} e^{i(m-n)x}dx$$

$$= \sum_{n=-\infty}^{\infty} A_n \int_0^{2\pi} \left[\cos\left((m-n)x\right)+\right.$$

$$\left.+i\sin\left((m-n)x\right)\right]dx \qquad (2.3.6)$$

$$= \sum_{n=-\infty}^{\infty} A_n 2\pi\delta_{mn} = 2\pi A_m$$

where $\delta_{mn}$ is the Kronecker delta (see section 2.1). Equivalently, we are free to define the Fourier Series expansion with the opposite sigh convention, replacing $n$ with $-n$ in equations (2.3.4)-(2.3.5) above,

$$f(x) = \sum_{n=-\infty}^{+\infty} A_n e^{-inx} \qquad (2.3.7)$$

with

$$A_n = \frac{1}{2\pi}\int_0^{2\pi} f(x)e^{inx}dx. \qquad (2.3.8)$$

Fourier series follow a periodic behaviour by definition since trigonometric functions are periodic. Therefore, they are especially useful for the description of periodic phenomena and for solving problems such as differential equations with periodic boundary conditions. If the function $f(x)$ is periodic then its Fourier series expansion describes the value of $f(x)$ for any x. We can also change the interval of integration in (2.3.2,2.3.3) to fit the period of the function $f(x)$, T, using a change of variables. Then, the Fourier series of a periodic function $f(x)$ is

$$f(x) = \frac{a_0}{2} + \sum_{n=1}^{\infty} a_n \cos\left(\frac{2n\pi x}{T}\right) + \sum_{n=1}^{\infty} b_n \sin\left(\frac{2n\pi x}{T}\right) \qquad (2.3.9)$$

where

$$a_n = \frac{2}{T}\int_0^T f(x)\cos\left(\frac{2n\pi x}{T}\right)dx \qquad (2.3.10)$$

for $n \in \{0, \mathbb{N}\}$ and

$$b_n = \frac{2}{T} \int_0^T f(x) \sin\left(\frac{2n\pi x}{T}\right) dx \tag{2.3.11}$$

for $n \in \mathbb{N}$. We can define the frequency for $n = 1$ as the fundamental frequency, $f_1 = 1/T$, and frequencies for $n > 1$ as harmonics of the fundamental frequency, $f_n = nf_1$. Respectively, for the exponential Fourier series of a periodic function we have

$$f(x) = \sum_{n=-\infty}^{+\infty} A_n e^{i\frac{2n\pi}{T}x} \tag{2.3.12}$$

with

$$A_n = \frac{1}{T} \int_0^T f(x) e^{-i\frac{2n\pi}{T}x} dx \tag{2.3.13}$$

and $n \in \mathbb{Z}$. For example, in the case of a wave moving in space in the $+x$ direction, the function which describes it is a Fourier Series with argument $f = f(x - ut)$, where $u$ is the phase velocity of the wave. For a wave moving in the $-x$ direction, the corresponding function describing it will be $f = f(x + ut)$. In the case of light the phase velocity is the speed of light.

We can generalize the Fourier series for $T \to \infty$. In this limit, we can replace the discrete $A_n$ of equation (2.3.12) with a continuous function $f(k)$ and the summation with integration to reach the expansion

$$f(x) = \int_{-\infty}^{\infty} dk f(k) e^{2\pi ikx} \tag{2.3.14}$$

with

$$f(k) = \int_{-\infty}^{\infty} dx f(x) e^{-2\pi ikx}. \tag{2.3.15}$$

This is the Fourier transform of $f(x)$. If we want to write the Fourier transform in terms of the angular frequency $\omega = 2\pi/T$, then

$$f(t) = \frac{1}{2\pi} \int_{-\infty}^{\infty} d\omega f(\omega) e^{i\omega t} \tag{2.3.16}$$

with

$$f(\omega) = \int_{-\infty}^{\infty} dt f(t) e^{-i\omega t}. \tag{2.3.17}$$

We will adopt the Fourier transform convention of equations (2.3.16) and (2.3.17) throughout this thesis. Note that this is not a universal convention for the Fourier transform. Another common convention is a "symmetric" Fourier transform where both formulas (2.3.16) and (2.3.17) have a $1/\sqrt{2\pi}$ factor [64].

We can also define the logarithmic Fourier transform

$$\tilde{f}(k) = |k|f(k) = |k| \int_{-\infty}^{\infty} dt f(t) e^{-2\pi i k t}, \qquad (2.3.18)$$

which provides the advantage that $\tilde{f}(k)$ and $f(t)$ have the same physical dimensions.

## 2.4 The Dirac delta function and Heaviside step function

The Dirac delta function, introduced by the physicist Paul Dirac, is used to describe quantities that are zero everywhere except from a single point, whose integral over the entire space has finite value. The Dirac delta function is defined as [64, 68]

$$\delta(x) = 0, \text{ for } x \neq 0 \qquad (2.4.1)$$

along with the property

$$\int_{-\infty}^{\infty} f(x)\delta(x-a)dx = f(a), \qquad (2.4.2)$$

where $f(x)$ is a continuous function. For $f(x) = 1$, we find that

$$\int_{-\infty}^{\infty} \delta(x)dx = 1 \qquad (2.4.3)$$

From the above equation we conclude that the delta function is infinitely high at $x = 0$ and zero everywhere else. No function with such properties exists, but it can be approximated by the limit of a sequence of functions. The property

$$\int_{b}^{c} f(x)\delta(x-a)dx = f(a), \qquad (2.4.4)$$

also holds with the point $a$ included in the interval of integration. The discrete analogue of the delta function is the Kronecker delta, defined in 2.1. The delta function is an even function, $\delta(-x) = \delta(x)$.

In the case of more than one dimension we can define the delta function such that it carries unit integrated weight, i.e. the property (2.4.2) holds, but instead of $dx$ we have the volume element of the space we integrate over. For example, in 3 dimensional Euclidean space we have

$$\int_{-\infty}^{\infty} \int_{-\infty}^{\infty} \int_{-\infty}^{\infty} f(\vec{r}(x,y,z))\delta(\vec{r}(x,y,z) - \vec{r_i})dxdydz = f(\vec{r_i}), \qquad (2.4.5)$$

where $\delta(\vec{r}) = \delta(r_x)\delta(r_y)\delta(r_z)$. We use the notation $\vec{r}$ to denote vectors and $\vec{r}(x,y,z) = r_x\vec{i} + r_y\vec{j} + r_z\vec{k}$, where $(\vec{i}, \vec{j}, \vec{k})$ is the orthonormal Cartesian basis. In spherical coordinates we have

$$\int_0^\infty \int_0^\pi \int_0^{2\pi} f(\vec{r}(r,\theta,\phi))\delta(\vec{r}(r,\theta,\phi) - \vec{r_i})dV = f(\vec{r_i}), \qquad (2.4.6)$$

where $dV = r^2\sin(\theta)drd\theta d\phi$ the volume element in spherical coordinates and $\delta(\vec{r}(r,\theta,\phi) - \vec{r_i}) = \delta(r - r_i)\delta(\theta - \theta_i)\delta(\phi - \phi_i)/r^2\sin(\theta)$, where $\theta_i \in [0,\pi]$ and $\phi_i \in [0,2\pi)$.

One can prove that the integral of the delta function is given by [64]

$$H(x) = \int_{-\infty}^x \delta(x')dx' = \begin{cases} 0, & x < 0, \\ 1, & x > 0. \end{cases} \qquad (2.4.7)$$

The above function is called the Heaviside step function, also known as Heaviside theta function or Heaviside unit step function, $H(x)$.

In the upcoming chapters we will use the above definition of the delta function when calculating integrals in Fourier space, and the definition of the Heaviside theta function to denote restrictions to the domain of physical quantities. Note that by domain of values of a physical quantity we mean the set of values that serve as input for the quantity.

## 2.5  Probability and statistics

In this section we will discuss the properties of random variables. A random variable is defined as a function $X(\Omega) = E$ that obtains values from a domain set of events, $\Omega$, to a set $E$ with a given probability. We will also define the probability distribution, which is a function of the probability of the random variable to obtain a certain value, $P(X) \in [0, 1]$, where $X : \Omega \to E$. The value $P(X) = 0$ means that an event is impossible, while the value $P(X) = 1$ means that an event is certain. We will provide as example the problem of a coin toss. In this problem, the set of events has two elements, tails or heads, which define the random variable with two values, $x_1, x_2$, both with probability of occurring $P(x_i) = 0.5$, with $i = 1, 2$. The domain set is $\Omega = \{\text{heads}, \text{tails}\}$ and

$$X(y) = \begin{cases} x_1, & \text{if } y = \text{heads}, \\ x_2, & \text{if } y = \text{tails}. \end{cases} \tag{2.5.1}$$

Then, $E = \{x_1, x_2\}$, where the value $x_1$ corresponds to the event of tails, while the value $x_2$ corresponds to the event of heads. This is the case of a discrete random variable. In the case where the set $E$ is uncountable, the random variable is continuous.[4]

We will also discuss the statistical properties of data sets. These data sets can be obtained through some experiment or through probing numerically a problem with different values of the problem parameters, each time selected randomly, due to lack of an analytic approach that could solve the problem or in order to compare with an analytic approach. In both of these cases, after a sufficiently large number of data is collected, we can look into the distribution of the data and infer some conclusions about their behaviour. Quantities that are usually useful to analyse the data are the mean value of the data, the standard deviation and the error. It is also helpful to plot a histogram of the data, to visualize their distribution. A histogram is constructed by defining bins that divide the entire set of values of the data into non-overlapping intervals which are, commonly but non necessarily, of equal size. By counting the number of times a value of the data set falls into each bin, we can place on the horizontal axis the bins and draw a rectangle with height corresponding to the total number of events belonging to each bin.

---

[4]A set is countable when the number of its elements is some subset of the set of natural numbers. Therefore, it can be a finite set or a countably finite set. For example, the set of integer numbers is countable, while the set of real numbers is uncountable.

By applying the theory of probability to statistics, we can test hypotheses about the statistical data compared to a probability distribution, such as the normal distribution, the uniform distribution, the Poisson distribution, etc. For a sufficiently large number of data one can test the confidence level of the hypothesis and generalize the properties tested for the given sample of data to the whole set of events. A method for comparing the data to a distribution is the Kolmogorov–Smirnov test [69].

Finally, in the case of sets of data in pairs $(x_i, y_i)$, we can apply the least square method to find the curve which best fits the data. This method can be applied for the mean values, $\bar{x}_\alpha$, of several data sets that differ to one another by the value of a parameter $\alpha$ of the problem we study. Then, by finding the least square fit of $(\alpha_i, \bar{x}_{\alpha_i})$ we can understand the effect of the parameter $\alpha$ on the system.

### 2.5.1 Probability distributions

In this section we will discuss examples of probability distributions and their properties. A probability distribution, $P(X)$, is defined as a function that gives the probability of occurrence of different outcomes of an experiment. We will focus on discrete probability distributions, i.e. distributions that take a countable number of values, since we will use them in upcoming chapters. In particular, we will define the discrete uniform distribution, the geometric distribution and the Poisson distribution. Note that the total probability of all the possible outcomes of the experiment should always equal to unity, $\sum_{i=1}^{k} P(X_i) = 1$, where $k$ is an integer number that corresponds to all of the possible outcomes of the experiment.

The discrete uniform distribution is defined as a probability distribution where a finite number of variables are equally likely to be observed. If a random variable $X$, which obeys the discrete probability distribution, obtains integer values $X_1, X_2, ...., X_n$ in the interval $[a, b]$, where $a$ and $b$ are integers with $a < b$ and $n = b - a + 1$, then its probability distribution is given by $P(X = X_i) = 1/n$. A simple situation that can be described by the discrete uniform distribution is the throw of a dice.

The geometric distribution is defined as [69]

$$P(X = k) = (1 - p)^k p \qquad (2.5.2)$$

for $k = 0, 1, 2, ...$. The above gives the probability that there are $k$ failures until the first success of the trial, when the probability of the event to occur at each trial is constant and equal to $p$. For $X = 0$, the event has occurred with zero failures. An example is the throw of a dice until the result "6" occurs, where $p = 1/6$, and for $X = 0$ the probability of success with zero failures is $1/6$. This distribution can model problems where each trial is independent of the previous trials, the probability $p$ is constant for each trial, and there are two possible outcomes of the trial which we name success and failure. The probabilities $P(k)$ follow a geometric sequence with $k$. The geometric sequence is defined as a sequence of powers of a fixed number $a$, i.e. $b, ba, ba^2, ba^3, ...$, where $b$ is the sequence start value and $a$ is the common ratio, such that $b_k = ab_{k-1}$ for all $k \in \mathbb{N}$. When the common ratio is positive, all values $a_n$ are positive as well, while when it is negative the values of $a_n$ alternate between positive and negative. Also, when the common ratio is greater than unity the values $a_n$ increase exponentially towards infinity as $n \to \infty$, while when it is less than unity the values of $a_n$ converge to zero. In section 5.7.8, we will use this property of the geometric distribution to describe the distribution of a discrete quantity that decreases exponentially. There, we will denote the geometric distribution as $G(p, k)$, to differentiate it from other distributions, and because we will consider $p$ as a variable, rather than a fixed value.

The Poisson distribution is a discrete probability distribution defined as [69]

$$P(X = k) = \frac{\lambda^k e^{-\lambda}}{k!} \qquad (2.5.3)$$

where $\lambda > 0$, $\lambda \in \mathbb{R}$, $k = 0, 1, 2, ...$, $e$ is Euler's number and $k!$ denotes the factorial of $k$. If the average number of events in an interval is $\lambda$, then the above expresses the probability, $P(X = k)$, of observing $k$ events in the interval. It is straight forward that the distribution peaks at $k = \lambda$. The interval can describe a fixed interval in time or a fixed interval in space. This distribution can be used to describe events that are independent of one another, but appear with a constant rate over a given interval. It is also assumed that there exists a small enough interval (of time or space), where we can always count one or zero events, and no more. The Poisson distribution describes the fact that the probability of observing $k$ events, which has an average constant rate $\lambda$, peaks at $k = \lambda$, but there is always the chance that a much smaller or much higher number of events can occur. One example where the Poisson distribution can be applied is the number of meteors that hit the Earth's atmosphere per year. However, it cannot be applied for the number of people that enter an airport per hour, since the rate of people changes greatly during the day and during the night, and it can vary during the holiday months. Another example is the case of the air raids of London during World War II. During that period, it was observed by people that some regions in London were more impacted by attacks compared to others. It was then assumed that the attacks could be targeted towards these regions. However, by dividing London into equal sized squares, counting the number of attacks in each square and creating a histogram with the number of squares hit by a given number of attacks, it was found that the frequency distribution followed a Poisson distribution. This meant that the attacks occurred with a constant average rate across the squares. It was therefore concluded that the attacks were not focused towards specific regions, but exhibited a probability $P(k)$ that some regions faced almost zero attacks while others were more devastatingly hit. We will follow a similar procedure to determine the way that cusps are distributed on cosmic string loops in section 5.5.2.

Finally, Benford's law, or the first digit law, is a probability distribution which describes the distribution of the first digits in the numbers of large data sets. In sets that obey the law, the digit 1 is most likely to appear, while the digit 9 is the least likely. This law is obeyed by many sets of data in real life, such as electricity bills, death and birth rates, the area of rivers, constants in physics and mathematics. It arises naturally and can be used to detect cases of fraud in data, which were manually produced with the human perception that all numbers are distributed evenly [70]. We will discuss this distribution in more detail in section 5.7.8.

### 2.5.2   Properties of data sets

Let us assume a set of data $(x_1, x_2, ..., x_n)$, which have been gathered from an experiment or are results of a numerical simulation. The mean value of the set is defined by

$$\bar{x} = \frac{1}{n} \sum_{i=1}^{n} x_i. \tag{2.5.4}$$

A measure of the spread of the data with respect to the mean value is given by the standard deviation

$$\sigma = \sqrt{\frac{1}{n} \sum_{i=1}^{n} (x_i - \bar{x})^2}. \tag{2.5.5}$$

A small value of $\sigma$ implies that the data are distributed close to the mean value, while a large value implies that the individual data are in general broadly spread around the mean value. We can then write each of the data as

$$x_i = \bar{x} + e_i, \tag{2.5.6}$$

where $e_i$ is their deviation from the mean value of the data set. The above satisfy

$$\sum_{i=1}^{n} e_i = 0 \tag{2.5.7}$$

by definition.

Suppose that we would like to calculate the average value of a function $f(x)$ based on the measurements $(x_1, x_2, ..., x_n)$. By applying a Taylor expansion of $f(x)$ and using (2.5.6), we find that [64]

$$\bar{f} = f(\bar{x}) + \frac{1}{2} \sigma^2 f''(\bar{x}) + ... \tag{2.5.8}$$

It can also be proved that the spread of values of $f(x_i)$ is

$$\sigma^2(f) = (f'(\bar{x}))^2 \sigma^2 \tag{2.5.9}$$

where we have approximated $f(x_i) = \bar{f} + f'(\bar{x}) e_i$. Therefore, the error propagation for a function of one variable is

$$f(\bar{x} \pm \sigma) = f(\bar{x}) \pm f'(\bar{x})\sigma. \tag{2.5.10}$$

In the above, we have approximated $\bar{f} \simeq f(\bar{x})$ from equation (2.5.8).

Error propagation can be generalized to functions that depend on $n$ variables. An application of that is to find the error in the average value of a data set. We can regard the mean value $\bar{x}$ of the measurements $(x_1, x_2, ..., x_n)$ as a function of $n$ variables

$$\bar{x} = f(x_1, x_2, ..., x_n) = \frac{1}{n}\sum_{i=1}^{n} x_i. \tag{2.5.11}$$

Then, one can prove that the variance of $\bar{x}$ is given by

$$\sigma^2(\bar{x}) = \sigma^2/n, \tag{2.5.12}$$

which indicates that the spread of values decreases as the number of data increases, tending to zero for $n \to \infty$. Let us now assume that the exact value for the mean value of our experiment is $\mu$, and the relation $\bar{x} = \mu + \alpha$ holds. We can calculate the variance of the data set with respect to the value $\mu$, which we will denote $s$. By estimating that $\alpha \simeq s/\sqrt{n}$, which is a reasonable assumption if we compare with equation (2.5.12), we find that

$$s = \sqrt{\frac{\sum_{i=1}^{n}(x_i - \bar{x})^2}{n-1}}. \tag{2.5.13}$$

The above is called the sample standard deviation, and it describes the variance of the data set with respect to the average $\mu$. A detailed proof of the above will not be presented here, but it can be found in [64].

Given a set of n experimental data pairs $(x_i, y_i)$, we can assume that $y_i$ depend on $x_i$ according to the relation

$$y_i = f(x_i; a_1, a_2, ..., a_m) \tag{2.5.14}$$

where $a_1, a_2, ..., a_m$ are unknown parameters. The least square method provides a way of determining the unknown parameters by minimizing a sum expression.

In section 5.5 we will use the built-in method in Mathematica to fit our data in curves.

In many cases, we would like to have an intuition of the behavior of a data set, and to extend this behavior from the sample set of data to the rest of the potential events of the experiment, for a sufficiently large $n$. To achieve this, one can evaluate the mean value and the variance of the distribution from the data set, and assume a reference probability distribution. An assumption about an unknown probability distribution is called a statistical hypothesis. The Kolmogorov–Smirnov test is a goodness-of-fit test that can be used to compare the sample data set with an assumed probability distribution $P(X)$, to a confidence level. If we call $H_0$ the hypothesis that the sample comes from the distribution $P(X)$ and $H_1$ that the sample does not come from the distribution $P(X)$, the test provides a value $p$. A small value of $p$ suggests that $H_1$ is likely to hold, while a large $p$ value suggest that $H_0$ holds, i.e. that the data can be described from the probability distribution $P(X)$. The confidence level is defined by a value $\alpha$ that determines whether the value of $p$ is small or large. A typical value for $\alpha$ is 0.05, such that the null hypothesis is rejected if $p < \alpha$. In section 5.5.2 where we use the Smirnov-Kolmogorov test, we will use the built-in symbol in Mathematica to apply it. It is not the focus of this thesis to explain the theory behind the Smirnov-Kolmogorov test, but more on its derivation can be found in [69].

## 2.6 Numerical methods

In this section, we will give a short account of the numerical methods that we will use in the upcoming sections. One is Simpson's rule which is used to numerically evaluate integrals of functions that we are otherwise unable to calculate analytically. The other is the Monte Carlo method which is a stochastic numerical method based on random sampling of a given set of values and it can be used to solve problems of probabilistic nature.

### 2.6.1 Simpson's rule

Simpson's rule is a numerical method to approximate the value of definite integrals. For an interval $[x_0, x_n]$ which we split using a grid spacing $h$, Simpson's formula for numerical integration of the function $f(x)$ in the aforementioned interval is

$$\int_{x_0}^{x_n} f(x)dx = \frac{h}{3} \sum_{i=0}^{n-2} \left( f(x_i) + 4f(x_{i+1}) + f(x_{i+2}) \right) + O\left(h^5\right), \qquad (2.6.1)$$

where

$$h = \frac{x_n - x_0}{n} \qquad (2.6.2)$$

and $x_i = x_0 + ih$. The total error of the above method is

$$-\frac{x_n - x_0}{180} h^5 f^4(\xi) \qquad (2.6.3)$$

where $\xi \in [x_0, x_n]$ and $f^{(4)}(\xi)$ is the fourth derivative of the function $f(x)$ in terms of $x$ and evaluated at $\xi$. Note that $O(h^5)$ describes the total asymptotic error. It means that there is a positive constant $C$ such that for a sufficiently small value of $|h^5|$, it satisfies $|O(h^5)/h^5| < C$ [71].

We will use Simpson's rule to numerically evaluate the integrals of functions whose values are only known in a list form, i.e. they can be evaluated at specific points but we have no means to obtain their expression for any $x$. In particular we will apply this method in section 5.5.2.

### 2.6.2 Monte Carlo method

The Monte Carlo method can be used when we wish to evaluate a result in a probabilistic manner over a given domain of values. For example, one can evaluate the value of a definite integral by randomly generating a large amount of points and checking their properties. In particular, we define a domain that includes the limits of integration and the values of the function $f(x)$. Then, by generating random points in this domain, one can check for each point whether it lies under the curve of $f(x)$ or above it. Knowing the total area of the domain and the fraction of points under and over the curve, we can obtain the integral of $f(x)$.

Another example we will mention here is Monte Carlo decay. We will assume a number $N_0$ of particles at time $t$ and that they have a constant chance of decaying over an interval of time $dt$,

$$P = \frac{dN(t)}{N(t)} = -\lambda dt \qquad (2.6.4)$$

where $N(t)$ is the number of particles at time $t$ and $\lambda$ is the decay rate. To find the number of particles with time, $N(t)$, using the Monte Carlo method, we follow the following procedure. We define an amount of time $dt$, and we set time to take the values $t_i = t_0 + idt$, where $i$ is a counter number obtaining integer values and $t_0$ is the initial time, such that $N(t_0) = N_0$. At each time $t_i$, we find a random real number in the interval $[0, 1]$, using a random number generator, and we compare it with $\lambda dt$. If it is smaller than that, then the number of particles decreases by one, and otherwise it remains the same. We repeat this for the total number of particles $N(t_i)$, until we have determined the fate of each particle. Then, we repeat the same method for the next moment in time $t_{i+1}$, until we reach a moment $t_f$, where most of the particles have decayed. We compare the result with the analytic solution of the decay problem, which is the exponential decay of the particles $N(t) = N_0 e^{-\lambda t}$. The comparison is depicted in Figure 2.6.1, where the red line is the exponential decay and the blue line is the Monte Carlo decay, which was obtained with the above mentioned method.

We will use the Monte Carlo method in section 5.7.6 to compare our results for the toy model presented in 5.7 with the analytic approach.

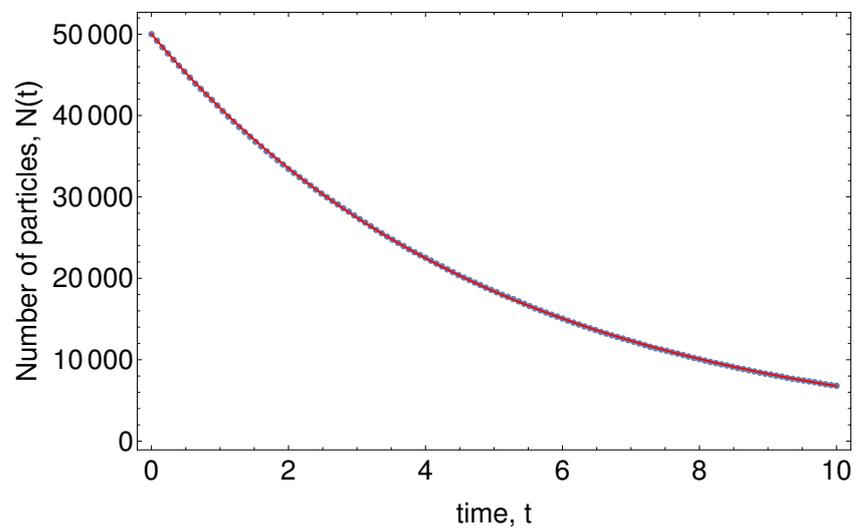

Figure 2.6.1: Comparison of the exponential decay (red curve) with the Monte Carlo decay (blue dots).

Gravitational waves

In this chapter, we will present the Einstein-Hilbert action, which is the action of the general theory of relativity, and derive from it the Einstein field equations, that describe the geometry of spacetime, given its matter content. Furthermore, we will show that general relativity supports radiative solutions, similar to electromagnetism, called gravitational waves. In the context of general relativity, gravitational waves are disturbances of the curvature of spacetime, caused by an accelerated mass, which corresponds to the source of the radiation and is described from the energy momentum tensor. Gravitational waves propagate in spacetime with the speed of light and cause a distortion of the spacetime, which causes distances to increase and decrease periodically as the gravitational wave passes through a region of spacetime. They have been the most elusive of the predictions of general relativity to observe, until their direct detection was announced by the LIGO and Virgo Collaboration on February 2016 [72].

Note that when describing gravitational waves one can use the metric description or the covariant approach, where the Weyl tensor is used instead of the metric [62], and the description uses quantities that do not depend on coordinate choice. In this thesis, we will use the metric description, where spacetime is assumed to be close to $\mathbb{R}^4$, i.e. our metric is a perturbation of the Minkowski metric. Since our aim is to calculate the amplitude of the gravitational wave (in chapter 6), a scalar quantity that does not depend on the coordinate choice, we are not worried about the coordinates that we will use. The theory outlined here is discussed in more detail in [73], [61].

Hereafter, we will work in units where the speed of light is dimensionless and equal to unity, $c = 1$.

## 3.1 The Einstein-Hilbert action

We can derive the Einstein field equations of General Relativity in vacuum (i.e. in empty spacetime) by using the variational principle on the Einstein-Hilbert action, given by

$$S[g] = \frac{1}{2\kappa} \int_\Omega R \sqrt{-g} \, d^4x \tag{3.1.1}$$

where $g = \det g$, the determinant of the metric tensor, $\Omega$ denotes integration over all spacetime, $R = R_{ab} g^{ab}$ is the Ricci scalar, which is the contraction of the metric tensor and the Ricci tensor, and $\kappa = 8\pi G$ where $G$ is Newton's constant. The above can be varied with respect to the metric tensor, to extremize the action $S[g]$ according to section 2.2,

$$\int_\Omega \delta \left( R \sqrt{-g} \right) d^4x = 0. \tag{3.1.2}$$

To simplify the above integral, we first need to calculate the variation of the square root of the metric tensor determinant, $\delta g$, and the variation of the Ricci scalar, $\delta R$.

To find the variation of the metric tensor determinant, let us perform the calculation for any symmetric tensor. For a non-degenerate matrix, $A_{ab}$, with

$A = \det A < 0$, Jacobi's formula holds

$$\frac{\partial A}{\partial A_{ab}} = [adj(A_{ab})]^T. \tag{3.1.3}$$

which expresses the derivative of the determinant of $A_{ab}$ with respect to the adjugate of the matrix, $adj(A_{ab})$, i.e. the transpose of its cofactor matrix. By the superscript $T$ we denote the transpose of a matrix. For an invertible matrix the identity $adj(A_{ab}) = AA^{ab}$ holds. If we specify $A_{ab}$ to be a metric tensor, i.e. it is symmetric, $A_{ab} = A_{ba}$, then we reach the relation

$$\frac{\partial A}{\partial A_{ab}} = AA^{ab}. \tag{3.1.4}$$

Using (3.1.4), we find that

$$\frac{\partial \sqrt{-A}}{\partial A_{ab}} = \frac{\sqrt{-A}}{2} A^{ab}, \tag{3.1.5}$$

which can be applied to the metric tensor.

Therefore, we find that

$$\frac{\partial \sqrt{-g}}{\partial g_{ab}} = \frac{\sqrt{-g}}{2} g^{ab}. \tag{3.1.6}$$

From the above, and using the chain rule (2.2.4), we conclude that the variation of the square root of the determinant is

$$\delta \left( \sqrt{-g} \right) = \frac{\partial \sqrt{-g}}{\partial g_{ab}} \delta g_{ab} = \frac{1}{2} \sqrt{-g} g^{ab} \delta g_{ab} \tag{3.1.7}$$

We can now take the variation of equation (2.1.1), to find that

$$\delta \left( g^{ab} g_{bc} \right) = 0 \tag{3.1.8}$$

since the variation of the Kronecker delta is zero. The above expands to

$$\begin{aligned}
&\delta g^{ab} g_{bc} + g^{ab} \delta g_{bc} = 0 \Rightarrow \delta g^{ab} g_{bc} g^{cd} + g^{ab} \delta g_{bc} g^{cd} = 0 \Rightarrow \\
&\delta g^{ad} = -g^{ab} g^{cd} \delta g_{bc} \Rightarrow \delta g_{ad} = -g_{ab} g_{cd} \delta g^{bc}.
\end{aligned} \tag{3.1.9}$$

Note that $\delta g_{ab}$ is a tensor quantity.

### 3.1.1 Variation of the Ricci scalar

To calculate the variation of the Ricci scalar, we first need to find the variation of the Riemann tensor, defined in (2.1.11). Let us start by varying the Christoffel symbols, defined in equation (2.1.10),

$$\delta\Gamma^a_{bc} = \frac{1}{2}\delta g^{ad}\left(\partial_c g_{db} + \partial_b g_{cd} - \partial_d g_{bc}\right) + \frac{1}{2}g^{ad}\left(\partial_c \delta g_{db} + \partial_b \delta g_{cd} - \partial_d \delta g_{bc}\right) \quad (3.1.10)$$

Substituting equation (3.1.7) in the above, we find that

$$\begin{aligned}
\delta\Gamma^a_{bc} =& \frac{1}{2}\left(-g^{ae}g^{fd}\delta g_{ef}\right)\left(\partial_c g_{db} + \partial_b g_{cd} - \partial_d g_{bc}\right) + \\
&+ \frac{1}{2}g^{ad}\left(\partial_c \delta g_{db} + \partial_b \delta g_{cd} - \partial_d \delta g_{bc}\right).
\end{aligned} \quad (3.1.11)$$

Using the definition of the Christoffel symbol (equation (2.1.10)), we notice that the above can be written as

$$\begin{aligned}
\delta\Gamma^a_{bc} =& - g^{ae}\Gamma^f_{bc}\delta g_{ef} + \\
&+ \frac{1}{2}g^{ad}\left(\partial_c \delta g_{db} + \partial_b \delta g_{cd} - \partial_d \delta g_{bc}\right).
\end{aligned} \quad (3.1.12)$$

Rearranging the above, we find

$$\delta\Gamma^a_{bc} = \frac{1}{2}g^{ad}\left(\partial_c \delta g_{bd} + \partial_b \delta g_{cd} - \partial_d \delta g_{bc} - 2\Gamma^e_{bc}\delta g_{de}\right). \quad (3.1.13)$$

We would like to write the above in terms of the covariant derivative of $\delta g_{ab}$. From equation (2.1.9), we find that

$$\nabla_c \delta g_{ab} = \partial_c \delta g_{ab} - \Gamma^f_{ac}\delta g_{fb} - \Gamma^f_{bc}\delta g_{fa}. \quad (3.1.14)$$

We can add and subtract in equation (3.1.13) the two quantities, $2\Gamma^e_{cd}\delta g_{eb}$ and $2\Gamma^e_{bd}\delta g_{ec}$, to find

$$
\begin{aligned}
\delta\Gamma^a_{bc} =& \frac{1}{2}g^{ad}\left(\partial_c\delta g_{bd} + \partial_b\delta g_{cd} - \partial_d\delta g_{bc} - 2\Gamma^e_{bc}\delta g_{de} + 2\Gamma^e_{bd}\delta g_{ec} + \right.\\
&+ 2\Gamma^e_{cd}\delta g_{eb} - 2\Gamma^e_{bd}\delta g_{ec} - 2\Gamma^e_{cd}\delta g_{eb}\right)\\
=& \frac{1}{2}g^{ad}\left(\partial_c\delta g_{db} - \Gamma^e_{bc}\delta g_{de} - \Gamma^e_{cd}\delta g_{eb} + \right.\\
&+ \partial_b\delta g_{dc} - \Gamma^e_{bc}\delta g_{de} - \Gamma^e_{bd}\delta g_{ec} - \\
&- \partial_d\delta g_{bc} + \Gamma^e_{bd}\delta g_{ec} + \Gamma^e_{cd}\delta g_{eb} + \\
&+ \Gamma^e_{cd}\delta g_{eb} + \Gamma^e_{bd}\delta g_{ec} - \Gamma^e_{cd}\delta g_{eb} - \Gamma^e_{bd}\delta g_{ec}\right).
\end{aligned}
\tag{3.1.15}
$$

In the above equation, the four last terms cancel with each other, while if we combine equation (3.1.14), we conclude that the variation of the Christoffel symbols is given by

$$
\delta\Gamma^a_{bc} = \frac{1}{2}g^{ad}\left(\nabla_c\delta g_{bd} + \nabla_b\delta g_{dc} - \nabla_d\delta g_{bc}\right).
\tag{3.1.16}
$$

Since $\nabla_a\delta g_{bc}$ is a tensor and the sum of tensors is a tensor, the above equation implies that the variation of the Christoffel symbols is a tensor.

When it comes to the variation of the Riemann tensor, defined in equation (2.1.11), we will follow the method described in [74]. At any arbitrary point $P_0$ in spacetime, we can calculate spacetime quantities in a local coordinate system called normal coordinates, where the Christoffel symbols vanish $\Gamma^a_{bc} = 0$ (but not their derivatives, since they are non zero at points around $P_0$), and the metric tensor is constant $\partial_c g_{ab} = 0$, at $P_0$. Then, the Riemann tensor in normal coordinates at $P_0$ can be written as

$$
R^a_{\ bcd} = \partial_c\Gamma^a_{bd} - \partial_d\Gamma^a_{bc}.
\tag{3.1.17}
$$

and its variation is

$$
\delta R^a_{\ bcd} = \partial_c\delta\Gamma^a_{bd} - \partial_d\delta\Gamma^a_{bc}.
\tag{3.1.18}
$$

In the above we have used the property of commutation between $\delta$ and $\partial$, as we recall from (2.2.3). Note that the partial derivative of a quantity is equal to its covariant derivative in normal coordinates, which can be proved by applying $\Gamma^a_{bc} = 0$, which holds for normal coordinates, to the definition of the covariant derivative in equation (2.1.9). Furthermore, the variation of the Christoffel symbols is a tensor, as we proved in equation (3.1.16). Therefore, we conclude that equation (3.1.18) is a covariant expression, and it holds at any arbitrary point in spacetime and in any coordinate system, in the tensorial expression

$$\delta R^a_{\ bcd} = \nabla_c \delta \Gamma^a_{bd} - \nabla_d \delta \Gamma^a_{bc}. \tag{3.1.19}$$

By contracting the two indices of the variation of the Riemann tensor (i.e. by multiplying the left- and the right-hand-side of the equation with $\delta^c_a$), we find

$$\delta R_{bd} \equiv \delta R^c_{\ bcd} = \nabla_c \delta \Gamma^c_{bd} - \nabla_d \delta \Gamma^c_{bc}. \tag{3.1.20}$$

The above is called the Palatini identity, and corresponds to the variation of the Ricci tensor. Note that the Kronecker delta intercommutes with $\nabla$, since its covariant derivative is zero.

We can now calculate the variation of the Ricci scalar, which is

$$\begin{aligned}
\delta R &= \delta \left( g^{ab} R_{ab} \right) = \delta g^{ab} R_{ab} + g^{ab} \delta R_{ab} \\
&= \left( -g^{ad} g^{bc} \delta g_{dc} \right) R_{ab} + \sqrt{-g} g^{ab} \left( \nabla_c \Gamma^c_{ab} - \nabla_b \Gamma^c_{ac} \right) \\
&= -R^{ab} \delta g_{ab} + \sqrt{-g} \, \nabla_c \left( g^{ab} \Gamma^c_{ab} - g^{ac} \Gamma^d_{ad} \right),
\end{aligned} \tag{3.1.21}$$

where we have used the fact that $\nabla_c g_{ab} = 0$.

## 3.1.2 The Einstein field equations

We can now apply the results of the previous section to the calculation of the Einstein-Hilbert action variation from equation (3.1.2). We find that

$$\delta \left( R \sqrt{-g} \right) = \delta R \sqrt{-g} + R \delta \left( \sqrt{-g} \right). \tag{3.1.22}$$

Substituting in equation (3.1.22) the results from equation (3.1.7) and (3.1.21), we find that

$$\delta \left( R \sqrt{-g} \right) = \sqrt{-g} \left[ -R^{ab} \delta g_{ab} + \nabla_c \left( g^{ab} \Gamma^c_{ab} - g^{ac} \Gamma^d_{ad} \right) + \frac{1}{2} R g^{ab} \delta g_{ab} \right] \tag{3.1.23}$$

Therefore, we find that the variation of the action is

$$\int_\Omega \left( -R^{ab}\delta g_{ab} + \frac{1}{2}Rg^{ab}\delta g_{ab} \right)\sqrt{-g}d^4x +$$
$$+ \int_\Omega \nabla_c \left( g^{ab}\Gamma^c_{ab} - g^{ac}\Gamma^d_{ad} \right)\sqrt{-g}d^4x = 0 \tag{3.1.24}$$

From the divergence theorem, which was mentioned in (2.2.6) and applied for the derivation of the Euler-Lagrange equations in section 2.2, we find that the second integral in the variation of the action (3.1.24) is a boundary term at infinity which vanishes, and we find that the Einstein field equations in vacuum are

$$G^{ab} \equiv R^{ab} - \frac{1}{2}Rg^{ab} = 0. \tag{3.1.25}$$

The left-hand-side of the above equation is called the Einstein tensor, $G_{ab}$. It can be proved from the contracted Bianchi identity, an identity that is derived from the geometric properties of the spacetime manifold, that $\nabla_a G^{ab} = 0$ [75]. Note that equations (3.1.25) are the Einstein field equations when no matter is present in spacetime, called Einstein field equations (EFE) in vacuum. The above can be further simplified if we notice that the trace of $G^{ab}$, i.e. $G = g_{ab}G^{ab}$, is equal to $-R$. Since $G^{ab} = 0$, its trace is also equal to 0. Then we conclude that the Einstein field equations are simply $R^{ab} = 0$.

If we would like to include matter (or equivalently energy) in the description of spacetime, we add terms to the Lagrangian of the system. Then, the total action of the system would be the Einstein-Hilbert action, plus a Lagrangian, $\mathcal{L}_m$, corresponding to the matter-energy sources of the system,

$$S_{tot} = \int_\Omega \left[ \frac{1}{2\kappa}R + \mathcal{L}_m \right]\sqrt{-g}\,d^4x. \tag{3.1.26}$$

Varying $S_{tot}$ with the metric tensor, $g_{ab}$, as we did for the action of vacuum, we find that the Einstein field equations with matter-energy sources are

$$R^{ab} - \frac{1}{2}Rg^{ab} = 8\pi G T^{ab}, \tag{3.1.27}$$

which are the equations of motion of the metric tensor in a spacetime with a matter source of Lagrangian $\mathcal{L}_m$. The tensor $T^{ab}$ is called the stress-energy-momentum tensor, also called the energy-momentum tensor, which is a symmetric tensor that corresponds to the variation of the matter Lagrangian with the metric field,

$$T^{ab} = \frac{-2}{\sqrt{-g}} \frac{\delta\left(\sqrt{-g}\mathcal{L}_m\right)}{\delta g_{ab}}. \tag{3.1.28}$$

This tensor represents the matter-energy content of spacetime. One of the postulates of general relativity is that the energy-momentum tensor is conserved

$$\nabla_a T^{ab} = 0. \tag{3.1.29}$$

The contracted Bianchi identities, that imply the conservation of $G^{ab}$, ensure the consistency of the vanishing divergence on both sides of the equations (3.1.27).

## 3.2 The linearized Einstein field equations

The Einstein field equations are non-linear partial differential equations, which makes them very difficult to solve for general cases. In this section, we will derive the linearized expression of the Einstein field equations, by assuming that the spacetime metric is described as a small deviation from the Minkowski metric, $h_{\mu\nu}$, which vanishes at infinity. This simplified form of the Einstein field equations corresponds to scenarios where gravity is weak. In practice, this is a quite frequently occurring scenario, since any gravitational signal that is likely to be observed will have weak intensity.

This assumption allows us to linearize the Einstein equations by assuming that our spacetime is close to $\mathbb{R}^4$ and there exist global coordinates $x^\mu$ for which the metric is written as

$$g_{\mu\nu} = \eta_{\mu\nu} + h_{\mu\nu}, \tag{3.2.1}$$

where $\eta_{\mu\nu} = \text{diag}(-1, 1, 1, 1)$, the Minkowski metric. The weakness of gravity in this linearized approach is reflected in the fact that $|h_{\mu\nu}| \ll 1$. Note that we use Greek letters to denote the indices in the above equation, since we describe our quantities in Cartesian coordinates (i.e. $\mathbb{R}^4$ coordinates) where the Minkowski metric takes the above mentioned components. Recall from section 2.1, that we used Latin indices for tensorial quantities only. Since we restrict ourselves to first order terms, hereafter we will lower or raise indices using the Minkowski tensor, instead of $g^{\mu\nu}$, which agrees with our first order approach neglecting higher order terms. Also, note that $h_{\mu\nu}$ is assumed to transform as a tensor under Lorentz transformations, i.e. as a tensor would transform in the context of special relativity.

Let us now calculate the inverse of $g_{\mu\nu}$. We assume that $g^{\mu\nu} = \eta^{\mu\nu} + k^{\mu\nu}$, which is reasonable since the inverse of the metric will also be a small perturbation from flat spacetime. Then we find

$$g^{\mu\nu}g_{\nu\rho} = (\eta^{\mu\nu} + k^{\mu\nu})(\eta_{\nu\rho} + k_{\nu\rho}) = \delta^\mu_\rho + k^{\mu\nu}\eta_{\nu\rho} + \eta^{\mu\nu}h_{\nu\rho} + k^{\mu\nu}h_{\mu\rho}. \qquad (3.2.2)$$

The last term of the above result is ignored since it is a second order term. Using equation (2.1.1) and substituting in the above result, we find that

$$\begin{aligned}
\delta^\mu_\rho + k^{\mu\nu}\eta_{\nu\rho} + \eta^{\mu\nu}h_{\nu\rho} &= \delta^\mu_\rho \Leftrightarrow \\
k^{\mu\nu}\eta_{\nu\rho} + \eta^{\mu\nu}h_{\nu\rho} &= 0 \Leftrightarrow \\
\eta^{\rho\tau}k^{\mu\nu}\eta_{\nu\rho} + \eta^{\mu\nu}\eta^{\rho\tau}h_{\nu\rho} &= 0 \Leftrightarrow \\
k^{\mu\tau} &= -\eta^{\mu\nu}\eta^{\rho\tau}h_{\nu\rho} = -h^{\mu\tau}.
\end{aligned} \qquad (3.2.3)$$

Therefore,

$$g^{\mu\nu} = \eta^{\mu\nu} - h^{\mu\nu}. \qquad (3.2.4)$$

In order to linearize the Einstein field equations, we should first calculate the linearized Einstein tensor. From the definition of the Christoffel symbols (2.1.10), we find that for $g_{\mu\nu}$ given by equation (3.2.1)

$$\Gamma^\mu_{\nu\rho} = \frac{1}{2}g^{\mu\tau}\left(\partial_\rho h_{\tau\nu} + \partial_\nu h_{\rho\tau} - \partial_\tau h_{\nu\rho}\right), \qquad (3.2.5)$$

since $\partial_\mu\eta_{\nu\rho} = 0$. Also, since the quantity in the parenthesis is of first order,

$$\Gamma^\mu_{\nu\rho} = \frac{1}{2}\eta^{\mu\tau}\left(\partial_\rho h_{\tau\nu} + \partial_\nu h_{\rho\tau} - \partial_\tau h_{\nu\rho}\right), \qquad (3.2.6)$$

which is the expression of the linearized Christoffel symbols. We can now see from the definition of the Riemann tensor, in equation (2.1.11), that the two last terms on the right-hand-side are second order terms and should be ignored. Therefore,

$$
\begin{aligned}
R_{\mu\nu\rho\sigma} =& \eta_{\mu\nu}\left(\partial_\rho \Gamma^\tau_{\nu\sigma} - \partial_\sigma \Gamma^\tau_{\nu\rho}\right) = \\
=& \frac{1}{2}\left(\partial_\rho\partial_\nu h_{\mu\sigma} + \partial_\mu\partial_\sigma h_{\nu\rho} - \partial_\mu\partial_\rho h_{\nu\sigma} - \partial_\nu\partial_\sigma h_{\mu\rho}\right).
\end{aligned}
\tag{3.2.7}
$$

Then, the linearized Ricci tensor is given by

$$
\begin{aligned}
R_{\mu\nu} = \eta^{\tau\rho} R_{\tau\mu\rho\nu} =& \eta^{\tau\rho}\frac{1}{2}\left(\partial_\mu\partial_\rho h_{\tau\nu} + \partial_\tau\partial_\nu h_{\sigma\mu} - \partial_\tau\partial_\rho h_{\mu\nu} - \partial_\mu\partial_\nu h_\tau h_{\nu\rho}\right) \\
=& \frac{1}{2}\left(\partial^\tau\partial_\mu h_{\tau\nu} + \partial^\tau\partial_\nu h_{\mu\tau} - \partial^\rho\partial_\rho h_{\mu\nu} - \partial_\mu\partial_\nu h\right),
\end{aligned}
\tag{3.2.8}
$$

where $h$ is the trace of the metric perturbation, $h = h_\mu{}^\mu$. The Ricci scalar, which is the trace of the Ricci tensor, is found to be

$$
R = \eta^{\mu\nu} R_{\mu\nu} = -\frac{1}{2}\partial^\rho\partial_\rho h - \frac{1}{2}\partial_\rho\partial^\rho h + \frac{1}{2}\partial^\tau\partial^\nu h_{\tau\nu} + \frac{1}{2}\partial^\tau\partial^\nu h_{\nu\tau}.
\tag{3.2.9}
$$

By adding the terms that are equal, the above can be simplified to the following expression for the linearized Ricci scalar

$$
R = -\partial^\rho\partial_\rho h + \partial^\tau\partial^\nu h_{\tau\nu}.
\tag{3.2.10}
$$

Finally, we have all the components required to calculate the linearized Einstein tensor. From the definition of the Einstein tensor in equation (3.1.25), we find that to linear order it equals

$$
\begin{aligned}
G_{\mu\nu} =& \frac{1}{2}\left[-\partial^\rho\partial_\rho h_{\mu\nu} - \partial_\mu\partial_\nu h + \partial^\tau\partial_\mu h_{\tau\nu} + \partial^\tau\partial_\nu h_{\tau\nu} - \right. \\
& \left. - \left(\partial^\tau\partial^\nu h_{\tau\nu} - \partial^\rho\partial_\rho h\right)\right].
\end{aligned}
\tag{3.2.11}
$$

It is useful to define the trace-reversed metric perturbation

$$
\bar{h}_{\mu\nu} = h_{\mu\nu} - \frac{1}{2}h\eta_{\mu\nu},
\tag{3.2.12}
$$

where $h = h^\mu{}_\mu$. The trace of the Minkowski metric $\eta^{\mu\nu}\eta_{\mu\nu} = 4$, by definition. Therefore, $\bar{h} = -h$, and

$$
h_{\mu\nu} = \bar{h}_{\mu\nu} - \frac{1}{2}\bar{h}\eta_{\mu\nu}.
\tag{3.2.13}
$$

The linearized Einstein tensor in terms of $\bar{h}_{\mu\nu}$ is given by

$$
\begin{aligned}
G_{\mu\nu} = &-\frac{1}{2}\partial^\rho\partial_\rho\left(\bar{h}_{\mu\nu} - \frac{1}{2}\bar{h}\eta_{\mu\nu}\right) - \frac{1}{2}\partial_\mu\partial_\nu(-\bar{h}) + \\
&+\frac{1}{2}\partial^\tau\partial_\mu\left(\bar{h}_{\tau\nu} - \frac{1}{2}\bar{h}\eta_{\tau\nu}\right) + \frac{1}{2}\partial^\tau\partial_\nu\left(\bar{h}_{\mu\tau} - \frac{1}{2}\bar{h}\eta_{\mu\tau}\right) - \\
&-\frac{1}{2}\left[-\partial^\rho\partial_\rho(-\bar{h}) + \partial^\tau\partial^\sigma\left(\bar{h}_{\tau\sigma} - \frac{1}{2}\bar{h}\eta_{\tau\sigma}\right)\right]\eta_{\mu\nu}
\end{aligned}
\tag{3.2.14}
$$

If we expand the above, we find that all terms including the trace, $\bar{h}$, cancel with one another, and we finally reach the expression

$$
G_{\mu\nu} = \frac{1}{2}\left(-\partial^\rho\partial_\rho\bar{h}_{\mu\nu} + \partial^\tau\partial_\mu\bar{h}_{\tau\nu} + \partial^\tau\partial_\nu\bar{h}_{\mu\tau} - \eta_{\mu\nu}\partial^\tau\partial^\sigma\bar{h}_{\tau\sigma}\right).
\tag{3.2.15}
$$

Therefore, the linearized Einstein field equations, given in equation (3.1.27), are

$$
-\frac{1}{2}\partial^\rho\partial_\rho\bar{h}_{\mu\nu} + \partial^\rho\partial_{(\mu}\bar{h}_{\nu)\rho} - \frac{1}{2}\eta_{\mu\nu}\partial^\rho\partial^\sigma\bar{h}_{\rho\sigma} = 8\pi G T_{\mu\nu}.
\tag{3.2.16}
$$

## 3.2.1 Gauge fixing

Equation (3.2.16) does not yield unique solutions due to the gauge invariance of the Einstein field equation. We will prove that any coordinate transformation of the form

$$
x^\mu \to x'^\mu = x^\mu + \epsilon^\mu(x^\mu),
\tag{3.2.17}
$$

leaves solutions of (3.2.16) unchanged. Note that to maintain the weakness of the field $h_{\mu\nu}$, $\partial\epsilon^\mu/\partial x^\nu$ should be a first order quantity, but $\epsilon^\mu$ is otherwise an arbitrary function of the coordinates. The metric transforms under (3.2.17) as

$$
g'^{\mu\nu} = \frac{\partial x'^\mu}{\partial x^\kappa}\frac{\partial x'^\nu}{\partial x^\lambda}g^{\kappa\lambda}
\tag{3.2.18}
$$

as we can see from the definition of tensor transformations in equation (2.1.3). Combining the above result with (3.2.1) and lowering the indices, we find that

$$
\eta^{\mu\nu} - h'^{\mu\nu} = \frac{\partial x'^\mu}{\partial x^\kappa}\frac{\partial x'^\nu}{\partial x^\lambda}\left(\eta^{\kappa\lambda} - h^{\kappa\lambda}\right) \Leftrightarrow
\tag{3.2.19}
$$

$$\eta^{\mu\nu} - h'^{\mu\nu} = \left(\frac{\partial x^\mu}{\partial x^\kappa} + \frac{\partial \epsilon^\mu}{\partial x^\kappa}\right)\left(\frac{\partial x^\nu}{\partial x^\lambda} + \frac{\partial \epsilon^\nu}{\partial x^\lambda}\right)\left(\eta^{\kappa\lambda} - h^{\kappa\lambda}\right) \tag{3.2.20}$$

Removing all second order terms (keep in mind that we chose $\epsilon^\mu$ such that $\partial \epsilon^\mu / \partial x^\nu$ is a first order quantity) and using the identity $\partial x^\mu / \partial x^\nu = \delta^\mu_\nu$, the above reduces to

$$\eta^{\mu\nu} - h'^{\mu\nu} = \left(\delta^\mu_\kappa \delta^\nu_\lambda + \delta^\mu_\kappa \frac{\partial \epsilon^\nu}{\partial x^\lambda} + \delta^\nu_\lambda \frac{\partial \epsilon^\mu}{\partial x^\kappa}\right)\left(\eta^{\kappa\lambda} - h^{\kappa\lambda}\right) \Leftrightarrow \tag{3.2.21}$$

$$\eta^{\mu\nu} - h'^{\mu\nu} = \eta^{\mu\nu} + \eta^{\mu\lambda}\frac{\partial \epsilon^\nu}{\partial x^\lambda} + \eta^{\kappa\nu}\frac{\partial \epsilon^\mu}{\partial x^\kappa} - h^{\mu\nu} - h^{\mu\lambda}\frac{\partial \epsilon^\nu}{\partial x^\lambda} - h^{\nu\kappa}\frac{\partial \epsilon^\mu}{\partial x^\kappa}. \tag{3.2.22}$$

Ignoring the last two terms on the right-hand-side of the above, because they are second order terms, we finally find that $h_{\mu\nu}$ transforms under the coordinate transformation (3.2.17) as

$$h'_{\mu\nu} = h_{\mu\nu} - \frac{\partial \epsilon_\mu}{\partial x^\nu} - \frac{\partial \epsilon_\nu}{\partial x^\mu}. \tag{3.2.23}$$

If we substitute the above into the linearized Einstein equations (3.2.16), by using the definition (3.2.12), we can verify that it leaves (3.2.16) unchanged. Therefore, if $\bar{h}_{\mu\nu}$ is a solution to (3.2.16) then $\bar{h}'_{\mu\nu}$ is also a solution. This property of the linearized general relativity is called the gauge invariance of the field equations. The above implies that under coordinate transformations of the type defined in (3.2.17), $h_{\mu\nu}$ remains unchanged, which indicates a redundancy in our description. However, if we are happy to identify the redundant coordinate systems then the solutions of (3.2.16) become unique. Since any choice of $\epsilon^\mu$ leaves the solutions of the linearized Einstein equations unchanged, a particular choice for $\epsilon^\mu$ is of no physical meaning, and we can identify different coordinate systems, without any loss in our description. Another way to state this is that any coordinate system can be reached via the coordinate transformation (3.2.17) and they are identified because of our gauge symmetry that $h_{\mu\nu}$ remains unchanged regardless of the coordinate system. Note here the similarity to electromagnetism in flat spacetime, where we introduce the vector potential $A_\mu$, which is related to the Faraday tensor via $F_{\mu\nu} = 2\partial_{[\mu}A_{\nu]}$. Then, under the transformation

$$A'_\mu(x) = A_\mu(x) + \partial_\mu \Lambda(x) \tag{3.2.24}$$

the Faraday tensor remains unchanged, due to the intercommutation of the partial derivative. This implies that $A_\mu$ and $A_\mu + \partial_\mu \Lambda$ correspond to the same physical state (i.e. the same electric and magnetic field), regardless of the choice of $\Lambda$. To fix the gauge and produce unique choices of $A_\mu$ a simple gauge choice is $\partial_\mu A^\mu = 0$, called the Lorentz gauge. This choice implies that $\partial_\mu \partial^\mu \Lambda = 0$, which is a wave equation. We will follow a similar reasoning to impose a gauge condition for the linearized Einstein equations.

Since we are free to fix $\epsilon^\mu$, we will now proceed to a convenient choice for our coordinate system, which will remove the gauge ambiguity. It is a matter of simplification of our calculations what condition for $\epsilon^\mu$ we will choose, and it can vary depending on the approach. To make a convenient gauge choice, note first that $\partial^\nu \bar{h}_{\mu\nu}$ transforms under (3.2.17) as

$$\left(\partial^\nu \bar{h}_{\mu\nu}\right)' = \partial^\nu \bar{h}_{\mu\nu} + \partial^\nu \partial_\nu \epsilon_\mu. \tag{3.2.25}$$

Motivated from the above, we will fix $\epsilon_\mu$ such than the condition

$$\left(\partial^\nu \bar{h}_{\mu\nu}\right)' = 0, \tag{3.2.26}$$

is satisfied. This is called the harmonic gauge or harmonic coordinate system. For the above to hold, we choose $\epsilon_\mu$ to satisfy the wave equation

$$\partial^\nu \partial_\nu \epsilon_\mu = -\partial^\nu \bar{h}_{\mu\nu}, \tag{3.2.27}$$

in order to impose the gauge (3.2.26). The $\epsilon_\mu$ condition (3.2.27) is a wave equation which supports non trivial solutions and can be solved using Green's functions. For an application of Green's functions to the inhomogeneous wave equation see section 3.2.3.

Using the gauge condition (3.2.26), we can show the linearized Einstein equations (3.2.16) take the form

$$\Box \bar{h}_{\mu\nu} = -16\pi G T_{\mu\nu}, \tag{3.2.28}$$

where $\Box = \partial_\kappa \partial^\kappa$. We notice that the partial differential equation (3.2.28) is an inhomogeneous wave equation, with the right-hand-side serving as the source of the waves. It is known that differential equations of this form can be solved us-ing Green's functions. Therefore, we conclude that in the harmonic gauge the components of $\bar{h}_{\mu\nu}$ satisfy the wave equation, the solution of which is interpreted as gravitational radiation produced by the energy-momentum tensor, which corresponds to the gravitational wave source. In this way, it is proved that general relativity supports radiative solutions, similarly to the radiative solutions of electromagnetism, which is the electromagnetic wave. In the same way that the plane wave solutions of Maxwell's equations lead to an equivalent interpretation in terms of the photon particle, we can deduce the particle of the gravitational plane waves, the graviton.

The study of gravitational radiation in a more general context is complicated by the non-linearity of the Einstein field equations. The gravitational wave carries momentum and energy which contributes to its own gravitational field. In the above, i.e. the linearization of the Einstein equations, we essentially simpified the gravitational wave description by assuming that the gravitational wave is weak and it does not affect its own propagation, and therefore removing the aforementioned complication. Special cases of the exact Einstein equations have been studied producing non-linear radiative solutions in spacetime [73]. However, the linearized approach is a satisfactory approach for weak signals arriving to an observer on Earth from a distant source. It is also a way to describe the graviton, when it is far away from sources and other particles, and when the effects of its own field on itself is neglected, which is a fundamental description for an elementary particle.

In the homogeneous case (i.e. when the right-hand side of (3.2.28) is zero), the solutions are interpreted as gravitational radiation coming from infinity, as we will discuss in the following section.

## 3.2.2 The homogeneous linearized Einstein field equations

In the case of vacuum, i.e. spacetime with no matter content, the linearized Einstein equations reduce to the homogeneous wave equation

$$\Box \bar{h}_{\mu\nu} = 0. \tag{3.2.29}$$

The general solution of this equation is a linear superposition of plane waves

$$\bar{h}_{\mu\nu} = e_{\mu\nu}e^{ik_\lambda x^\lambda} + e^*_{\mu\nu}e^{-ik_\lambda x^\lambda}, \tag{3.2.30}$$

where the superscript $*$ indicates the complex conjugate of a quantity. The tensor $e_{\mu\nu}$ is a symmetric tensor which describes the polarization of the plane wave, and we will call it the polarization tensor, and $k^\mu$ is the waveform, which indicates the direction of propagation of the plane wave. The plane wave satisfies (3.2.29) if $k_\mu k^\mu = 0$ and the gauge condition (3.2.26) if

$$k_\mu e^\mu_\nu = \frac{1}{2}k_\nu e^\mu_\mu. \tag{3.2.31}$$

A symmetric $4 \times 4$ tensor has $16 - 6 = 10$ independent components. From the 4 conditions (3.2.31), the independent components of $e_{\mu\nu}$ are reduced to 6. Let us now take a coordinate change $x'^\mu = x^\mu + \epsilon^\mu(x^\mu)$ which we choose to be

$$\epsilon^\mu(x) = i\epsilon^\mu e^{ik_\lambda x^\lambda} - i\epsilon^{*\mu}e^{-ik_\lambda x^\lambda}. \tag{3.2.32}$$

Then, $\bar{h}'_{\mu\nu}$ is given by (3.2.23),

$$\begin{aligned}\bar{h}'_{\mu\nu} =& \bar{h}_{\mu\nu} - i\epsilon_\mu ik_\nu e^{ik_\lambda x^\lambda} + i\epsilon^*_\mu + i\epsilon^*_\mu(-ik_\nu)e^{-ik_\lambda x^\lambda} - i\epsilon_\nu ik_\mu e^{ik_\lambda x^\lambda} + \\ &+ i\epsilon^*_\nu(-ik_\mu)e^{-ik_\lambda x^\lambda}.\end{aligned} \tag{3.2.33}$$

Expanding the above, we find that

$$\bar{h}'_{\mu\nu} = \left(e_{\mu\nu} + \epsilon_\mu k_\nu + \epsilon_\nu k_\mu\right)e^{ik_\lambda x^\lambda} + \left(e^*_{\mu\nu} + \epsilon^*_\mu k_\nu + \epsilon^*_\nu k_\mu\right)e^{-ik_\lambda x^\lambda}. \tag{3.2.34}$$

Therefore, the polarization tensor of $\bar{h}'_{\mu\nu}$ is

$$e'_{\mu\nu} = e_{\mu\nu} + \epsilon_\mu k_\nu + \epsilon_\nu k_\mu. \tag{3.2.35}$$

From the above we conclude that since $e_{\mu\nu}$ and $e'_{\mu\nu}$ correspond to the same physical state for our problem, and they are related with the arbitrary 4-component vector $\epsilon^\mu$, the polarization tensor has $6 - 4 = 2$ independent components. An example of the above for a wave moving in the z-direction can be found in [73].

We can define the helicity, $\lambda$, such that

$$\psi' = e^{i\lambda\theta}\psi, \qquad (3.2.36)$$

where $\psi$ is a plane wave and $\psi'$ is its transformation by a rotation of angle $\theta$ about the direction of propagation. It can be proved that the polarization tensor of a gravitational wave can be decomposed into parts with helicity $\pm 2$, $\pm 1$, 0. However, the parts with helicity $\pm 1$ and 0 can vanish under appropriate coordinate transformations and the parts that are physically significant are those with helicity $\pm 2$. In comparison, note that for an electromagnetic wave the physically significant parts have helicity $\pm 1$. An equivalent way to express the above is that gravitational waves have spin 2, while electromagnetic waves have spin 1. A proof of the above can be found in [73].

### 3.2.3 Solutions in the local wave zone for a periodic source

We will now look into the inhomogeneous linearized Einstein equations, where the source of the gravitational waves appears on the right-hand-side of equation (3.2.16), under the local wave zone assumption.

In the local wave zone approximation, we assume that the observer is at a distance from the source $T_{\mu\nu}$ much larger than the size of the source. We will also assume that our source is periodic, since we will apply our results for cosmic string loops, which are periodic sources. We can then decompose the time variation of the energy momentum tensor $T_{\mu\nu}$ into a Fourier series

$$T_{\mu\nu}(\vec{x}, t) = \sum_{n=-\infty}^{\infty} T_{\mu\nu}(\vec{x}, \omega_n)e^{-i\omega_n t}, \qquad (3.2.37)$$

where the Fourier series coefficients are given by

$$T_{\mu\nu}(\vec{x}, \omega_n) = \frac{1}{T}\int_0^T dt\, e^{i\omega_n t}T_{\mu\nu}(\vec{x}, t), \qquad (3.2.38)$$

as we have shown is section 2.3. In the above, $T$ is the period of the energy momentum tensor and $\omega_n = n\omega = 2\pi n/T$. Note that for a non-periodic source we should have expanded the energy-momentum tensor into a Fourier transform. Also, we use a minus sign in the exponential for the transformation of the time coordinate, and a plus sign in the exponential for the transformation of the space coordinate both for the Fourier series and the Fourier transform, to follow the convention of [4].

To solve (3.2.28), we will start by using the Green's function $G(\vec{x}, \vec{x}')$, defined by $L(\vec{x})G(\vec{x}, \vec{x}') = -4\pi\delta(\vec{x} - \vec{x}')$ where $L(\vec{x})$ is a linear differential operator. First, we will transform the time component of the equation to momentum space by applying Fourier series on both sides of the differential equation. Then, (3.2.28) is written as

$$(\Delta + \omega_n^2)\bar{h}_{\mu\nu}(\vec{x}, \omega_n) = -16\pi G T_{\mu\nu}(\vec{x}, \omega_n),\qquad(3.2.39)$$

which is the Helmholtz equation. In the above, the operator $\Delta$ is used to denote $\nabla^a\nabla_a$. The Green's function $G(\vec{x}, \vec{x}')$ of the Helmholtz equation is known to be

$$G(\vec{x}, \vec{x}') = \frac{e^{i\omega|\vec{x} - \vec{x}'|}}{|\vec{x} - \vec{x}'|},\qquad(3.2.40)$$

and therefore the solution is

$$\bar{h}_{\mu\nu}(\vec{x}, \omega_n) = 4G \int d^3\vec{x}' \frac{e^{i\omega_n|\vec{x} - \vec{x}'|}}{|\vec{x} - \vec{x}'|} T_{\mu\nu}(\vec{x}', \omega_n),\qquad(3.2.41)$$

which is integrated over the source.

Since we have assumed that the observer is in the local wave zone of the source, i.e. at a distance $r$ from the source which is much larger than the size of the source,[1] it holds that $|\vec{x}| \gg |\vec{x}'|$. Therefore, we can estimate $|\vec{x} - \vec{x}'|$ by $r - \vec{n} \cdot \vec{x}'$ in the phase factor of (3.2.41) and by r in the denominator, where $\vec{n} = \frac{\vec{x}}{r}$. Let us now define $\vec{k} = \omega\vec{n}$, such that $k^\mu = (\omega, \vec{k})$ is the 4-frequency of the gravitational waves in the $\vec{n}$ direction. Then, (3.2.41) can be written as

$$\bar{h}_{\mu\nu}(\vec{x}, \omega_n) = 4G\frac{e^{i\omega_n r}}{r} \int d^3\vec{x}' e^{-i\omega_n \vec{n} \cdot \vec{x}'} T_{\mu\nu}(\vec{x}', \omega_n) = 4G\frac{e^{i\omega_n r}}{r} T_{\mu\nu}(\vec{k}, \omega_n),\quad(3.2.42)$$

by Fourier transforming the spatial part of the energy momentum tensor. Follow-ing [4], we use the Fourier transform convention $S(\vec{k}, t) = \int d\vec{x} e^{-ik_i x^i} S(\vec{x}, t)$ for the spatial part of a vector, and $S(\vec{x}, \omega) = \int dt e^{i\omega t} S(\vec{x}, t)$, for the temporal part of a vector. Finally, by transforming back to the time domain using the Fourier series, we arrive at the expression

$$\bar{h}_{\mu\nu}(\vec{x}, t) = \frac{\kappa_{\mu\nu}(t - r, \vec{n})}{r}, \tag{3.2.43}$$

where

$$\kappa_{\mu\nu}(t - r, \vec{n}) = 4G \sum_{n=-\infty}^{\infty} e^{-i\omega_n(t-r)} T_{\mu\nu}(\vec{k}, \omega_n) \tag{3.2.44}$$

is the asymptotic waveform. We will use the above result to calculate the gravitational waveform emitted from a cusp on a cosmic string in section 6.2. In chapter 6 we will use the expression of the asymptotic waveform to calculate the gravitational waves from cusps observed away from a cosmic string loop.

---

[1] The distance $r$ is also assumed to be much smaller than the horizon distance. As we will see in section 6.4, on cosmological scales we will use the FLRW metric to evolve the amplitude of the gravitational wave.

# The standard cosmological model

Cosmology is a branch of astronomy that deals with the dynamics of radiation, galaxies and other astronomical objects at large scales, of the order of 100 Mpc or larger. It is a theory that aims to describe the history of the universe as a whole; its beginning, its current state and its future.

A flat and static universe which is both temporally and spatially infinite was one of the first cosmological assumptions to be adopted, proposed by Thomas Digges and supported by Newton as well [76]. However, it was soon found that such a universe could under some reasonable conditions contradict our observations on Earth. For example, Olber's paradox states that in an infinite universe, populated by an infinite number of stars which are distributed evenly, any line of sight from an observer on Earth should lead to the surface of a star. Hence, the night sky should appear bright during the night from starlight [77]. Shortly after the development of general relativity, the prevailing cosmological theory was a static universe, temporally infinite but spatially finite. This model was proposed in 1917 by Albert Einstein, it is called Einstein's static universe, and it was developed by adding a cosmological constant, $\Lambda$, to the Einstein field equations. With this addition, the Einstein field equations, which were shown in (3.1.27), obtain the form

$$R^{ab} - \frac{1}{2}Rg^{ab} + \Lambda g^{ab} = 8\pi G T^{ab}, \qquad (4.0.1)$$

which is a modification consistent with the theory of relativity. The purpose of this cosmological constant is to counteract the gravitational attractive forces, which would alone cause a spatially infinite static universe to collapse [78]. Soon after the development of this model, observational evidence from Leavitt, Hubble and others showed that the universe is not a static universe but an expanding one, which contained many distant galaxies. An observational law was developed that related the distance of a galaxy from Earth and its speed of recession, called Hubble-Lemaître law [62]. Friedmann, Lemaître, Robertson and Walker had shown independently that an expanding universe could be described by an isotropic cosmological model which obeys the Einstein field equations, called the FLRW universe. Friedmann's equations are a special case of the Einstein field equations which describe the expansion of the universe with the FLRW model assumptions [73]. The Einstein universe is the only non-trivial static solution of the Friedmann equations. It was also proposed by Lemaître that an expanding universe could be tracked backwards in time to a single point, its origin. He called this theory the primeval atom, although the name Big Bang theory became more prevalent. George Gamow, Ralph Alpher and Robert Herman developed the Big Bang model further and predicted the nuclei content of the early universe (Big Bang nucleosynthesis) and the presence of a cosmic background radiation filling all space, the CMB, both of which were confirmed observationally [79]. By this time most cosmologists considered the cosmological constant to be zero. This suggests that the expansion of the universe would be decelerating. However, observation of type Ia supernovae in 1988 indicated the unexpected, that the expansion of the universe was in fact accelerating [80]. Further evidence to this direction was found from baryon acoustic oscillations [81]. Therefore, the concept of the cosmological constant was revived to describe the accelerating expansion of the universe. In equation (4.0.1), if we move the $\Lambda$ term to the right-hand-side of the equation, then the term can be interpreted as an intrinsic energy of space, which is called the dark energy or vacuum energy because it is the energy of empty space.

The FLRW model sets the foundation for the development of the standard model of modern cosmology. A cosmological model that attempts to describe inhomogeneities of the universe, such as the clustering of galaxies, the physics of galaxy formation etc, and accounts the real universe as a perturbation of the FLRW universe is called the standard cosmological model. By perturbation of the FLRW universe we mean a model that follows the FLRW metric apart from primordial density fluctuations, which appear as higher order terms in the FLRW metric. Another means of perturbing the FLRW model is by using gauge invariant quantities, without choosing a coordinate system, and calculating their perturbations, as in the 1+3 covariant approach [71]. The primordial density fluctuations are defined as density fluctuations in the early universe that are considered as the seeds of all structure in the universe, since they set the appropriate conditions for galaxy formation. These anisotropies in the density in the early universe have been observed in detail in the CMB. The most widely accepted theory to explain the cause of the primordial anisotropies is the phenomenological model of inflation, which lies in the realm of particle cosmology. This is a model that describes a very short exponential expansion of spacetime in the early universe driven by a single scalar field, called the inflaton [79, 82].

A common choice for the standard cosmological model is the $\Lambda$CDM model with inflation, which is successful in describing the existence and structure of the CMB, the observed abundances of elements in the universe, the accelarating expansion of the universe and galaxy formation [79]. This model assumes three types of matter, first dark energy, mentioned above, second cold dark matter (CDM), a postulated type of matter which is non-baryonic, moves with a speed much less than the speed of light and does not emit radiation, and finally baryonic matter and radiation. The presence of the postulated cold dark matter is necessary to describe the observation of flat rotation curves of galaxies (which indicate the presence of unseen matter) and the observation of enhanced galaxy clustering. Other types of standard cosmology models include modifications to some of the $\Lambda$CDM assumptions, such as alternatives to dark matter (such as modified Newtonian dynamics [83]) and modified gravity models [84], or models which are intrinsically different such as the conformal cyclic cosmology, which assumes that the universe iterates through a countable sequence of cycles of FLRW spacetimes each beginning with a Big Bang [85, 86].

## 4.1 The FLRW metric

We will assume that the universe on large scales ($\geq 100$ Mpc) is isotropic, which means that the distribution of the galaxies is the same in every direction. This is called the cosmological principle and it implies that we are not in a special position in the universe, but it looks isotropic to any observer in the universe. The isotropy assumption at all points implies that the universe is also homogeneous, i.e. translationally invariant. These two assumptions are strongly supported by the observed CMB signature which has a uniform temperature distribution with tiny fluctuations that are, to a first approximation, isotropic. We will also assume that the universe can be described by the theory of general relativity, which is known to describe well the dynamics on large scales. We also know that the universe is expanding from observation of the recession velocities of distant galaxies described by Hubble's law, as was discussed in the introduction of this chapter.

The above assumptions force the line element, $ds^2$ (defined in equation (2.1.4)), to take the form

$$ds^2 = -dt^2 + a^2(t)dl^2. \tag{4.1.1}$$

In the above, $t$ is the coordinate time, $a(t)$ is the scale factor and $dl$ is a constant curvature three-dimensional line element. In cylindrical coordinates it becomes

$$dl^2 = \frac{d\hat{r}^2}{1 - k\hat{r}^2} + \hat{r}^2 d\Omega^2, \tag{4.1.2}$$

where

$$d\Omega^2 = d\theta^2 + \sin^2\theta d\phi^2 \tag{4.1.3}$$

and $\hat{r}$ is the metric distance. The parameter $k$ corresponds to the constant spatial curvature of the universe. For $k = 0$, the line element corresponds to flat space, for $k = 1$, it corresponds to spherical space and for $k = -1$ it corresponds to hyperbolic space. All three cases satisfy the isotropy and homogeneity conditions [79]. The line element (4.1.1) is invariant under the rescaling $a \to \lambda a$, $\hat{r} \to \hat{r}/\lambda$ and $k \to k/\lambda^2$. The above allow us to fix $a_0 = 1$, the value of the scale factor today. From here onwards, we will use the subscript 0 to denote scalar quantities evaluated in the present time. In the case of tensor quantities we will use a parenthesis to avoid confusion with the index notation. With this choice, the scale factor becomes dimensionless and $r$ and $k^{-1/2}$ have units of length. Note that the symmetries of the universe have reduced the free parameters of the metric to two, the scale factor, which is a function of time, and the curvature parameter $k$. We do not allow for non-diagonal components in the metric, $g^{0i}$, because this would break the isotropy assumption. Also, a non-trivial time component can always be reabsorbed by a coordinate transformation $dt' = \sqrt{g^{00}}dt$. The coordinates of the spatial part of the metric are called comoving coordinates and they are related to the physical coordinates via

$$r_{ph} = r(t) = a(t)\hat{r} \tag{4.1.4}$$

and

$$k_{ph} = k/a^2. \tag{4.1.5}$$

The line element of equation (4.1.2) is called the FLRW metric and it describes the geometry of the FLRW universe. In the local wave zone approximation, the physical distance defined in equation (3.2.43) corresponds to $r_{ph}$ as defined in (4.1.4).

The physical velocity in terms of the comoving velocity is given by

$$v_{ph} = \frac{dr_{ph}}{dt} = \frac{d(a\hat{r})}{dt} = \dot{a}\hat{r} + a\dot{\hat{r}} = Hr_{ph} + v_{pec}, \tag{4.1.6}$$

where the dot above quantities denotes $d/dt$. Therefore, the physical velocity of an object in the universe decomposes into two terms. One is the Hubble flow, $Hr_{ph}$, where we have defined the Hubble parameter

$$H(t) = \frac{\dot{a}(t)}{a(t)}. \tag{4.1.7}$$

This is the velocity of the object due to the expansion of the universe. We can define comoving observers, i.e. the observers who follow the Hubble flow. Then, the other term of the physical velocity, $v_{pec} = \dot{a}\hat{r}$, is the peculiar velocity of the object and it corresponds to its velocity as measured by a comoving observer.

For studying the propagation of light, it is useful to define the conformal time

$$d\eta = \frac{dt}{a(t)}. \tag{4.1.8}$$

Then, the line element (4.1.1) becomes

$$ds^2 = a^2(\eta)\left(-d\eta^2 + dl^2\right), \tag{4.1.9}$$

and the metric in the brackets is static. Since light propagates on null geodesics, $ds^2 = 0$, we find that the change of conformal time equals the change in the metric distance $\hat{r}$ if we drop radial motion ($d\Omega = 0$)

$$\Delta\eta = \Delta\hat{r}, \tag{4.1.10}$$

for $k = 0$, which implies that photons move on straight lines in flat space.

From equations (4.1.1)-(4.1.2), we find that an infinitesimal distance in a Euclidean ($k = 0$) FLRW universe is

$$ds^2 = -dt^2 + a^2(t)(d\hat{r}^2 + \hat{r}^2 d\Omega^2) = a^2(\eta)\left[-d\eta^2 + d\hat{r}^2 + \hat{r}^2 d\Omega^2\right] \tag{4.1.11}$$

expressed both in coordinate and conformal time. In the case of $k = 0$, the metric distance coincides with the comoving distance $r(t)$ (or proper distance or cosmic distance) defined as

$$r(t) = \int_{t_0}^{t_1} \frac{dt}{a(t)} \tag{4.1.12}$$

which is the distance between the observer at time $t_1$ and an object emitting a photon at time $t_0$, normalized by the scale factor, and it remains constant as the universe expands given that radial motion is insignificant.

## 4.2 The energy-momentum tensor

Let us now specify the matter content of the universe. For a comoving observer our assumptions from section 4.1, that the universe is isotropic and homogeneous, imply that the energy momentum tensor should have the form

$$T_{00} = \rho(t), \quad T_{0i} = 0, \quad T_{ij} = p(t)g_{ij}, \tag{4.2.1}$$

which is the energy-momentum tensor of a perfect fluid. In the above, we use the indices $i, j, k, ...$ to denote the three spatial components of the tensor. These take the values $i, j, k, ... = 1, 2, 3$. The energy-momentum tensor can be written in the covariant form

$$T_{\mu\nu} = (\rho(t) + p(t)) \, u_\mu u_\nu + p(t)g_{\mu\nu}. \tag{4.2.2}$$

In the above, $u^\mu$ is the relative 4-velocity between the perfect fluid and an observer,

$$u^\mu = \frac{dx^\mu}{d\tau} \tag{4.2.3}$$

where $\tau$ is the proper time of the observer, normalized such that $g^{\mu\nu}u_\mu u_\nu = -1$. Also, $\rho(t)$ and $p(t)$ are the density and pressure of the fluid in its rest frame, respectively. With respect to a comoving observer, $u_\mu = (-1, 0, 0, 0)$, the fluid is at rest and the energy-momentum tensor takes the diagonal form $T^\mu_\nu = \text{diag}(-\rho(t), p(t), p(t), p(t))$.

From the conservation of the energy-momentum tensor

$$\nabla^\mu T_{\mu\nu} = 0 \tag{4.2.4}$$

we obtain a conservation law for the fluid in an expanding universe

$$\dot{\rho}(t) + 3H \left( \rho(t) + p(t) \right) = 0, \tag{4.2.5}$$

called the continuity equation. Assuming a constant equation of state for the fluid

$$\rho = \omega p, \tag{4.2.6}$$

with $\omega$ a constant, we can write the continuity equation in terms of the density

$$\frac{\dot{\rho}}{\rho} + 3 \left( 1 + \omega \right) H = 0. \tag{4.2.7}$$

The above differential equation can be solved by integration, to find that

$$\rho \propto a^{-3(1+\omega)}. \tag{4.2.8}$$

In terms of the present day scale factor, $a_0$, and the energy density, $\rho_0$,

$$\rho(t) = \rho_0 \left( \frac{a(t)}{a_0} \right)^{-3(1+\omega)}. \tag{4.2.9}$$

Hereafter, when $a_0$ appears we will assume that it obtains the value $a_0 = 1$.

### 4.2.1 Matter sources

The three most popular equations of state that describe the matter content of the universe are $\omega = 0$, $\omega = 1/3$ and $\omega = -1$. The case of $\omega = 0$ corresponds to the so-called pressureless matter.[1] This category includes all types of matter for which the pressure is negligible compared to the energy density. This is the case for a gas consisting of non-relativistic particles, whose energy content is dominated by mass. Setting $\omega = 0$ in (4.2.9) we find that for matter

$$\rho(t) \propto a^{-3}, \tag{4.2.10}$$

which implies that for this type of matter the only reason there is dilution of the energy density is the expansion of the universe. Types of matter with the above equation of state in the universe are baryons, a term used in cosmology to refer to non-relativistic ordinary matter (which consists of atoms and electrons), and cold dark matter, a postulated type of matter consisting of non-relativistic heavy particles of unknown nature, as described above.

The case of $\omega = 1/3$ corresponds to radiation, which describes relativistic particles the energy density of which is dominated by kinetic energy. This type matter content includes photons, neutrinos and gravitons. Setting $\omega = 1/3$ in (4.2.9) we

---

[1]Note that we use the term matter in cosmology to specifically refer to types of energy/matter in the universe whose equation of state is $\omega = 0$. It should not be confused with the term matter content which is more general and covers all types of energy/matter in the universe.

find that for radiation

$$\rho(t) \propto a^{-4}, \tag{4.2.11}$$

which shows that the dilution of the energy density is due to the redshifting of the energy along with the dilution caused from the expansion of the universe.

Finally, the case of $\omega = -1$ corresponds to dark energy, which has a negative pressure effect and explains the accelerated expansion of the universe. This appears as the $\Lambda$ term in (4.0.1). Here, we will assume that the term is absorbed into the energy-momentum tensor and we will treat it as a fluid. This choice has to do purely with the nature of $\Lambda$ and it brings no loss of generality to our calculations for the cosmic strings in the upcoming chapters. Setting $\omega = -1$ in (4.2.9) we find that for the dark energy

$$\rho(t) \propto a^{0}, \tag{4.2.12}$$

which implies that the density of dark energy is constant. This does not contradict energy conservation because it is a solution of the continuity equation.

## 4.3   Dynamics of the universe

To find the evolution of the universe the Einstein field equations (3.1.27) need to be solved. From the FLRW universe assumptions we have found that the only unknown function that is needed to describe our system is the scale factor, $a(t)$, which appears in the FLRW metric (4.1.1). Having specified our metric, we can evaluate the Einstein tensor. We find that

$$G_{00} = -3\left(H^2 + \frac{k}{a^2}\right), \tag{4.3.1}$$

$$G_{0j} = 0, \tag{4.3.2}$$

$$G_{ij} = \left(2\frac{\ddot{a}}{a} + H^2 + \frac{k}{a^2}\right)\delta_{ij}. \tag{4.3.3}$$

### 4.3.1   The Friedmann equations

We can write the Einstein field equations for an FLRW universe by combining equations (4.3.1)-(4.3.3) with the expression of the energy-momentum tensor (4.2.1) to find that

$$H(t)^2 = \frac{8\pi G}{3}\rho(t) - \frac{k}{a(t)^2}, \qquad (4.3.4)$$

$$\frac{\ddot{a}}{a} = -\frac{4\pi G}{3}\left(\rho(t) + 3p(t)\right). \qquad (4.3.5)$$

The above are called the first and the second Friedmann equations, respectively. The expressions of $\rho(t)$ and $p(t)$ are perceived as the sum of all the matter sources in the universe.

From the first Friedmann equation (4.3.4), we can define the critical energy density of the universe by setting $k = 0$,

$$\rho_{crit} = \frac{3H^2}{8\pi G}. \qquad (4.3.6)$$

The critical energy density today is estimated to be $\rho_{crit,0} = 8 \times 10^{-26}\,g\,cm^{-3}$. Using this quantity we can define the fractional density for each matter source

$$\Omega_X = \frac{\rho_{X,0}}{\rho_{crit,0}}. \qquad (4.3.7)$$

The index $X$ corresponds to the matter source component. We will use the notation $\rho_M$ for the energy density of matter, $\rho_R$ for the energy density of radiation and $\rho_\Lambda$ for the energy density of dark energy. Then, for $k = 0$,

$$\Omega_M + \Omega_R + \Omega_\Lambda = 1, \qquad (4.3.8)$$

by definition.

Using the above notation and equations (4.2.9)-(4.2.12) the total energy density can be written as

$$\rho(t) = \sum_X \rho_{X,0}\,a(t)^{-3(\omega+1)}, \qquad (4.3.9)$$

while the Hubble constant can be written as

$$H^2(t) = H_0^2 \sum_X \left[ \Omega_{X,0}\, a(t)^{-3(\omega+1)} \right] + H_0^2 \Omega_{k,0} a(t)^{-2}. \tag{4.3.10}$$

Summation over $X$ implies summation over all the different matter sources in the universe. We have also introduced a curvature fractional density, $\Omega_{k,0} = -k/H_0^2$, for ease of notation. We can calculate the fractional density of each matter source in terms of the scale factor, $a(t)$, by using the solutions (4.2.9)-(4.2.12),

$$\Omega_M = \frac{8\pi G}{3H_0^2} \rho_M = \frac{\Omega_{m,0}}{a(t)^3}, \tag{4.3.11}$$

$$\Omega_R = \frac{8\pi G}{3H_0^2} \rho_R = \frac{\Omega_{r,0}}{a(t)^4}, \tag{4.3.12}$$

$$\Omega_\Lambda = \frac{8\pi G}{3H_0^2} \rho_\Lambda = \Omega_{\Lambda,0}. \tag{4.3.13}$$

Then, the Friedmann equations can be written in terms of the $\Omega_X$'s instead of $\rho$,

$$\dot{a}^2 = H_0^2 \left( \frac{\Omega_{M,0}}{a} + \frac{\Omega_{R,0}}{a^2} + \Omega_{\Lambda,0} a^2 \right) - k, \tag{4.3.14}$$

$$\ddot{a} = -H_0^2 \left( \frac{\Omega_{M,0}}{2a^2} + \frac{\Omega_{R,0}}{a^3} - \Omega_{\Lambda,0} a \right). \tag{4.3.15}$$

Hubble's constant (4.3.10) can be written in the more explicit form

$$H^2(t) = H_0^2 \left[ \Omega_{M,0} a^{-3} + \Omega_{R,0} a^{-4} + \Omega_{\Lambda,0} + \Omega_{k,0} a^{-2} \right]. \tag{4.3.16}$$

### 4.3.2 Possible FLRW universe outcomes

Having substituted the contribution of each fluid element into the Friedmann equations, we can now solve them to find the expression for $a(t)$, the only unknown in the FLRW metric (4.1), given that we input the values of the densities and curvature today. From equations (4.3.14)-(4.3.15) we observe that each of the fluid components has a different scaling with $a(t)$ in the Friedmann equations, for matter it is $a^{-3}$, for radiation $a^{-4}$ and for dark energy $a^0$. This implies that it is a reasonable assumption to regard each of the fluid component terms separately when solving the Friedmann equations. The physical interpretation of this approach is that in different eras of the universe evolution, it is dominated by a different fluid component.

First of all, let us discuss the evolution of the universe from the properties of $\dot{a}$ and $\ddot{a}$ in equations (4.3.14)-(4.3.15). In the following, we will assume $k$ to be a constant parameter that obtains real values. Let us start by setting $\Omega_\Lambda = 0$, i.e. include only matter and radiation in our universe. From (4.3.15), we see that $\ddot{a} < 0$ should always hold. Therefore, in this case the universe expansion is always decelerating. Also, it does not allow any static solutions since $\ddot{a} < 0$. Since the scale factor is a positive quantity by definition, the above implies that for $t = 0$ it should approach the value $\lim_{t \to 0} a(t) = 0$. For the different values of $k$ we have the following outcomes:

- For $k > 0$, which corresponds to spherical topology, the universe expands with time ($\dot{a} > 0$ is the only option for $a(t)$ to obtain positive values) until it reaches $\dot{a} = 0$. Then, $\dot{a}$ becomes negative, being a monotonically decreasing function due to $\ddot{a} \leq 0$, and this point corresponds to a maximum for $a(t)$. After this, $a(t)$ decreases and the universe recolapses back to itself. This is called the closed universe.

- For $k = 0$, which corresponds to Euclidean topology, we find that $\dot{a} > 0$ always. Therefore, $a(t)$ will expand forever, but $\dot{a}$ will tend to 0 as time approaches infinity. This is called the flat universe. In the case where we set $\Omega_r = 0$ this is called the Einstein-de Sitter universe.

- For $k < 0$, which corresponds to hyperbolic topology, we have that at infinity, $\lim_{t \to \infty} \dot{a}(t) = 1$. This corresponds to the open universe.

We do not discuss the case where $k < 0$ and $a(t)$ evolves from $\infty$ at $t = 0$ and reaches 0 asymptotically, since it does not correspond to an expanding universe.

Let us now include the effect of dark energy. For $\Omega_\Lambda \neq 0$ we have that $\ddot{a} < 0$ when the matter and radiation terms in (4.3.15) dominate, while $\ddot{a} > 0$ when the dark energy terms dominate. This corresponds to the following possibilities

- For $k > 0$ the universe can expand from $a = 0$ until it reaches a maximum and recollapse back to $a = 0$. In this case, the value of the curvature is such that the universe never reaches the dark energy domination era. This is the closed universe. Equivalently, the universe can collapse from $a = \infty$, reach a minimum and expand back to $\infty$. In this case, the universe is always in the dark energy domination era. This is called the bouncing universe.

- For $k > 0$, we can obtain the special case where $\ddot{a} = 0$ and $\dot{a} = 0$, which corresponds to the Einstein static universe. In this case, $a(t)$ is constant and finite. Since this is an unstable case, one can assume a universe that starts from $a = 0$ and crosses the Einstein static universe arbitrarily close, until it continues to expand again forever. This is called the loitering universe [87].

- For $k = 0$, $\dot{a} \geq 0$. In this case, the universe starts from $a = 0$ and expands indefinitely, but with an accelerated rate at late times due to dark energy domination. This is the flat universe.

- For $k < 0$ the universe starts from $a = 0$ and it expands forever, with dark energy dominating at late times, causing an accelerated expansion. This corresponds to an open universe.

## 4.4  Redshift

The light we receive is stretched due to the expansion of the universe. This effect is referred to as redshifting of the electromagnetic waves. If a wave is emitted at time $t_{em}$ and with wavelength $\lambda_{em}$ and is received on Earth at time $t_0$ and with wavelength $\lambda_{rec}$, then

$$\lambda_{rec} = \frac{a_0}{a_{em}}\lambda_{em} = \frac{1}{a_{em}}\lambda_{em}. \tag{4.4.1}$$

Because of the expansion of the universe $a_0 > a_{em}$, which implies that the wave is stretched, $\lambda_{rec} > \lambda_{em}$.

Motivated from the above, we define the redshift as the difference of the wave-length from emission to observation, divided by its value at emission,

$$z(t_{em}) \equiv \frac{\lambda_{rec} - \lambda_{em}}{\lambda_{em}} = \frac{1}{a(t_{em})} - 1. \tag{4.4.2}$$

The above supplies a one-to-one relation of the scale factor with the redshift. Therefore, for any object observed at redshift $z(t)$ on Earth, we can find the scale factor $a(t)$.

In the frequency domain it holds that

$$f_{rec}a_{rec} = f_{em}a_{em}. \tag{4.4.3}$$

We can also write

$$f_{em} = (1+z)f_{rec} \tag{4.4.4}$$

where $f_{rec}$ is the observed frequency at Earth, and $a_{rec} \equiv a_0 = 1$ is the present cosmological scale factor.

Let us Taylor expand $a(t)$ around the present time

$$a(t) = 1 + H_0(t - t_0) + \frac{1}{2}\ddot{a}(t_0)(t - t_0). \tag{4.4.5}$$

The distance that light travels is given by (4.1.10) and it equals $d = t_0 - t_1$ for small distances. If we also Taylor expand the right-hand-side of equation (4.4.2) around $a_0 = 1$, we find to first order,

$$z = -(a - 1). \tag{4.4.6}$$

Combining equations (4.4.5) and (4.4.6), we find that

$$z = H_0 d, \tag{4.4.7}$$

for observations close enough to Earth. This is Hubble's law, which implies that objects redshift away from the Earth at a constant rate with their distance from us, and it confirmed by supernovae observations.

Having defined the redshift, we can now express a relation between $dt$ and $dz$.

Note that

$$H(z) = \frac{1}{a}\frac{da}{dt} = \frac{da}{dz}\frac{dz}{dt}(1+z). \tag{4.4.8}$$

Differentiating (4.4.2),

$$da = -\frac{dz}{(1+z)^2}. \tag{4.4.9}$$

Therefore,

$$dt = -\frac{dz}{H(z)(1+z)}. \tag{4.4.10}$$

Given the above, we can express the age of the universe in terms of the redshift as

$$t_0 = \int_0^{t_0} dt = \int_0^{\infty} \frac{dz}{H(z)(1+z)}, \tag{4.4.11}$$

while the coordinate time at redshift $z$ is

$$t(z) = \int_z^{\infty} \frac{dz}{H(z)(1+z)}. \tag{4.4.12}$$

We can also write the Hubble parameter from (4.3.10) in terms of the redshift,

$$H^2 = H_0^2 \sum_X \Omega_{X,0} \, (1+z)^{3(\omega+1)}. \tag{4.4.13}$$

Finally, the comoving distance (or cosmic distance) in a flat universe $k = 0$, is given by

$$r(z) = \int_{t_1}^{t_0} \frac{dt}{a(t)} = \int_0^z \frac{dz}{H(z)}. \tag{4.4.14}$$

We will use the above definitions of the cosmological functions in section 5.6.1.

## 4.5   The ΛCDM universe

Observational results from the 2015 Planck results [22] indicate that the spatial curvature of our universe is very close to zero with the curvature density parameter satisfying,

$$|\Omega_k| < 0.005. \tag{4.5.1}$$

This implies that our universe is very close to the flat universe described in section 4.3.2 with $k = 0$. It is also found that the universe consists of matter, radiation and dark energy in the following proportions

$$\Omega_M = 0.308, \ \Omega_R = 9.1476 \times 10^{-5}, \ \Omega_\Lambda = 1 - \Omega_m - \Omega_r = 0.692. \qquad (4.5.2)$$

The value of the Hubble constant today is evaluated to be $H_0 = 100hkms^{-1}Mpc^{-1}$ where $h = 0.678$. The equation of state for the dark energy is found to be very close to $\omega = -1$, which means that it is well described by the concept of the cosmological constant.

Given the above, we will consider the effects of curvature to be negligible, $\Omega_k \simeq 0$. Also, as discussed in 4.3.2, the universe during its evolution is dominated by a single fluid, first by radiation, then by matter and finally by dark energy. We call these stages of the universe evolution as the radiation era, the matter era and the dark energy era, respectively. We can solve the Friedmann equations for a flat universe and assuming a single-component fluid to find the solution for the scale factor in each era. Equation (4.3.4), can be written for a single component as

$$\frac{\dot{a}(t)}{a(t)} = H_0 \sqrt{\Omega_X} a(t)^{-\frac{3}{2}(1+\omega_X)}. \qquad (4.5.3)$$

By integrating the above we find that during the radiation era, for $\omega = 1/3$,

$$a(t) \propto t^{1/2} \qquad (4.5.4)$$

and in conformal time

$$a(\eta) \propto \eta. \qquad (4.5.5)$$

Similarly, for the matter era, $\omega = 0$, we find that

$$a(t) \propto t^{2/3} \qquad (4.5.6)$$

and in conformal time

$$a(\eta) \propto \eta^2. \qquad (4.5.7)$$

Finally, during the accelerated expansion of the dark energy era, we find that

$$a(t) \propto e^{Ht} \qquad (4.5.8)$$

and in conformal time

$$a(\eta) \propto (-\eta)^{-1}. \tag{4.5.9}$$

In the evolution of the universe, there is also the transition from one era to another. During this period, the universe is best described by a two fluid approach. The moment of equality of radiation and matter, i.e. the transition from the radiation to the matter era, is called the matter-radiation equality

$$\rho_m = \rho_r, \tag{4.5.10}$$

and it occurred at redshift $z_{eq} \simeq 3366$. One can solve the above issue either by solving the Friedmann equations for a two component universe or by introducing an interpolating function between the two eras. We will follow the latter approach in Model I discussed in section 5.6.1. In Model II of the same section, we will use a numerical approach, by solving the integrals $t(z)$ and $r(z)$, which cannot be solved analytically for any $z$. We will also derive their asymptotic behaviour. Note that we will use the above values of the cosmological parameters in section 5.6.1.

# CHAPTER 5

---

# Cosmic strings

---

In this chapter, we will look into cosmic strings under the assumption that their width is infinitely small, approaching zero with respect to their length, which is known as the wire approximation [14]. Indeed, a cosmic string has width less than the electron radius, but its length can extend to the limits of the visible Universe, in general. In this approach, we can visualize the string as a line of particles of infinitesimal mass, interacting with each other via a string-like tension. With this approach, one can derive the classical relativistic equations of motion for the cosmic string. The string is allowed to be either open, understanding that it stretches beyond the limits of the visible universe, or closed, meaning that it forms a loop, called a cosmic string loop.

A point of view other than the wire approximation is the field theory approach, where the width of the cosmic string is taken into account, and the field theoretic properties of the cosmic strings can be derived. Of this approach there will be little discussion in this thesis, but more analysis can be found in [9].

In this chapter, we will derive the classical equation of motion of the cosmic string, using the notions introduced in section 2.2. Following that, we will focus on the closed cosmic string embedded in a flat spacetime background. We will also discuss the gauge issue and the solution of the equation of motion, as well as possible string trajectories. An important feature relevant to the gravitational wave emission from cosmic strings are the cusps, points on the string that move at the speed of light. Therefore, we will give some additional focus on the properties of the string at the cusps.

## 5.1 Basic elements of cosmic strings in spacetime

The cosmic string is a one-dimensional object which traces a two-dimensional surface in spacetime in the course of its evolution, called the worldsheet [88, 14, 9]. Cosmic strings can be of two types; these are the open strings, i.e. strings with two ends, and the closed strings, i.e. strips with no ends. The former will trace an open surface in spacetime, while the latter will trace a closed curve. In the following we will focus on closed cosmic strings, also called cosmic string loops. The worldsheet requires two parameters $(\xi^1, \xi^2)$ to be described, which we can also denote in index notation as $\xi^A$, with $A = 1, 2$.[1] The parametrized curve traced out by the string in spacetime, namely the worldsheet, can therefore be described by the mapping functions

$$X^\mu(\xi^1, \xi^2), \tag{5.1.1}$$

which map from the two-dimensional parameter space $\{\xi^A\}$ to spacetime $\{x^\mu\}$, i.e. $x^\mu(\xi^A) = X^\mu(\xi^A)$.

An infinitesimal distance on the worldsheet is (recall equation (2.1.4))

$$ds^2 = g_{ab} dX^a(\xi^1, \xi^2) dX^b(\xi^1, \xi^2) = g_{ab} \frac{\partial X^a}{\partial \xi^A} \frac{\partial X^b}{\partial \xi^B} d\xi^A d\xi^B. \tag{5.1.2}$$

---

[1] We use capital indices $A$, $B$... to denote the coordinates in the two-dimensional parameter space $(\xi^1, \xi^2)$.

The set of the two spacetime vectors $\partial X^a / \partial \xi^A$ spans the tangent space of the worldsheet. The motion of the string is assumed to be non-tachyonic, i.e. at any point on the worldsheet the tangent space is spanned by a timelike and a spacelike vector. This rule is only broken at points where the string moves momentarily at the speed of light, called cusps, where the tangent space collapses to a null line. This implies that at the cusps both the vectors $\partial X^a / \partial \xi^A$ are null. From equation (5.1.2) we realise that we can define a two-dimensional induced metric on the worldsheet, or worldsheet metric (see [88]),

$$\gamma_{AB} = g_{ab} \frac{\partial X^a}{\partial \xi^A} \frac{\partial X^b}{\partial \xi^B},$$ (5.1.3)

Note that the determinant of $\gamma_{AB}$ is zero whenever the vectors $\partial X^a / \partial \xi^A$ are linearly dependent. This happens only at the cusps, where the tangent space is null. It can be proved that away from the cusps for a non-tachyonic moving strings, the determinant of the induced metric is negative (see [88, 14]). Therefore, we conclude that

$$\left( \frac{\partial X}{\partial \xi^1} \right)^2 \left( \frac{\partial X}{\partial \xi^2} \right)^2 - \left( \frac{\partial X}{\partial \xi^1} \cdot \frac{\partial X}{\partial \xi^2} \right)^2 \leq 0$$ (5.1.4)

at any point on the worldsheet (recall the relativistic product notation from section 2.1). The induced metric has mixed signature, at any point on the worldsheet apart from the cusps, where all the components of the induced metric tend to zero (for a proof in a general curved background see section 2.6 of [14]) and the induced metric becomes degenerate.

The area A that a string sweeps in spacetime is

$$A = \int d\tau d\sigma \sqrt{\left( \frac{\partial X}{\partial \xi^1} \cdot \frac{\partial X}{\partial \xi^2} \right)^2 - \left( \frac{\partial X}{\partial \xi^1} \right)^2 \left( \frac{\partial X}{\partial \xi^2} \right)^2}.$$ (5.1.5)

From equation (5.1.4), we conclude that the quantity under the square root is positive or zero. Using the worldsheet metric $\gamma_{AB}$ we can write the surface swept by the string as

$$A = \int d\xi^1 d\xi^2 \sqrt{-\gamma},$$ (5.1.6)

where $\gamma = \det(\gamma_{AB})$ is the determinant of the induced metric.

## 5.2   The equation of motion of a cosmic string

To study the dynamics of the relativistic string, we will start by formulating its action, which reflects its physical and geometrical nature and we will derive the equations Euler-Lagrange for a cosmic string.

The Nambu-Goto action, which describes the motion of the relativistic cosmic string, is proportional to the area the worldsheet sweeps in spacetime, given in equation (5.1.5),

$$S = -\mu \int d\xi^1 d\xi^2 \sqrt{-\gamma},$$ (5.2.1)

and it describes a zero thickness string in spacetime. In the above, the multiplication factor, $\mu$, is the string tension, i.e. the mass per unit length.

This action was formulated by Nambu and Goto [89, 90], and it is a unique solution under the following assumptions:

- the action is invariant under spacetime transformations $x^\mu \to \tilde{x}^\mu$ and under reparametrizations of the worldsheet coordinates $\xi^A \to \tilde{\xi}^A$.

- the action involves only first derivatives of $X^\mu$.

Note the similarity of the string action with the action for a relativistic particle which is proportional to the proper length (i.e. the elapsed proper time) it traces in spacetime, and thus the analogy of the terms "worldline" and "worldsheet". A comparison of the relativistc string dynamics with the nonrelativistic one can be found in [88].

The relativistic string action can be generalized to include terms with the curvature of the string as well, by violating the second of the aforementioned conditions and including higher derivative terms [91]. A typical string has such a large radius that its curvature has a negligible effect on its dynamics in general, but it can be significant to describe the string dynamics around cusps and kinks, where the string curvature obtains very large values.

### 5.2.1 Derivation of the equation of motion

We can apply the variational principle, which we discussed in section 2.2, to the action (5.2.1) to derive the equation of motion of a Nambu-Goto cosmic string.

We derive the Euler-Lagrange equations, by extremizing the action $S[X^\mu]$ in equation (5.2.1) with respect to the functions $X^\mu$. It follows that (using the reasoning of section 2.2)

$$
\delta S = 0 \Leftrightarrow \int d^2\xi \left[ \frac{\partial \sqrt{-\gamma}}{\partial X^a} \delta X^a + \frac{\partial}{\partial \xi^A} \left( \frac{\partial \sqrt{-\gamma}}{\partial (\partial_A X^a)} \delta X^a \right) - \right.
$$
$$
\left. \frac{\partial}{\partial \xi^A} \left( \frac{\partial \sqrt{-\gamma}}{\partial (\partial_A X^a)} \right) \delta X^a \right] = 0 \tag{5.2.2}
$$

The second term of the above equation is a boundary term and it vanishes. Therefore, we find that the string satisfies the Euler-Lagrange equations

$$
\frac{\partial \sqrt{-\gamma}}{\partial X^a} - \frac{\partial}{\partial \xi^A} \left( \frac{\partial \sqrt{-\gamma}}{\partial (\partial_A X^a)} \right) = 0. \tag{5.2.3}
$$

Using the above results and keeping also in mind (5.1.3), we can expand,

$$
\frac{\partial \sqrt{-\gamma}}{\partial X^a} = \frac{1}{2} \sqrt{-\gamma} \gamma^{AB} \frac{\partial g_{bc}}{\partial X^a} \partial_A X^b \partial_B X^c \tag{5.2.4}
$$

and

$$
\frac{\partial}{\partial \xi^C} \left( \frac{\partial \sqrt{-\gamma}}{\partial (\partial_C X^a)} \right) = \partial_C \left( \frac{1}{2} \sqrt{-\gamma} \gamma^{AB} g_{bc} \frac{\partial (\partial_A X^b \partial_B X^c)}{\partial (\partial_C X^a)} \right) =
$$
$$
\partial_C \left[ \sqrt{-\gamma} \gamma^{BC} g_{ab} \partial_B X^b \right]. \tag{5.2.5}
$$

To reach the above result, we have used the formula

$$
\frac{\partial (\partial_A X^a \partial_B X^b)}{\partial (\partial_C X^c)} = \delta^a_c \delta^A_C \partial_B X^b + \delta^b_c \delta^B_C \partial_A X^a. \tag{5.2.6}
$$

Taking into account (5.2.5-5.2.6), the Euler-Lagrange equations (5.2.3) for a curved spacetime can be written as

$$
\partial_B \left( \sqrt{-\gamma} \gamma^{AB} g_{ab} \partial_A X^a \right) - \frac{1}{2} \sqrt{-\gamma} \gamma^{AB} \partial_b g_{ac} \partial_A X^a \partial_B X^c = 0. \tag{5.2.7}
$$

## 5.2.2 Derivation of the energy-momentum tensor

The variation of the Nambu-Goto action with respect to the metric gives us the energy-momentum of the cosmic string

$$T^{ab} = \frac{-2}{\sqrt{-g}} \frac{\delta S}{\delta g_{ab}} \tag{5.2.8}$$

The above can be written as

$$T^{ab} = \frac{-2}{\sqrt{-g}} \frac{\delta S}{\delta \left(\sqrt{-\gamma}\right)} \frac{\delta \left(\sqrt{-\gamma}\right)}{\delta g_{ab}} \tag{5.2.9}$$

Using equations (5.1.3) and (3.1.5), we find that

$$\frac{\delta \left(\sqrt{-\gamma}\right)}{\delta g_{ab}} = \frac{1}{2\sqrt{-\gamma}} \frac{\delta \left(-\gamma\right)}{\delta g_{ab}} = \tag{5.2.10}$$

$$\frac{1}{2\sqrt{-\gamma}} \frac{\delta \left(-\gamma\right)}{\delta \gamma_{AB}} \frac{\delta \gamma_{cd}}{\delta g_{ab}} = \tag{5.2.11}$$

$$\frac{1}{2\sqrt{-\gamma}} \left(-\gamma \gamma^{AB}\right) \frac{\partial X^a}{\partial \xi^A} \frac{\partial X^b}{\partial \xi^B}. \tag{5.2.12}$$

Therefore, the energy-momentum tensor of the cosmic string is

$$T^{ab} = \frac{\mu}{\sqrt{-g}} \int \int d^2\xi \sqrt{-\gamma} \gamma^{AB} \frac{\partial X^a}{\partial \xi^A} \frac{\partial X^b}{\partial \xi^B} \delta \left(x^c - X^c(\xi^C)\right). \tag{5.2.13}$$

In the above, the delta function ensures that the energy-momentum tensor is zero at any point $x^c$ in spacetime that does not lie on the worldsheet surface $X^c(\xi^C)$.

# 5.3 Closed strings in flat spacetime

In the following chapters, our main focus is to look into the gravitational waves emitted from a certain family of cosmic string loops, as well as develop a toy model to describe the loop fragmentation. We will focus on closed cosmic string loops which are embedded in a fixed Minkowski background, which we will describe in this section. This implies that the loops are far away from any gravitational field and that the gravitational back-reaction from the string on itself is negligible.

Under these assumptions, the equation of motion of a cosmic string (5.2.7) in flat spacetime simplifies to

$$\partial_B \left( \sqrt{-\gamma} \gamma^{AB} \partial_A X^a \right) = 0, \tag{5.3.1}$$

since $\partial_b \eta^{ac} = 0$. The problem of solving this differential equation to find the four functions $X^\mu$ has two degrees of freedom which appear due to the reparametrization freedom of the worldsheet coordinates $\xi^A$ [14]. The equation of motion of the cosmic string can be simplified by making the appropriate worldsheet coordinate choice.

## 5.3.1 Gauge fixing

The induced metric (5.1.3) is the 2-by-2 matrix

$$\gamma_{AB} = \begin{pmatrix} \frac{\partial X}{\partial \xi^1} \cdot \frac{\partial X}{\partial \xi^1} & \frac{\partial X}{\partial \xi^1} \cdot \frac{\partial X}{\partial \xi^2} \\ \frac{\partial X}{\partial \xi^1} \cdot \frac{\partial X}{\partial \xi^2} & \frac{\partial X}{\partial \xi^2} \cdot \frac{\partial X}{\partial \xi^2} \end{pmatrix} \tag{5.3.2}$$

It can be proved that on a 2-dimensional surface with mixed signature metric we can choose a set of real coordinates $(u, v)$ such that the line element on the surface obtains the form [92]

$$ds^2 = f(u, v) du dv. \tag{5.3.3}$$

We will call this choice of worldsheet coordinates the light-cone gauge. In this gauge, the induced matrix has zero diagonal elements

$$\gamma_{AB} = \begin{pmatrix} 0 & \frac{\partial X}{\partial u} \cdot \frac{\partial X}{\partial v} \\ \frac{\partial X}{\partial u} \cdot \frac{\partial X}{\partial v} & 0 \end{pmatrix} = \sqrt{-\gamma} \begin{pmatrix} 0 & 1 \\ 1 & 0 \end{pmatrix}, \tag{5.3.4}$$

since $\gamma = -[(\partial X/\partial u) \cdot (\partial X/\partial v)]^2$ and equation (5.3.3) is equivalent to the conditions

$$\eta_{\mu\nu} \frac{\partial X^\mu}{\partial u} \frac{\partial X^\nu}{\partial u} = \eta_{\mu\nu} \frac{\partial X^\mu}{\partial v} \frac{\partial X^\nu}{\partial v} = 0. \tag{5.3.5}$$

The action (5.2.1) in this gauge becomes

$$S = -\mu \int du dv \, \eta_{\mu\nu} \frac{\partial X^\mu}{\partial u} \frac{\partial X^\nu}{\partial v}. \tag{5.3.6}$$

By varying the action with respect to $X^\mu$, we find that in the light-cone gauge the equation of motion is

$$\frac{\partial^2 X^\mu}{\partial u \partial v} = 0. \tag{5.3.7}$$

For any set of light-cone gauge coordinates, we can always define another set of gauge coordinates; a timelike coordinate [2]

$$\tau = \frac{v - u}{2} \tag{5.3.8}$$

and a spacelike coordinate

$$\sigma = \frac{v + u}{2}. \tag{5.3.9}$$

By timelike (spacelike) coordinate we mean that the tangent vector in the direction of the coordinate is a timelike (spacelike) vector. Rewriting the line element (5.3.3) in terms of $(\tau, \sigma)$, we find that

$$ds^2 = \frac{1}{2} \left[ -\left(\frac{\partial X}{\partial \tau}\right)^2 + \left(\frac{\partial X}{\partial \sigma}\right)^2 \right] \left( -d\tau^2 + d\sigma^2 \right). \tag{5.3.10}$$

Therefore, this gauge choice converts the induced metric into a diagonal form

$$\gamma_{AB} = \begin{pmatrix} \left(\frac{\partial X}{\partial \tau}\right)^2 & 0 \\ 0 & \left(\frac{\partial X}{\partial \sigma}\right)^2 \end{pmatrix} = \sqrt{-\gamma} \begin{pmatrix} -1 & 0 \\ 0 & 1 \end{pmatrix}, \tag{5.3.11}$$

since $\gamma = (\partial X/\partial \tau)^2 (\partial X/\partial \sigma)^2 = -[(\partial X/\partial \tau) \cdot (\partial X/\partial \tau)]^2 = -[(\partial X/\partial \sigma) \cdot (\partial X/\partial \sigma)]^2$. Our line element choice implies the conditions

$$\frac{\partial X}{\partial \tau} \cdot \frac{\partial X}{\partial \sigma} = 0, \ \left(\frac{\partial X}{\partial \tau}\right)^2 + \left(\frac{\partial X}{\partial \sigma}\right)^2 = 0, \tag{5.3.12}$$

called the Virasoro conditions. We will call this choice of worldsheet coordinates the conformal gauge because the induced metric obtains the conformally flat form

$$\gamma_{AB} = \sqrt{-\gamma}\, \eta_{AB}, \ \gamma^{AB} = \frac{1}{\sqrt{-\gamma}}\, \eta^{AB}, \tag{5.3.13}$$

where $\eta_{AB} = \text{diag}(-1, 1)$.

---

[2]Note here a possible confusion between the notation of the timelike coordinate and the proper time. Because the choice of the character $\tau$ is classic for both of these quantities in their own fields, we will keep it for both, but we will be careful to specify which one we refer to in every case.

Transforming the action (5.3.6) in $(\tau, \sigma)$ coordinates we find

$$S = -\mu \int d\tau d\sigma \, \frac{1}{2} \eta_{\mu\nu} \left( \frac{\partial X^\mu}{\partial \sigma} \frac{\partial X^\nu}{\partial \sigma} - \frac{\partial X^\mu}{\partial \tau} \frac{\partial X^\nu}{\partial \tau} \right). \tag{5.3.14}$$

In the above we have used the formula for change of coordinates in a 2-dimensional integral [66]

$$d\tau d\sigma = \left| \det \left( \frac{\partial(\sigma, \tau)}{\partial(u, v)} \right) \right| du dv = \frac{1}{2} du dv, \tag{5.3.15}$$

where

$$\frac{\partial(\sigma, \tau)}{\partial(u, v)} = \begin{pmatrix} \partial\sigma/\partial u & \partial\tau/\partial u \\ \partial\sigma/\partial v & \partial\tau/\partial v \end{pmatrix}. \tag{5.3.16}$$

Therefore, we find that the equation of motion for this gauge choice is

$$\left( \frac{\partial^2}{\partial \tau^2} - \frac{\partial^2}{\partial \sigma^2} \right) X^\mu (\tau, \sigma) = 0, \tag{5.3.17}$$

which is a 2-dimensional wave equation.

Unlike open strings, where the worldsheet coordinates have an $\mathbb{R}^2$ domain, the worldsheet coordinates for closed strings feature a periodic identification along the spacelike direction, due to the cylindrical topology of the closed string. Under a particular choice of the $(u, v)$ coordinates (see section 2.3 of [14]), we can align the coordinate system such that lines of constant $\tau$ form closed curves on the worldsheet, with a fixed period $\Delta\sigma = l$, such that

$$X^\mu(\tau, \sigma) = X^\mu(\tau, \sigma + l). \tag{5.3.18}$$

The parameter $\sigma$ describes position on the string, and therefore, a reasonable choice is $l$ to have units of length. However, we are also free to assume that $\sigma$ is dimensionless, and another convenient choice for its period is $2\pi$. The choice of units and period does not carry any physical significance. For instance, if we choose $\sigma$ to have units of length, then $\partial X^a/\partial \xi^A$ is dimensionless and in equation (5.1.2) $ds^2$ will obtain its units of length squared from the product $d\xi^A d\xi^B$. Otherwise, if we choose $\sigma$ to be dimensionless, then $\partial X^a/\partial \xi^A$ has units of length, and $ds^2$ will obtain its units from the product $\partial X^a/\partial \xi^A \partial X^b/\partial \xi^B$ in (5.1.2).

We can define the invariant length of the loop, $l_{inv}$, which is the length of the loop measured in its rest frame when the loop is momentarily stationary. This length is invariant under gauge transformations and it is a conserved quantity [93]. We can choose the gauge period $l$ to be equal to the invariant length, which implies a lower bound for the period of the loop trajectory. For ease of notation, we will denote the invariant length by $l$ and we will always choose it to coincide with the gauge period hereafter.

### 5.3.2 String dynamics in the conformal time gauge

The conditions of the conformal gauge are not enough to specify a unique choice of worldsheet coordinates [14]. Note that a solution to (5.3.17) for the temporal part, with index value $\mu = 0$, is $X^0 = \tau$, which we will choose to fix the remaining gauge invariance. This time gauge identifies the parameter $\tau$ with the time coordinate t. We still have freedom of reparametrizing the parameter $\sigma$. Hereafter, we will use the notation $\dot{X}^\mu = \partial X^\mu / \partial \tau$ and $X^{\mu\prime} = \partial X^\mu / \partial \sigma$ to simplify our expressions. Also, the derivative of any one-parameter function $f(x)$ will be denoted as $f'(x)$ meaning differentiation with $x$. Using this notation, the equation of motion (5.3.17) becomes

$$\ddot{X}^\mu(\tau, \sigma) - X^{\mu\prime\prime}(\tau, \sigma) = 0 \tag{5.3.19}$$

and the Virasoro conditions (5.3.12) become

$$\dot{X} \cdot X' = 0 \tag{5.3.20}$$

$$\dot{X}^2 + X'^2 = 0. \tag{5.3.21}$$

The general solution to the equation of motion (5.3.19) in the conformal time gauge is a superposition of left- and right-moving waves

$$X^0 = \tau, \; X^i(\tau, \sigma) = \frac{1}{2} \left( a^i(u) + b^i(v) \right), \tag{5.3.22}$$

where $u = \sigma - \tau$ and $v = \sigma + \tau$. The functions $\vec{a}(u)$ and $\vec{b}(v)$ are called the left- and right-moving wave functions respectively (or left- and right-movers). In the above, the Latin characters $i, j, k...$ denote the space indices of spacetime quantities. Also, the notation $\vec{r}$ is used for three-vector functions, i.e. $\vec{r} = (r^1, r^2, r^3)$. We can also define the 4-vectors $a^\mu = (a^0, \vec{a}^i)$ and $b^\mu = (b^0, \vec{b}^i)$ by decomposing $X^\mu$ as

$$X^\mu = \frac{1}{2} \left( a^\mu(u) + b^\mu(v) \right). \tag{5.3.23}$$

Since $X^0 = \tau$, we find that $a^0 = -u$ and $b^0 = v$. The Virasoro conditions (5.3.20-5.3.21) simplify to

$$\left( a^{\mu\prime} \right)^2 = 0, \ \left( b^{\mu\prime} \right)^2 = 0. \tag{5.3.24}$$

The above conditions yield

$$\vec{a}^{\prime 2} = \vec{b}^{\prime 2} = 1. \tag{5.3.25}$$

Therefore, the left- and right-movers both lie on a unit sphere, which is called the Kibble-Turok sphere.

From equation (5.3.18), it follows that

$$\vec{a}(-\tau) + \vec{b}(\tau) = \vec{a}(l - \tau) + \vec{b}(l + \tau) \tag{5.3.26}$$

Also, since the derivative of a periodic function is periodic, $\vec{X}'$ is periodic in $\sigma$, satisfying

$$\vec{a}'(-\tau) - \vec{b}'(\tau) = \vec{a}'(-\tau + l) - \vec{b}'(\tau + l) \tag{5.3.27}$$

These equations imply that the left- and right-movers are each the sum of a vector function that has period $l$, $\vec{a}_1$ and $\vec{b}_1$ respectively, and a time dependent vector function, which corresponds to the motion of the string as a whole in spacetime [50]. Therefore,

$$\vec{a}(\sigma - \tau) = \vec{a_1}(\sigma - \tau) + \vec{V}\tau, \tag{5.3.28}$$

$$\vec{b}(\sigma + \tau) = \vec{b_1}(\sigma + \tau) + \vec{V}\tau, \tag{5.3.29}$$

where $\vec{V}$ is the string's bulk velocity, which is a constant vector.

We can now prove that the period in time of the string trajectory is $l/2$. If we calculate $\vec{X}$ at $(\tau + l/2, \sigma + l/2)$, we find that

$$X^\mu(\tau + l/2, \sigma + l/2) = \frac{1}{2}\vec{a_1}(-\tau + \sigma) + \frac{1}{2}\vec{b_1}(\tau + \sigma + l) + \vec{V}(\tau + \frac{l}{2}). \tag{5.3.30}$$

In the centre-of-momentum frame, the bulk velocity is zero, and the above ex-

pression becomes

$$\vec{X}(\tau + l/2, \sigma + l/2) = \frac{1}{2}\vec{a_1}(-\tau + \sigma) + \frac{1}{2}\vec{b_1}(\tau + \sigma) = \vec{X}(\tau, \sigma). \qquad (5.3.31)$$

For specific families of cosmic string solutions it is possible for the period to be shorter that $l/2$, when the string trajectory is invariant under a translation of a fraction of $l/2$,

$$(\tau, \sigma) \rightarrow (\tau + l/2n, \sigma + l/2n) \qquad (5.3.32)$$

where $n$ is a positive integer. An example of such a string loop is the non-planar $N/M$ single harmonic loop, where $N$ is the harmonic order of the left-mover and $M$ is the harmonic order of the right-mover [14].

Using equation (5.2.13) along with (5.3.11) and (5.3.13), we find that the energy-momentum tensor of the string in this gauge is

$$T^{\nu\kappa} = \mu \int \int d\tau d\sigma \left( -\frac{\partial X^\nu}{\partial \tau}\frac{\partial X^\kappa}{\partial \tau} + \frac{\partial X^\nu}{\partial \sigma}\frac{\partial X^\kappa}{\partial \sigma} \right) \delta\left( x^\lambda - X^\lambda \right). \qquad (5.3.33)$$

The limits of integration for the worldsheet coordinate $\sigma$ are from 0 to $l$, the invariant length, and the limits of integration for $\tau$ are from 0 to $l/2$, the period of the loop. Since $X^0 = \tau = t$, the above can be written as

$$T^{\nu\kappa} = \mu \int d\sigma \left( -\frac{\partial X^\nu}{\partial \tau}\frac{\partial X^\kappa}{\partial \tau} + \frac{\partial X^\nu}{\partial \sigma}\frac{\partial X^\kappa}{\partial \sigma} \right) \delta\left( \vec{x} - \vec{X}(\tau, \sigma) \right) \qquad (5.3.34)$$

The total energy of the string is

$$E = \mu \int d^3\vec{x} T^0_0, \qquad (5.3.35)$$

which is a conserved quantity [9]. We find that

$$T^0_0 = \eta_{0\kappa} T^{0\kappa} = \mu \int d\sigma \delta\left( \vec{x} - \vec{X}(\tau, \sigma) \right) \qquad (5.3.36)$$

and therefore the total energy of the string is

$$E = \mu \int d\sigma = \mu l. \qquad (5.3.37)$$

The general solution of the equation of motion of a closed cosmic string loop (5.3.19) can be expressed as an infinite series of harmonic modes. Since we will be looking at the movement of $\vec{a}'$ and $\vec{b}'$ on the Kibble-Turok sphere, we will parametrize the solution in terms of the derivatives of the left- and right-movers

$$\vec{a}'(u) = -\vec{V} + \sum_{n=1}^{\infty} \vec{A}_n \cos(\frac{2\pi nu}{l}) + \sum_{n=1}^{\infty} \vec{B}_n \sin(\frac{2\pi nu}{l}), \qquad (5.3.38)$$

$$\vec{b}'(v) = \vec{V} + \sum_{n=1}^{\infty} \vec{C}_n \cos(\frac{2\pi nv}{l}) + \sum_{n=1}^{\infty} \vec{D}_n \sin(\frac{2\pi nv}{l}), \qquad (5.3.39)$$

where $\vec{V}$ is the bulk velocity of the cosmic string loop. In this way, the string loop trajectory will be of the form

$$\begin{aligned}
\vec{X}(t,\sigma) = \vec{X}_0 + \vec{V}t + \sum_{n=1}^{\infty} &\left[ \vec{a_n^s} \sin\left(\frac{2\pi nu}{l}\right) + \vec{a_n^c} \cos\left(\frac{2\pi nu}{l}\right) \right] + \\
\sum_{n=1}^{\infty} &\left[ \vec{b_n^s} \sin\left(\frac{2\pi nv}{l}\right) + \vec{b_n^c} \cos\left(\frac{2\pi nv}{l}\right) \right]
\end{aligned} \qquad (5.3.40)$$

and

$$\vec{a}(u) = \vec{X}_0 - \vec{V}u + \sum_{n=1}^{\infty} \left[ \frac{l}{2\pi n} \vec{A}_n \sin(\frac{2\pi nu}{l}) - \frac{l}{2\pi n} \vec{B}_n \cos(\frac{2\pi nu}{l}) \right], \qquad (5.3.41)$$

$$\vec{b}(v) = \vec{X}_0 + \vec{V}v + \sum_{n=1}^{\infty} \left[ \frac{l}{2\pi n} \vec{C}_n \cos(\frac{2\pi nv}{l}) - [\frac{l}{2\pi n} \vec{D}_n \sin(\frac{2\pi nv}{l}) \right], \qquad (5.3.42)$$

where

$$\vec{a_n^s} = \frac{l}{2\pi n} \vec{A}_n, \ \vec{a_n^c} = -\frac{l}{2\pi n} \vec{B}_n \qquad (5.3.43)$$

and

$$\vec{b_n^s} = \frac{l}{2\pi n} \vec{C}_n, \ \vec{b_n^c} = -\frac{l}{2\pi n} \vec{D}_n. \qquad (5.3.44)$$

### 5.3.3 The Kibble-Turok sphere

In the conformal time gauge with $t = \tau$, we have seen that the position vector of a cosmic string loop is described by two wave functions $\vec{a}$ and $\vec{b}$. The derivatives of these functions move on the Kibble-Turok sphere as we can see from equa-tion (5.3.25). We have also found that the functions $\vec{a}'$ and $\vec{b}'$ are periodic and form closed loops on the Kibble-Turok sphere in the centre-of-momentum frame. Therefore, they satisfy

$$\int_0^l \vec{a}' du = \int_0^l \vec{b}' dv = \vec{0}. \qquad (5.3.45)$$

The above condition implies that neither of the curves $\vec{a}'$ and $-\vec{b}'$ can lie completely in a single hemisphere of the Kibble-Turok sphere, and it requires contrived conditions to keep them from crosssing. Therefore, the curves $\vec{a}'$ and $-\vec{b}'$ will in general cross within a single period $l$ [94, 9], given that $\vec{a}'$ and $\vec{b}'$ are smooth. At the crossing points, $\vec{a}'(u_1) = -\vec{b}'(v_1)$, or equivalently $\vec{a}'(u_1) \cdot \vec{b}'(v_1) = -1$, [3] we find that

$$\dot{\vec{X}}(t, \sigma) = \frac{1}{2}\left(\frac{\partial \vec{a}}{\partial t}(u) + \frac{\partial b(v)}{\partial t}\right) = \frac{1}{2}\left(-\vec{a}'(u) + \vec{b}'(v)\right) \Rightarrow |\dot{\vec{X}}| = 1. \qquad (5.3.46)$$

Therefore, the crossing points are points where the string moves at the speed of light, which we defined as cusps (see section 5.1). This cusplike behaviour of the string was first introduced in [94]. The cusps will always appear in pairs, given that the left- and right-movers are smooth, since two closed continuous curves on the sphere will cross an even number of times necessarily. Also, since the motion of the Nambu-Goto string is periodic, cusps will also appear periodically over the motion of the string.

Since the curves $\vec{a}'(u)$ and $\vec{b}'(v)$ move on the sphere it is also convenient to describe them in spherical polar coordinates $(\theta, \phi)$. In Figure 5.3.1 (a), we depict $\vec{a}'(u)$ and $\vec{b}'(v)$ on the Kibble-Turok sphere.[4] We notice that the curves cross at two points, where cusps occur. In Figure 5.3.1 (b), we plot the $(\theta(u), \phi(v))$ coordinates of the curves $\vec{a}'(\theta(u), \phi(u))$ and $\vec{b}'(\theta(v), \phi(v))$. It is possible for cusps to appear momentarily as in Figure 5.3.1 or to be permanent, which occurs for example on planar loops. Permanent cusps propagate on the string with the speed of light with respect to the spacetime, and instead of being generic, as the instantaneous cusps, they require specific string configurations.

---

[3]To see that $\vec{a}' = -\vec{b}'$ and $\vec{a}' \cdot \vec{b}' = -1$ are equivalent, we should note that in Euclidean space $\vec{a}' \cdot \vec{b}' = |\vec{a}'||\vec{b}'|cos(\theta)$ where $\theta$ is the angle between the two vectors. Therefore, for unit vectors we have that $\theta = \pi$ in this case and the vectors are antiparallel $\vec{a}' = -\vec{b}'$. The inverse also holds under the same reasoning.

[4]Note that for this example plot we have used an exact cosmic string solution, which we will discuss in section 5.4.

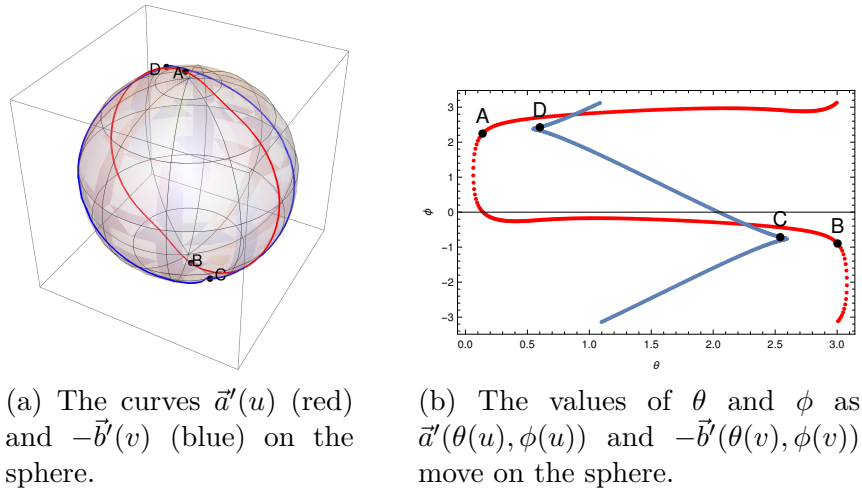

(a) The curves $\vec{a}'(u)$ (red) and $-\vec{b}'(v)$ (blue) on the sphere.

(b) The values of $\theta$ and $\phi$ as $\vec{a}'(\theta(u), \phi(u))$ and $-\vec{b}'(\theta(v), \phi(v))$ move on the sphere.

Figure 5.3.1: Plots of $\vec{a}'(u)$ and $-\vec{b}'(v)$. The points depicted in the plots correspond to $A = (\theta_{\vec{a}'}(u = 0), \phi_{\vec{a}'}(u = 0))$, $B = (\theta_{\vec{a}'}(u = \pi), \phi_{\vec{a}'}(u = \pi))$, $C = (\theta_{\vec{b}'}(v = 0), \phi_{\vec{b}'}(v = 0))$, $D = (\theta_{\vec{b}'}(v = \pi), \phi_{\vec{b}'}(v = \pi))$. By $\theta_{\vec{a}'}$ and $\theta_{\vec{b}'}$ we denote the $\theta$ coordinates of the curves $\vec{a}'$ and $-\vec{b}'$ respectively. Similarly for $\phi$.

Relaxing the condition of continuity, which we assumed upon writing the general solution of the string motion in Fourier Series (equations (5.3.38) and (5.3.39)), we can allow the curves $\vec{a}'$ and $\vec{b}'$ to be discontinuous. A simple solution with discontinuous movers has been constructed by Garfinkle and Vachaspati [53]. Realistically, these discontinuities appear on daughter loops that have formed from an initial loop which self-intersected, and are called kinks. In particular, at the point where the initial loop self-intersects, it chops into two daughter loops and a set of four kinks are produced. Each daughter loop then possesses two of the kinks, which are permanent effects. More on this process will be discussed in section 5.7. A kink corresponds to a discontinuity on either of the left- and right-movers. For a loop with kinks, cusps are no longer a generic feature since they are suppressed by the fact that the curves $\vec{a}'$ and $-\vec{b}'$ are less likely to cross if they support many kinks. Also, loops with kinks don't necessarily have an even number of cusps. However, kinks are only an approach since on a realistic string there will not be actual discontinuities of the string movers, but instead regions on the string where $\vec{a}''$ or $\vec{b}''$ obtain very large values.

### 5.3.4 Cusps in flat spacetime

Let us assume that a cusp occurs at a point $(u^{(c)}, v^{(c)})$ on the string loop. We can define the 4-vector

$$l^\mu = \left(1, \vec{n}^{(c)}\right) = -a^{\mu\prime}(u^{(c)}) = b^{\mu\prime}(v^{(c)}). \tag{5.3.47}$$

Since

$$\dot{X}^\mu(u^{(c)}, v^{(c)}) = \frac{1}{2}\left(-a^{\mu\prime}(u^{(c)}) + b^{\mu\prime}(v^{(c)})\right) = -a^{\mu\prime}(u^{(c)}) = b^{\mu\prime}(v^{(c)}), \tag{5.3.48}$$

the 4-vector $l^\mu$ lies on the direction of the luminal string velocity at the cusp. Since the tangent vectors to $\vec{a}'$ and $\vec{b}'$ are proportional to $\vec{a}''$ and $\vec{b}''$ respectively, the vector $(\vec{b}'' \times \vec{a}'')$ is normal to the surface of the Kibble-Turok sphere. We can place cusps into two categories based on the relative orientation of the left- and the right-movers. If $(\vec{b}''_{(c)} \times \vec{a}''_{(c)}) \cdot \vec{b}'_{(c)} > 0$ the cusp is called a procusp, while if $(\vec{b}''_{(c)} \times \vec{a}''_{(c)}) \cdot \vec{b}'_{(c)} < 0$ the cusp is called an anti-cusp. For a kinkless loop the procusps and anti-cusps alternate on the Kibble-Turok sphere and appear in pairs [95].

Without any loss of generality, we can shift the origin of the worldsheet coordinates such that $(u^{(c)}, v^{(c)}) = (0, 0)$ and the origin of $X^\mu$ such that $X^\mu_{(c)} = 0$. We can apply a Taylor expansion to the left- and right-movers around the cusp

$$a^\mu(u) = -l^\mu u + \frac{1}{2}a^{\mu\prime\prime}_{(c)}u^2 + \frac{1}{6}a^{\mu\prime\prime\prime}_{(c)}u^3 + \cdots, \tag{5.3.49}$$

$$b^\mu(v) = l^\mu v + \frac{1}{2}b^{\mu\prime\prime}_{(c)}v^2 + \frac{1}{6}b^{\mu\prime\prime\prime}_{(c)}v^3 + \cdots, \tag{5.3.50}$$

and also

$$a^{\mu\prime}(u) = -l^\mu + a^{\mu\prime\prime}_{(c)}u + \frac{1}{2}a^{\mu\prime\prime\prime}_{(c)}u^2 + \cdots, \tag{5.3.51}$$

$$b^{\mu\prime}(v) = l^\mu + b^{\mu\prime\prime}_{(c)}v + \frac{1}{2}b^{\mu\prime\prime\prime}_{(c)}v^2 + \cdots, \tag{5.3.52}$$

where the subscript c implies evaluation at the cusp [52, 4]. Also, differentiating the Virasoro conditions (5.3.24) we obtain

$$a^{\mu\prime}a^{\prime\prime}_\mu = 0, \ a^{\mu\prime\prime}a^{\prime\prime}_\mu + a^{\mu\prime}a^{\prime\prime\prime}_\mu = 0, \tag{5.3.53}$$

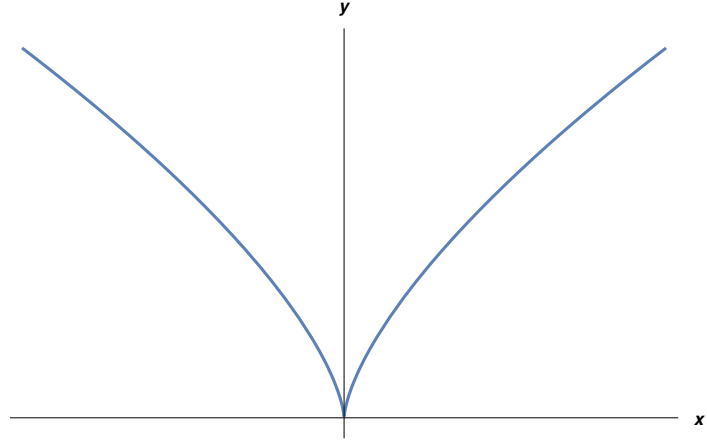

Figure 5.3.2: The shape of the cosmic string around a cusp.

$$b^{\mu'}b^{''}_\mu = 0, \ b^{\mu''}b^{''}_\mu + b^{\mu'}b^{'''}_\mu = 0, \tag{5.3.54}$$

or with respect to $l^\mu$,

$$l^\mu a^{''}_\mu = 0, \ l^\mu a^{'''}_\mu = \left(a^{\mu''}\right)^2, \tag{5.3.55}$$

$$l^\mu b^{''}_\mu = 0, \ l^\mu b^{'''}_\mu = -\left(b^{\mu''}\right)^2. \tag{5.3.56}$$

The shape of the string around the cusp at $t = 0$ is given by [9, 14]

$$X^\mu(0, \sigma) = \frac{1}{4}\left(a^{\mu''}_{(c)} + b^{\mu''}_{(c)}\right)\sigma^2 + \frac{1}{6}\left(a^{\mu'''}_{(c)} + b^{\mu'''}_{(c)}\right)\sigma^3. \tag{5.3.57}$$

Also, the magnitude of the string velocity around the cusp is

$$\dot{\vec{x}}^2 = 1 - \frac{1}{4}\left[(\sigma - t)\vec{a}^{''}_{(c)} + (\sigma + t)\vec{b}^{''}_{(c)}\right]^2. \tag{5.3.58}$$

It follows from equations (5.3.53)-(5.3.54) that the direction of the cusp $\left(a^{\mu''}_{(c)} + b^{\mu''}_{(c)}\right)$ is vertical to the luminal velocity $l^\mu$. Under an apropriate choice of coordinates we find that the string in the region of a cusp has the characteristic shape $y \propto x^{2/3}$, as depicted in Figure 5.3.2 [9].

Contracting equations (5.3.49)-(5.3.50) with $l^\mu$ and using equations (5.3.53)-(5.3.54) we find that around the cusp

$$l_\mu a^\mu(u) = \frac{1}{6}\left(a^{\mu''}_{(c)}\right)^2 u^3. \tag{5.3.59}$$

Similarly,

$$l_\mu b^\mu(v) = -\frac{1}{6}\left(b^{\mu}_{(c)}{}''\right)^2 v^3. \qquad (5.3.60)$$

The above result implies that it was crucial to keep terms of third order in the Taylor expansion of $X^\mu$.

If we would like to find the total number of cusps occurring on a string loop, we need to solve the set of equations

$$\vec{a}'(u) = -\vec{b}'(v) \Leftrightarrow \vec{a}'(u) \cdot \vec{b}'(v) = -1. \qquad (5.3.61)$$

Using the Fourier series (5.3.41)-(5.3.42), we can fix the parameters $\vec{A}_n$, $\vec{B}_n$, $\vec{C}_n$, $\vec{D}_n$, and solve it for any exact solution of the cosmic string loop. The set of equations (5.3.61) is a non-linear system of equations, and although it can be solved analytically for particular string solutions, it requires a numerical approach for more general cases. We will discuss a numerical method in Section 5.5.1.

### 5.3.5  Kinks in flat spacetime

Kinks are defined as points on the worldsheet where the tangent plane to the worldsheet is discontinuous. Unlike the cusp which is in principle a momentary event, a kink on a cosmic string loop moves on the Kibble-Turok sphere on a curve $s = \sigma_{(k)}(\tau)$, which has period $T_l$. At the region around the kink $(\tau, \sigma_{(k)})$, the string can be described by a piecewise function

$$h^\mu = \begin{cases} h_1^\mu(\tau, \sigma) & \text{for } \sigma \leq \sigma_{(k)}(\tau) \\ h_2^\mu(\tau, \sigma) & \text{for } \sigma \geq \sigma_{(k)}(\tau) \end{cases}$$

which satisfies $h_1^\mu(\tau, \sigma_{(k)}) = h_2^\mu(\tau, \sigma_{(k)})$. Note that $h(\tau, \sigma)$ is used to denote the right-mover or the left-mover. The above implies that the requirement for $a^{\mu\,\prime}$ and $b^{\mu\,\prime}$ to be continuous is relaxed when a kink is present on the loop. A simple loop which supports four kinks, two on each of the loop movers, is the loop solution introduced by Garfinkle and Vachaspati in [53]. It can be proved that kinks propagate with the speed of light [9].

Kinks appear on the Kibble-Turok sphere as gaps in the $\vec{a}'(u)$ and $\vec{b}'(v)$ curves of Figure 5.3.1(a). This implies that our argument for the intrinsic presence of cusps on loops (see section 5.3.3) no longer holds when at least a kink is present either on $a^{\mu\,\prime}$ or on $b^{\mu\,\prime}$. Therefore, their presence on the loop is suppressed by the kink presence. It is also evident that cusps no longer appear necessarily in pairs.

It should be mentioned that the nature of kinks is perceived in a different way for the case of a loop that has width different from zero, a case closer to reality than the Nambu-Goto string approach. For a real string, the kink does not correspond to an actual discontinuity but to a very steep change of the values of the left-mover or the right-mover over a very small interval of $\sigma$, i.e. there is a very sudden "jump" in the values of $\Delta a^{\mu\,\prime} = a^{'\mu}(u+du) - a^{'\mu}(u)$ and $\Delta b^{\mu\,\prime} = b^{'\mu}(v+dv) - b^{'\mu}(v)$ at the region of the kink.

Note that kinks appear when a string segment crosses another string segment, where the segments can belong to the same string or two different strings, open or closed. The string dynamics of two crossing segments has been approached using numerical methods, since the Nambu-Goto approach cannot simulate the result of their interaction. It was found in [96, 97, 98, 99], that at the crossing the segments intercommute, i.e. they exchange parts, as can be seen in Figure 5.3.3. At the crossing, two kinks are created on each of the newly formed string segments, one moving on the left-mover and one moving on the right-mover. Note that in the case of cosmic super-strings there can be other outcomes of the string crossing, which are the case that the segments do not interact and the case of entanglement [9].

### 5.3.6 Self-intersections of cosmic string loops

Cosmic string loops can self-intersect whenever two string segments that belong to the loop cross. The crossing of these two segments has as a result their inter-commutation and exchange of parts, as we described in section 5.3.5. After the self-intersection, the original loop, called the parent loop, breaks into two daughter loops. Each of the daughter loops will support two kinks one moving one the left-mover and one on the right-mover. Note that in flat spacetime if the invariant length of the parent loop is $l_0$ and the invariant length of the two daughter loops is $l_1$ and $l_2$, then the inequality $l_0 > l_1 + l_2$ necessarily holds. The difference of the parent loop length and the sum of the daughter loop lengths contributes to the kinetic energy of the daughter loops after the loop fragmentation [100].

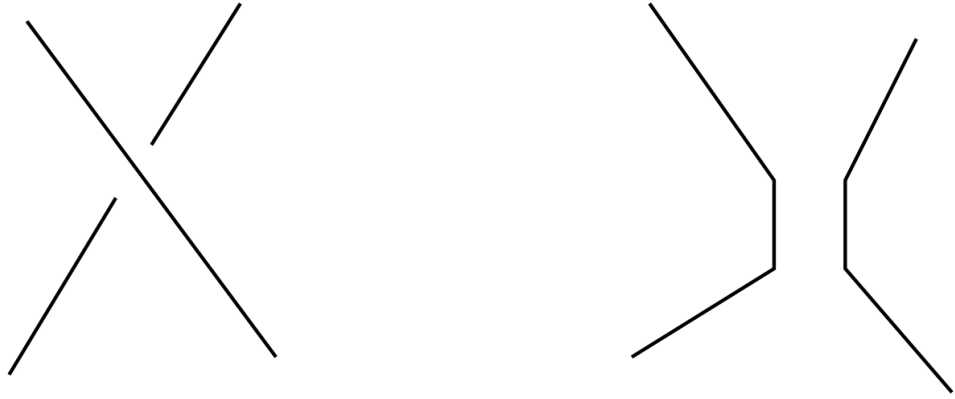

Figure 5.3.3: Crossing of string segments. The figure on the left depicts the segments before the crossing and the figure on the right depicts the segments after they have intercommuted. Adopted from [1].

We can find the number of times in a period $T_l = l/2$ the equation

$$\vec{x}(\sigma, t) = \vec{x}(\sigma', t) \qquad (5.3.62)$$

is satisfied. In the above, $\sigma$ and $\sigma'$ both belong in the interval $[0, l)$. The above can be written in terms of the Fourier Series introduced in (5.3.41)-(5.3.42). This yields three non-linear equations in terms of $\sigma$, $\sigma'$ and t, given that we have fixed all the vectors $\vec{A}_n$, $\vec{B}_n$, $\vec{C}_n$, $\vec{D}_n$, i.e. that we are solving for a specific loop. The system can be solved analytically for particular loops, however it requires a numerical approach for more general loop solutions, as we will see in Section 5.4.6. Note that a cosmic string loop will chop into two loops at the point of self-intersection, i.e. at a solution of the above equation. However, finding the total number of solutions of equation (5.3.62) can be useful for improving the toy model of section 5.7, as we discuss in section 5.7.9. Also, in the case of cosmic superstrings, the chopping of the loop at the point of self-intersection is not certain. Therefore, the total number of solutions of equation (5.3.62) can be used to determine the chance of self-intersection.

One could try to describe the loop fragmentation by using full binary trees, that would describe the splitting of the loop into two parts. We have applied this method and we discuss it in section 5.7. Note that a similar method has been used in [101], while in [102] the problem was tackled for a particular string family.

Note that in the case of cosmic superstrings there are more potential outcomes after an intersection of string segments compared to cosmic strings, as we can see in Figure 1 of [103].

## 5.4 Exact loop solutions in flat spacetime

Although it is very simple to solve the differential equation (5.3.19), which describes a cosmic string in the conformal time gauge in flat spacetime, finding the solutions that satisfy the Virasoro conditions (5.3.25) is a non-trivial problem.

In particular, the vectors $\vec{V}$, $\vec{A}_n$, $\vec{B}_n$ are constrained from the Virasoro condition $\vec{a}'^2 = 1$, which gives $4N+1$ separate conditions on these $2N+1$ vectors, for Fourier modes from 1 to $N$. Similarly for $\vec{b}'$. We can aim to solve these conditions for specific low harmonic loops where simple smooth solutions can be found. We will also discuss a general approach to tackle this constraint problem, the spinor representation, as well as the family of odd-harmonic strings.

Recall that any term of equations (5.3.38) and (5.3.39) satisfies the equation of motion (5.3.19). Therefore, we can seek for specific solutions that have different combinations of harmonics. By the term harmonic $N$, we mean the Fourier Series terms of $\cos{(2\pi N x/l)}$ and/or $\sin{(2\pi N x/l)}$ (see also section 2.3). We will call the harmonics of a string mover $N_1 N_2 N_3...$ if the mover contains harmonics of these orders. An $N_1 N_2 N_3.../M_1 M_2 M_3...$ harmonics string will have a right-mover $\vec{a}'$ with harmonics $N_1 N_2 N_3...$ and a left-mover with harmonics $M_1 M_2 M_3...$. Note that there is no restriction between the harmonics of the left-mover and of the right-mover. We will also identify a string mover by its harmonic order, i.e. by its highest harmonic. Therefore, the $N/M$ string refers to a string with the right-mover having $N$ as its highest harmonic, and the left-mover having $M$ as its highest harmonic.

If we describe the loop in its centre-of-momentum frame, then we can drop the terms $\vec{X}_0 + \vec{V}t$ in the string loop trajectory (5.3.40). This implies that the term $\vec{X}_0 + \vec{V}u$ in $\vec{a}(u)$ and the term $\vec{X}_0 + \vec{V}v$ in $\vec{b}(v)$ are also dropped in the centre-of-momentum frame. We will call these terms 0-order terms.

Exact solutions of the string equations of motion can be found in [14, 104, 105, 53, 106]. In this section, we will present the single harmonic string, the Burden string, the Kibble-Turok string, and the odd-harmonic string.

## 5.4.1   The 01-harmonics loop

Assume a cosmic string with terms of zero order and first order harmonics. In the calculations we will absorb the $2\pi/l$ term in the $u$ and $v$ coordinates for ease of notation, and restore it back at the end. Then, the 01-harmonics string is

$$\vec{b}'(v) = \vec{V} + \vec{C}_1 \cos\left(v\right) + \vec{D}_1 \sin\left(v\right).$$  (5.4.1)

The gauge condition $|\vec{b}'|^2 = 1$ imposes that

$$\left(\vec{V} + \vec{C}_1 \sin\left(v\right) + \vec{D}_1 \cos\left(v\right)\right) \cdot \left(\vec{V} + \vec{C}_1 \sin\left(v\right) + \vec{D}_1 \cos\left(v\right)\right) = 1 \Leftrightarrow$$
$$V^2 + C_1^2 \cos^2\left(v\right) + D_1^2 sin^2\left(v\right) + 2\vec{V} \cdot \vec{C}_1 \cos\left(v\right) + 2\vec{V} \cdot \vec{D}_1 \sin\left(v\right) +$$  (5.4.2)
$$2\vec{C}_1 \vec{D}_1 \cos\left(v\right) sin\left(v\right) = 1$$

Since we want the above expression to hold at all times, the following conditions should hold

$$\vec{V} \cdot \vec{C}_1 = 0, \ \vec{V} \cdot \vec{D}_1 = 0, \ \vec{C}_1 \cdot \vec{D}_1 = 0, \ \vec{C}_1^2 = \vec{D}_1^2, \ \vec{V}^2 + \vec{C}_1^2 = 1.$$  (5.4.3)

These conditions imply that the vectors $\vec{V}$, $\vec{C}_1$, $\vec{D}_1$ are vertical to one another. Therefore, if $\vec{C}_1$ is parallel to the unit vector $\vec{n}_1$ and $\vec{D}_1$ is parallel to the unit vector $\vec{n}_2$, then $\vec{n}_1 \cdot \vec{n}_2 = 0$ must hold and $\vec{V}$ is forced to lie along the direction of $\vec{n}_1 \times \vec{n}_2$. Also, $|\vec{C}_1| = \sin\phi$ and $|\vec{V}| = \cos\phi$, so that $\vec{V}^2 + \vec{C}_1^2 = 1$ is satisfied for any value of the angle $\phi$. Note that we can choose the magnitude of $\vec{C_1}$ to be $\sin \phi$ and of $\vec{V}$ to be $\cos \phi$ since the parameter $\phi$ is free to take any value from 0 to $2\pi$. If we move to the centre-of-momentum frame of the loop then $\cos \phi = 0$, so that the zero order term is dropped. We can also align the string on the $(x, y, z)$ plane by choosing $\vec{n_1} = \vec{i}$ and $\vec{n_2} = \vec{j}$, where $\vec{i}$ and $\vec{j}$ are the unit vectors along the $x$ axis and $y$ axis respectively. Then $\vec{n_1} \times \vec{n_2} = \vec{k}$, where $\vec{k}$ is the unit vector along the z axis.

Our gauge fixing has left one freedom, which allows us to reparametrize the coordinate $\sigma$. Therefore, we can in general introduce a phase parameter $\theta$ in our arguments. The solution aligned to the $(x, y, z)$ coordinates is written as

$$\vec{b}'(v) = \cos\left(\frac{2\pi}{l}(u - \theta)\right) \sin\phi\, \vec{i} + \sin\left(\frac{2\pi}{l}(u - \theta)\right) \sin\phi\, \vec{j} + \cos\phi\, \vec{k}. \quad (5.4.4)$$

We can follow the same procedure to obtain the solution $\vec{a}'(u)$. We reach the result

$$\vec{a}'(u) = \cos\left(\frac{2\pi}{l}(u - \theta)\right) \sin\phi\, \vec{e_1} + \sin\left(\frac{2\pi}{l}(u - \theta)\right) \sin\phi\, \vec{e_2} - \cos\phi\, \vec{k}. \quad (5.4.5)$$

Note that we did not align $\vec{a}'$ to the $(x, y, z)$ plane in order to maintain generality, apart from keeping the condition that the zero order terms of $\vec{a}'$ and $\vec{b}'$ are antiparallel. The vectors $(\vec{e_1}, \vec{e_2})$ are orthonormal unit vectors.

## 5.4.2 Loop with one harmonic

We can write the solutions (5.4.4) and (5.4.5) for an observer in the centre-of-momentum frame. Dropping the 0-order terms in (5.4.5) and (5.4.4), we obtain the first order harmonic string

$$\vec{a}'(u) = \cos\left(\frac{2\pi}{l}(u - \theta)\right) \vec{e_1} + \sin\left(\frac{2\pi}{l}(u - \theta)\right) \vec{e_2} \quad (5.4.6)$$

and

$$\vec{b}'(v) = \cos\left(\frac{2\pi}{l}(u - \theta)\right) \vec{i} + \sin\left(\frac{2\pi}{l}(u - \theta)\right) \vec{j}. \quad (5.4.7)$$

If the movers lie in the same plane, i.e. $\vec{a}'$ in (5.4.6) has no z-component, the loop will either have 2 permanent cusps or collapse in a cusp momentarily, depending on the relative orientation of $\vec{a}'$ and $\vec{b}'$ in the x-y plane. If the movers are non-planar, then we can write equation (5.4.6), without loss of generality, as

$$\begin{aligned}
\vec{a}'(u) = &\cos\left(\frac{2\pi}{l}(u-\theta)\right)\vec{i} + \cos\phi\sin\left(\frac{2\pi}{l}(u-\theta)\right)\vec{j} + \\
&+ \sin\phi\sin\left(\frac{2\pi n}{l}(u-\theta)\right)\vec{k}.
\end{aligned} \tag{5.4.8}$$

For the rest of this section we will fix $\theta = 0$, and also for ease of notation we can reparametrize $2\pi u/l \to u$ and $2\pi v/l \to v$, such that $u$ and $v$ will range from 0 to $2\pi$. We can restore the units later.[5] Therefore, we can write the single harmonic non-planar loop as

$$\vec{a}'(u) = \cos(u)\vec{i} + \cos\phi\sin(u)\vec{j} + \sin\phi\sin(u)\vec{k} \tag{5.4.9}$$

and

$$\vec{b}'(v) = \cos(v)\vec{i} + \sin(v)\vec{j}, \tag{5.4.10}$$

with $\phi \in (0, 2\pi)$. This loop supports two simultaneous cusps at $\tau = \pi/2$ (see Appendix A.1). It is easy to find the cusps analytically in this case. Recalling that the cusp condition is $\vec{a}' = -\vec{b}'$, we reach the conditions,

$$\cos(u) = -\cos(v), \tag{5.4.11}$$

$$\cos\phi\sin(u) = -\sin(v), \tag{5.4.12}$$

$$\sin\phi\sin(u) = 0. \tag{5.4.13}$$

The third condition is satisfied for $u = 0$ or $u = \pi$. The second for $v = 0$ or $v = \pi$. From the first condition, we conclude that there are two solutions, $(u = 0, v = \pi)$ and $(u = \pi, v = 0)$. In terms of the $\tau - \sigma$ coordinates, the cusps appear at $(\pi/2, \pi/2)$ and $(\pi/2, 3\pi/2)$.

We can generalize the above procedure and derive the single harmonic string, where both $\vec{a}'$ and $\vec{b}'$ have the $N^{th}$-harmonic only. To achieve this, let's choose from equations (5.3.38) and (5.3.39) the $N^{th}$ term of the Fourier Series. We are allowed to do this, since each term of the Fourier Series is a solution of the Nambu-Goto equation of motion of the string. Our function has the same functional form as for the first-harmonic loop. Therefore, the general form of the $N^{th}$ harmonic string in terms of its left- and right-movers is

$$\vec{a}'(u) = \cos{(Nu)}\ \vec{e_1} + \sin{(Nu)}\ \vec{e_2}, \tag{5.4.14}$$

$$\vec{b}'(v) = \cos{(Nv)}\ \vec{i} + \sin{(Nv)}\ \vec{j}. \tag{5.4.15}$$

---

[5]Note that this reparametrization is equivalent to defining the dimensionless variables $u' = 2\pi u/l$ and $v' = 2\pi v/l$.

This solution describes the most general string which contains only the $N^{th}$-harmonic in its centre-of-momentum frame.

### 5.4.3 The Burden loop

We will now look into the cosmic string loop where $\vec{a}'$ has one harmonic of order $N$ and $\vec{b}'$ has one harmonic of order $M$, with N and M being relatively prime. The loop of this form, with the phase $\theta = 0$,

$$\vec{a}'(u) = \cos{(Nu)}\ sin\phi\ \vec{e_1} + \sin{(Nu)}\ \vec{e_2}, \tag{5.4.16}$$

$$\vec{b}'(v) = \cos{(Mv)}\ sin\phi\ \vec{i} + \sin{(Mv)}\ \vec{j}, \tag{5.4.17}$$

is called the Burden loop, introduced by C. Burden and L. Tassie [107, 51].

Regarding the cusp occurence of the Burden loops, it differs depending on the relative position of $\vec{a}'$ and $\vec{b}'$. In the case of planar $\vec{a}'$ and $\vec{b}'$ (i.e. $\vec{a}'$ has no z-component), the Burden loop supports permanent cusps. If $\vec{a}'$ and $\vec{b}'$ are non-planar, then the loop exhibits a total number of $4|NM|$ cusps in a period [14].

### 5.4.4 The Kibble-Turok string

An example of string solutions where the occurence of cusps can be found analytically is the Kibble-Turok family of strings [50]

$$\vec{a}'(u) = [(1 - \alpha)\cos(u) + \alpha\cos(3u)]\,\vec{i} + \tag{5.4.18}$$

$$+ [(1 - \alpha)\sin(u) + \alpha\sin(3u)]\,\vec{j} + \tag{5.4.19}$$

$$+ 2\sqrt{\alpha(1 - \alpha)}\sin(u)\,\vec{k} \tag{5.4.20}$$

and

$$\vec{b}'(v) = \cos(v)\vec{i} + \sin(v)\vec{j}. \tag{5.4.21}$$

In the above, the parameter $\alpha$ belongs in the interval $[0, 1]$. This family of string solutions supports two simultaneous cusps per period $\pi$ at the points $(\tau, \sigma) = (\pi/2, \pi/2)$ and $(\tau, \sigma) = (\pi/2, 3\pi/2)$. We can show this by solving the cusp conditions for the Kibble-Turok string. These are

$$(1 - \alpha)\cos(u) + \alpha\cos(3u) = -\cos(v), \tag{5.4.22}$$

$$(1 - \alpha)\sin(u) + \alpha\sin(3u) = -\sin(v), \tag{5.4.23}$$

$$2\sqrt{\alpha(1 - \alpha)}\sin(u) = 0. \tag{5.4.24}$$

The third condition is satisfied for $u = 0$ or $u = \pi$. The second condition is satisfied for $v = 0$ or $v = \pi$. From the third condition we conclude that the two solutions of the system are $(u = 0, v = \pi)$ and $(u = \pi, v = 0)$.

### 5.4.5 A general loop solution in the conformal time gauge

A method to solve the Virasoro conditions was introduced by R. Brown et al., who used the spinor representation method to describe the loop trajectory in terms of a matrix product [108, 109]. Following the spinor representation description, we find that the derivatives of the cosmic string movers of harmonic order $N$ are of

the functional form

$$\vec{h}'_N(v) = \rho_{N+1}R_z(u)\dots\rho_3R_z(u)\,\rho_2R_z(u)\,\rho_1\vec{k} \tag{5.4.25}$$

where

$$\rho_i = \rho(\theta_i,\,\phi_i) = R_z(-\theta_i)\,R_x(\phi_i)\,R_z(\theta_i)\,, \tag{5.4.26}$$

such that

$$0 \le \theta_i \le \pi,\, 0 \le \phi_i \le 2\pi, \tag{5.4.27}$$

and

$$R_z(\omega) = \begin{pmatrix} \cos\omega & -\sin\omega & 0 \\ \sin\omega & \cos\omega & 0 \\ 0 & 0 & 1 \end{pmatrix}, \tag{5.4.28}$$

$$R_x(\omega) = \begin{pmatrix} 1 & 0 & 0 \\ 0 & \cos\omega & -\sin\omega \\ 0 & \sin\omega & \cos\omega \end{pmatrix}. \tag{5.4.29}$$

Note that we have assumed a 3 dimensional Cartesian coordinate system $(x, y, z)$, with unit vectors $(\vec{i}, \vec{j}, \vec{k})$. We have introduced a matrix notation, where $R_x(\phi)$ denotes a rotation of angle $\phi$ around the x axis. In expression (5.4.25), we chose the $\vec{k}$ unit vector, without loss of generality; any other unit vector can be chosen and apply the same rule or equivalently we can change the coordinate system alignment.

## 5.4.6 The odd-harmonic string

The spinor representation method provides a general expression for the cosmic string solution up to any order of harmonics as a product of matrices. However, if we wish to describe the cosmic string in its centre-of-mass frame, the cosmic string solution must also satisfy the condition (5.3.45), which implies that $\vec{a}'(u)$ and $\vec{b}'(v)$ both have period $2\pi$. Although a general expression of any order N has not been found analytically that solves (5.3.45), we can still take a set of solutions that satisfy the above condition. This is the odd-harmonic string, introduced by X. Siemens and T. Kibble in [2], which allows only for odd-harmonics in the cosmic string solution. The loss of generality is the price we have to pay in order to move to the centre-of-mass frame of the cosmic string. The function that describes the derivatives of the left- and right-movers of the N-harmonic odd-harmonic string is

$$h'_N(u) = \rho_{N+2} R_z(2u) \rho_N R_z(2u) \dots R_z(2u) \rho_3 R_z(u) \rho_1 \vec{k}, \qquad (5.4.30)$$

with $\theta_1 = -\pi/2$ and $\phi_1 = \pi/2$ for both $\vec{a}'$ and $\vec{b}'$. A generalization of the above solution can be found in [110], where an explanation of the reasoning behind its construction is explained as well.

Given the aim of our study we can further simplify the above solution, since we are interested only in the relative orientation freedom of $a'(u)$ and $b'(v)$. In particular, we obtain results that can be used for the study of the gravitational wave effects from the cosmic string loops. These are the number of cusps that occur per period on the string, the relative positions of the points where the cusps occur and the magnitude of the second derivatives of the cosmic string loop movers evaluated at the cusps. Therefore, we can eliminate parameters from the orientation freedom of the string by removing the $\theta_{N+2}$ parameter from $a'(u)$ and the $\theta_{N+2}$ and $\phi_{N+2}$ parameters from $b'(v)$. Leaving the parameter $\phi_{N+2}$ in $a(u)$ ensures the relative freedom of the string movers. The solutions for the derivatives of the right- and left-movers that we obtain are

$$\vec{a}'_N(u) = R_x(\phi_{N+2}) R_z(2u) \rho_N R_z(2u) \dots R_z(2u) \rho_3 R_z(u) \vec{i} \qquad (5.4.31)$$

and

$$\vec{b}'_M(v) = R_z(2v) \rho_M R_z(2v) \dots R_z(2v) \rho_3 R_z(v) \vec{i}, \qquad (5.4.32)$$

respectively. We will call the string appearing in equations (5.4.31) and (5.4.32), the N/M odd harmonic string, where N refers to the order of harmonic we have chosen for $a'(u)$ and M to the order of harmonic we have chosen for $b'(v)$. In our study we will use this parametrization of the cosmic strings to obtain results about their behaviour. The angles that appear in the matrix $\rho_i(\theta_i, \phi_i)$ will be denoted as $\theta_{ia}$ and $\phi_{ib}$ for the $\rho_i$ matrices appearing in $\vec{a}'(u)$ and $\vec{b}'(v)$ respectively. By moving to the centre-of-mass frame, we have induced an extra symmetry to the cosmic string loop, which now satisfies $\vec{X}(\tau, \sigma + \pi) = -\vec{X}(\tau, \sigma)$.

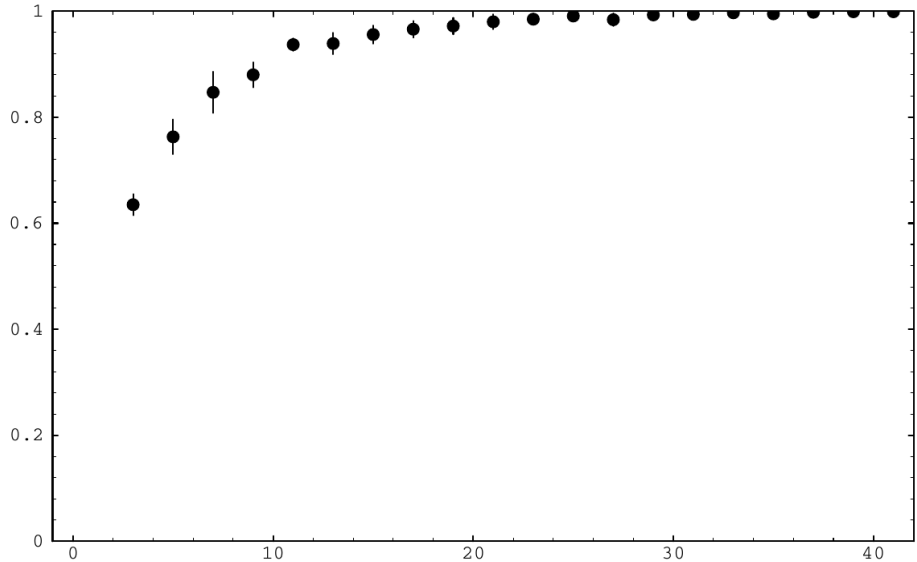

Figure 5.4.1: The probability of self-intersection vs the string harmonic order. Adopted from [2].

### 5.4.6.1 Self-intersections of the odd-harmonic string

In [2], the self-intersection of the odd-harmonic for high harmonics was also studied. In particular, the condition for the self-intersection of the cosmic string loops, presented in equation (5.3.62), was solved numerically. To achieve this, they split the string in timesteps $t_i$, with t ranging from 0 to $\pi$, which is the string effective period. At each timestep, they checked for the presence of solutions of (5.3.62) around the points $(u_j, v_j) = (\sigma_j - t_i, \sigma_j + t_i)$ and $(u_k, v_k) = (\sigma_k + t_i, \sigma_k + t_i)$, with $u$ and $v$ ranging from 0 to $2\pi$. To avoid trivial solutions, i.e. $\sigma = \sigma'$, a cutoff value $\delta\sigma$ was used, such that when an intersection occurred for $|\sigma_j - \sigma_k| < \delta\sigma$, it was ignored. Under these assumptions, the probability of a self-intersection occurring was calculated for $N/N$ odd-harmonic strings up to harmonic order 25. The values of the probabilities are depicted in Figure 5.4.1, with $\delta\sigma = 0.126$ radians. Note that the probability of self-intersections was calculated only for strings that had the same number of harmonics in each mover. We will use the probabilities found with this method in section 5.7 to develop a toy model for the chopping of a cosmic string loop.

### 5.4.6.2 First and third order odd-harmonic string cases

The odd harmonic string for harmonic order N=1 can be evaluated from equations (5.4.31)-(5.4.32), and we can find that it coincides with the single harmonic non-planar loop presented in equations (5.4.9)-(5.4.10).

For the case of the third order N=3, if we set

$$\theta_{3a} = 0,\ \phi_{3a} = 0,\ \theta_{3b} = 0,\ \phi_{3b} = \pi, \tag{5.4.33}$$

we find the following single parameter expression for the odd-harmonic loop

$$\vec{a}'(u) = \left[\sin^2\left(\frac{\phi_{3a}}{2}\right)\cos(u) + \cos^2\left(\frac{\phi_{3a}}{2}\right)\cos(3u)\right]\vec{i}+ \tag{5.4.34}$$

$$+ \left[\sin^2\left(\frac{\phi_{3a}}{2}\right)\sin(u) + \cos^2\left(\frac{\phi_{3a}}{2}\right)\sin(3u)\right]\vec{j}+ \tag{5.4.35}$$

$$+ \sin(\phi_{3a})\sin(u)\vec{k} \tag{5.4.36}$$

and

$$\vec{b}'(v) = \cos(v)\vec{i} + \sin(v)\vec{j}. \tag{5.4.37}$$

We notice that for

$$\alpha = \cos^2\left(\frac{\phi_{3a}}{2}\right), \tag{5.4.38}$$

the Kibble-Turok string, defined in equations (5.4.20)-(5.4.21), coincides with the above set of equations. Therefore, the family of Kibble-Turok strings can be produced from the above single parameter N=3 odd harmonic family of strings.

In following sections, we will use our results from analytic solutions for the above families of loops, in particular their cusp occurrence, to compare as a check with our numerical findings. Another quantity we will use to test our results is the maximum magnitude of the second derivatives of the string movers, which can also be calculated analytically. It is obvious from equations (5.4.9)-(5.4.10), that the second derivatives of the first harmonic loop have constant unit magnitude. When it comes to the third order odd-harmonic string, we can reproduce it through equations (5.4.31)-(5.4.32), differentiate them and calculate their magnitudes to

be

$$|\vec{a}''|^2(u; \theta_{3a}, \phi_{3a}) = 4 + \cos\left[2\left(u + \theta_{3a}\right)\right] + 4\cos\left(\phi_{3a}\right) + \\ + 2\cos\left(2\phi_{3a}\right)\sin^2\left(u + \theta_{3a}\right),$$ 
(5.4.39)

$$|\vec{b}''|^2(v; \theta_{3b}, \phi_{3b}) = 4 + \cos\left[2\left(v + \theta_{3b}\right)\right] + 4\cos\left(\phi_{3b}\right) + \\ + 2\cos\left(2\phi_{3b}\right)\sin^2\left[v + \theta_{3b}\right].$$ 
(5.4.40)

In equation (5.4.39), we can see that the parameter $\theta_{3a}$ serves as a phase factor, and therefore does not affect the maximum value of (5.4.39), as we can easily infer by making the coordinate transformation $u \to u - \theta_{3a}$. This observation leaves us with one free parameter, $\phi_{3a}$. This only appears as an argument on the cosine function and in positive only terms, which implies that it maximizes the expression (5.4.39) for $\phi_{3a} = 0$. Finally, having set the values of the parameters $\theta_{3a}$ and $\phi_{3a}$ it is easy to conclude that the function obtains its maximum value for $u = \pi/2$, which can also be deduced by plotting the function in the interval $[0, 2\pi)$. We find that the maximum value of the second derivative of the right string mover is max $(|\vec{a}''|) = 3$. Following the same reasoning, we find that max $\left(|\vec{b}''|\right) = 3$. We will compare the above analytical results with our numerical results in section 5.5.2.

## 5.5   Cusp occurence on the odd-harmonic string

The occurrence of cusps on strings not only informs us of the properties of the strings, but is also useful for the study of gravitational waves from cosmic strings. For this purpose, our aim is to study the occurrence of cusps in the odd-harmonic family of strings containing high orders of harmonics. It is clear if we insert the equations producing the odd-harmonic string (5.4.31)-(5.4.32) to the cusp conditions (5.3.61), the system cannot be solved analytically in general. In order to solve it we should use a numerical approach. However, there are particular choices of parameters that lead to a set of equations which can be solved analytically for the presence of cusps, as we discussed in section 5.4.6.2.

Another quantity of relevance for gravitational waves emanating from cusps on cosmic strings is the second derivatives of the right- and left-movers at the points on the worldsheet where cusps occur, $\vec{a}''(u^{(c)})$ and $\vec{b}''(v^{(c)})$ respectively. Once we find numerically the points where cusps occur, $(u^{(c)}, v^{(c)})$, we can then easily calculate the values of the second derivatives, by differentiating equations (5.4.31) and (5.4.32), and calculating them at $(u^{(c)}, v^{(c)})$.

## 5.5.1 Numerical method

We will now describe the structure of the code we used to solve the non-linear system of equations (5.3.61). It consists of two parts, one to obtain the loops and another to solve the non-linear system as follows:

- Part I

    1. We choose the harmonic order $N$, which is the same for the left- and right-movers and the number of loops M we will analyse.

    2. We do an iterative process to produce the $N$-order harmonic loop in $(N-1)/2$ steps, using the definition of the odd-harmonic loop, (5.4.31) and (5.4.32). In each step of the iteration we choose randomly 2 angles for the $a_N(u)$ and 2 angles for $b_N(v)$ and in the last step we randomly choose one extra angle for each of the left- and right-movers, which are needed to eventually build the N-harmonic loop with $2N - 1$ random angles in total. In this way we aim to sample the plane of angles for a large number of different parametrizations.

    3. We append the angles chosen for the string and the left- and right-movers of the string and their derivatives to lists. The process of step 2 is repeated M times, to eventually obtain in lists all the information for the parametrization of the loops we produced.

- Part II

    1. We enter an iterative process where we choose the ith element of the lists we have produced in Part I, where i takes integer values from 1 to M.

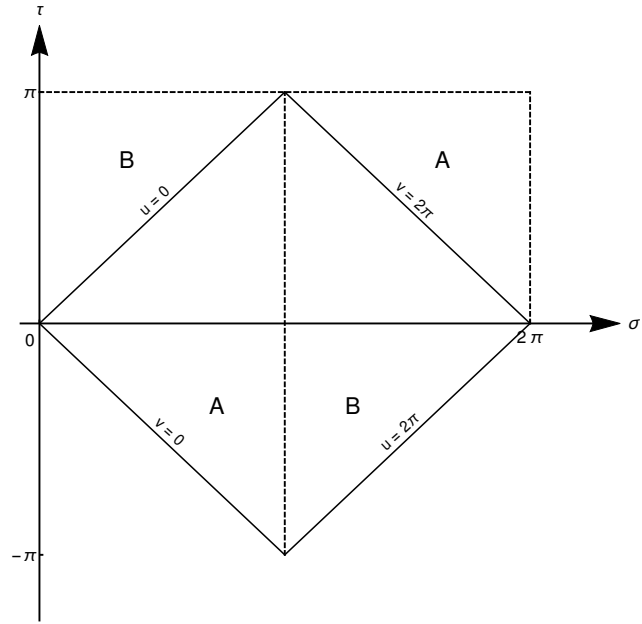

Figure 5.5.1: The domain in terms of the $(\tau, \sigma)$ and $(u, v)$ coordinates.

2. We split the $(u, v)$ plane in an equally sized grid. We use FindRoot, a built-in process in Mathematica to treat non-linear equations, to solve the system. Then we append the cusp solutions $(u^{(c)}, v^{(c)})$ in a list, and we test that they are indeed solutions, i.e. that they satisfy (5.3.47) by using the analytic expressions we saved from Part I, up to $10^{-5}$ accuracy. We decrease the size of the grid. We repeat this process until no new $(u^{(c)}, v^{(c)})$ are found. After we obtain the list with the cusp points, we calculate $|\vec{a}''(u^{(c)}, v^{(c)})|$, $|\vec{b}''(u^{(c)}, v^{(c)})|$.

3. This process is repeated until we analyse all loops from Part I.

Note that the harmonic order of both the left- and the right-mover is assumed to be the same. However, there is no restriction that $\vec{a}(u)$ and $\vec{b}(v)$ should satisfy this condition, which we only assume for simplicity. There are unlimited combinations for the harmonics of the string movers, and it would be an interesting subject to test what happens under different assumptions and how they vary from the one we assumed.

Since the general equation of motion for the string comprises of the periodic functions $\vec{a}(u)$ and $\vec{b}(v)$, each with a period $2\pi$, the domain in u-v space $[0, 2\pi) \times$ $[0, 2\pi)$ contains all the information about the string motion. In Figure 5.5.1, we can see that the $[0, 2\pi) \times [0, 2\pi)$ $(u, v)$ domain can be mapped to the $[0, \pi) \times [0, 2\pi)$ $\tau - \sigma$ domain, as expected. Indeed, the domains A are equivalent to each other, due to the periodicity of the string, $\vec{X}(\tau, \sigma) = \vec{X}(\tau + \pi, \sigma + \pi)$. The same holds for the domains B.

The numerical analysis of the strings can be quite time consuming. To reduce the time required we take advantage of the extra symmetry of the string, $\vec{X}(\tau, \sigma + \pi) = -\vec{X}(\tau, \sigma)$, which is equivalent to $\vec{X}(u, v) = \vec{X}(u + \pi, v + \pi)$. This implies that if a cusp occurs at $(u^{(c)}, v^{(c)})$, it also occurs at $(u^{(c)} + \pi, v^{(c)} + \pi)$, and we only need to look for cusp solutions in the domain $[0, \pi) \times [0, 2\pi)$ of the $u - v$ space, and then map them to the full domain $[0, 2\pi) \times [0, 2\pi)$ to obtain the full space of cusp solutions.

### 5.5.2 Distribution of results in terms of the harmonic order

In order to check on the accuracy of our simulations, we have compared our numerical results with the analytic cases mentioned in sections 5.4.2 and 5.4.4, namely the one harmonic loop and the Kibble-Turok loop, which can both be obtained from the odd-harmonic loop as we have shown in section 5.4.6.2. In both cases we find the same number and positions for the cusps, with an accuracy of $10^{-5}$, and the same values for $|a^{(c)''}|$, $|b^{(c)''}|$ evaluated at the cusp. As a further check to our numerical method, we find that the Turok solution [111], a generalization of the Kibble-Turok string with two free parameters, exhibits two or six cusps per period, with two of them always occuring at $(\tau, \sigma) = (\pi/2, \pi, 2)$ and $(\tau, \sigma) = (\pi/2, 3\pi, 2)$, as expected. The Turok string is not a subfamily of the odd-harmonic string. Therefore, we input the equation of the left- and right-movers of the Turok string (given in [9]) into the Part II of our code (described in section 5.5.1). Note that the Turok string can also exhibit 4 cusps per period for a specific choice of its two free parameters. Since we choose these parameters randomly, it is unlikely that we will come across this case. The cusp structure of the Turok string with respect to the two-parameter space is provided in [14]. Finally, for the third harmonic order odd harmonic string, we find that the maximum values of $|\vec{a}''|$ and $|\vec{b}''|$ are indeed 3, as we found in section 5.4.6.2.

Having confirmed the consistency of our approach with known analytic solutions, we can confidently go on to look at the case for more general odd-harmonic strings, going up to harmonic order 21. The first question we wish to address is what is the average value of the second derivative of the left- and right-movers? Recall this is an important contributor to the gravitational wave power emitted by cusps, and the assumption being made when calculating the associated gravitational wave power emerging from the cusp region is that in units of $2\pi/l$, the average number is of order 1 [53, 4, 3]. If it was substantially smaller, or if there are a significant number of cusps on a loop producing such small values, this would increase the associated power. Our key results are presented in Figures 5.5.2 and 5.5.3. In Figure 5.5.2 we have calculated the mean value for the second derivative of the left- and right-movers on the string defined through $g_1$

$$g_1 = \left( |\vec{a}''(u^{(c)})||\vec{b}''(v^{(c)})| \right)^{-\frac{1}{3}}, \tag{5.5.1}$$

for many loops as a function of the harmonic order ranging from N=1 to 21. The error bars indicate the usual standard error associated with the mean, defined in equation 2.5.13. They are vanishingly small for low harmonics and increase for larger because we are not able to analyse as many large harmonic loops because of time constraints associated with analysing such complex loops. Of particular note is the fact that $\langle g_1 \rangle$ decreases rapidly initially as the harmonic order increases but eventually plateaus for large $N$. We also note that it is indeed consistent with the claims in [4], namely that it is a number of order unity, given that it ranges from 1 for small $N$ down to 0.4 for large $N$.

What about the range of possible values of $g_1$ for a given harmonic? In particular how large can it go, and how frequent are these large values? We address this question in Figure 5.5.3, where each of the four plots represents the frequency distribution of the parameter $g_1$ produced from 1000 loops of a specific harmonic order N. On the horizontal axis we have the values of $g_1$ and on the vertical axis the number of times the parameter $g_1$ obtains a value which lies in the corresponding bin. Note that the first plot, depicting the frequency distribution of $g_1$ for N=3 harmonic loops, decreases almost monotonically from its initial high value in the $[0.45, 0.5)$ bin, except for a secondary subsidiary peak in the $[0.95, 1)$ bin. As we increase the number of harmonics $N$ on the loop, the histograms becomes more peaked around small values of $g_1 \sim 0.4$, a feature that is particularly noticeable for the cases $N = 11$ and $N = 19$. We also notice that the total number of counts increases as we increase the harmonic order of the loops N. This occurs since the average number of cusps per period increases as we increase the harmonic order of the loops. Given that each value of $g_1$ corresponds to a cusp, the total number of $g_1$ counts obtained over a large sample of loops, will therefore increase with N. We can understand the typical smallest value $g_1$ takes. For the case of the $N = 3$ harmonic order loop, we can show analytically (from Eqns. (5.4.31) and (5.4.32)) that it is 0.5, matching the numerical result. Unfortunately analytic approaches break down for higher orders, as the function $g_1(u, v)$ becomes complicated, hence more difficult to find its maximum analytically. It is clear from the four cases depicted in Figure 5.5.3 that the number of loops with large values of $g_1 \gg 1$ become negligible, hence increasing the number of harmonics does not apparently have a significant impact in the range of possible values of $g_1$, they remain close to the assumed value of order unity.

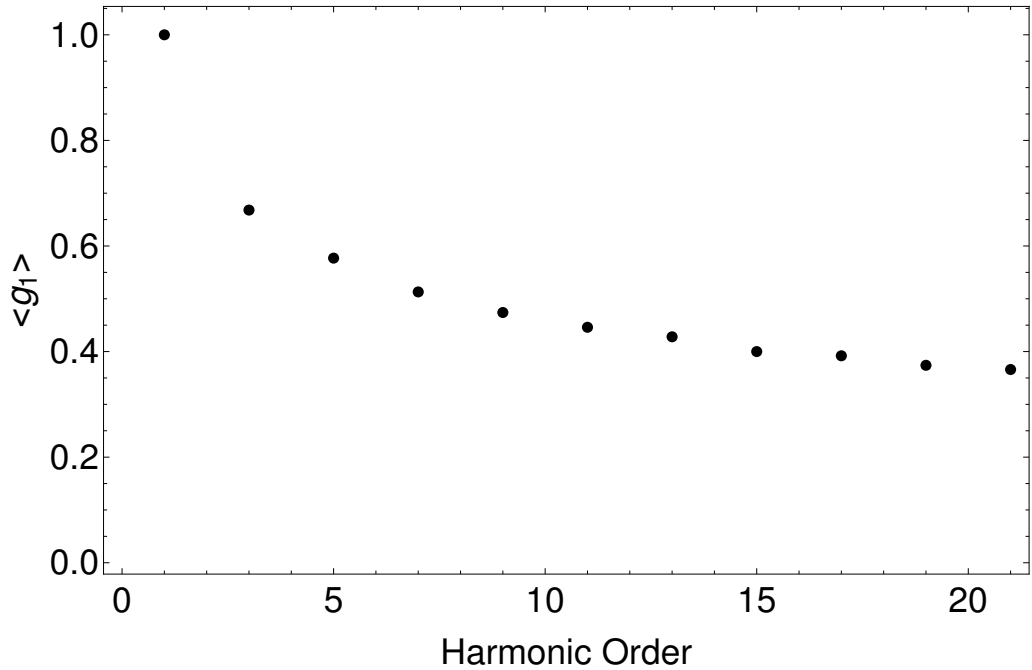

Figure 5.5.2: Our numerical results for the mean value of $g_1$, with harmonic order from 1 to 21. Note that the harmonic order takes only odd values.

In Figure 5.5.4 we plot the mean value of yet another quantity that is related to the

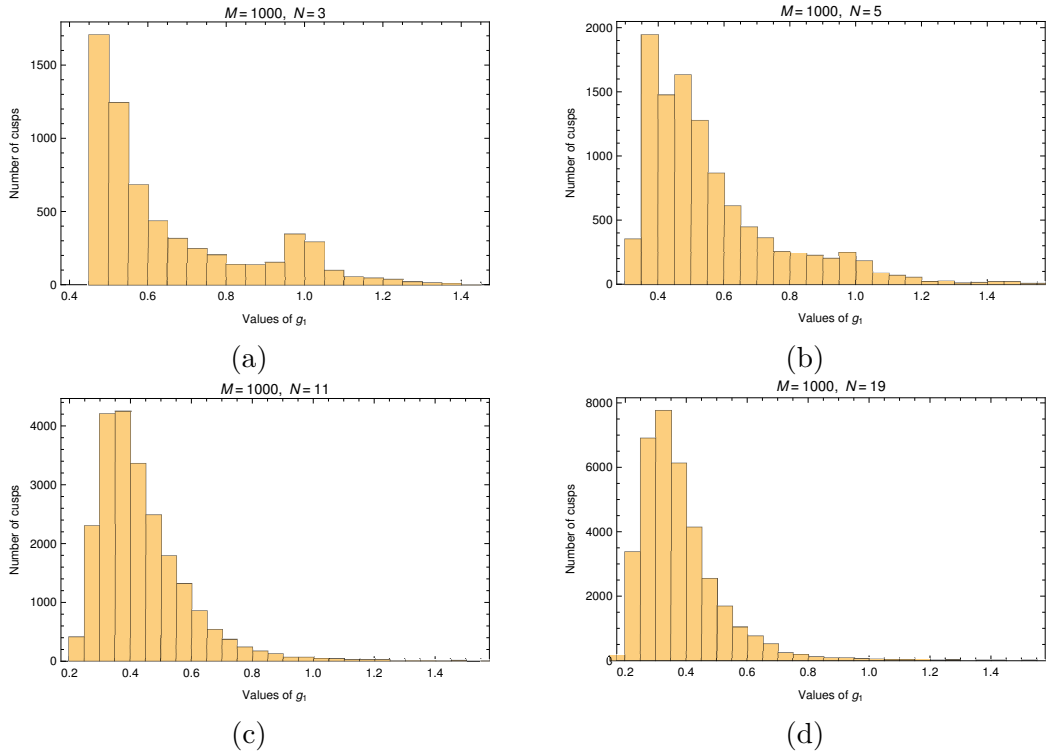

Figure 5.5.3: The frequency distribution of the parameter $g_1$ calculated for different harmonic orders. Note that the data originate from M=1000 cosmic string loops for each plot. We can see that the total number of cusps increases with the harmonic order, which is expected since the average number of cusps per period for each loop also increases with the harmonic order. Note that we have cut some rare higher values of $g_1$ from the histograms. The maximum value of $g_1$ in the data is given in Table 5.5.9.

gravitational wave signal emitted from cusps on cosmic strings. The parameter $g_2$ is defined as

$$g_2 = \left( \min \left( |\vec{a}''(u^{(c)})|, |\vec{b}''(v^{(c)})| \right) \right)^{-1}, \qquad (5.5.2)$$

and it is inversely proportional to the beaming angle of the cusp $\theta^{div}$. In particular, we define the angle $\theta^{div}$ to be the angle that divides the observation angles of a cusp into two sets, one where the gravitational wave signal is roughly the same as it is along the direction of the cusp emission ($\theta < \theta^{div}$), and one where the signal is smoothed ($\theta > \theta^{div}$), which corresponds to the gravitational wave signal away from the cusp [4]. The observer receives the gravitational wave which has emanated from the cusp on the cosmic string if and only if the observation angle with respect to the direction of the cusp satisfies $\theta < \theta^{div}$. As $g_2$ decreases, the angle $\theta^{div}$ increases, which implies that the cusp signal can be received from a broader range of observation angles, and leads to an enhanced overall gravitational wave signal from cosmic strings on Earth. We will derive this angle in greater detail in section 6.2.3. From Figure 5.5.4, we notice that the average value of $g_2$ for each harmonic order follows a pattern similar to the one in Figure 5.5.2, starting from an average value of unity at the first order harmonic string, and gradually decreasing in value until it plateaus at just below 0.4. Once again we note that the values of $g_2$ obtained using the odd-harmonic string do not deviate from the usual assumption that its value is equal to unity [4].

Turning our attention to the number of cusps of harmonic order $N$, $c_N$ appearing per period, we see from Figure 5.5.5 the interesting result that the average value $\langle c_N \rangle$ shows a linear behaviour (for $N$ ranging from 1 to 21), satisfying $\langle c_N \rangle \simeq 2N$. We note that this result differs from that predicted in [1], where they suggested $c_N \propto N^2$. The argument for $N^2$ is based on the fact that each mode roughly corresponds to a great circle on the Kibble-Turok sphere, so the number of intersections (i.e. the number of occasions a cusp forms) is proportional to $N^2$. It isn't obvious to us at the moment why we are differing and is something that is worthy of further investigation. Figure 5.5.6 depicts the frequency distribution of the cusps per period produced from the class of loops represented in Figure 5.5.3. In the histograms depicted in Figure 5.5.6 the bins chosen are the intervals $[0.5, 1.5), [1.5, 2.5), [2.5, 3.5)....$ Figure 5.5.6 allows us to determine the general idea of how the histogram changes with harmonic order. From the symmetries of the odd-harmonic string we conclude that the number of cusps per period, $c_N$,

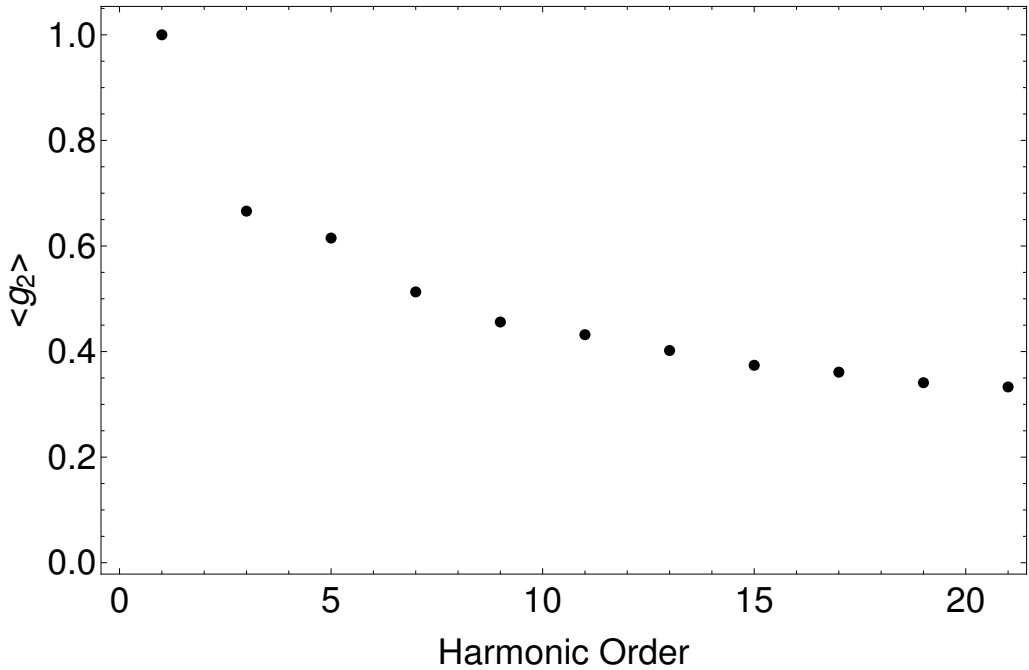

Figure 5.5.4: Our numerical results for the mean value of $g_2$, with harmonic order from 1 to 21. Note that the harmonic order takes only odd values.

has to be an even number. Also, note that in low harmonics it is far more likely to have values of $c_N$ that are not multiplies of 4, as is clear from Figure 5.5.6(a). This can be compared with the Turok solution [94], where $c_N$ obtains the value of 4 only for very specific choices of the string parameters, and otherwise it takes the values 2 or 6 (see [14]).[6] As the harmonic order increases, we notice that it becomes more likely to have values of $c_N$ that are multiples of 4. It is worth noting a few things from Figures 5.5.5 and 5.5.6. First of all note that even for low harmonic loops with $N = 3$ there are on average 6 cusps per period, going up to approximately 40 for the case $N = 19$. For the 1000 $N = 3$ loops shown in Fig. 5.5.6, 650 of them have more than 6 cusps per period, for the N=5 case, that number rises to 850, and for the $N = 19$ case, almost all the loops satisfy that condition. It raises the obvious question, what is the influence of these very cuspy loops when it comes to estimating the gravitational wave beaming from them? At the very least, it suggests that the effective number of cusps per period could well be significantly more than has been assumed to date. Hereafter, we will denote $c_N$ by $c$ as well, to follow the notation of [4], and to denote the cusps

___________

[6]Do not confuse the number of cusps, c, with the indexed notation $c_k^i$, or with the speed of light c, which we have set equal to one from the beginning of this thesis.

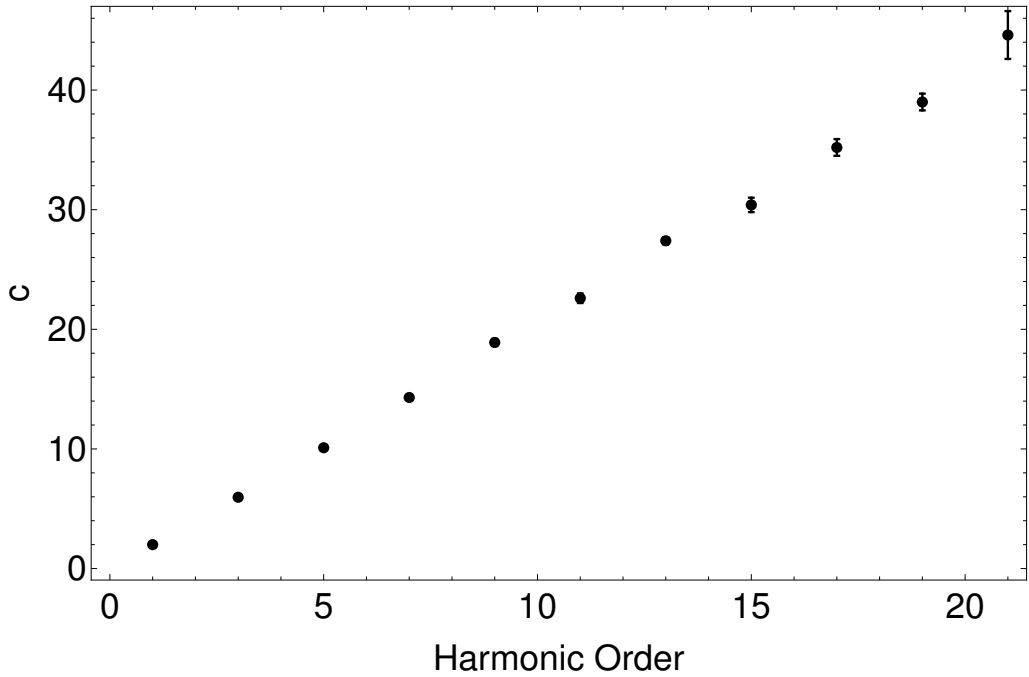

Figure 5.5.5: Our numerical results for the mean value of the cusps per period $c$, with harmonic order from 1 to 21.

per period from the cosmic strings contained in a spacetime volume.

Let us recall that we sample the spherical angles $\theta$ and $\phi$ of the odd-harmonic string, given in equations (5.4.31)-(5.4.32), in a uniform manner by using a random number generator as we discussed in section 5.7.9, to produce a large number of cosmic string loops with different parameters. Another choice for the sampling of the angles could be to sample uniformly $\cos\theta$ and $\phi$. This would ensure isotropy of the points on the sphere. In particular, by sampling the angles $\theta$ and $\phi$ uniformly we would gather more points around the poles of the sphere rather than around the equator of the sphere. However, by sampling $\cos\theta$ and $\phi$, we do not sample the angle $\theta$ uniformly, but we choose more values of $\theta$ around $\pi/2$, and less values around 0 and $\pi$. In the last paragraph of this section, we provide a discussion of the distribution of the cusp solutions on the Kibble-Turok sphere when we use the uniform sampling of the angles, which we plot in Figure 5.5.8. As we discussed, one could also produce this distribution using a uniform distribution of points on the sphere when producing the cosmic string loops and see the effect of this change in the distribution on the location of the cusps on the Kibble-Turok sphere. Also, the effect of this on the average cusps per period could be tested.

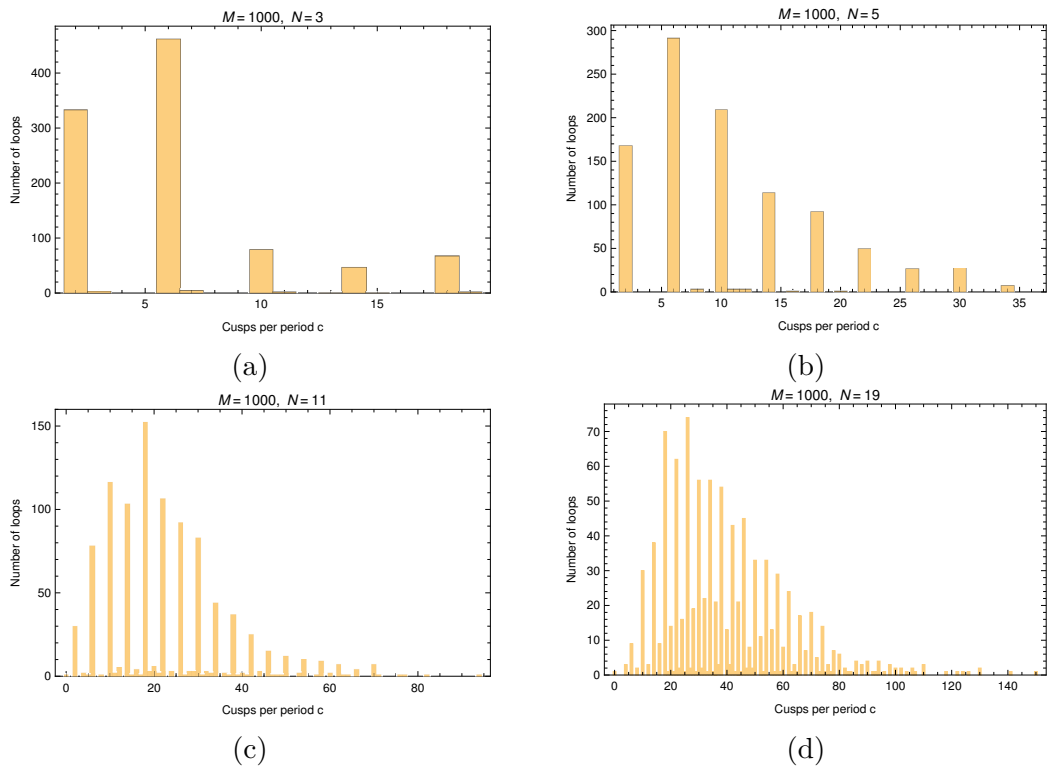

Figure 5.5.6: The frequency distribution of the cusps per period. Each plot is produced from 1000 odd-harmonic string loops of the same harmonic order, N=3, N=5, N=11, N=19.

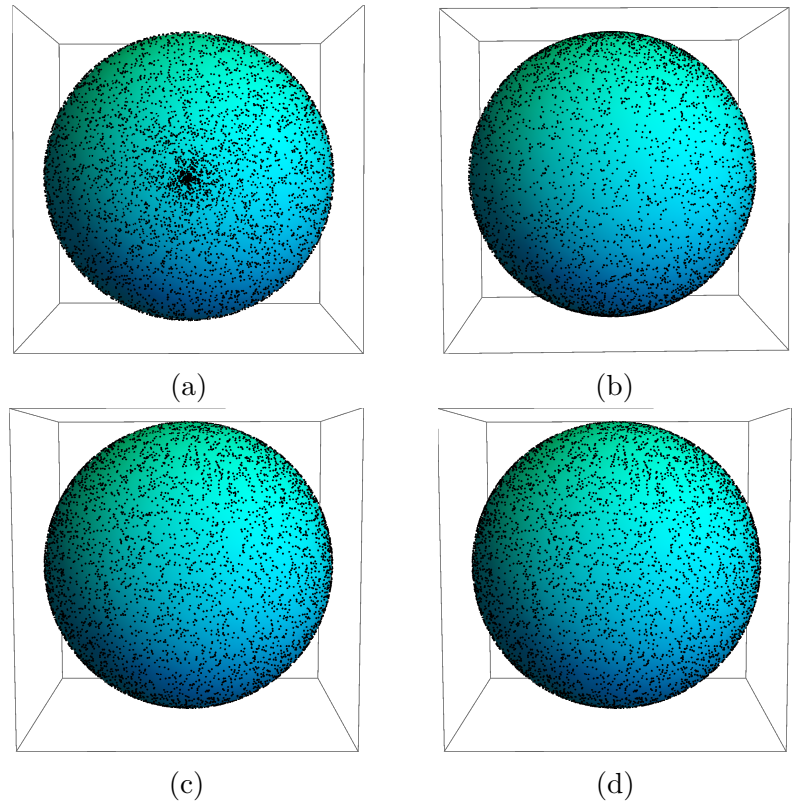

(a)                    (b)

(c)                    (d)

Figure 5.5.7: The distribution of points on the unit sphere depending on the choice of distribution for the angles $\theta$ and $\phi$. If we choose the angles $\theta$ and $\phi$ uniformly, the points on the sphere appear as in Figure (a) around the poles and as in Figure (b) around the equator. If we choose $\cos\theta$ and the angle $\phi$ uniformly, the points on the sphere appear as in Figure (c) around the poles and as in Figure (d) around the equator.

The values of the second derivatives are not expected to be affected by such a change, since their values are obtained locally around the cusp.

Table 5.5.9, provides an elegant summary of the main results we have obtained. In it we show the mean number of cusps per period for different harmonic order, as seen in Fig. (5.5.5). We also show the average values of $g_1$ and $g_2$ (including the maximum value we obtain for $g_1$) for each harmonic order. The take home message is pretty clear, they are fairly closely distributed around unity, as has been assumed in the literature, but at least we can now say it is a justified assumption. In fact if anything they are slightly lower than unity, indicating that the amplitude of gravitational signals will be less strong from these effects than has been assumed.

There is a nice formal mathematical aspect to the distribution of the $(u^{(c)}, v^{(c)})$ pairs on the plane. In Figure 5.5.8 we can see that the pairs of harmonic order N=3 follow a pattern, which does not persist for the higher harmonic orders, as we can see for the N=19 case. We can quantify this using the two-dimensional Kolmogorov-Smirnov test (discussed in section 2.5.2), and doing it we find that for large harmonic orders, up to N=19, the hypothesis that the distribution of the $(u^{(c)}, v^{(c)})$ pairs follows the two-dimensional uniform distribution is not rejected at the 5 percent level, hence is consistent with it appearing to be uniform. However, for harmonic order N=13 and smaller, we find that the above-mentioned hypothesis is rejected at the 5 percent level, indicating there is some underlying structure present. Another way to test the behaviour of the $(u^{(c)}, v^{(c)})$ pairs is to convert the two-dimensional distribution to a one-dimensional one. One way to achieve this is to split the domain $[0, 2\pi) \times [0, 2\pi)$ in u-v space into equal sized squares. The number of the squares is taken to be of the order of the number of $(u^{(c)}, v^{(c)})$ pairs. We can then make a distribution of the number of $(u^{(c)}, v^{(c)})$ pair counts in each square, and compare it with the Poisson distribution of the same mean value. Recall our discussion of the Poisson distribution in section 2.5.1. Using the Kolmogorov-Smirnov test to compare these two distributions we find again that the null hypothesis is rejected for N=13 or smaller at the 5 percent level, while it is true for N=15 to N=19, which is the maximum harmonic order we have tested.

## 5.6 Cosmic string network

In this section, we shall look into the evolution of the cosmic string loop network after the end of the friction-dominated period [9], [112]. Straight cosmic strings after their formation tend to intersect and create loops. Therefore, a cosmic string network of loops forms which evolves with time on cosmological scales and loses energy via gravitational wave emission in the wire approach.[7] The cosmic string network, in the scale of a Hubble volume and at cosmic time $t$, consists of a number of straight strings that stretch across the Hubble volume and a significantly larger number of closed loops. Regarding the evolution of the string network, there are different approaches to its behaviour and many unanswered questions, regarding the length of the loops, the number of cusps and kinks per period, the effect of gravitational backreaction, the initial conditions of the loops formed such as their length and number of harmonics.

---

[7]If the strings are modeled as Abelian-Higgs strings, there are also claims in the literature that the dominant form of decay is via the fields themselves and not gravitationally [31].

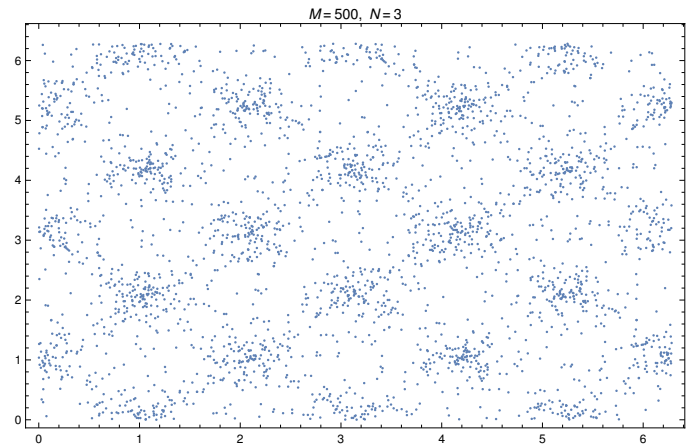

(a)

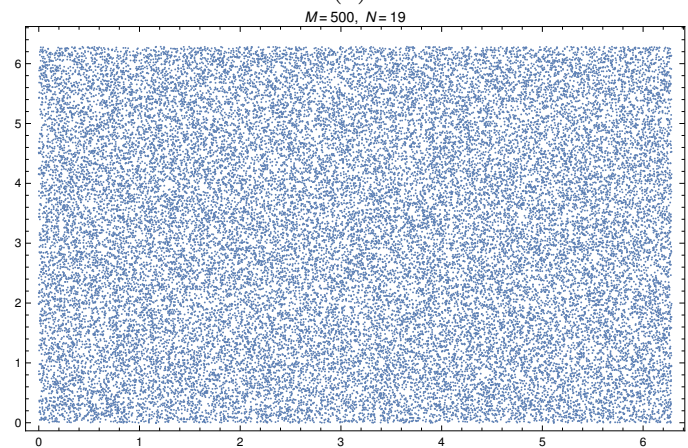

(b)

Figure 5.5.8: The values of the $(u^{(c)}, v^{(c)})$ pairs on the u-v plane. Note that for N=3 the pairs are more orderly placed, which is not observed for N=19.

| Harmonic order | c | $\langle g_1 \rangle$ | Max ($\{g_1\}$) | $\langle g_2 \rangle$ |
|:---:|:---:|:---:|:---:|:---:|
| 1 | 2.00 | 1.00 | 1.00 | 1.00 |
| 3 | 5.96 | 0.668 | 5.18 | 0.666 |
| 5 | 10.1 | 0.577 | 3.15 | 0.615 |
| 7 | 14.3 | 0.513 | 3.17 | 0.513 |
| 9 | 18.9 | 0.474 | 3.27 | 0.456 |
| 11 | 22.6 | 0.446 | 3.01 | 0.432 |
| 13 | 27.4 | 0.428 | 2.04 | 0.402 |
| 15 | 30.4 | 0.400 | 3.35 | 0.374 |
| 17 | 35.2 | 0.392 | 2.14 | 0.361 |
| 19 | 39.0 | 0.375 | 1.91 | 0.341 |

Table 5.5.9: Mean values of the cusp number per period and the average values of $g_1$ and $g_2$ evaluated at the cusp events for each harmonic order. We also provide the maximum value in the list of the values of $g_1$ we have obtained, $\{g_1\}$, for each harmonic order.

Three candidate models to explain the above complex evolution of a cosmic string network appear in the recent LIGO Collaboration publication for the detection of gravitational wave emission from cosmic string loops, [3]. These are labelled as Model 1, Model 2 and Model 3. In all three of these models the loops produced from the long string network are assumed to be large loops. We will define the meaning of small and large loops in the following section. The Model 1 corresponds to the "one-scale" model of string evolution, discussed in [9, 113]. This is an analytic model that assumes that all the loops produced from the long string network at time $t$ have the same size. The Model 2, which is presented in [46], follows a numerical approach to determine the size of the loops produced at time $t$, as opposed to Model 1 that assumes the size of the loops as a postulate. In Model 3, the expected number density of the cosmic string loops is semi-analytically derived, from creation to later times, taking into account the gravitational backreaction of the cosmic strings, as presented in [114]. The above analytic approach is fit for the numerical results of [42]. Another model, used in the early work for the calculation of GWBs from cosmic strings of [54, 4], is a "one-scale" model with small loop size.

In the following sections, we will focus on the "one-scale" model described in two ways. One requires numerical integration of quantities that follows the approach of Model 1 of [3] for large loops. We will call this Model I in the following. The other way is the case where the quantities are described analytically using interpolating functions that follows the approach of [54, 4] for small loops. We will call this Model II.

### 5.6.1 The "one-scale" model

We will focus on the simplest analytical model, called the "one-scale" model. In this approach, the cosmic string network evolution reaches a scaling regime. This stage of evolution is called scaling because the energy density of the long cosmic strings divided by the total energy density of the universe is a constant. In [9], the reasoning behind this assumption is explained. The long strings are losing energy through the production of cosmic string loops, whose size is significantly smaller than the Hubble radius, $(aH)^{-1}$. The production mechanism of closed loops is via the chopping of the long strings. This model assumes that all loops that chopped-off the long string network at a cosmic time $t$ formed with the same length

$$l \simeq \alpha t \qquad (5.6.1)$$

where $\alpha$ is a dimensionless parameter determined by the gravitational backreaction of the cosmic string loop. The value of the $\alpha$ parameter is still uncertain because the length of the loops chopped off the long network is yet to be determined. There are models that suggest a small value for $\alpha$, which corresponds to small loops, and models that suggest a large value, which corresponds to large loops, as we discussed above. We also find that the number density of loops formed at time $t$ is

$$n \simeq \alpha^{-1} t^{-3}. \qquad (5.6.2)$$

Following the calculation of [115], the gravitational backreaction scale is

$$\Gamma G \mu \qquad (5.6.3)$$

where the dimensionless constant $\Gamma \simeq 50$ appears in the calculation of the average power radiated via gravitational radiation of the loop

$$P = \Gamma G \mu^2. \tag{5.6.4}$$

Given that the rate the loop decays is

$$\frac{dl}{dt} = -\Gamma G \mu, \tag{5.6.5}$$

the length of the loops with respect to time is

$$l(t) = \alpha t_i - \Gamma G \mu (t - t_i), \tag{5.6.6}$$

The above calculations have been achieved with the assumption that the loops do not self-intersect once formed. The lifetime of the loop is given by solving $l(t) = 0$.

### 5.6.1.1 Large loops

For the case of large loops we will follow the approach of [3]. We will use the following functions for cosmic time and proper distance. The cosmic time is defined as

$$t(z) = \frac{\varphi_t(z)}{H_0} \tag{5.6.7}$$

where

$$\varphi_t(z) = \int_z^\infty \frac{dz'}{\mathcal{H}(z')(1 + z')}, \tag{5.6.8}$$

the proper distance (or cosmic distance) is defined as

$$r(z) = \frac{\varphi_r(z)}{H_0} \tag{5.6.9}$$

where

$$\varphi_r(z) = \int_0^z \frac{dz'}{\mathcal{H}(z')}, \tag{5.6.10}$$

and the proper spatial volume between redshifts $z$ and $z + dz$ is

$$dV(z) = \frac{\varphi_V(z)}{H_0^3} dz \qquad (5.6.11)$$

where

$$\varphi_V(z) = \frac{4\pi\varphi_r^2(z)}{(1+z)^3 \mathcal{H}(z)}, \qquad (5.6.12)$$

where $\mathcal{H}(z)$ is related to the Hubble parameter via

$$H(z) = H_0 \mathcal{H}(z) \qquad (5.6.13)$$

where

$$\mathcal{H}(z) = \sqrt{\Omega_\Lambda + \Omega_M(1+z)^3 + \Omega_R \mathcal{G}(z)(1+z)^4}. \qquad (5.6.14)$$

The quantity $\Omega_\Lambda$ satisfies $\Omega_\Lambda = 1 - \Omega_M - \Omega_R$, as we discussed in section 4.3.1. Also, the radiation matter equality redshift is $z_{eq} = 3366$. We will use the same values of the cosmological quantities as used in [3]. These are the Planck 2015 results presented in [22], with $H_0 = 100h\,km\,s^{-1}\,Mpc^{-1}$, $h = 0.678$, $\Omega_M = 0.308$, $\Omega_R = 9.1476 \times 10^{-5}$. Equation (5.6.14) differs from the definition of the Hubble parameter given in (4.4.13), by the inclusion of the function $\mathcal{G}(z)$. This function is related to the entropy released from particle species which become non-relativistic as the universe cools. It varies mainly during the electron-positron annihilation and the QCD phase transition [112]. Therefore, we can describe it as the piecewise-function

$$\mathcal{G}(z) = \begin{cases} 1 & z < 10^9, \\ 0.83 & 10^9 < z < 2 \times 10^{12}, \\ 0.39 & 2 \times 10^{12} < z. \end{cases} \qquad (5.6.15)$$

The theory behind the above mentioned phenomenon is beyond the scope of this thesis. However, we will include it in our calculations for a more accurate approach, following [3].

The integrals that yield $\varphi_t(z)$ and $\varphi_r(z)$ cannot be solve analytically for any $z$, and need to be evaluated numerically. However, we can calculate their asymptotic behaviour. For $z \gg 1$, we can simplify the Hubble constant

$$\mathcal{H}(z) \simeq \sqrt{\Omega_R} z^2. \tag{5.6.16}$$

Note that the above holds given that $z \gtrsim \Omega_M / \Omega_R \mathcal{G}(z) \Rightarrow z \gtrsim 3000$, such that the second term under the square root of equation (5.6.14) is smaller than the third term. Then, using (5.6.16), the integral (5.6.8) can easily be calculated

$$\varphi_t(z) = \frac{1}{2\sqrt{\Omega_R \mathcal{G}(z)}} \frac{1}{z^2}, \tag{5.6.17}$$

for $z \gg 1$. We can not use the above method to calculate the large $z$ limit of $\phi_r$, because its limits of integration vary over all the range of the $z$ values (note that in the above case they only obtained large $z$ values). Instead, we can calculate it by splitting the integral into two parts, since we will approximate its limits of integration to extend from 0 to $z \simeq \infty$. The first part is the one where the first two terms under the square root of (5.6.14) are significant and the second where the second and third term under the square root of (5.6.14) are significant. Then, we find that

$$\varphi_r(z) = \int_0^{0.31} \frac{dz'}{\sqrt{\Omega_\Lambda + \Omega_M(1+z)^3}} + \\ + \int_{0.31}^\infty \frac{dz'}{\sqrt{\Omega_M(1+z)^3 \Omega_R \mathcal{G}(z)(1+z)^4}}. \tag{5.6.18}$$

The integral limit that we introduced corresponds to the solution of

$$\Omega_\Lambda = \Omega_M(1+z)^3 \Rightarrow z \simeq 0.31. \tag{5.6.19}$$

Using the above, we find that

$$\varphi_r(z) \simeq 3.39, \tag{5.6.20}$$

for $z \gg 1$. Note that the above result differs from the one calculated in [3] at the second significant digit.

For $z \ll 1$, we can estimate $\varphi_t$ to be

$$\varphi_t(z) = \int_0^\infty \frac{dz'}{\mathcal{H}(z')(1+z')}, \tag{5.6.21}$$

i.e. it is roughly constant for small $z$ values. We find that the value of the above integral is $\varphi_t(z) \simeq 0.9566$. To calculate it we split the integral into 4 integrals. The first had limits of integration from 0 to 3366, and it was calculated numerically, with Mathematica's NIntegrate. The other three parts were calculated analytically, by simplifying the $\mathcal{H}(z)$ function and splitting into the intervals where the piece-wise function $\mathcal{G}(z)$ changes values. We find that all the contribution comes from the first integral, and the remaining three are negligible. Finally, to calculate the small z behaviour of $\varphi_r(z)$, we will calculate the Taylor series around $z = 0$ of the integrand function in (5.6.10). We follow this method, since the integrand function is integrated from 0 to $z \ll 1$, i.e. in a region of the variable close to 0. We find that the first order Taylor series of the integrand function is $1 - 0.46z'$. Keeping only the zero order term, we find that

$$\varphi_r(z) = z, \tag{5.6.22}$$

for $z \ll 1$.

For the large loop case we will assume that the parameter $\alpha$ in equation (5.6.1) takes the value $\alpha = 0.1$ [112],[3]. The rate of length loss is the same as in the small loop case, i.e. $dl/dt = -\Gamma G \mu$. This has as a result that the large loops have a larger life expectancy than the small loops, and they survive for longer than a Hubble time. Their lifetime is given by

$$\tau = \frac{\alpha}{\Gamma G \mu} t. \tag{5.6.23}$$

Let us define the relative size of the loops

$$\gamma = l/t \tag{5.6.24}$$

and the loop distribution

$$\mathcal{F}(\gamma, t) = n(l, t) t^4 \tag{5.6.25}$$

where $n(l, t)$ is the number density of loops of length $l$ and at cosmic time $t$. A significant change is that the loop length was considered constant in the small loop approach (only depending on the time of the loop creation), while in the large loop approach we will take into account its decrease due to gravitational wave emission. We will also assume that the loops do not self-intersect, as we did in the small loop approach. Therefore, the length of a loop formed at cosmic time $t_i$ with respect to time is given by (5.6.6), with $\alpha = 0.1$.

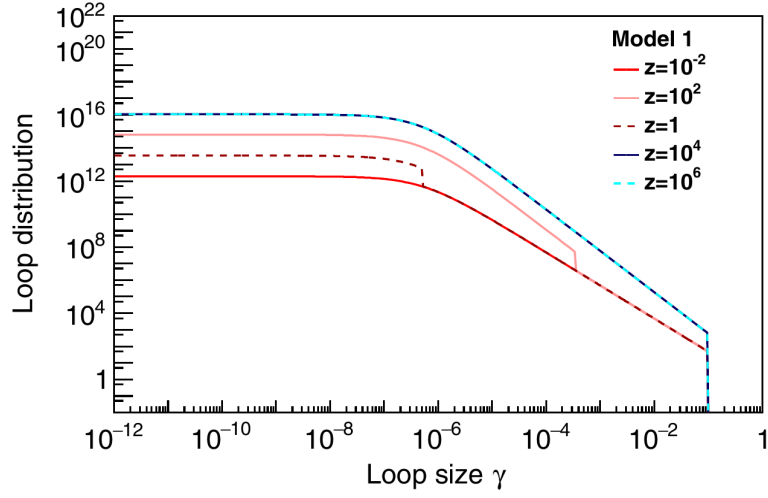

Figure 5.6.1: The loop distribution $\mathcal{F}(\gamma, z)$ ploted against the loop size $\gamma$ and for $G\mu = 10^{-8}$. The figure is adopted from [3].

The loop distribution of large loops in the radiation era ($z > 3366$) is [9], [3]

$$\mathcal{F}_{rad}(\gamma) = \frac{C_{rad}}{(\gamma + \Gamma G\mu)^{5/2}} \Theta(\alpha - \gamma).$$ (5.6.26)

The Heaviside step function ensures that $\gamma < \alpha$, which is always true since all the loops are formed with a length $\alpha t$ and after that decay. In the matter era the loop distribution function consists of two different contributions. One is from the loops of the radiation era that survive into the matter era. These loops will have length at cosmic time $t$ which is less than the length of a loop formed at $t_{eq}$, i.e. $l(t) < \alpha t_{eq} - \Gamma G\mu(t - t_{eq})$, since any loop with length larger than this would have formed in the matter era. If we define the function

$$\beta(t) = \alpha t_{eq} - \Gamma G\mu(t - t_{eq})$$ (5.6.27)

we can then write the loop distribution function which consists of the radiation era loops that survive into the matter era

$$\mathcal{F}_{mat}^{(1)}(\gamma) = \frac{C_{rad}}{(\gamma + \Gamma G \mu)^{5/2}} \left( \frac{t_{eq}}{t} \right)^{1/2} \Theta(-\gamma + \beta(t)). \qquad (5.6.28)$$

The loop distribution for loops formed in the matter era is

$$\mathcal{F}_{mat}^{(2)}(\gamma) = \frac{C_{mat}}{(\gamma + \Gamma G \mu)^2} \Theta(\alpha - \gamma)\Theta(\gamma - \beta(t)). \qquad (5.6.29)$$

In the above, the function $\Theta(\alpha - \gamma)$ ensures that all loops considered have length less than the formation length at time t, and the function $\Theta(\gamma - \beta(t))$ ensures that no loops surviving from the radiation era are taken into account in the matter era loop distribution. Therefore, the total matter era loop distribution function is

$$\mathcal{F}_{mat}(\gamma) = \mathcal{F}_{mat}^{(1)}(\gamma) + \mathcal{F}_{mat}^{(2)}(\gamma) \qquad (5.6.30)$$

The constants appearing in equations (5.6.26)-(5.6.29) have not been determined analytically, but can be calculated through numerical simulations of Nambu-Goto strings. Following [9], [113], [3], we set these parameters to the following values

$$C_{rad} \simeq 1.6, \qquad C_{mat} \simeq 0.48. \qquad (5.6.31)$$

### 5.6.1.2 Small loops

In the case of small values for the $\alpha$ parameter, we will follow the assumptions made in [4]. In particular, our aim will be to obtain analytic expressions for the cosmic time and cosmic distance, and obtain functions that interpolate between the matter era and the radiation era. We will also assume a small size for the loops created from the long string network. Following the calculation of [115], the small value of $\alpha$ is given by the gravitational backreaction scale

$$\alpha \simeq \Gamma G \mu \qquad (5.6.32)$$

and the lifetime of the loops, which can be found by solving $l(t) = 0$, given in (5.6.6), is

$$\tau_l \simeq \frac{l}{\Gamma G \mu} \simeq t. \qquad (5.6.33)$$

The above calculations have been achieved with the assumption that the loops do not self-intersect once formed. These imply that the small loops are short lived, with lifespan within a Hubble time.

Let us define the redshift $z_{eq}$ at which the matter density is equal to the radiation density

$$z_{eq} \simeq 2.410^4 \times \Omega_{M,0} h^2 \simeq 10^{3.9}. \tag{5.6.34}$$

We will also set the Hubble parameter to be $H_0 \simeq 65 km s^{-1} Mpc^{-1}$ and the age of the universe to be $t_0 \simeq 10^{10} yr \simeq 10^{17.5} s$. The above is derived from

$$t_0 = \frac{2}{3H_0}, \tag{5.6.35}$$

which holds for an Einstein-de Sitter universe with $\Omega_M = 1$. Notice how these values differ from those presented in section 5.6.1.1 for the case of large loops. It is because in both the large loop approach, which follows [3], and in the small loop approach, which follows [4], we want to directly compare our results with those in the afore-mentioned publications in sections 6.6.2 and 6.6.1. Therefore, we choose the cosmological parameters to have the same values as in the publications.

During matter domination ($z < z_{eq}$), the scale factor is connected to the redshift as

$$(1+z)^{-1} = \frac{a(t)}{a_0} = \left(\frac{t}{t_0}\right)^{2/3} \tag{5.6.36}$$

and during radiation domination ($(z > z_{eq})$), the scale factor is connected to the redshift as

$$(1+z)^{-1} = \frac{a(t)}{a_0} = \frac{a_{eq}}{a_0} \left(\frac{t}{t_0}\right)^{1/2}. \tag{5.6.37}$$

Therefore, we find that

$$t = t_0 (1+z)^{-3/2} \tag{5.6.38}$$

in the matter era, and

$$t = t_0 (1+z_{eq})^{1/2} (1+z)^{-2} \tag{5.6.39}$$

in the radiation era (see section 4.5). To combine the matter and radiation expressions into one we define a function that interpolates between the two eras

$$\varphi_l(z) = (1+z)^{-3/2} \left(1 + \frac{z}{z_{eq}}\right)^{-1/2}. \tag{5.6.40}$$

In Appendix A.3 we show the derivation of the above interpolating function. We will use the above to write the cosmic string equations in terms of the redshift. From equations (5.6.1) and (5.6.4), we find that the length of loops formed at redshift $z$ is

$$l \simeq \alpha t_0 \varphi_l(z) \tag{5.6.41}$$

and their number density is

$$n_l \simeq \alpha^{-1} t_0^{-3} \varphi_l^{-3}. \tag{5.6.42}$$

Also, the cosmic time is

$$t \simeq t_0 \varphi_l(z). \tag{5.6.43}$$

Note that since the lifetime of small loops is less than one Hubble time (defined as $H^{-1}$, which is roughly the time in which the scale factor doubles, depending on the era), equations (5.6.41)-(5.6.42) can determine the length distribution and the number density distribution of the loops. Finally, we will use the expression for the proper spatial volume in the redshift interval $(z, z + dz)$

$$dV = 54\pi t_0^3 \left[ (1+z)^{1/2} - 1 \right]^2 (1+z)^{-11/2} dz \tag{5.6.44}$$

in the matter era, and in the radiation era

$$dV = 72\pi t_0^3 (1+z_{eq})^{1/2} (1+z)^{-5} dz. \tag{5.6.45}$$

We will also use a simplified expression of the cosmic distance, defined in (5.6.9),

$$r(z) = 3t_0 \left( 1 - \frac{1}{\sqrt{1+z}} \right) \tag{5.6.46}$$

with

$$\varphi_r(z) = 3t_0 H_0 \left( 1 - \frac{1}{\sqrt{1+z}} \right). \tag{5.6.47}$$

The above expression for the cosmic distance corresponds to the cosmic distance in a spatially flat matter dominated universe, $\Omega_m = 1$. This approximation relies on the fact that for large redshifts, the value of $\varphi_r$ tends to the finite limit 3.39, as we found in equation (5.6.20), while the large redshift limit of (5.6.47) is 2.

We will further simplify the above to obtain the result [4]

$$r(z) = \frac{t_0 z}{1 + z} \tag{5.6.48}$$

with

$$\varphi_r(z) = t_0 H_0 \frac{z}{1 + z}. \tag{5.6.49}$$

We use the above simplifications to obtain analytical results and follow the calculations of [4] exactly.

## 5.7 A toy model for the loop evolution

In this section, we will introduce a toy model that estimates the number of cusps per period $c$ of a cosmic string loop with analytical arguments only. The value of $c$, a quantity relevant to the detection of the GWB signal emitted from cusps on cosmic strings, is an elusive one due to the many unknown parameters of the cosmic string loop initial conditions and evolution.

For this toy model we will apply assumptions regarding the initial conditions of the loops and the way the loops change after they self-intersect. Our aim is to build a model that provides the average cusp number produced per fundamental period $T = l/2$ from a cosmic string network of loops in a unit spacetime volume. We will not use the results of the toy model to change any of the quantities of the "one-scale" model of section 5.6 in the following sections. The purpose of this section is to develop an intuitive understanding for the number of cusps that can be created as a loop self-intersects. This will allow us to provide a new improved estimate for the crucial value $c$, the cusps per period of a cosmic string network introduced in section 5.5.2, and is currently assumed to be unity. It will also allow us to compare the number with the number of cusps we found forming in the higher harmonic loops in section 5.5.2. Therefore, in some sense, we will use the properties of string loops as provided in section 5.6, but change the estimation of the cusps per period they provide by using an effective value, which will be calculated by the toy model. We should note that the toy model includes assumptions of the evolution of loops more complicated than in section 5.6, such as allowing loops to self-intersect.

This model also provides a means of estimating the number of stable loops originating from a parent loop. By stable loops we define loops that no longer self-intersect (i.e. there is not one instance in their periodic motion where two points on the loop will cross) and by parent loop we define a cosmic string loop which has been produced from the cosmic string network and has not yet self-intersected [1]. Note that in the "one-scale" model of section 5.6, all loops are parent loops since they do not self-intersect.

Note that the toy model strings will be assumed to be odd-harmonic strings, and their chance to self-intersect will be a function of their harmonic order, as estimated in [2]. We will also ignore the effect of kinks, which will be discussed in section 5.7.9.

Our motivation for developing this model is to estimate the contribution to the GWB signal from high harmonic cosmic string loops that chop off the long string network. These loops could potentially support a large number of cusps leading to an enhanced GWB signal. In particular, the cosmic string model of [4] and Model 1 of the LIGO publication [3] calculate the gravitational wave signal from cosmic string loops assume that the cosmic string network follows the one scale model, which assumes that loops do not self intersect, while at the same time it is assumed that the loops contain roughly one cusp per period, $c = 1$. Therefore, in this way, a large signal that could be emitted from high harmonic cosmic string loops chopping from the long string network is ignored. With the toy model we present we aim to take into account this effect of cusp events from high harmonic order loops allowing for the possibility that they might self-intersect, and providing an integrated effect for the value of $c$, while at the same time keeping all the other key assumptions of the one scale model.

To achieve the calculation of this integrated effect, we need to assume the distribution of harmonics of the loops that chop off the long string network. We then calculate the number of cusps that each produce during their lifetime, given the probability that they might chop into smaller loops. The chopping process is modeled using the binary tree evolution described below. At each level of the tree, the probability that a loop will chop or not is given by Siemens and Kibble [2], which provides the probability of an odd-harmonic string of harmonic order $N$ to self-intersect. Then, the cusps per period produced from each loop chopped from the long string network is averaged over all its lifetime, by averaging over all the possible evolutions it might have (i.e. different binary trees, each with an assigned probability). Having calculated this quantity for odd-harmonic strings for several harmonic orders, we can calculate the cusps per period from a distribution of harmonics of the parent loops, which provides us with the final value of c, for a network of strings that chop off following a binary tree evolution. Given the many unknown quantities this calculation involves, we will make, as a first approach, the simplest and the most general assumptions that could be made, with a stochastic reasoning, that render this approach a toy model. In section 5.7.9, we discuss how these assumptions could be improved.

Note that in [101], it was also assumed that the evolution of a string can be imitated with a binary tree with nodes corresponding to loops that chop with a given probability. In that model the probability of chopping was constant regardless of the harmonic order of the initial loop, or the tree level, and the maximum number of tree levels (which they called generation) was also not restricted from the harmonic order of the initial loop. It was found that if the probability of self-intersection is larger than 1/2 there is a probability that the loops would chop infinitely, i.e. binary trees occurred with an infinite number of tree levels. A similar approach was assumed in [116], where the production and absorption of cosmic string loops from the long string network was studied. There, the loop self-intersection was important to be studied, because the self-intersection would mean smaller loops, whose reabsorption from the long string network became more unlikely. They modelled the loop self-intersection by assuming that for each loop there was a probability to split in two equally sized daughter loops. It was also assumed that the two daughter loops would oscillate with half the period of their parent loop and, therefore, they would split faster into two equally sized loops, compared to their parent loop.

In [117], the problem of fragmentation was tackled in a numerical manner. Two different families of cosmic string loops were assumed, and each family of strings was tested for self-intersections, their evolution tracked using a numerical method until stable loops were produced, i.e. loops that do not self-intersect. It was found that the probability of chopping was not a constant but it reduced with each loop generation. It was also found that the splitting of the loops did not necessarily occur in half, but with any other manner in a rather uniform distribution, i.e.

the production of very small daughter loops was also observed. It was also found that if one assumes the splitting of loops into two with a probability that would decrease at each generation, then this analytic approach matched the numerical approach well, when it comes to the daughter loops produced at each generation. Note that in this publication it was found that there is a correlation between the harmonic order of the parent loop, but it was not studied whether the maximum number of generations also depended on the harmonic order of the parent loop.

Finally, we should comment on the fact that the total number of daughter loops described using a binary tree model increases exponentially with the tree height. In particular, a fully expanded binary tree of height $n$, i.e. one where all of its nodes split, has $2^n$ daughter loops. By fully expanded binary tree we mean that any node at any level of the tree splits, until it reaches height $n$. In the above, we imply that a single loop is a binary tree of height 0. In any intermediate situation, i.e. at any tree where the nodes split or do not split with some probability, the increase will be exponential but with a basis less than 2. Therefore, it is reasonable to anticipate that any quantity that has a linear relation with the number of daughter loops will inherit an exponential behaviour.

### 5.7.1 Assumptions of the model

First of all, we will assume that the loops maintain their Nambu-Goto nature and can be described via any type of Nambu-Goto loop solution. In section 5.4.6, we presented a specific Nambu-Goto loop set of solutions called the odd-harmonic string, introduced in [2], and we tested their behaviour at cusp points in section 5.5. We will assume that at any stage of evolution the loops belong to this set of solutions.

We will call the initial loop the parent loop, and all the loops produced through self-intersections starting from the parent loop, daughter loops. Recalling our knowledge of self-intersections (see section 5.3.6), we know that the parent loop will self-intersect if equation (5.3.62) has at least one solution. After the self-intersection happens, the parent loop will split into two daughter loops which will have different initial conditions than the parent loop. If equation (5.3.62) has no solution then the parent loop is a stable loop, i.e. it will evolve without any self-intersection occuring in its lifetime. Note that the motion of the string is periodic. Therefore, if it does not self-intersect within one period (i.e. if equation (5.3.62) has no solutions in the region $(\tau.\sigma) \in [0, \pi) \times [0, 2\pi)$), then it will not self-intersect at all until it dissolves. We will assume that each loop will self-intersect at one point only, producing two daughter loops, excluding the case of a loop self-intersecting at two points simultaneously. The daughter loops produced will be checked for self-intersections as we did in the case of the parent loop, and they may produce or not produce further daughter loops. Eventually, the system will reach an equilibrium state where all possible self-intersections have happened, and it will consist of a number of stable loops, all produced from the initial parent loop. At any stage of evolution, when we refer to the system at some time t we will mean all the daughter loops produced from the initial parent loop that exist at the time (or in the case that the parent loop does not self-intersect the term system will refer simply to the parent loop). The above-mentioned terminology can also be found in [101, 117, 1].

Every time a loop self-intersects its length is reduced and divided between the two daughter loops. If $l$ is the length of the initial loop, and $l_1$ and $l_2$ the lengths of the daughter loops, then it does not necessarily hold that $l = l_1 + l_2$, since some of the initial loop energy turns into kinetic energy of the daughter loops. However, we will assume that the above equality holds and the kinetic energy is negligible. The simplest scenario for how the length is divided is to assume that it is halved, i.e. that the daughter loops have equal length $l_1 = l_2$. We will suppose that this is the case for any self-intersection that occurs. Since the length is halved, the fundamental harmonic of the loop that chops, which is the one with the longest wavelength, will no longer be supported from the daughter loops, which will therefore necessarily have a smaller total number of harmonics compared to the parent loop. We will also fix how the harmonics transition after a chopping, assuming that the daughter loops have harmonic order $N_i - 2$ given that they were produced from a loop of harmonic order $N_i$. This ensures that the total harmonics are reduced as the loops chop, and they maintain the odd number of harmonics format. We will also take that any self-intersection occurs after the loop has oscillated for half of its period. This choice is also based on taking an average value for the time it takes for a loop to self-intersect, since the actual value can vary from an infinitesimal time of oscillation to an almost complete oscillation of the loop that splits. In this toy model, there will be no need to determine the size of the loops, which is a subject of debate as we discussed in section 5.6.

The splitting of the parent loop forms a full binary tree.[8] We will call internal nodes the points (in our case loops) that are linked to points at the next level of the tree, and leaves the points of the tree that are not linked with points at the next level. The height of the tree $h$ is the number of levels it has. A tree that consists of a single leaf has total height zero. The top level of the tree is the level at height zero, while the bottom level is the level at maximum height. We will denote by $N$ the highest order loop, which occurs at height $h = 0$. Then, a loop of order $N_i$ occurs at level of height

$$h = \frac{N - N_i}{2},$$
(5.7.1)

where $N$ and $N_i$ take odd integer values. The index $i$ corresponds to the height of the tree level.

We can assign a function $P(N_i, N_j)$ between any two levels of the tree, which is the probability of the loop of order $N_i$ splitting to loops of harmonic $N_j$. Note that the harmonics reduce always by 2 because our loops support odd-harmonics only, and both $N_i$ and $N_j$ obtain odd values. Also, we assume that for any parent loop and daughter loop, the left-moving and right-moving functions have an equal number of harmonics, which we have assumed in the previous sections as well.

## 5.7.2 The stable loop number and cusp production

In this section, we will display our method for calculating the splitting of an initial odd-harmonic string, and evaluating the average number of stable loops from an initial parent loop and the average number of cusps emitted from the system, with respect to the harmonic order of the parent loop.

---

[8]It is called tree because of its structure, full because each point (in our case loop) splits into a number from 0 to n daughter points, and binary because the number of possible daughter points is necessarily two in our case.

At each harmonic order we have a total number of binary trees, given by (5.7.5), defined below. To find the number of stable loops for a harmonic order, we calculate the number of stable loops for each of the binary trees and then we average over all the binary trees. The number of cusps is calculated in a similar manner. We calculate for each binary tree the value of the total number of cusps produced divided by the total number of periods of the system. By total number of cusps emitted we mean the sum of all cusp events that occurred in the system of loops until all loops in the system have vanished, i.e. in the lifetime of the system. By total number of periods we mean the number of periods of the parent loop (which is equal to $T_l = l/2$) that have occurred in its lifetime $\tau_l = l/\Gamma G \mu$. In this way, we can use the above value of the cusps per period as an estimation of the cusps per period emitted from an N order harmonic string in its lifetime. Instead of using the number of cusps that it would produce if it did not evolve, we use a different approach that aims to take its evolution into account. Once more, averaging over the values of all binary trees we find the final result of the average cusps per period of an N harmonic order loop.

### 5.7.3 Calculation for any harmonic order

For the probability of self-intersection of an $N$ order odd-harmonic loop, we will use the values obtained from the plot 5.4.1, which is obtained from [2]. In table 5.7.1, we present these probability values in terms of the harmonic order. The probability value of an $N_i$ harmonic order loop not to self-intersect is $[1 - P(N_i, N_i - 2)]$. The number $\varepsilon$ is a much smaller than unity, which indicates the very low probability of a high-harmonic loop not to self-intersect. We will set it to be $\varepsilon = 0.01$. The first order odd-harmonic loop is always stable and does not self-intersect, $P(1, 1) = 1$.

For the values of the cusps per period of a loop of a given harmonic order $N_i$, $c_{N_i}$, we will use the values obtained for the odd-harmonic loops presented in Table 5.5.9.

Regarding the total number of periods, in terms of the period of the parent loop, this will be calculated by dividing the total lifetime of the tree by the period of the parent loop. The lifetime of stable loops is significantly longer than the lifetime of a loop that chops off, since $\Gamma G\mu$ is of the order $10^{-6}$ or smaller. Also, the lifetime of a stable loop is larger the smaller its tree height. Therefore, the total lifetime of the tree is given by the sum of the lifetimes of the stable loops with the smallest tree height plus the lifetime of the loops that precede it.

| Harmonic order $N_i$ | Probability of self-intersection |
|:---:|:---:|
| 3 | P(3,1)=0.6 |
| 5 | P(5,3)=0.8 |
| 7 | P(7,5)=0.9 |
| 9 | P(9,7)=1-0.05 |
| ≥11 | P($N_i$,$N_i$-2)=1-$\varepsilon$ |

Table 5.7.1: In the table, we present the probability value of a loop of harmonic order $N_i$ to split into two loops of harmonic order $N_i - 2$.

Let us denote by $a(h)$ the number of trees of a given height $h$, where $h$ takes positive integer values, including zero. Then, the recurrence relation [118]

$$a(h+1) = a(h)^2 + 2a(h)\left[a(h-1) + a(h-2) + \cdots + a_0\right], \quad a_0 = 1 \quad (5.7.2)$$

holds. If we also denote by $b(h)$ the cumulative number of trees up to height $h$ we know that it is also expressed by the following recurrence relation [119]

$$b(h+1) = b(h)^2 + 1, \quad b_0 = 1. \quad (5.7.3)$$

Note that the zero height values are obvious, since they correspond to a single point. The simplicity of relation (5.7.3), allows us to easily find the total number of all trees that can exist and have height from 0 to $h$. This describes the total number of configurations we can potentially have when a loop of harmonic order $N$ self-intersects, which corresponds to height $h = N - 1$. We notice that the number of trees increases in a recurrence power law manner. This implies that the value of height $h$ is the square of the previous value, which is the square of the value before that, and so on so forth, which corresponds to a very rapid increase of the total cummulative number of trees with respect to height [120].

We will now define the trees with same type, i.e. the trees that have the same number of leaves and internal nodes at each level. Note that this definition implies that these trees have the same value of height too. An example of trees of the same type are the trees that appear in Figures 5.7.3(c) and 5.7.3(d), that we will examine in more detail in the following sections. We will also define the multiplicity (also called cardinality) $d(h)$, which is the number of different types of trees with height $h$. For example, from Figure 5.7.3 we can see that the $d(2) = 2$. The recurrence series of the tree multiplicity is given in [121]. We can also define the degeneracy of a tree type $N(h, i)$, which is the number of trees of height $h$ that belong in the same tree type. The index $i$ corresponds to the number of different tree types at height $h$ and obtains values $1, \ldots, d(h)$. For example, from Figure 5.7.3, it is obvious that for $h = 2$ the degeneracy is $N(2, 1) = 2$, since we have one tree type and two degenerate loops. Given the above, we can also write the number of trees of a given height in terms of the tree degeneracy

$$a(h) = \sum_{i=1}^{d(h)} N(h, i) \qquad (5.7.4)$$

and the cumulative number of trees up to height $h$ is written as

$$b(h) = \sum_{j=0}^{h} a(j) = \sum_{j=0}^{h} \sum_{i=1}^{d(j)} N(j, i). \qquad (5.7.5)$$

### 5.7.3.1 Calculation of the average stable loops

In our model we deal with loops of harmonic order $K$, where $K$ is odd. This loop will self-intersect producing odd-harmonic order loops. This process can continue and loops of harmonic orders $K - M$ are produced, where $M$ takes even values and satisfies $0 \leq M \leq K - 1$. Then, the height is given by $h = M/2$.

We will define the final harmonic order of a string that evolves to be the smallest harmonic order of any of the stable loops of the system, $K - M$. Note that this does not prevent the system to also include stable loops with harmonic order greater than $K - M$.

Let us now produce the formula that calculates the average stable loops from a parent loop of harmonic order $K$. We should start with the calculation of the stable loops for trees with fixed tree height, between harmonic order $K$ and $K - M$. This is described by the quantity

$$f_{K,K-M} = \sum_{i=1}^{d_{K,K-M}} c_{K,K-M}^i P(K, K-M)^i N_{K,K-M}^i, \qquad (5.7.6)$$

and includes all trees of fixed height $M/2$. In the above, $d_{K,K-M}$ is $d(M/2)$, which is the number of different types of trees with height $M/2$. The quantity $c_{K,K-M}^i$ is the number of stable loops (leaves) of trees with height $M/2$ and of the same type $i$. Also, $P(K, K-M)^i$ is the total probability of the trees of type $i$ and height $M/2$ to occur. As we mentioned above, each tree configuration has a probability of occurring, and the total probability of a type $i$ of trees is the summation of all the probabilities of the type $i$ trees. Finally, the quantity $N_{K,K-M}^i$ is the degeneracy of trees of same type $i$ and with height $M/2$. [9]

Given the above, we can calculate the average number of stable loops of a parent loop of harmonic order $N$ which evolves (defined in section 5.7.2), if we sum the above quantity $f_{K,K-M}$ over all possible tree heights, i.e. ranging from 0 (which corresponds to the case where the parent loop does not self-intersect) to $(N-1)/2$, which is the maximum tree height for an $N$ order harmonic loop that is allowed to self-intersect up to first order harmonic loops. If we denote the average stable loops of an $N$ order harmonic string by $sl_N$, we find that it can be calculated as

$$sl_N = \sum_{i=0}^{\frac{N-1}{2}} f_{N,N-2i}, \qquad (5.7.7)$$

where $N$ is odd. Note that the subscript where $N$ appears twice (i.e. $N, N$) implies an $N$ harmonic order loop that does not self-intersect. Since the minimum harmonic order is $1 = N - (N - 1)$, the maximum height is $(N - 1)/2$. The summation of all tree configuration probabilities over all possible tree heights is equal to 1 by definition (see section 5.7.5).

---

[9]Confusion should be avoided with the capital indices $A, B, ...$ in section 5.1, used to denote the worldsheet parameters, and the capital indices in (5.7.6) and in the following expressions used to denote the harmonic order. The former are dummy indices that imply summation while the latter are not.

### 5.7.3.2  Calculation of the average cusps per period

For the calculation of the average cusp number, we will need to define the periods of the daughter loops, as well as explain the concept of number of periods in the lifetime of the system (see section 5.7.2).

As we discussed in section 5.7.1, a loop that self-intersects will split into two equal sized loops. Since the period of a loop is half the length of the loop this implies that the period of the two half length loops will be equal and also half compared to the loop that they originated from. Therefore, we can express the period of a daughter loop as

$$T_{K-M}^{(K)} = \frac{T_{K-M+2}}{2} \tag{5.7.8}$$

where $2 \leq M \leq K - 1$. The superscript $(K)$ indicates the harmonic order of the parent loop of the system. The subscript $K - M$ indicates the harmonic order of the loop. The above relation is a recurrence relation between the period of a $K - M$ harmonic order loop compared to a loop with harmonic order $K - M + 2$. The former loop emerged from the self-intersection of the latter loop. It is also useful to know how long a loop lives. If the loop self-intersects, it will then live for half its period, according to our assumptions in section 5.7.1. Therefore, the expression for its lifetime is

$$\mathcal{T}_{K-M+2,K-M}^{(K)} = \frac{T_{K-M+2}^{(K)}}{2}, \tag{5.7.9}$$

where $2 \leq M \leq K - 1$. If the loop does not self-intersect, then its lifetime is given by equation (5.6.33), which is calculated with the rate of energy loss due to gravitational radiation. In this case, the lifetime of the loop is

$$\mathcal{T}_{K-M,K-M}^{(K)} = \frac{T_{K-M+2}^{(K)}}{\Gamma G \mu}. \tag{5.7.10}$$

Note that in the above expressions, the subscript $K - M, K - M$ indicates a loop of order $K - M$ that does not self-intersect, while the subscript $K - M + 2, K - M$ indicates a loop of order $K - M + 2$ that will self-intersect into two $K - M$ loops.

The total lifetime of the system will be the total time from the moment the parent loop of harmonic order $K$ is created until all of the loops that were created via self-intersections have evaporated. Note that the lifetime of a stable loop is significantly larger than a loop that self-intersects, because $\Gamma G \mu$ is of the order $10^{-6}$ or smaller. From equation (5.7.10), we can see that the larger the harmonic order $K - M$ the longer the stable loop lives. Therefore, the total lifetime of a system with maximum leaf having harmonic order $K - M$ will be given by

$$\mathcal{T}_{K-M}^{(K)} = \mathcal{T}_{K,K-2}^{K} + \mathcal{T}_{K-2,K-4}^{K} + \cdots + \mathcal{T}_{K-M+2,K-M}^{(K)} + T_{K-M,K-M}^{(K)}. \qquad (5.7.11)$$

Note that the above describes the total lifetime of a system where $K - M$ is the highest order harmonic of any of its stable loops. The highest order harmonic leaf does not necessarily correspond to a single loop of the system, since there can be multiple stable loops of the same harmonic order. The above expression summarizes the lifetimes of the loops from harmonic order $K$ to harmonic order $K - M$, which corresponds to a stable loop.

For a parent loop with harmonic order $K$ we need $M/2$ steps in the tree (i.e. $M/2$ differences in height) to reach a loop of harmonic order $K - M$. Therefore, the period of the $K - M$ harmonic order loop can be written as

$$T_{K-M}^{(K)} = 2^{-\frac{M}{2}} T_K^{(K)} = 2^{-\frac{M}{2}} \frac{\alpha t_i}{2} = 2^{-\frac{M+2}{2}} \alpha t_i. \qquad (5.7.12)$$

In the above, we used the fact that at each step the period of the loops is half of the period of the loops of the previous step. The harmonic order $M$ is given by the integer values in the interval $[0, K - 1]$. The lifetime of a loop of harmonic order $K - M + 2$ which splits (into a $K - M$ harmonic order loop) is

$$\mathcal{T}_{K-M+2,K-M}^{(K)} = \frac{T_{K-M+2}^{(K)}}{2} = \mathcal{T}_{K-M}^{(K)}, \qquad (5.7.13)$$

and we can write it in terms of the period of the parent $K$ harmonic loop as

$$\mathcal{T}_{K-M+2,K-M}^{(K)} = 2^{-\frac{M+2}{2}} \alpha t_i \qquad (5.7.14)$$

where $2 \leq M \leq K - 1$. If we apply the parameter transformation $M \to M + 2$, we find that equation (5.7.14) is written as

$$\mathcal{T}_{K-M,K-M-2}^{(K)} = 2^{-\frac{M+4}{2}} \alpha t_i, \qquad (5.7.15)$$

where $0 \leq M \leq K-3$. Combining equations (5.7.12) and (5.7.15), we find that,

$$\frac{\mathcal{T}_{K-M,K-M-2}^{(K)}}{T_{K-M}^{(K)}} = \frac{1}{2} \tag{5.7.16}$$

with $0 \leq M \leq K-3$. For the case of a stable loop, combining the equations (5.7.8) and (5.7.10), we find that

$$\frac{\mathcal{T}_{K-M,K-M}^{(K)}}{T_{K-M}^{(K)}} = \frac{T_{K-M+2}^{(K)}}{\Gamma G \mu T_{K-M}^{(K)}} = \frac{2T_{K-M}^{(K)}}{\Gamma G \mu T_{K-M}^{(K)}} = \frac{2}{\Gamma G \mu}. \tag{5.7.17}$$

Given the above results, we can calculate the total number of cusp events produced by a tree with initial harmonic order $K$ and final harmonic order $K-M$

$$g_{K,K-M} = \sum_{i=1}^{d_{K,K-M}} P(K,K-M)^i \tilde{c}_{K,K-M}^i N_{K,K-M}^i, \tag{5.7.18}$$

using the same reasoning as in equation (5.7.6). The upper sum limit $d_{K,K-M}$ is $d(M/2)$, i.e. the number of different types of trees with height $M/2$. The probability $P_{K,K-M}^i$ is the total probability of trees of type $i$ and height $M/2$. Also, $N_{K,K-M}^i$ is the degeneracy of trees of the same type $i$ and with height $M/2$. Finally, the quantity $\tilde{c}_{K,K-M}^i$ corresponds to the total cusps produced from the nodes and the leaves of the tree of type $i$

$$\tilde{c}_{K,K-M}^i = \sum_{j=0}^{M/2-1} \frac{\mathcal{T}_{K-2j,K-2j-2}^{(K)}}{T_{K-2j}^{(K)}} c_{K-2j} n_{(K,K-M,K-2j)}^i +$$
$$+ \sum_{j=0}^{M/2} \frac{\mathcal{T}_{K-2j,K-2j}^{(K)}}{T_{K-2j}^{(K)}} c_{K-2j} l_{(K,K-M,K-2j)}^i. \tag{5.7.19}$$

Note that by total cusps we mean all the cusp events produced from the creation until the evaporation of the loop of type $i$ and height $\lambda/2$. In the above, the first sum on the right-hand side of the equation summarizes the cusp contribution from the nodes (i.e. loops that split) of the tree and $n_{(K,K-M,K-2j)}^i$ is the number of nodes at tree height $K-2j$ for a loop with initial harmonic order $K$ and final $K-M$. The second sum on the right-hand side of the equation above summarizes the cusp contribution from the leaves (i.e. stable loops) of the tree and $l_{K-2j}$ is the number of leaves at tree height $K - 2j$ for a loop with initial harmonic order $K$ and final $K - M$. The nodes contribute cusp events for half of their period, $\mathcal{T}_{K-2j,K-2j-2}^{(K)}$, after which they split. The leaves contribute for much longer since they do not split, $\mathcal{T}_{K-2k,K-2k}^{(K)}$. The parameter $c_{K-2j}$ are the cusps per period of an odd-harmonic loop of harmonic order $K - 2j$, the values of which are given in Table 5.5.9. The fractions in equation (5.7.19) give the number of periods for which the corresponding loop oscillates until it chops or dissolves.

Having calculated the total number of cusps produced for a tree configuration of a given height, we will summarize over the possible tree heights. We will also normalize the result in terms of the total number of periods the $K$-th order harmonic parent loop oscillates, $2/\Gamma G\mu$, to finally obtain the average cusps per period of a $\kappa$-th order harmonic loop

$$c^{(K)} = \frac{\Gamma G\mu}{2} \sum_{i=0}^{\frac{K-1}{2}} g_{K,K-2i}. \tag{5.7.20}$$

Note that in deriving the above formula we followed the same reasoning in our calculations as in section 5.7.3.1 where we calculated the stable loops number.

## 5.7.4 Application for low harmonic loops

In this section we will provide an example of the toy model calculations presented in 5.7.3 for cosmic string loops of harmonic order $N = 1$, $N = 3$ and $N = 5$.

The simplest case is the $N = 1$ odd-harmonic string, which does not self-intersect at all and does not produce any daughter loops. For this case we know that the stable loops produced will always correspond to a single loop with average cusps per period $c_1 = 2$ (as we found in section 5.4.2).

### 5.7.4.1 The $N = 3$ parent loop case

For the case of the $N = 3$ odd-harmonic string, we have two possible states of evolution, i.e. the total number of binary trees is 2. The one case is that the

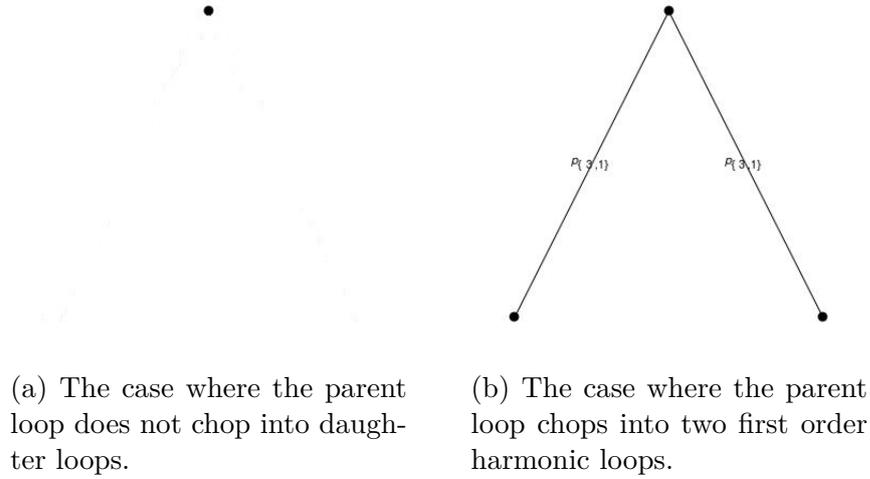

(a) The case where the parent loop does not chop into daughter loops.

(b) The case where the parent loop chops into two first order harmonic loops.

Figure 5.7.2: All the possible cases of evolution for a third order harmonic string. The dot represents the loop.

parent loop does not self-intersect, and the other is that the parent loop self-intersects, producing two $N = 1$ daughter loops. The former case corresponds to Plot 5.7.2(a) and the latter corresponds to Plot 5.7.2(b). To calculate the number of stable loops, we note that the binary tree in plot 5.7.2(a) has one leaf, i.e. one stable loop, and its probability of occurring is $[1 - P(3, 1)] = 0.4$. The binary tree in plot 5.7.2(b) has two leaves and its probability of occurring is $P(3, 1) = 0.6$. Averaging over the two possible cases we find that the average number of stable loops for $N = 3$ is

$$n_s^{(3)} = 2P(3, 1) + [1 - P(3, 1)] = 1.6. \qquad (5.7.21)$$

The lifetime of a loop was estimated in equation (5.6.33), $\tau_l = l_i/\Gamma G\mu$. Note that we do not specify whether the value of $\alpha$ corresponds to small or large loops since its value cancels in the calculations. The period of the $N = 3$ loop is $T_3^{(3)} = l/2 = \alpha t_i/2$. If it splits into two equal sized loops, each will have period $T_1^{(3)} = (l/2)/2 = T_3^{(3)}/2$ (see equation (5.7.8)). The two daughter loops produced will be of harmonic order $N = 1$ and they will not split further. This means that they will live for $\tau_l = l_i/4\Gamma G\mu$, since they lose energy with rate $\Gamma G\mu$ (see equation (5.7.10)). Therefore, the average number of cusps emitted from the system per period of the initial loop defined in equations (5.7.18)-(5.7.20) consists of two terms, one with weight $P(3, 3) = 0.4$, which corresponds to case (a) of plot 5.7.2 and one with weight $P(3,1) = 0.6$, which corresponds to case (b) of 5.7.2. In case (a), the total number of cusps emitted from the system on average is

$$c_3 \frac{\frac{\alpha t_i}{\Gamma G \mu}}{\frac{\alpha t_i}{2}} = c_3 \frac{2}{\Gamma G \mu}, \tag{5.7.22}$$

while the total number of periods is

$$\frac{\frac{\alpha t_i}{\Gamma G \mu}}{\frac{\alpha t_i}{2}} = \frac{2}{\Gamma G \mu}, \tag{5.7.23}$$

as we calculated in the previous section. Therefore, in the case that the loop does not split it produces $c_3$ cusps per period, as expected. In case (b), the total number of cusps emitted from the system is

$$c_3 \frac{\frac{\alpha t_i}{4}}{\frac{\alpha t_i}{2}} + 2c_1 \frac{\frac{\alpha t_i}{2\Gamma G \mu}}{\frac{\alpha t_i}{4}} = \frac{1}{2}c_3 + \frac{4}{\Gamma G \mu}c_1. \tag{5.7.24}$$

The first term in the right-hand side of (5.7.24), corresponds to the average number of cusp events that occured from the parent $N = 3$ loop, in the half period interval before it splits, and the second term on the right-hand side of (5.7.24) corresponds to the average number of cusp events that occured from the two $N = 1$ loops. Note that $c_3/2 \ll 4c_1/\Gamma G \mu$ since $\Gamma G \mu$ is of the order $10^{-6}$ or smaller. We can now calculate the average number of cusps emitted from the system of the $N = 3$ loop per period of the initial loop, which is

$$c^{(3)} = P(3,3)c_3 + P(3,1) \left( \frac{1}{2}c_3 + \frac{4}{\Gamma G \mu}c_1 \right) \frac{\Gamma G \mu}{2} \simeq 4.8. \tag{5.7.25}$$

In the above, we have used the values $c_1 = 2$ and $c_3 = 5.96$ from table 5.5.9.

### 5.7.4.2 The $N = 5$ parent loop case

In the case of the fifth order odd-harmonic loop we have 5 configurations, as we can see in Figure 5.7.3. Two of the configurations, the ones that appear in the Figures 5.7.3(c) and 5.7.3(d), are identical. They account for the case where only one of the two daughter loops at height $h = 2$ self-intersects. Since labelling the loops is not of importance in our calculations, we can account for these two

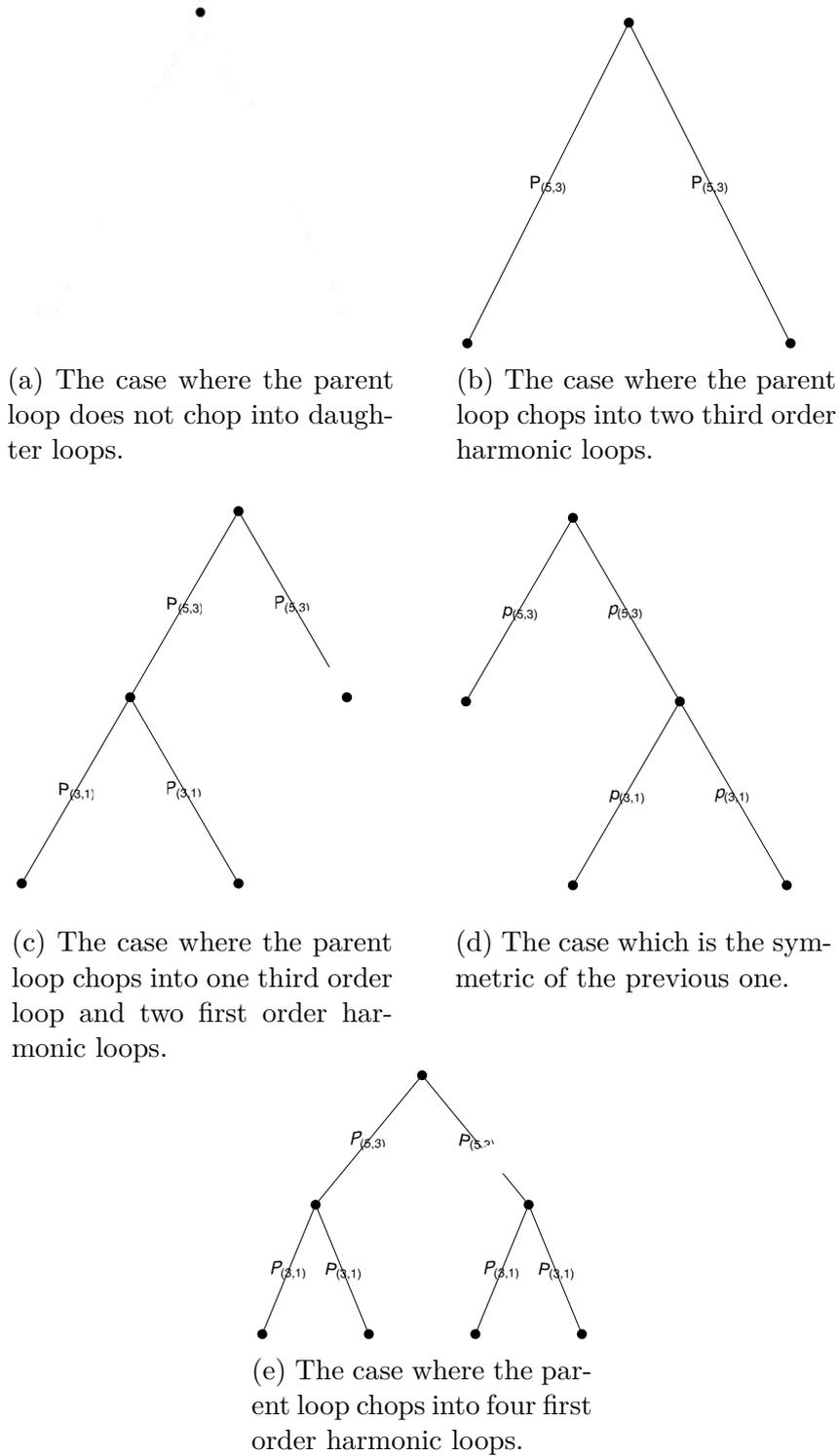

(a) The case where the parent loop does not chop into daughter loops.

(b) The case where the parent loop chops into two third order harmonic loops.

(c) The case where the parent loop chops into one third order loop and two first order harmonic loops.

(d) The case which is the symmetric of the previous one.

(e) The case where the parent loop chops into four first order harmonic loops.

Figure 5.7.3: All the possible cases of evolution for a fifth order harmonic string.

configurations as one configuration of multiplicity two.

The probability of configuration (a) of plot 5.7.3 to occur is $P(5,5)$. The probability of configuration (b) is the probability of the 5 order loop to self-intersect, and none of the daughter loops to do so, which corresponds to $P(5,3)\left[1-P(3,1)\right]^2$. Using the same reasoning, the probability of configurations (c) and (d) is $P(5,3)$ $P(3,1)\left[1-P(3,1)\right]$, and of (e) is $P(5,3)P(3,1)^2$.

We find that the average number of stable loops is

$$
\begin{aligned}
n_s^{(5)} =&P(5,5) + 2P(5,3)\left[1-P(3,1)\right]^2 + 6P(5,3)P(3,1)\left[1-P(3,1)\right] + \\
&4P(5,3)P(3,1)^2 = 2.76.
\end{aligned}
\tag{5.7.26}
$$

The total number of cusps per period produced on average from the configuration (a) of plot 5.7.3 coincides with $c_5$ from table 5.5.9, similarly to the $N = 3$ case. Its lifetime is also the same as the lifetime of an $N = 5$ string loop that does not self-intersect, $\alpha t_i / \Gamma G \mu$. The total number of cusps on average produced from configuration (b) is

$$
c_5 \frac{\frac{\alpha t_i}{4}}{\frac{\alpha t_i}{2}} + 2c_3 \frac{\frac{\alpha t_i}{2\Gamma G\mu}}{\frac{\alpha t_i}{4}} = \frac{1}{2}c_5 + \frac{4c_3}{\Gamma G\mu} \simeq \frac{4c_3}{\Gamma G\mu}.
\tag{5.7.27}
$$

The configurations (c) and (d) each produce total cusps

$$
c_5 \frac{\frac{\alpha t_i}{4}}{\frac{\alpha t_i}{2}} + c_3 \frac{\frac{\alpha t_i}{8}}{\frac{\alpha t_i}{4}} + c_3 \frac{\frac{\alpha t_i}{2\Gamma G\mu}}{\frac{\alpha t_i}{4}} + 2c_1 \frac{\frac{\alpha t_i}{4\Gamma G\mu}}{\frac{\alpha t_i}{8}} \simeq \frac{2c_3 + 4c_1}{\Gamma G\mu}.
\tag{5.7.28}
$$

Finally, the configuration (e) produces the following total cusps per period

$$
c_5 \frac{\frac{\alpha t_i}{4}}{\frac{\alpha t_i}{2}} + 2c_3 \frac{\frac{\alpha t_i}{8}}{\frac{\alpha t_i}{4}} + 4c_1 \frac{\frac{\alpha t_i}{4\Gamma G\mu}}{\frac{\alpha t_i}{8}} \simeq \frac{8c_1}{\Gamma G\mu}.
\tag{5.7.29}
$$

Given the above, we find that an $N = 5$ loop, which is allowed to self-intersect, will produce on average

$$
\begin{aligned}
&P(5,5)c_5 + P(5,3)\left[1-P(3,1)\right]^2 2c_3 + 2P(5,3)P(3,1) \\
&\left[1-P(3,1)\right](c_3 + 2c_1) + 4P(5,3)P(3,1)^2 c_1 = 9.7
\end{aligned}
\tag{5.7.30}
$$

total number of cusps per period of the initial loop.

## 5.7.5 Implementation with Mathematica

The number of different tree configurations increases rapidly with the tree height. As we can see from the recurrence relation (5.7.5), the number of configurations is 1 for $h = 0$, 2 for $h = 1$, 5 for $h = 2$, 26 for $h = 3$, 677 for $h = 4$, and so on so forth. Since we can not proceed with calculations as we did in the previous section as the height increases, it is necessary to obtain an automated means of calculating the values of the stable loops and cusps per period for $h > 2$.

The flow chart of the method we used to calculate higher order harmonic cases is the following:

1. Set the value of "treeheight" (which is equal to the longest tree height minus one) to the value that we are interested to calculate. Note that the harmonic order of the parent loop $K$ is related to the value of "treeheight" according to

$$K = 2 \,\text{``treeheight''} + 3 \qquad (5.7.31)$$

Then, we initialize the value of a list for the tree of height 0 and the tree of height 1, which will be needed to calculate trees of higher height. This list is defined to include the following elements; the tree height, the number of leaves, the number of equivalent trees with the given characteristics, the probability of each tree configuration, and a list called "newconttreedata" which contains all the information for the structure of the tree, i.e. its internal nodes and leaves at every level of the tree, and calculates their cusp contribution according to equation (5.7.19). Note that the initialization values are given in section 5.7.4. We also need a method to produce the trees of the next level, given that we have calculated all possible trees of a given level. For this, we use a list "conttree", which contains all the information needed for this purpose. We initialize the value of "conttree" for the smallest value (i.e. for transitioning from the level 0 tree to 1). The list includes the probability of the parent loop splitting, which is $P(K, K - 2)$, and its cusp contribution. We also initialize the list "AllContTrees", which contains all trees used to produce next level trees, and the list "AllTrees", which contains all trees.

2. For the values between 1 and "treeheight" repeat the following nested loops

   (a) Initialize the list that saves the new trees types, produced in this iteration. Also, set the number of tree categories, given by the length of "AllContTrees".

   (b) For all the tree types previously found (i.e. up to the previously calculated trees) repeat the following

      i. Set the variables (of the given tree category) for the height "h", for the number of total leaves ("leaves"), for the number of leaves at the bottom layer ("bleaves"), for the number of trees of this type ("n") and the probability of this tree type ("P"). Also keep in a list (called "conttreedata") the number of leaves and internal nodes at every level of this type of tree.

      ii. For all the possible values of leaves on the bottom layer (i.e. values between 1 and "bleaves") repeat the following

         A. For the value of the bottom leaves "i" of the previous tree, calculate the values of "h", "leaves", "bleaves", "n" and "P" and the information for leaves and internal nodes for the new tree type (the list "newconttreedata"). The height of the new tree will be $(h + 1)$, the number of bottom leaves will be $2i$, the number of total leaves will be leaves $+ 1$, the number of trees of this type will be

$$n\binom{\text{bleaves}}{i}. \tag{5.7.32}$$

Note that in the above we have used the binomial coefficient

$$\binom{n}{k} = \frac{n!}{k!(n-k)!}, \tag{5.7.33}$$

defined for positive integer values of $n$ and $k$. The probability of the new tree is

$$
\begin{aligned}
&P \times P(K-2h, K-2h)^{\text{bleaves}-i} \times \\
&P(K-2h, K-2h-2)^{i} \times \\
&P(K-2h-2, K-2h-2)^{2i},
\end{aligned}
\tag{5.7.34}
$$

where $P$ is the probability of the tree type of "treeheight"$= h$. We also create the list used to create the $h+2$ trees.

    B. Add the above calculated tree to the list of trees. The sum of all probabilities found is equal to 1.

3. Use the formulas (5.7.7)-(5.7.6) to calculate the average number of stable loops and the formulas (5.7.18)-(5.7.20) to calculate the average number of cusps.

4. Print the analytic formulas for the stable loops and the cusps per period and then their numerical values by using the values for the cusps per period from table 5.5.9 and the probabilities $P(M, M-2)$ from table 5.7.1.

## 5.7.6 Monte Carlo method

We can also obtain the results of section 5.7.5 with a numerical method. The method that can be applied for the specific problem is the Monte Carlo method, described in section 2.6.2.

The flow chart of the method we used to calculate higher order harmonic cases numerically with the Monte Carlo method is the following:

1. Set the maximum loop order ("maxlooporder") and the numerical values transition probabilities p(x,x). Set the values for c(x) and $\Gamma G\mu$. Also set the amount of random numbers ("Nran") used for the calculation of the trees.

2. Initialize the variables counting the loops and cusps at zero value.

3. For the amount of random numbers repeat the following

   (a) Initialize variables starting with a single node (i.e. a single leaf).

   (b) While the tree keeps growing and the current loop order ("loopord") is greater or equal to 1 repeat the following:

      i. For the amount of bottom leaves repeat the following:

         A. Calculate a pseudo-random real number "ran0" between 0 and 1 from uniform distribution.

         B. If "ran0" greater or equal to p("loopord","loopord") the current bottom leaf will become an internal node on the next iteration and also the tree increases by one layer, else it remains a bottom leaf.

      ii. If at any point all bottom leaves remain bottom leaves then the current tree has ended.

   (c) Sum the variables for the calculation of loops and cusps.

4. Divide the sum of the variables for loops and cusps by "Nran" to find the average stable loops and average stable cusps.

## 5.7.7 Number of stable loops and cusps per period with harmonic order

In this section we will present the results we reached using the methods described in sections 5.7.5 and 5.7.6. We will also compare the results from the different methods, to check that they are consistent and to understand the differences of the two methods on the computation execution time.

In Table 5.7.5, we present the results that were obtained using the analytic approach that was implemented with Mathematica. This method allows us to obtain results for harmonic order of the parent loop ranging from 1 to 17, this upper limit being due to the fact that the computational time increases rapidly with harmonic order. The value of cusps per period at harmonic order 17 was not possible to calculate due to the computational time required. In Table 5.7.6 we present the results we obtained using the Monte Carlo method. This method is much faster than the previous one and allows us to calculate data for higher harmonics loops. We notice that the data in both methods are exactly the same up to two significant digits, and they only deviate a little in the third digit.

We can compare the data for the average stable loops per period presented in Tables 5.7.5-5.7.6 with the data we presented in Table 5.5.9, which are the average number of cusps produced from an odd-harmonic loop with harmonic order. We notice that at the third order harmonic, the average number of cusps per period calculated with the toy model is smaller than the value we found in Table 5.5.9. At the fifth harmonic order, the average number of cusps becomes very close with the value in Table 5.5.9. At higher harmonics, the number of cusps become much larger, since the data exhibit exponential growth, while the ones in Table 5.5.9 increase linearly. By applying a linear fit to the logarithm of the data presented in Tables 5.7.5-5.7.6 for the number of cusps per period, we find that they grow with an exponential base of 1.94 with harmonic order. Note that the base is the same up to the third significant digit for both the data in Table 5.7.5, obtained with the analytic method, and the data in Table 5.7.6, obtained with the Monte Carlo method.

We notice that at this rate of growth, the cusps per period rapidly reach large values for high harmonic loops. For example, as we calculated with the Monte Carlo method, in Table 5.7.6 we can see that at harmonic order $N = 51$ the average cusps per period become of the order $10^7$. We believe that the algorithm we have adopted accurately represents the number of cusps produced for the low harmonic loops, but tends to overestimate the number produced for the high harmonic orders. In particular, one of our assumptions was that each of the loops of the system that splits, does so at half its period, and splits into two equal sized loops with harmonic order smaller by two. This assumption however, might not follow the behaviour of high harmonic loops, which tend to break up rapidly and into many small fragments, as was simulated and discussed in [1]. For this reason, in the following sections we will focus our attention on the results regarding the lower harmonic parent loops, from $N = 1$ to $N = 7$, which we believe provides a more realistic outcome of the evolution of a network of cosmic strings. We discuss this issue further and possible improvements of the toy model in section 5.7.9.

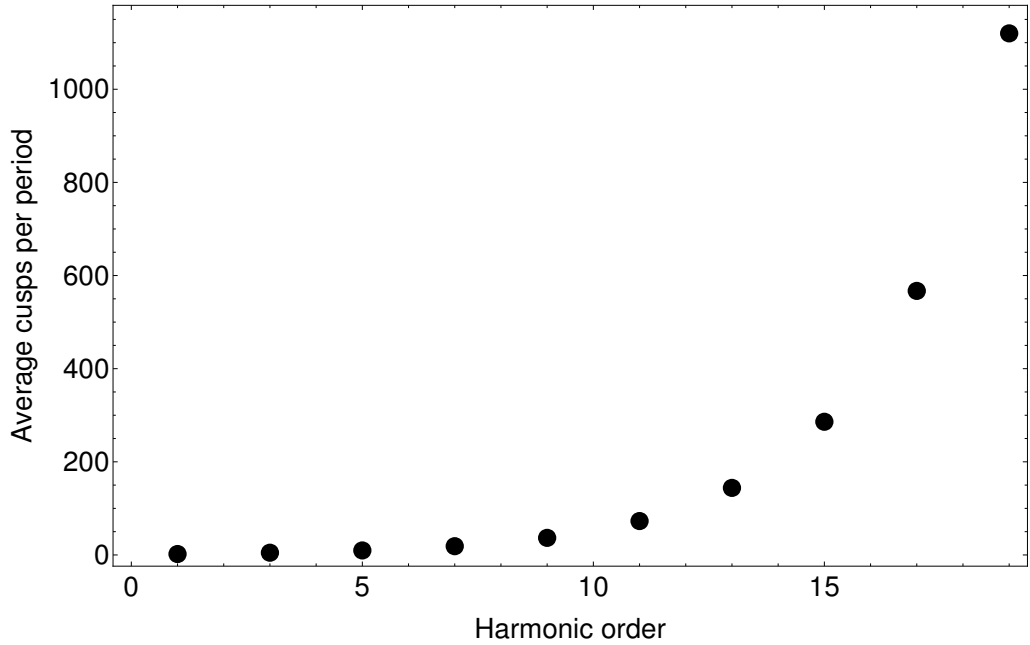

Figure 5.7.4: The average cusps per period vs the harmonic order of the parent loop. This Figure is obtained from the results of the algorithms in sections 5.7.5 and 5.7.6.

Using the least squares method, we find that the cusp numbers are related to the harmonic order via

$$\ln(c) = a + \ln(\beta)N \Rightarrow c = e^a \beta^N \tag{5.7.35}$$

where $a = 0.348$ and $\ln(\beta) = 1.42$. As expected from our discussion in the last paragraph of section 5.7, the behaviour is exponential and the exponential base is less than 2.

| Harmonic order | average stable loops | average cusps per period |
|:---:|:---:|:---:|
| 1 | 1 | 2.00 |
| 3 | 1.6 | 4.8 |
| 5 | 2.76 | 9.68 |
| 7 | 5.07 | 18.8 |
| 9 | 9.68 | 36.7 |
| 11 | 19.2 | 72.8 |
| 13 | 38.0 | 144 |
| 15 | 75.2 | 286 |
| 17 | 154 | Unknown |

Table 5.7.5: The average number of stable loops and cusps per period calculated using the analytical method described in section 5.7.5, for loops of harmonic order 1 to 17. Note that the unknown value at harmonic order 17 is due to very long execution time.

| Harmonic order | average stable loops | average cusps per period |
|:---:|:---:|:---:|
| 1 | 1±0 | 2.00±0 |
| 3 | 1.60±0.49 | 4.8±0.98 |
| 5 | 2.76±1.1 | 9.68±1.2 |
| 7 | 5.06±2.0 | 18.8±2.3 |
| 9 | 9.69±3.4 | 36.7±5.3 |
| 11 | 19.2±5.1 | 72.8±9.1 |
| 13 | 38.0±8.1 | 144±17.5 |
| 15 | 75.1±13.6 | 286±35.7 |
| 17 | 149±24.2 | 567±73.1 |
| 19 | 295±45 | $1.12 \ 10^3 \ \pm150$ |
| 51 | $1.71 \times 10^7 \pm 1.6 \times 10^6$ | $6.51 \times 10^7 \pm 6.2 \times 10^6$ |

Table 5.7.6: The average number of stable loops and cusps per period calculated using the Monte Carlo method described in section 5.7.6, for loops of harmonic order 1 to 19. We also present a single very high harmonic calculation $N = 51$. Note the impressive agreement between this and the analytical method shown in Table 5.7.5 for those loops that can be compared. To obtain these results we produced 300000 trees using the Monte Carlo method at each harmonic order. We also provide indicative results from $N = 51$ harmonic order parent loop, by producing 10 trees. We reduce the number of trees for this case due to the very high computation time.

### 5.7.8 Calculation of cusps per period from a cosmic string network

When we calculate signals from the network of cosmic strings, we will need some additional assumptions regarding the harmonic order distribution of the strings in the network to implement our results. In particular, regarding the gravitational wave signal, the number of cusp events per unit spacetime volume [4]

$$\nu(t) = \frac{c n_l(t)}{T_l},$$

(5.7.36)

where $c$ is the average number of cusps per period, $T_l = l/2 = \alpha t$, and $n_l(t)$ is the number density of loops, defined in equation (5.6.42). It has been assumed in previous works that $c$ is roughly unity. To calculate the value of $c$ using our results from the toy model we studied in the previous sections, we will need to assume the distribution of the harmonic order of the loops in the unit volume. It appears that not much is known regarding the harmonic order distribution of the loops chopped from the long string network. Given this, we will follow a conservative approach and assume that low harmonic loops dominate over high harmonic loops when created from the long string network, and assume that the density of loops of a given harmonic order drops with the harmonic order. We will also take into account only loops with harmonic order from $N = 1$ to $N = 7$. The reason for this is that our results are more reliable for low order harmonic loops (see discussion in section 5.7.9).

Given the above mentioned assumptions, we will aim to split the harmonic order distribution of loops in a cosmic string network volume into first, third, fifth and seventh order harmonic loops, following a discrete distribution. The simplest and most straight forward way to achieve this is to assume a uniform distribution of the aforementioned harmonics. We also present this as a case where higher harmonic are more frequently encountered compared to the discussion above that higher order harmonics should be less frequent. We will compare this result with other distributions that follow the above reasoning. In this case, the average cusp number per period from a unit spacetime volume will be the average of the values for the cusps per period presented in Table 5.7.5, for harmonic order of the parent loop from $N = 1$ to $N = 7$. We will denote this as $c$, following [4],[10] and it obtains the value, $c = 8.82$. We will use this result in section 6.6, to modify the current assumption for the cusps per period when estimating the amplitude of the gravitational waves originating from cusps on cosmic strings. The current assumption for the cusps per period from a unit volume is taken to be $c = 1$ or $c = 0.1$ [4]-[3].

Since the way the loops are distributed in terms of their harmonic order is currently unknown, we should also take several cases for the distribution of loops of harmonic order $N = 1$ to $N = 7$, and see how the average number of cusps per period $c$ changes for a range of distributions. The only requirement for the discrete distribution that will describe the distribution of the harmonics is that it should drop quickly with the harmonic order. A discrete distribution that satisfies the above requirement is Benford's law. A set of numbers $P(d)$ (in our case the percentage of loops of a given harmonic) given by Benford's law satisfy

$$P(d) = \log_b \left( 1 + \frac{1}{d} \right).$$
(5.7.37)

Note that we do not take into account the physical meaning of Benford's law, mentioned in section 2.5.1, but only its property that the value of $P(d)$ drops with $d$, which is our preferred assumption for the behaviour of the harmonics of the parent loops. The parameter $b$ will be fixed to $b = 5$, since we are interested to take into account the harmonics $N = 1$ to $N = 7$ in the string network. Then, we have that $P(1) = 0.43$, $P(2) = 0.25$, $P(3) = 0.18$ and $P(4) = 0.14$, which satisfy

$$\sum_{d=1}^{4} P(d) = 1,$$
(5.7.38)

as expected. Then, we find that for this distribution of the parent loop harmonics the average cusps per period from a unit volume is $c = 6.43$. Note that this distribution provides us with only one choice of values for the distribution of harmonics of the parent loops.

Another distribution which provides us with possible values of the parent loop harmonics (that decrease as the harmonic order increases) is the geometric dis-

---

[10]Note that the quantity $c$ is denoted as $N_c$ in [3]

tribution

$$G(p, k) = (1 - p)^k p \tag{5.7.39}$$

where $k = \{0, \mathbb{Z}^+\}$ and $0 < p \leq 1$, which was discussed in section 2.5.1. In the above, we can fix the value of $p$ and obtain the values $G(p, k)$ for harmonics 1 to infinity, since $k$ can obtain the value of any natural number, with

$$\sum_{k=0}^{\infty} G(p, k) = 1. \tag{5.7.40}$$

This distribution has the advantage that gives us the choice of obtaining different values of the density of parent loops of a given harmonic, by using different values for $p$, unlike the even distribution and Benford's law. However, since we would like to focus on loops of harmonics up to $N = 7$, we will choose values of $p$, such that higher order parent loops are scarce in the unit volume. To choose such values, let us first produce the formula that calculates $c$ for this distribution. This is given by

$$c = \sum_{k=0}^{\infty} G(p, k) c_{2k+1} \tag{5.7.41}$$

where the subscript $2k+1$ corresponds to the harmonic order of the parent loop, which obtains odd values only, and the values of $c_{2k+1}$ are given in Table 5.7.5. Using equation (5.7.35), we find that

$$c = \sum_{k=0}^{\infty} (1 - p)^k p e^a \beta^{2k+1}. \tag{5.7.42}$$

By summing the above, we find that

$$c = p e^a \beta \frac{1}{1 + (p - 1)\beta^2}, \tag{5.7.43}$$

for $(1 - p)\beta^2 < 1$. Note that the sum is of the form of a geometric series, i.e. the sum of numbers in a geometric progression $x^k$, with $x = (1 - p)\beta^2$. This sum is known to converge if and only if the above inequality holds. Since the value of $\beta$ has been found using the least square method, we find that the above sum converges if and only if $p > 0.94$. For example, for $p = 0.95$, we find that $c = 38.9$. However, we are interested in the values of the sum from $k = 0$ (which corresponds to $N = 1$) to $k = 3$ ($N = 7$), and we would like to consider higher harmonic contributions negligible by minimizing their appearance in the distribution. Therefore, regardless of the sum convergence, we can examine the behaviour of the truncated sum for any value of $p$. As we can see in Figure 5.7.7, the values of $c$ with regards to $p$, calculated using the sum of the first 4 terms of (5.7.42), lie in the interval $[0, 5.03]$. The maximum value of the plot corresponds to $p = 0.35$, which has a relatively large percentage of high harmonic loops $\sum G(0.35, k > 3) = 0.18$, which are neglected. Preferably, we would like a value of $p$ that contains mainly lower harmonic order loops (up to $N = 7$), such that it follows the rule we have set, namely that the harmonic order loop distribution drops quickly with $N$. For $p = 0.6$, the percentages of the parent loops of each harmonic order are given by, $G(0.6, 0) = 0.6$, $G(0.6, 1) = 0.24$, $G(0.6, 2) = 0.1$, $G(0.6, 3) = 0.04$. Finally, the rest of the harmonic order loops sum to a total percentage of $\sum G(0.35, k > 3) = 0.02$, which we will consider negligible. For this case, the value of cusps per period is $c = 4.0$. Note that around this value of $p$, the value of $c$ is relatively stable, and the aforementioned value represents the distribution well.

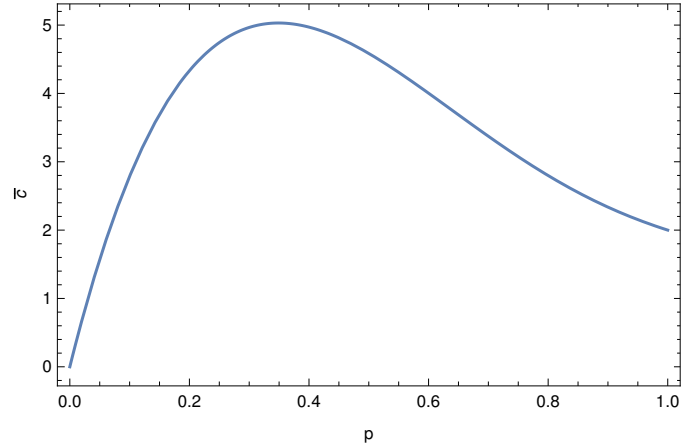

Figure 5.7.7: The values of the average cusps per period $c$ from a distribution of parent loops given by the geometric distribution $G(p, k)$, with respect to the value of $p$, and with $k$ ranging from 0 to 3.

Given the above calculations, the three different distributions we used (the uniform distribution, Benford's law and the geometric distribution) provide a value of $c$ evaluated over the lowest four harmonics that does not vary significantly from distribution to distribution. In the case of the uniform distribution the harmonic order of the loops is equally distributed, while in the other two cases the percentage of loops of a given harmonic drops as the harmonic order increases. Distributions where the percentage of loops increases as their harmonic order increases will not be taken into account, as intuition suggests that should not be the case, although they can not be rejected from our current knowledge of the cosmic string network.

### 5.7.9 Issues and possible improvements of the model

A main issue of the toy model is the fact that the emergence of kinks during the self-intersection process of the loops is ignored. As we discussed in section 5.3.3 (also discussed in [1],[9]) the occurence of cusps is supressed with the presence of kinks on the Kibble-Turok sphere. Therefore, this toy model gives an enhanced number of cusps per period output, which also becomes more intense as the harmonic order of the loops increases. A way to counter this enhancement of cusp events would be to add a suppresion factor which would account for the kink effect on the cusp production, see for example [1]. Another factor that suppresses cusps is the gravitational backreaction around cusps and kinks, which rounds off kinks and makes cusps weaker [122], [123]. This phenomenon could also pose a suppression factor for the cusp occurrence on the evolution of the loops. However, it could have a counter effect towards enhancement of the cusp number due to the rounding of the kinks.

Another issue is the percentage of the lifetime of the $N_i$ harmonic non stable loop, which we assumed to be half of the period. Could a better assumption work? An idea would be to check how many solutions equation (5.3.62) has, i.e. how many possible self-intersections could happen, and split the lifetime to that fraction of the period. Also, the assumption of the harmonic order of the daughter loops always following the rule to be minus 2 of the loop that chopped could change with this assumption, which would allow the loop to chop into smaller loops from earlier stages of evolution, which would affect also the assumption of the length of the daughter loops. For a numerical simulation of a loop chopping see [1]. In their website they provide a video of the evolution. The above comment explains why we used our results only to harmonic order $N = 7$ of the parent loop, and discarded higher harmonic order results. It is our belief that at higher harmonics a parent loop would quickly chop into small loops and it should be tested to what extend the results are impacted by the rule that harmonics decrease by 2 at each step of the evolution. This could lead to a significant change in the values of stable loops and cusps that would be calculated.

The restriction of the odd-harmonics only loops implies that the strings produced cannot obtain any even harmonic values, which fixes the evolution of the loops to a particular shape (although still with a large number of free parameters). Given the freedom of any harmonic order, the binary trees would have different branches. Furthermore, the results we find could be heavily dependent on the string family that one assumes. For example, in [117], two relatively similar families of strings are assumed, yet the difference in the results is significant.

Also, recall that we do not take into account the kinetic energy lost every time a loop self-intersects and forms two daughter loops. If this energy is a significant percentage of the system energy, it could cause the system to diminish faster.

Furthermore, the main essence of this model, that renders it analytical, is the stochastic behavior of the system of loops and their odd-harmonic string behavior at any stage of evolution. This kind of assumption could be quite restrictive for such a complicated system and it could prove inadequate compared to a potential numerical simulation that follows the exact motion of the strings on a grid at every time interval of the system evolution. Such models have been developed, see for example [117], but they are not directly comparable to our model, due to the different string configuration assumed. However, a numerical model developed using the odd-harmonic string and calculating the stable loops and the total cusps produced would be comparable with this toy model, and it could be a way of testing our assumptions and results.

The assumption of the harmonic distribution in the loops chopped-off the long string network in section 5.7.8, is completely fixed, due to the limitations of our model, that the maximum harmonic of the loops chopped is $N = 7$, and the fact that we assumed that the low harmonics would dominate, aiming to achieve a conservative approach. It is our belief that the actual value for $c$ could vary by orders of magnitude and it is sensitive to the assumptions made.

Finally, we have assumed that the loops can be described in flat spacetime. However, in reality they live in an FLRW spacetime, where their fate is expected to differ from the one predicted in Minkowski spacetime. Therefore, it would be important to check what the effect of curved spacetime is on the lifetime of the loops and their periods, since the stable loops are assumed to oscillate for much longer than one Hubble time in the calculations of the toy model.

In general, the aim of this toy model is to give a general idea of how the value of $c$, the cusps per period from a spacetime volume of loops, could be enhanced if we took into account the high-harmonic loop contribution to the cusp production in the "one-scale" string network, and to probe a stochastic evolution of loops. Although the amount of assumptions needed to achieve this and the fixed values required to calculate $c$ render it a toy model, we believe that a key feature has emerged which will hold in more realistic cases. Namely $c > 1$ appears a robust result and the fact we obtain $c \sim 4$ for the most conservative case we can suggests that in reality it could be considerably more. As we will see the value of $c$ has a big impact on the gravitational wave bursts from a network of strings.

---

# Gravitational waves from cosmic strings

---

It is known that cosmic string loops are efficient gravitational wave emitters, with a spectrum covering a wide range of frequencies (see chapter 7.5 in [9]). In this chapter we will focus on the gravitational waves emitted from sharp gravitational wave bursts (GWBs) emanating from cusps and kinks on cosmic string loops, as studied in [4].

In what follows we will restore the units of $u$ and $v$ so that they range from 0 to $l$, instead of 0 to $2\pi$, as was assumed in chapter 5 for our numerical computations. It was useful for the numerical computations to use normalized units, but here we will focus on natural quantities, and it is preferable to restore the units. However, we will use the normalized values for $|\vec{a}''|$ and $|\vec{b}''|$ that were calculated numerically in the previous chapter, and we will denote them with a hat when they appear.

# 6.1 Factorization of the energy momentum tensor

In section 3.2.3, we found that the energy momentum tensor of a periodic source (in our case a cosmic string loop) in Fourier space is

$$T^{\nu\kappa}(\omega_m, \vec{k}_m) = \frac{1}{T_l} \int_0^{T_l} dt \int d^3\vec{x}\, e^{i\left(\omega_m t - \vec{k}_m \cdot \vec{x}\right)} T^{\nu\kappa}(t, \vec{x}), \qquad (6.1.1)$$

where $T_l = l/2$ is the period of the loop. In the following, we assume that our observer is in the "local wave zone", i.e. at a distance from the source such that the spacetime can be approximated to be Minkowski, with metric $\eta_{\mu\nu}$. Recall that

$$k^\nu = (\omega_m, \vec{k}_m) = (\omega_m, \omega_m \vec{n}) \qquad (6.1.2)$$

is the 4-frequency of the gravitational waves moving in the $\vec{n}$ direction, which obtains discrete values over all the harmonics $\omega_m = m\omega_1$, with $m \in \mathbb{Z} - \{0\}$ and $\omega_1 = 2\pi/T_l$ the fundamental angular frequency of the string. We can assume $m$ to take only positive values, which implies that we must add the complex conjugate quantity in (6.1.1), as we did in section 3.2.2. Substituting in equation (6.1.1) the expression for the cosmic string loop energy-momentum tensor $T^{\nu\kappa}(t, \vec{x})$ in the conformal time gauge (5.3.34), we find that

$$T^{\nu\kappa}(\omega_m, \vec{k}_m) = \frac{\mu}{T_l} \int_0^{T_l} d\tau \int_0^l d\sigma \left( -\dot{X}^\nu \dot{X}^\kappa + \vec{X}^{\nu'} \vec{X}^{\kappa'} \right) e^{-ik_m \cdot X}, \qquad (6.1.3)$$

where $k_m \cdot X = \eta_{\mu\nu} k_m^\mu X^\nu$. We will write the energy momentum tensor (6.1.3) in terms of the left- and right-movers, defined in (5.3.22),

$$T^{\nu\kappa}(\omega_m, \vec{k}_m) = \frac{\mu}{T_l} \int_0^{T_l} d\tau \int_0^l d\sigma \frac{1}{2} \left( \partial_u a^\nu \partial_v b^\kappa + \partial_u a^\kappa \partial_v b^\nu \right) e^{-ik_m \cdot X}. \qquad (6.1.4)$$

To deduce the above result, we have used the chain rule

$$\partial_\tau = \partial_v - \partial_u, \; \partial_\sigma = \partial_v + \partial_u. \qquad (6.1.5)$$

We can also write equation (6.1.4) as

$$T^{\nu\kappa}(\omega_m, \vec{k}_m) = \frac{\mu}{Tl} \int_0^{Tl} d\tau \int_0^l d\sigma a^{(\nu'} b^{\kappa)'} e^{-\frac{i}{2}(k_m \cdot a + k_m \cdot b)}, \qquad (6.1.6)$$

using the notation introduced in definition (2.1.5). In the above, the primes denote the derivative with respect to the unique variable of the right- and left-movers, $u$ for $a^\mu$ and $v$ for $b^\mu$ respectively, unlike in equation (6.1.3) where the primes denote partial differentiation with respect to $\sigma$ of the two variable function $X^\mu$. We can further simplify the energy-momentum tensor by changing coordinates from the $\tau - \sigma$ plane to the $u - v$ plane. Using formula (5.3.15), we find that

$$T^{\nu\kappa}(\omega_m, \vec{k}_m) = \frac{\mu}{l} \int_0^l du \int_0^l dv a^{(\nu'} b^{\kappa)'} e^{-\frac{i}{2}(k_m \cdot a + k_m \cdot b)}. \qquad (6.1.7)$$

The above result can be written in the factorized form

$$T^{\nu\kappa}(\omega_m, \vec{k}_m) = \frac{\mu}{l} I_+^{(\mu} I_-^{\nu)}, \qquad (6.1.8)$$

where

$$I_+^\mu = \int_0^l dv \, b^{\mu'}(v) e^{-\frac{i}{2} k_m \cdot b} \qquad (6.1.9)$$

and

$$I_-^\mu = \int_0^l du \, a^{\mu'}(u) e^{-\frac{i}{2} k_m \cdot a}. \qquad (6.1.10)$$

Equation (6.1.8) gives us a very convenient form for the energy-momentum tensor that will help us look into its high-frequency behaviour [4], [54].

## 6.1.1 High-frequency behaviour of $T_{\mu\nu}$

It can be proved that integrals of the form $I(m) = \int_0^l d\sigma f^\mu(\sigma) e^{-im\phi(\sigma)}$ tend to zero faster than any negative power of m, as m goes to $\infty$, if the following conditions hold [4]

- Condition 1: $f^\mu(\sigma)$ and $\phi(\sigma)$ are smooth (i.e. $C^\infty$) in $[0, l]$.

- Condition 2: $\phi(\sigma)$ has no saddle points in $[0, l]$, i.e. $\phi'(\sigma) \neq 0$.

Taking this theorem into account, it is straightforward from equations (6.1.8)-(6.1.10), to show that $T_{\mu\nu}$ tends to zero faster than any negative power of n as $\omega_n = n\omega_1 \to \infty$ when the above conditions hold. However, the conditions for the suppression of the integral in high frequencies can be violated if the loop exhibits cusps and/or kinks.

Let us define $\phi_+ = k_1 \cdot b(v)$ and $\phi_- = k_1 \cdot a(u)$, where $k_1^\mu = (\omega_1, \vec{k}_1)$ is the fundamental 4-frequency of the gravitational waves emitted from the cosmic string. The saddle points in $\phi_-$ occur when $k_1 \cdot a(u)' = 0$, i.e. when the vectors $k_1$ and $a(u)'$ are orthogonal. The vectors $k_1$ and $a(u)'$ are both lightlike vectors, in particular $k_1 \cdot k_1 = 0$ from (6.1.2) and $a(u)' \cdot a(u)' = 0$ from the Virasoro conditions (5.3.25). It holds that two lightlike vectors are orthogonal if and only if they are proportional (see Corollary 1.1.5 in [124]). Therefore, the saddle points of $\phi_-$ occur when $k_1$ and $a(u)'$ are proportional. Similarly, $\phi_+$ exhibits saddle points when $k_1$ and $b(v)'$ are proportional.

A discontinuity in some derivative of $a(u)$ leads to a slow decay of $I_-$ as well, and similarly a discontinuity in derivatives of $b(v)$ leads to a slow decay of $I_+$. A frequently occurring cause of discontinuities on cosmic strings is the presence of kinks, i.e. the presence of discontinuities in the first derivatives of the left- and right-movers, $b^\mu(v)'$ and $a^\mu(u)'$ (we discussed kinks in section 5.3.5).

A slow decay of $T_{\mu\nu}$ at high frequencies occurs only if both $I_+$ and $I_-$ violate at least one of conditions 1 or 2. If one of the $I_\pm$ integrals does not violate any of the conditions, it will decay exponentially fast with m, and the product $T^{\mu\nu} \propto I_+^{(\mu} I_-^{\nu)}$ will also decay exponentially fast. We can see that condition 2 is violated for both $I_-$ and $I_+$ at the presence of cusps on the cosmic string. We have defined cusps as points on the cosmic string that momentarily reach the speed of light, $\dot{X}(t, \sigma) = 1$ (see section 5.3.3). By differentiating equation (5.3.22) with time, we see that $\dot{X}$ reaches the speed of light when $\vec{a}' = -\vec{b}'$. When the cusp condition holds both $a(u)'$ and $b(v)'$ are proportional to $k_1$, and a double saddle point occurs.

There are multiple ways to obtain a power law decay of $T_{\mu\nu}$ as $m \to \infty$. We will focus on the two cases which give the slowest decay of the energy-momentum tensor. These are (i) the case where a cusp occurs, and condition 2 is violated for both integrals $I_\pm$, and (ii) the case where $\vec{k}_1$ is proportional to $a(u)'$, i.e. condition 2 is violated for $I_-$, and $\vec{b}'$ has one or more kinks, i.e. condition 1 is violated for $I_+$ (or, equivalently, condition 2 violated for $I_+$ and condition 1 is violated for $I_-$) [4]. The gravitational waves produced when cases (i) and (ii) occur will be in the form of sharp bursts along certain directions. In case (i) these bursts will be along a finite set of directions, i.e. along the directions of the cusps formed on the cosmic string, while in case (ii) along the direction of the kink, as it propagates on the cosmic string with the speed of light.

Along these directions, the Fourier components of the asymptotic waveform, defined in (3.2.44) will decay as $m \to \infty$ in a power law manner. Since we will be looking into the high frequency behaviour of $T_{\mu\nu}$ along these directions, we can approximate the sum in the definition of the asymptotic waveform (3.2.44) by an integral

$$\sum_m \simeq \int dm = \frac{l}{2}\int df, \tag{6.1.11}$$

using the fact that the discrete frequency $f_m$ tends to a continuous frequency f as $N \to \infty$, and

$$m = \frac{f_m}{f_1} = f_m T_l = \frac{l}{2}f_m. \tag{6.1.12}$$

Therefore, the asymptotic waveform (3.2.44) becomes

$$\kappa_{\mu\nu}(t-r,\vec{n}) = 2Gl \int df e^{-i\omega(t-r)} T_{\mu\nu}(\vec{k},\omega) = \frac{Gl}{\pi}\int d\omega e^{-i\omega(t-r)} T_{\mu\nu}(\vec{k},\omega) \tag{6.1.13}$$

along the directions of cusps and kinks on cosmic string loops. The integral in the above expression ranges over negative frequencies as well, it being understood that if we restrict the frequencies to positive values only we should add the complex conjugate of $\kappa_{\mu\nu}$ in (6.1.13).

In terms of the logarithmic Fourier transform $\tilde{\kappa}(f) = |f|\kappa(f)$ (defined in (2.3.18)), we find that

$$\kappa(t-r,\vec{n}) = \int \frac{1}{|f|} df e^{-i\omega(t-r)} \tilde{\kappa}(f,\vec{n}). \tag{6.1.14}$$

Comparing the above with (6.1.13), we find that the logarithmic Fourier transform of the asymptotic waveform in terms of the energy-momentum tensor is

$$\tilde{\kappa}_{\lambda\nu}(f,\vec{n}) = 2Gl|f|T_{\lambda\nu}(\omega,\vec{k}). \tag{6.1.15}$$

Using (6.1.8), we reach the result

$$\tilde{\kappa}_{\lambda\nu}(f, \vec{n}) = 2G\mu|f|I_+^{(\lambda}I_-^{\nu)}.$$

(6.1.16)

Similarly, in terms of the Fourier transform, we find that

$$\kappa_{\lambda\nu}(f, \vec{n}) = 2G\mu I_+^{(\lambda}I_-^{\nu)}.$$

(6.1.17)

Assuming that this asymptotic approach can approximate the radiation emission of the cosmic string loop even to low harmonics, we can use it as a way to analytically determine the gravitational wave emission from a cusp and kink. In the following we will consider the high-frequency behaviour of the energy-momentum tensor in the case of the cusp. Our goal is to find the contribution of the amplitude of the gravitational wave emitted from cusps on the cosmic string loop.

## 6.2   The asymptotic waveform from cusps

In this section we will focus on the gravitational waves emitted from sharp GWBs emanating from cusps on cosmic string loops, studied in [4]. The calculations provided in this section follow their approach.

### 6.2.1   The asymptotic waveform in the direction of the cusp

In Section 5.3.4, we defined the luminal velocity of the string at the cusp

$$l^\nu = \left(1, \vec{n}^{(c)}\right) = -a^{\nu\prime}(u^{(c)}) = b^{\nu\prime}(v^{(c)}).$$

(6.2.1)

Substituting the Taylor expansion of $\dot{X}^\mu$ (derived in (5.3.51)-(5.3.52)) into equa-tions (6.1.9)-(6.1.10) we find

$$I_+^\nu = \int_{v_0}^{v_0+l} dv \ \left( l^\nu + b_{(c)}^{\nu''} v + \frac{1}{2} b_{(c)}^{\nu''} v^2 \right) e^{-\frac{i}{2} k_m \cdot b}, \tag{6.2.2}$$

$$I_-^\nu = \int_{u_0}^{u_0+l} du \ \left( -l^\nu + a_{(c)}^{\nu''} u + \frac{1}{2} a_{(c)}^{\nu''} u^2 \right) e^{-\frac{i}{2} k_m \cdot a}. \tag{6.2.3}$$

The quantities $a_{(c)}^{\nu''}$ and $b_{(c)}^{\nu''}$ denote the second derivatives of the left- and right-movers evaluated at the cusp. The constants $v_0$ and $u_0$ appear in the integral limits because of our shift of the cusp coordinates to the origin, $\left( u^{(c)}, v^{(c)} \right) = (0,0)$. It could be expected that the dominant term of $I_\pm^\nu$ would be $l^\nu$. However, this term can be removed by applying a coordinate transformation, and it does not correspond to a physical effect (see Appendix A.2). Following this and removing all gauge terms (i.e. terms that are removable via coordinate transformations), we find that,

$$I_+^{(\lambda} I_-^{(\nu)} = d_+^{(\lambda} d_-^{\nu)}, \tag{6.2.4}$$

where

$$d_+^\nu = b_{(c)}^{\nu''} \int_{v_0}^{v_0+l} dv \ v \, e^{-\frac{i}{2} k_m \cdot b} \tag{6.2.5}$$

and

$$d_-^\nu = a_{(c)}^{\nu''} \int_{u_0}^{u_0+l} du \ u e^{-\frac{i}{2} k_m \cdot a}. \tag{6.2.6}$$

Since we are calculating the GW that is emitted from the cusp, the GWB is centred around the 4-frequency

$$k_m^\mu = m \omega_1 l^\mu, \tag{6.2.7}$$

with $m \in \mathbb{Z} - \{0\}$. Using equations (5.3.59)-(5.3.60) we find that

$$k_m \cdot b = m \omega_1 l_\nu b^\nu(v) = -\frac{1}{6} m \omega_1 \left( b_{(c)}^{\nu''} \right)^2 v^3, \tag{6.2.8}$$

$$k_m \cdot a = l_\nu a^\nu(u) = \frac{1}{6} m \omega_1 \left( a_{(c)}^{\nu''} \right)^2 u^3. \tag{6.2.9}$$

Substituting the above in the expressions for $d_\pm^\nu$, we obtain

$$d_+^\nu = b_{(c)}^{\nu''} \int_{v_0}^{v_0+l} dv \ v \, \exp\left[ -\frac{i}{12} m \omega_1 \left( b_{(c)}^{''} \right)^2 v^3 \right] \tag{6.2.10}$$

and

$$d''_- = a^{\nu''}_{(c)} \int_{u_0}^{u_0+l} du\, u \exp\left[\frac{i}{12} m\omega_1 \left(a''_{(c)}\right)^2 u^3\right]. \tag{6.2.11}$$

We can introduce a change of variables

$$y = B_+ v \tag{6.2.12}$$

with

$$B_+ = \left[\frac{1}{12}|m|\omega_1 \left(b''_{(c)}\right)^2\right]^{1/3} \tag{6.2.13}$$

and

$$z = B_- u \tag{6.2.14}$$

with

$$B_- = \left[\frac{1}{12}|m|\omega_1 \left(a''_{(c)}\right)^2\right]^{1/3}. \tag{6.2.15}$$

Applying the above choice of variables, the integrals (6.2.10) and (6.2.11) become

$$d'_+ = b^{\nu''}_{(c)} \int_{y_1}^{y_2} dy\, \frac{y}{B_+^2} e^{\pm i y^3} \tag{6.2.16}$$

and

$$d'_- = a^{\nu''}_{(c)} \int_{z_1}^{z_2} dz\, \frac{z}{B_-^2} e^{\pm i z^3} \tag{6.2.17}$$

respectively. Note that the plus and minus sign in the above expressions arises because we take into account both positive and negative frequencies. It can be proved that for integrals of the form

$$I = \int_{x_1}^{x_2} dx\, e^{\pm i x^3} x \tag{6.2.18}$$

most of the contribution comes from an interval around zero.[1] Therefore, the integral $I$ can be calculated by setting $y_1 = -\infty$ and $y_2 = \infty$, which is evaluated analytically to have the exact pure imaginary value

$$I = \pm i \frac{2\pi}{3\Gamma(\frac{1}{3})}, \tag{6.2.19}$$

---

[1] More details on the calculation of the above integral can be found in [125]. In this paper, the case of $x_1$ or $x_2$ lying close to zero is also discussed, as well as the case where the interval between $x_1$ and $x_2$ does not include zero, which can occur for cosmic superstrings. Note that we have fixed the cusp to be at the point $(u^{(c)} = 0, v^{(c)} = 0)$.

where $\Gamma(x)$ is the Euler gamma function. The plus sign corresponds to positive frequencies and the minus sign to negative frequencies.

Therefore, the logarithmic Fourier transform of the asymptotic waveform in the direction of the cusp can be calculated

$$\tilde{\kappa}^{\lambda\nu}(f_m, \vec{n}_{(c)}) = 2G\mu|f_m|b_{(c)}^{(\lambda''} a_{(c)}^{\nu)''} \left( \int_{z_1}^{z_2} dz \, e^{\pm iz^3} \frac{z}{B_-^2} \right) \times \tag{6.2.20}$$

$$\times \left( \int_{y_1}^{y_2} dy \, e^{\pm iy^3} \frac{y}{B_+^2} \right) \tag{6.2.21}$$

$$= \frac{2G\mu|f|b_{(c)}^{(\lambda''} a_{(c)}^{\nu)''}}{B_+^2 B_-^2} I^2 = -\frac{2G\mu|f_m|}{\left(\frac{1}{12}|m|\omega_1\right)^{4/3}} \frac{b_{(c)}^{(\lambda''} a_{(c)}^{\nu)''}}{|\vec{b}_{(c)}''|^{4/3}|\vec{a}_{(c)}''|} I^2 \tag{6.2.22}$$

$$= -\frac{G\mu}{(2\pi|f_m|)^{1/3}} \frac{b_{(c)}^{(\lambda''} a_{(c)}^{\nu)''}}{|\vec{b}_{(c)}''|^{4/3}|\vec{a}_{(c)}''|} \frac{(12)^{4/3}4\pi}{\left(3\Gamma\left(\frac{1}{3}\right)\right)^2} \tag{6.2.23}$$

for positive and negative frequencies. Note that for our choice of gauge,

$$|\vec{a}''|^2 = \left(a^{\mu''}\right)^2 \geq 0, \tag{6.2.24}$$

since $a^0 = -u$. Similarly,

$$|\vec{b}''|^2 = \left(b^{\mu''}\right)^2 \geq 0. \tag{6.2.25}$$

If we define the constant

$$C = \frac{(12)^{4/3}4\pi}{\left(3\Gamma\left(\frac{1}{3}\right)\right)^2} \tag{6.2.26}$$

we reach the final expression

$$\tilde{\kappa}^{\lambda\nu}(f_m, \vec{n}_{(c)}) = -C \frac{G\mu}{(2\pi|f_m|)^{1/3}} \frac{b_{(c)}^{(\lambda''} a_{(c)}^{\nu)''}}{|\vec{b}_{(c)}''|^{4/3}|\vec{a}_{(c)}''|^{4/3}}. \tag{6.2.27}$$

In the calculations above, we have set the arrival time of the burst $t_c$ to zero. By taking the inverse logarithmic Fourier transform and reinserting $t_c \neq 0$,

$$\kappa(t) \propto \int df \frac{1}{|f|} \frac{1}{|f|^{1/3}} e^{-2\pi i f(t-t_c)}, \tag{6.2.28}$$

we find that in the time domain the asymptotic waveform at the direction of the

cusp is

$$\kappa(t) \propto |t - t_c|^{1/3}. \tag{6.2.29}$$

Although the high frequency waveform is zero at $t = t_c$, what distinguishes it from the low frequency waveform is its spiky shape. The curvature associated to the high frequency waveform is proportional to $\propto |t - t_c|^{-5/3}$ (see [4]), which also shows the spiky behaviour of the GWB from cusps.

## 6.2.2 The asymptotic waveform around the cusp

In the previous section we calculated the asymptotic waveform for an observer that lies exactly on the direction of the cusp. Here, we will assume that our observer does not lie on the direction of the GWB emission, but close to it. For the calculation of the asymptotic waveform around the cusp we follow [4].

Let us define the angle, $\theta$, between the direction of the observer, $\hat{k}^\mu = k^\mu/|k^\mu| = (1, \vec{n})$, and the 4-velocity of the cusp, $l^\mu = (1, \vec{n}_{(c)})$. We also define the difference between the direction of the observer and of the cusp velocity

$$\delta^\mu = l^\mu - \hat{k}^\mu = (0, \vec{n}^{(c)} - \vec{n}), \tag{6.2.30}$$

which is spacelike and has magnitude

$$\delta^2 = \delta^\mu \delta_\mu = \left(\vec{n}^{(c)}\right)^2 + \vec{n}^2 - 2\vec{n}^{(c)} \cdot \vec{n} = 2 - 2\cos\theta \simeq \theta^2. \tag{6.2.31}$$

To reach the above result we have used the fact that $\vec{n}$ and $\vec{n}^{(c)}$ are both unit vectors and $\cos\theta \simeq 1 - \theta^2/2$ for small angles. Recall that (see Appendix A.2)

$$I_\pm^\lambda = a_\pm k^\lambda + d_\pm^\lambda, \tag{6.2.32}$$

where $a_\pm$ are constants. We can remove the gauge terms (i.e. terms that are proportional to $l^\mu$, see Appendix A.2), and using equations (6.2.2)-(6.2.4), we find that

$$d_+^\lambda(\theta) = \int_{v_0}^{v_0+l} dv \left(\delta^\lambda + b_{(c)}^{\lambda''} v + \cdots\right) e^{-\frac{i}{2}k \cdot b}, \tag{6.2.33}$$

$$d_-^\lambda(\theta) = \int_{u_0}^{u_0+l} du \left( \delta^\lambda + a_{(c)}^{\lambda''} u + \cdots \right) e^{-\frac{i}{2} k \cdot a}. \tag{6.2.34}$$

We can also simplify the phase terms of the integrals (6.2.33)-(6.2.34). We use the Taylor expansions of $a^\mu$ and $b^\mu$ around the cusp (calculated in equations (5.3.49)-(5.3.50)) to find that,

$$k \cdot a = -(k \cdot l)u + \frac{1}{2} \left( k \cdot a_{(c)}'' \right) u^2 + \frac{1}{6} \left( k \cdot a_{(c)}''' \right) u^3 \tag{6.2.35}$$

and

$$k \cdot b = (k \cdot l)v + \frac{1}{2} \left( k \cdot b_{(c)}'' \right) v^2 + \frac{1}{6} \left( k \cdot b_{(c)}''' \right) v^3. \tag{6.2.36}$$

By definition, $k^2 = 0$ and $l^2 = 0$. Therefore,

$$k \cdot l = -\frac{1}{2}(k-l)^2 = -\frac{1}{2}\delta^2 \simeq -\frac{1}{2}\theta^2. \tag{6.2.37}$$

Taking into account the differentiated Virasoro conditions (5.3.55)-(5.3.55) as well, we find that

$$k \cdot a = \frac{1}{2}\theta^2 u - \frac{1}{2} \left( \delta \cdot a_{(c)}'' \right) u^2 + \frac{1}{6} \left[ \left( a'' \right)^2 - \delta \cdot a_{(c)}''' \right] u^3, \tag{6.2.38}$$

$$k \cdot b = -\frac{1}{2}\theta^2 v - \frac{1}{2} \left( \delta \cdot b_{(c)}'' \right) v^2 - \frac{1}{6} \left[ \left( b'' \right)^2 - \delta \cdot b_{(c)}''' \right] v^3, \tag{6.2.39}$$

In the above we can omit the terms $\delta \cdot a_{(c)}''' u^3$ and $\delta \cdot b_{(c)}''' v^3$ given their small magnitude, because $\delta$ is small and $u$ and $v$ are close to zero.

We can now see that the integrals $I_+(\theta)$ and $I_-(\theta)$, as $\theta$ tends to zero, tend to the values $I_+(0) = I_+$ and $I_-(0) = I_-$ (which were calculated along the direction of the cusp in subsection 6.2.1) as expected. In particular, since the magnitude of $\delta^\lambda$ is proportional to $\theta$ for small $\theta$, it approaches zero in the limit $\theta \to 0$. Also, the phase terms of the integrals (6.2.33)-(6.2.34) as $\theta \to 0$ become

$$k \cdot a = \frac{1}{6} \left( a_{(c)}'' \right)^2 u^2, \tag{6.2.40}$$

$$k \cdot b = \frac{1}{6} \left( b_{(c)}'' \right)^2 v^2, \tag{6.2.41}$$

which proves our statement, that $\lim_{\theta \to 0} I_+(\theta) = I_+$ and $\lim_{\theta \to 0} I_-(\theta) = I_-$, as we can see from equations (6.2.10)-(6.2.11).

To proceed with our discussion of the asymptotic waveform around the cusp, let us first recall the theorem mentioned at the beginning of the subsection 6.1.1, which implies that the integrals $I_{\pm}(\theta)$ are suppressed exponentially unless the first derivative of their phase terms becomes zero at one point at least (we do not consider the case of discontinuities). To check for solutions of the first derivative of the phase terms, we should differentiate them with respect to u for $I_{-}(\theta)$

$$(k \cdot a)' = \frac{1}{2}\theta^2 - \left(\delta \cdot a''_{(c)}\right) u + \frac{1}{2}\left(a''\right)^2 u^2, \qquad (6.2.42)$$

and with respect to v for $I_{+}(\theta)$

$$(k \cdot b)' = -\frac{1}{2}\theta^2 - \left(\delta \cdot b''_{(c)}\right) v - \frac{1}{2}\left(b''\right)^2 v^2. \qquad (6.2.43)$$

The equations

$$\left(k \cdot a_{(c)}\right)' = 0 \qquad (6.2.44)$$

$$\left(k \cdot b_{(c)}\right)' = 0 \qquad (6.2.45)$$

are both second degree polynomials. We can solve them by calculating their discriminant $\Delta$. The discriminant of equation (6.2.44) is $\Delta = \left(\delta \cdot a''_{(c)}\right)^2 - \theta^2 \left(a''\right)^2 = \delta^2 \left(a''\right)^2 (1 - \cos\phi) < 0$ in general. By $\phi$ we denote the angle between the 3-vectors $\vec{\delta}$ and $\vec{a}''$ (recall that both $\vec{\delta}$ and $\vec{a}''$ have zero time component). This implies that there are no saddle points around the cusp on the right-moving wave, and therefore $I_{-}$ is suppressed exponentially. Similarly, we find that there are no saddle points on the left-moving wave either, and the integral $I_{+}$ is also suppressed exponentially around the cusp.

We can also take a qualitative approach to provide intuition on the value of the angle around the cusp where the GW signal starts to decay exponentially. Following the same reasoning behind the reparametrization of subsection 6.2.1, we can define

$$\varepsilon_{-} = \frac{\theta B_{-}}{|a''_{(c)}|} = \theta \left(\frac{|m|\omega_l}{12|a''_{(c)}|}\right)^{1/3} \qquad (6.2.46)$$

$$\varepsilon_{+} = \frac{\theta B_{+}}{|b''_{(c)}|} = \theta \left(\frac{|m|\omega_l}{12|b''_{(c)}|}\right)^{1/3} \qquad (6.2.47)$$

Then we can see that the integrals (6.2.33)-(6.2.34) both take the form

$$d_{\pm}(\varepsilon_{\pm}) = \int_{x_1}^{x_2} dx \, (\varepsilon_{\pm} + x) \, e^{i\varphi_{\pm}(\varepsilon, x)} \qquad (6.2.48)$$

where

$$\varphi_{\pm}(\varepsilon, x) \propto x^3 + \varepsilon_{\pm} x^2 + \varepsilon_{\pm}^2 x. \qquad (6.2.49)$$

the phase of the integral. We can estimate that if $\varepsilon_{\pm} < 1$, $I_{\pm}(\varepsilon_{\pm}) \simeq I_{\pm}$, and if $\varepsilon_{\pm} > 1$, $I_{\pm}(\varepsilon_{\pm})$ is exponentially suppressed. This implies that for an angle $\theta$ satisfying

$$\theta^3 \geq \frac{1}{|f| T_l} \qquad (6.2.50)$$

the spiky cusp signal is smoothed. Therefore, the GWB from the cusp (6.2.23) is observed in a cone of opening $2\theta^{div}$ around the direction of the cusp velocity $\vec{n}^{(c)}$. We will choose a value for the angle $\theta^{div}$ in the next section.

## 6.2.3 An estimation of the waveform amplitude

Following the assumptions of [4], we will ignore any polarization effects of the waveform, i.e. the relative orientation of the vectors $a_{(c)}^{\lambda''}$ and $b_{(c)}^{\nu''}$. In particular, we can rewrite equation (6.2.27) as

$$\tilde{\kappa}^{\lambda\nu}(f_m, \vec{n}_{(c)}) = -C \frac{G\mu}{(2\pi |f_m|)^{1/3}} \frac{1}{\left(|\vec{a}_{(c)}''||\vec{b}_{(c)}''|\right)^{1/3}} \frac{b_{(c)}^{(\lambda''} a_{(c)}^{\nu)''}}{|\vec{b}_{(c)}''||\vec{a}_{(c)}''|}. \qquad (6.2.51)$$

We will remove the term

$$\frac{b_{(c)}^{(\lambda''} a_{(c)}^{\nu)''}}{|\vec{b}_{(c)}''||\vec{a}_{(c)}''|} \qquad (6.2.52)$$

which accounts for the polarization effects. Also, we define

$$a^{\mu''} = \frac{2\pi}{l} \hat{a}^{\mu''}, \, b^{\mu''} = \frac{2\pi}{l} \hat{b}^{\mu''}, \qquad (6.2.53)$$

which splits $a^{\mu''}$ into its magnitude part $2\pi/l$ and its Fourier Series part (see equation (5.3.41)), and similarly for $b^{\mu''}$. We can then take the absolute value of

(6.2.51) and find the amplitude of the asymptotic waveform to be

$$\tilde{\kappa}(f_m, \vec{n}_{(c)}) = C \frac{G\mu}{(2\pi|f_m|)^{1/3}} \left(\frac{l}{2\pi}\right)^{2/3} \frac{1}{\left(|\hat{\tilde{a}}''_{(c)}||\hat{\tilde{b}}''_{(c)}|\right)^{1/3}}. \tag{6.2.54}$$

Note that the constant coefficient of (6.2.54) has the value (calculated numerically in Mathematica)

$$\frac{C}{2\pi} = \frac{12^{4/3}4\pi}{(3\Gamma(1/3))^2} \frac{1}{2\pi} \simeq 0.8507. \tag{6.2.55}$$

If we also define the parameter

$$g_1 = \left(|\hat{\tilde{a}}''_{(c)}||\hat{\tilde{b}}''_{(c)}|\right)^{-1/3}, \tag{6.2.56}$$

we finally find that

$$\tilde{\kappa}^{cusp}(f_m, \vec{n}) = C g_1 \frac{G\mu l^{2/3}}{2\pi |f_m|^{1/3}}. \tag{6.2.57}$$

Note that $g_1$ was defined in equation (5.5.1), where we had normalized $\vec{a}''$ and $\vec{b}''$ by $2\pi/l$.

In section 6.2.2, we proved that the cusp GW signal is exponentially suppressed if the observation angle $\theta$ obtains a value such that $\varepsilon_\pm \geq 1$. We will define the angle $\theta^{div}$ which devides the observation angles into two sets, one where the signal is the same as it is along the direction of the cusp emission, and one where the signal is smoothed. We will assume that the division occurs when the 3rd order term of the Taylor expansion equals the first. From the definition of the Taylor expansion, this assumption implies that $\theta^{div}$ has such a value that the 3rd order term which is very small for small angles is large enough to compare to the first order term, which is the most dominant one for small angles. Under this reasoning we can conclude that this assumption qualifies as a divide for the small $\theta$'s and the large $\theta$'s. A different perspective behind this choice is provided in [126]. We will also set the values of $u$ and $v$ to be, $u \simeq B_-^{-1}$ and $v \simeq B_+^{-1}$.[2] Using

---

[2]This assumption is based on the discussion of section 6.2.1, where we concluded that the contribution to the value of the $I_\pm$ integrals is mainly from an interval of order $\Delta u \simeq B_-$ and $\Delta v \simeq B_+$.

these assumptions, we find that

$$\left(\theta_-^{div}\right)^2 = \frac{1}{3}\left(\vec{a}''_{(c)}\right)^2\left(\frac{1}{12}|\omega_m|\left(\vec{a}''_{(c)}\right)^2\right)^{-2/3} \tag{6.2.58}$$

and

$$\left(\theta_+^{div}\right)^2 = \frac{1}{3}\left(\vec{b}''_{(c)}\right)^2\left(\frac{1}{12}|\omega_m|\left(\vec{b}''_{(c)}\right)^2\right)^{-2/3}, \tag{6.2.59}$$

where $\theta_-^{div}$ is the "divide" angle of the emission from the right-moving wave and $\theta_+^{div}$ is the "divide" angle for the emission from the left-moving wave. Solving the above equations with respect to $\theta_\pm^{div}$, and substituting $f_m = \omega_m/2\pi$, we find

$$\theta_-^{div} = \left(\vec{a}''\right)^{1/3}\left(\frac{\sqrt{3}}{2}\pi|f_m|\right)^{-1/3} \tag{6.2.60}$$

and

$$\theta_+^{div} = \left(\vec{b}''\right)^{1/3}\left(\frac{\sqrt{3}}{2}\pi|f_m|\right)^{-1/3}. \tag{6.2.61}$$

It is obvious that if the angle of observation satisfies $\theta < \theta_-^{div}$ and $\theta < \theta_+^{div}$, then the cusp signal of equation (6.2.57) is observed. If $\theta \geq \theta_-^{div}$ and $\theta \geq \theta_+^{div}$, then a smoothed signal is observed. If we have the mixed case of $\theta < \theta_-^{div}$ and $\theta > \theta_+^{div}$, or $\theta > \theta_-^{div}$ and $\theta < \theta_+^{div}$, then we can conclude that a smoothed signal reaches the observer. In particular, because the signal is a multiplication of the integrals $d_+^\lambda(\theta)$ and $d_-^\lambda(\theta)$, defined in (6.2.33)-(6.2.34), if one is exponentially suppressed then the signal reaching the observer will also be suppressed (we follow here the same reasoning of section 6.1.1). Therefore, the observer receives the GWB emanated from the cusp on the cosmic string if and only if the observation angle with respect to the direction of the cusp satisfies $\theta < \theta_+^{div}$ and $\theta < \theta_-^{div}$.

Given the above, we finally calculate $\theta^{div}$, the "divide" angle between the observation of the spiky signal and the smoothed signal, to be

$$\theta^{div} = \min\left(\theta_+^{div}, \theta_-^{div}\right). \tag{6.2.62}$$

Alternatively, we can use the definitions (6.2.53) to express the above as

$$\theta_-^{div} = \left(\hat{\tilde{a}}''\right)^{1/3} \left(\frac{2\pi}{l}\right)^{1/3} \left(\frac{\sqrt{3}}{2}\pi|f_m|\right)^{-1/3} \qquad (6.2.63)$$

and

$$\theta_+^{div} = \left(\hat{\tilde{b}}''\right)^{1/3} \left(\frac{2\pi}{l}\right)^{1/3} \left(\frac{\sqrt{3}}{2}\pi|f_m|\right)^{-1/3}. \qquad (6.2.64)$$

Then, $\theta^{div}$ becomes

$$\theta^{div} = \left(\frac{4}{\sqrt{3}g_2|f_m|l}\right)^{1/3} \qquad (6.2.65)$$

where

$$g_2 = \left(min\left(|\hat{\tilde{a}}''|, |\hat{\tilde{b}}''|\right)\right)^{-1}. \qquad (6.2.66)$$

Using the above, we find that an estimation of the asymptotic waveform, which neglects the polarization effects, is

$$\tilde{\kappa}^{cusp}(f_m, \vec{n}) = C g_1 \frac{G\mu l^{2/3}}{2\pi|f_m|^{1/3}}\Theta\left(\theta^{div} - \cos^{-1}(\vec{n}\cdot\vec{n}^{(c)})\right), \qquad (6.2.67)$$

where $\Theta(x)$ is the Heaviside Theta Function, and $\theta = \vec{n}\cdot\vec{n}^{(c)}$ the angle between the direction of the cusp and the direction of observation.

We notice that in the above simplified result, we still have quantities that we have not specified a method for their calculation, namely $|\vec{a}''|$ and $|\vec{b}''|$. From section 5.3.2, we can see that $|\vec{a}''|$ and $|\vec{b}''|$ consist of the absolute value of a Fourier Series, the order of which depends on the harmonics of the loop, times the factor $2\pi/l$. In [4, 54], it was assumed that the Fourier Series part can be dropped, which leads to the following estimation of the second derivatives magnitude

$$|\vec{a}''| \sim \frac{2\pi}{l}, \; |\vec{b}''| \sim \frac{2\pi}{l}, \qquad (6.2.68)$$

and further simplifies the asymptotic waveform amplitude, with $g_1 = 1$ and $g_2 = 1$ always. However, the behaviour of $|\vec{a}''_{(c)}|$ and $|\vec{b}''_{(c)}|$ can in principle differ, and depending on how much it can vary, it could alter the observed GW amplitude (6.2.57) by orders of magnitude. For example, for values of $|\vec{a}''_{(c)}|$ and/or $|\vec{b}''_{(c)}|$ much smaller than unity (or equivalently $g_1 \gg 1$), the value of the strain (6.2.67) can become very large.

We will take a step further in these grounds by estimating the values of $|\vec{a}''_{(c)}|$ and $|\vec{b}''_{(c)}|$ using a particular family of strings, the odd-harmonic string, which we introduced in section 5.4.6. In the following, we will use our numerical calculation of the values of $|\vec{a}''_{(c)}|$ and $|\vec{b}''_{(c)}|$ to estimate their effects on $\kappa^{cusp}(f_m, \vec{n})$ for high-harmonic cosmic strings. This is the first time this has been done, to properly try to establish the values of the crucial $g_1$ and $g_2$ parameters.

### 6.2.4 Low-frequency limit

The result (6.2.57) comes with the assumption of a high frequency limit which implies almost continuous frequency values $f_m \to f$. Then, the cusp asymptotic waveform scales with frequencies as $\kappa^{cusp} \propto f^{-1/3} \propto m^{-1/3}$, where $m = f_m/f_1 \in \mathbb{N}$ the mode number. If we apply our high-frequency result to low frequencies, we can easily see that the magnitude of the asymptotic waveform (6.2.57) is of the order $G\mu l$. Therefore, the GWB from cusps is only a small correction to the low frequency background. However, its significance lies in its dominance at high-frequencies (since the asymptotic waveform is suppressed at high frequencies away from cusps and kinks), as well as providing us with a simple way to calculate the asymptotic waveform at both high frequencies, and as we approach the low frequency limit. The calculation of the GW emission of strings is non-linear in general, as we can see from the integral form of $T_{\mu\nu}$ in equation (6.1.3). Also, calculating the GW emission analytically for specific string cases can be very tedious. It also provides us with a solution for cosmic string loops of specific initial conditions. For example, see the calculation for the family of Burden loops in [51], as well as calculations for other families of strings in section 6.4 of [14]. Therefore, the result for the asymptotic cusp waveform of section 6.2.1 gives us a means of estimating the GW emission for any type of string in an analytic way.

We should mention that there is a second analytical method (i.e. not using numerical results) to reach the same result for the asymptotic waveform as we did with the high-frequency approach in equation (6.2.57) (up to an overall factor). This alternative method assumes that all the radiation is emitted from the $m = 1$ mode of the string, calculates the energy-momentum tensor and then extends the result to the higher harmonic modes, with $m > 1$. More details on this approach are included in [9],[112]. We should mention that compared to the low-frequency method, the high-frequency method introduced by Damour and Vilenkin in [4] gives an insight of the behaviour of cusps and kinks and provides the means to remove rare cusp events from the GW background.

### 6.2.5 An estimation of the key parameters $g_1$, $g_2$ and $c$

In this section we will estimate the values of $g_1$ and $g_2$ using our results for the second derivatives of the right- and left-movers at the cusp for odd-harmonic cosmic strings, $|\vec{a}''_{(c)}|$ and $|\vec{b}''_{(c)}|$, which we calculated in section 5.5.2.

We have concluded that the values of $|\vec{a}''_{(c)}|$ and $|\vec{b}''_{(c)}|$ are of order unity for the odd-harmonic family of strings, and their average value increases slowly as the harmonic order increases. In Figure 5.5.2, we have plotted the average value of the $g_1$ vs the harmonic order, where we can see that its value ranges from around 0.38 to 1.0. In Figure 5.5.4, we have plotted the average values of $g_2$ with harmonic order which range from 0.34 to 1. Since the values of the second derivatives do not differ a lot with respect to the harmonic order, an approach to estimate $g_1$ and $g_2$ would be to calculate them by averaging $|\vec{a}''_{(c)}|$ and $|\vec{b}''_{(c)}|$ over all the harmonic order values we obtained numerically.

If we calculate the average of $g_1$ over all odd harmonic order loops, from N=1 to N=23 (we discused our numerical method for producing these loops in section 5.5.1), we find that $g_1 = 0.489$. The above is averaged over a total of 278069 cusp events, ranging from first order odd-harmonic strings up to 21st odd-harmonic order loops, using the method described in section 5.5.1. In particular, we used 30000 cusp events from each harmonic order, except the 21st harmonic order where we used 6699 events and the 23rd harmonic order where we used 1370. This reduction in the number of events in higher harmonic order loops is because of the increased computation time they require to be analyzed. As a consequence of the new value for $g_1$, we find that it suppresses the GWB amplitude (6.2.67) by around half, compared to the estimation of [4]. For the same set of events we can also calculate the parameter $g_2$, which we find has an average value of $g_2 = 0.305$. This implies that the observation angle of the GWB from cusps, which is $2\theta^{div}$, will increase by a multiplication factor of $2g_2^{-1/3} = 1.87$ compared to [4]. Therefore, it is not clear for the moment whether the amplitude signal of the GWB will be amplified or suppressed when we set the new values of $g_1$ and $g_2$, compared to the values used in [4]. Thus, one of the set of values that we will use for the estimation of the GWB amplitude is $(g_1 = 0.489, g_2 = 0.305, c = 1)$, which we will call set 1. We will call set 0 the values $(g_1 = 1, g_2 = 1, c = 1)$, which is the set of values assumed in both [4] and [3].

Note that the weak dependence of the average values of $|\vec{a}''_{(c)}|$ and $|\vec{b}''_{(c)}|$ with respect to the harmonic order is not unexpected. We have seen in section 6.2, that the cusp events can be treated independently of the loop's initial conditions and locally around the cusp position to calculate the amplitude of the GWB from cusps. It does not come as a surprise that the second derivatives of the string movers at the cusp also seem to maintain rather stable values regardless of the loop's harmonic order.

To estimate values of $c$ different from one, we will use our assumptions of the harmonic order distribution of the parent loops in a spacetime volume $dV(z)$, which were presented in section 5.7.8. Using this distribution of harmonics, we will also calculate the corresponding average values of $g_1$ and $g_2$. Note that the distribution of the harmonics is assumed to include parent loops of harmonic orders from $N = 1$ to $N = 7$. For the uniform distribution case, we find that $(g_1 = 0.680, g_2 = 0.699, c = 8.82)$, for Benford's law we find $(g_1 = 0.773, g_2 = 0.779, c = 6.43)$ and for the geometric distribution, $(g_1 = 0.839, g_2 = 0.842, c = 4.0)$. We will call these set 2, set 3, and set 4, respectively.

## 6.3 The asymptotic waveform from kinks

As we discussed in section 6.1.1, emission of gravitational waves from kinks on cosmic string loops can occur when one of the integrals $I_\pm$ exhibits a saddle point, and the other has a kink, i.e. a discontinuity. Other combinations that produce an asymptotic decline of the energy-momentum tensor, $T^{\mu\nu}$, such as a kink-kink combination, will not be taken into account since they produce a weak signal.

It was proved in [53], that in the above mentioned scenario, of a saddle point and a kink appearing in the integration of $I_+$ and $I_-$, the gravitational waves are emitted in the direction of the kink velocity in pulses. The pulses from the moving kink are emitted on a one-dimensional curve (because of the curve $s = \sigma_{(k)}(\tau)$ that the kink moves on, see section 5.3.5) in a fan-like manner. Following the approach of [4], in this section we will calculate its GWB amplitude.

Let us assume that the integral $I_-$ is the one that exhibits the saddle point, without loss of generality. This integral is solved in the same manner as the integral of section 6.2.1 in the cusp scenario. Therefore,

$$I_-^\mu = a^{\mu\,\prime\prime}(u_s) \int_{u_0}^{u_0+l} u e^{-\frac{i}{2}k_m \cdot b} du \tag{6.3.1}$$

and its value is given by equations (6.2.17) and (6.2.19). The integral $I_+$ which exhibits the kink should be treated in a different manner due to the discontinuity that appears in the integrand. If the kink is at $v = v_{(k)}$, then

$$\begin{aligned}
I_+^\mu &= \int_0^l b^{\mu\,\prime}(v_s) u e^{-ik_m \cdot b(v)} dv = \\
&= \int_0^{v_{(k)}} b^{\mu\,\prime}(v_s) u e^{-ik_m \cdot b(v)} dv + \int_{v_{(k)}}^l b^{\mu\,\prime}(v_s) u e^{-ik_m \cdot b(v)} dv.
\end{aligned} \tag{6.3.2}$$

Integrating the above by parts, we find

$$\begin{aligned}
I_+^\mu &= \left[ \frac{b^{\mu\,\prime}(v)}{-ik \cdot b'} \right]_0^{v_{(k)}} - \\
&\quad - \int_0^{v_{(k)}} i\frac{b^{\mu\,\prime\prime}(v)(k_m \cdot b'(v)) - b^\mu(v)(k_m \cdot b''(v))}{k_m \cdot b'(v)^2} e^{-ik_m \cdot b} dv + \\
&\quad + \left[ \frac{b^{\mu\,\prime}(v)}{-ik \cdot b'} \right]_{v_{(k)}}^l - \\
&\quad - \int_{v_{(k)}}^l i\frac{b^{\mu\,\prime\prime}(v)(k_m \cdot b'(v)) - b^\mu(v)(k_m \cdot b''(v))}{k_m \cdot b'(v)^2} e^{-ik_m \cdot b} dv.
\end{aligned} \tag{6.3.3}$$

In the above, we will ignore the integrals since they give a contribution of $1/\omega^2$, and also the terms calculated at 0 and at $l$ in the brackets cancel with each other, due to the periodicity of $b^{\mu\,\prime}(u)$. Finally, we find that

$$I_+^\mu = \frac{2i}{\omega_m} \left[ \frac{b^{\mu\,\prime}(v_{(k)})}{1 - \vec{n} \cdot \vec{b}'(v_{(k)})} \right] e^{-i\frac{\omega_m}{2}(v_{(k)} + \vec{n}\cdot\vec{b}(v_{(k)}))}. \tag{6.3.4}$$

Note that the above method could not have been applied for the cusp case, because $k \cdot b$ cannot appear in the denominator. For the kink, $1 - \vec{n} \cdot \vec{b}(v_{(k)}) \neq 0$, since the kink is not a saddle point. We find that the kink GWB scales with $m \to \infty$ as $m^{-1}$, i.e. it drops faster than the cusp result by a factor $m^{-1/3}$.

To calculate the kink waveform, we will need to make assumptions regarding the cusp discontinuity. We assume the discontinuity in $b^{\mu\,\prime}(u)$ to be of the order $\omega_l |b^{mu\,\prime\prime}| \simeq 1$, and we estimate the quantity in the brackets to be of order 1 [4]. Then, we find that the waveform, defined in equation (3.2.44), takes the form

$$\tilde{\kappa}^{kink}(f, \vec{n}) \simeq \frac{G\mu l^{1/3}}{|f|^{2/3}}. \tag{6.3.5}$$

The above has been obtained by following the same reasoning as in section 6.2.3.

Since the gravitational wave amplitude emitted from kinks declines faster with frequency than the amplitude from cusps, it is obvious that a GWB event from cusps will be easier to observe than one from kinks. However, the kink emits GWBs in a more continuous manner than a cusp which has a momentary pulse of gravitational emission. One should study the complete contribution from a unit spacetime volume, $dV(z)$, of string loops to determine which signal is stronger. This was calculated in [4], and it was shown that the effect from kinks arriving at an observer at Earth is in fact weaker than the one from cusps. However, in recent studies it was found that, in the case of cosmic superstrings with Y-junctions, it is possible that the number of kinks increases enough with time on each loop, via a mechanism called kink proliferation [125]. It was therefore concluded that their overall signal might in fact have been underestimated. As it was shown in [39], even for Nambu-Goto strings a high number of kinks can imply an even higher number of collisions between the left- and right-moving kinks, an effect that can produce an enhanced gravitational wave emission which can potentially dominate the gravitational wave radiation from cosmic strings. It is also a fact that the number of cusps and the number of kinks on a loop can vary depending on the approach one assumes, and this greatly affects the signal received from each source. It is still an open question whether our focus should be on cusps or kinks when it comes to gravitational wave emission. In this study, our focus will be on the cusp signal and we will focus on that in the following sections. Our motivation was its slower decline with frequency, as well as the synergy cusps can have with our results of the odd-harmonic string, which calculate the values of the second derivatives of the left- and right-movers at the cusps and the number of cusps per period, as well as our toy model introduced in section 5.7. However, it could be possible to include the effect of kinks as an improvement of our approach, for example by including the kink effect in the toy model.

## 6.4 Propagation of GWBs in FLRW spacetime

In section 3.2.3, we calculated the GWB emission from a cosmic string loop at distances that were large compared to the wavelength of the gravitational waves (or equivalently distances much larger than the source), but small compared to the Hubble radius, $(aH)^{-1}$, i.e. negligible compared to cosmological scales. This allowed us to calculate the distance independent asymptotic waveform $\kappa_{\mu\nu}$ in a flat spacetime background, which characterized the GWB amplitude $\bar{h}_{\mu\nu}$ in flat spacetime around the source (see equation (3.2.43)). In [4], they found that the logarithmic Fourier transform of the propagated GWB amplitude in an FLRW flat spacetime, which is observed at a redshift $z$, a distance $r(z)$ and at frequency $f = f_{rec}$ is given by

$$\tilde{h}^{cusp}(f, z) = C g_1 \frac{G\mu l^{2/3}}{(1+z)^{1/3} f_{rec}^{1/3} r(z)}. \tag{6.4.1}$$

In the above, we used the result (6.2.57) and also dropped the subscript of the observed frequency for simplicity, i.e. write $f_{rec}$ simply as $f$, since we will only refer to the observed frequency from now on. We also removed the absolute value of the frequency, and assumed the value of the frequency to be positive hereafter. We are free to enforce this without any loss, given our current purposes in this section. In particular, the negative frequencies appeared from the definition of the exponential Fourier transform and series. This approach proved useful to sim-plify our calculations (instead of using trigonometric series or taking the complex conjugate of any Fourier expression) and it allowed for a simple way to study the presence of the cuspy signal in equations (6.2.27)-(6.2.29). However, at this stage we will not consider the frequency as a variable anymore, but fix it to its observed value, which is positive. The quantity $\tilde{h}$ is the logarithmic Fourier transform of the GWB amplitude. Using (2.3.18), we find that the Fourier transform of the GWB amplitude

$$h^{cusp}(f, z) = C g_1 \frac{G\mu l^{2/3}}{(1+z)^{1/3} f_{rec}^{4/3} r(z)},$$ (6.4.2)

which coincides with the result in [3].

We can use the cosmic distance approximation (5.6.48) to simplify the expression for the GWB amplitude. This approach is very useful since the integral of $r(z)$ cannot be calculated for any $z$, and requires a numerical calculation. Using the above approximation, we find that the logarithmic Fourier transform of the amplitude of the cusp waveform observed at redshift $z$ and frequency $f$ is

$$\tilde{h}^{cusp}(f, z) = C g_1 \frac{G\mu l^{2/3}}{(1+z)^{1/3} f^{1/3}} \frac{1+z}{t_0 z},$$ (6.4.3)

which is the result obtained in [4].

We can also propagate the value of the angle $\theta^{div}$, defined in equation (6.2.62), to its observed value

$$\theta^{div}(f, z) = \left(\frac{4}{\sqrt{3} g_2 (1+z) fl}\right)^{1/3} \simeq (g_2 (1+z) fl)^{-1/3},$$ (6.4.4)

by substituting the frequency with the observed frequency. We should also recall the restriction that a cusp GWB (6.4.1) is only observed if the angle between the velocity of the cusp and the direction of the observer is less than $\theta^{div}$.

Also, the result (6.4.1) was calculated for high frequencies, but if we extend the result to low harmonic orders, i.e. $f_{em} = (1+z)f \simeq T_1^{-1}$ for $m \simeq 1$, we find that the amplitude would be of order

$$\tilde{h}^{cusp}_{LF} \simeq C g_1 G\mu l \frac{1+z}{t_0 z},$$ (6.4.5)

and therefore, the high frequency amplitude compared to the low frequency one is $\tilde{h}^{cusp}(z) \simeq \theta_m \tilde{h}^{cusp}_{LF}$, where

$$\theta_m \simeq ((1+z)fl)^{-1/3} \simeq m^{-1/3}. \tag{6.4.6}$$

Since the cosmic string loop emits at frequencies $f_m$ with $|m| \geq 1$, the condition $\theta_m \leq 1$ should also hold. This restriction is necessary to make sure that in our calculations we do not take into account non-existing modes with $|m| \leq 1$.

## 6.5 Rate of GWBs from a cosmic string network

We will calculate the GWBs observed on Earth using two different cosmic string network models, one analytic and one which combines analytic and numerical results, both presented in section 5.6. The analytic model follows the approach of Damour and Vilenkin, presented in [4],[54], while the model that combines analytic and numerical results follows Model 1 of the more recent work from the LIGO Collaboration, presented in [3]. We will call the former Model I, the analytic model, or the small loops model, since it assumes small cosmic string loops. We will call the latter Model II, the numerical model, or the large loop model, since it assumes large loop size.

### 6.5.1 Model I

We defined the cosmic time in equation (5.6.7), the cosmic distance in equation (5.6.9) and the proper volume in equation (5.6.11). We also calculated their asymptotic behaviour. We can simplify the above quantities to obtain analytic approximations, as we did in section 5.6.1.2. We will use the expressions (5.6.38)-(5.6.39) for the cosmic time in the radiation and matter era, the expressions (5.6.48)-(5.6.49) for the proper distance and the expressions (5.6.44)-(5.6.45) for the proper volume in the radiation and matter era, for an analytic approximation.

Substituting equation (5.6.41) into (6.4.3), we obtain the following expression for the GWB amplitude from cusps

$$
\begin{aligned}
\tilde{h}^{cusp}(f,z) &= Cg_1 \frac{G\mu \left(\alpha t_0 \varphi_l(z)\right)^{2/3}}{(1+z)^{1/3} f^{1/3}} \frac{1+z}{t_0 z} \Theta(1-\theta_m) \\
&= Cg_1 \frac{G\mu \alpha^{2/3}}{t_0^{1/3} f^{1/3}} z^{-1} (1+z)^{-1/3} \left(1 + \frac{z}{z_{eq}}\right)^{-1/3} \Theta(1-\theta_m).
\end{aligned}
\tag{6.5.1}
$$

If we define

$$
\varphi_h(z) = z^{-1} (1+z)^{-1/3} \left(1 + \frac{z}{z_{eq}}\right)^{-1/3},
\tag{6.5.2}
$$

then the above can be written as

$$
\tilde{h}^{cusp}(f,z) = Cg_1 \frac{G\mu \alpha^{2/3}}{t_0^{1/3} f^{1/3}} \varphi_h(z) \Theta(1-\theta_m).
\tag{6.5.3}
$$

Note that since we calculate the amplitude from gravitational waves emitted from cusps only, we can emit the index cusp in $h$ hereafter. We can also calculate the function for $\theta_m$ with $\alpha$, i.e. the size of the loops, by substituting equation (5.6.41) into equation (6.4.6),

$$
\begin{aligned}
\theta_m(\alpha, f, z) &= [\alpha f t_0 (1+z) \varphi_l(z)]^{-1/3} \\
&= (\alpha f t_0)^{-1/3} (1+z)^{1/6} (1 + \frac{z}{z_{eq}})^{1/6}.
\end{aligned}
\tag{6.5.4}
$$

The rate of GWBs emanating from cusps at frequency f in the redshift interval $dz$ was estimated in [4] to be

$$
dR \simeq \frac{1}{4} \left(\theta^{div}\right)^2 (1+z)^{-1} \nu(z) dV(z).
\tag{6.5.5}
$$

In the above, the term $(\theta^{div})^2/4$ accounts for the fraction of the sky which is covered by the radiation beam of the cusp. As we proved in section 6.2.2, the cusp beams only towards the direction of the cusp velocity and around it. The whole sphere has $4\pi$ steradians of solid angle, while in the case of a small angle $\theta^{div}$ the solid angle of the cone is

$$
\Omega = 4\pi \sin^2 \left(\frac{\theta^{div}}{2}\right) \simeq 4\pi \left(\frac{\theta^{div}}{2}\right)^2 = (\theta^{div})^2 \pi.
\tag{6.5.6}
$$

Therefore, the fraction of the solid angle of the cone compared to the solid angle of the whole sphere is $\pi(\theta^{div})^2/4\pi = (\theta^{div})^2/4$. The factor of $(1 + z)^{-1}$ emerges from the transformation between the time the signal is received, $t_{rec}$, and the time it is emitted, $t_{em}$, which is given by $dt_{rec} = (1 + z)dt_{em}$. Since in the calculation of $dR$ the derivative is calculated in terms of the time received $dt_{rec}$, the factor of $(1 + z)^{-1}$ appears on the right hand side of the equation (6.5.5). The quantity $\nu(t)dt_{em}dV(z)$, was defined in section 5.7.8, and it corresponds to the total number of cusp events emitted from a unit spacetime volume in the redshift interval $[z, z + dz]$ and in the time interval $dt_{em}$. Note that equation (6.5.5) has been derived by initially "constructing" an approximate expression for the quantity

$$dR \sim \Omega\, \nu(z)dV(z)dt_{em}, \qquad (6.5.7)$$

which corresponds to the total number of cusps emitted from a volume $dV(z)$ in a time period of $dt_{em}$ and multiplied by the fraction of the sky that the cusps are observed in $\Omega$.

In the following, we will substitute $dV(z)$ with equations (5.6.44)-(5.6.45), which are analytic approximations of the proper volume in the matter and radiation era respectively. Hereafter, we will also use the logarithmic density of $dR$, which is defined as $R(f, z) \equiv dR/d\ln z$ [4]. If we substitute $dR$ from equation (6.5.5), we find that

$$R(f, z) = \frac{dR}{d(\ln z)} = \frac{dR}{dz}\frac{dz}{d(\ln z)} = z\frac{dR}{dz}. \qquad (6.5.8)$$

with

$$R(f, z) = \frac{1}{2}c\alpha^{-8/3}g_2^{-2/3}\left(ft_0\right)^{-2/3}(1 + z)^{-2/3}\left(1 + \frac{z}{z_{eq}}\right)^{1/3}t^{-4}z\frac{dV(z)}{dz}, \qquad (6.5.9)$$

where $\alpha = \Gamma G\mu$. We can write $R(f, z)$ in terms of the interpolating function $\varphi_n(z)$,

$$R(f, z) = 10^2ct_0^{-1}\alpha^{-8/3}g_2^{-2/3}\left(ft_0\right)^{-2/3}\varphi_n(z) \qquad (6.5.10)$$

where

$$\varphi_n(z) = z^3(1 + z)^{-7/6}\left(1 + \frac{z}{z_{eq}}\right)^{11/6}. \qquad (6.5.11)$$

In Appendix A.3, we show how to derive $\varphi_n(z)$.

We can now define the rate of GWBs from cusps that arrive to an observer on Earth over all redshifts, $R$ [4]. Since the function $\varphi(z)$ increases with the redshift in a power law manner, we will approximate the value of the GWB rate to be dominated by the large redshifts, i.e.

$$R = \int_0^{z_m} R(f, z) d\ln z \simeq R(f, z_m), \qquad (6.5.12)$$

where $z_m$ is the maximum observed redshift. The above is assumed by [4] to achieve an analytic approach to the problem. We can calculate $h^{cusp}$ in terms of the frequency and $R$, which will give us an estimate of the amplitude of bursts emanating from cusps observed on Earth, given that we know the rate of observation of the bursts. We need first to invert the function $R$ to $z_m = z_m(R, f)$. To achieve this, first of all define (for ease of notation) the parameter

$$y = 10^{-2} \frac{R(f, z)}{c} t_0 \alpha^{8/3} g_2^{-2/3} (ft_0)^{2/3} \qquad (6.5.13)$$

which is essentially $\varphi_n(z)$ defined in (6.5.11). Then for $y < 1$, the most significant factor in $z^3 (1+z)^{-7/6} (1 + z/z_{eq})^{11/6}$ is $z^3$, since $z < 1$. For $1 < y < y_{eq} \equiv z_{eq}^{11/6}$, the most significant term is $z^3 z^{-7/6} = z^{11/6}$. Finally, for $y > y_{eq}$, the most significant term is $z^{11/3}$. Thus,

$$z = y^{1/3} (1+y)^{7/33} \left(1 + \frac{y}{y_{eq}}\right)^{-3/11} \qquad (6.5.14)$$

Now, let us also write the function $\varphi_h$, defined in (6.5.2), in terms of $y$. For $z < 1$, which also implies that $y < 1$, $\varphi_h = z^{-1} = z^{-1} \simeq y^{1/3}$. For $1 < z < z_{eq}$, $\varphi_h(z) = z^{-4/3} \simeq y^{-8/11}$. For $z > z_{eq}$, $\varphi(z) = z^{-5/3} \simeq y^{-5/11}$. Therefore,

$$\varphi_h(y) = y^{-1/3} (1+y)^{-13/33} \left(1 + \frac{y}{y_{eq}}\right)^{3/11} \qquad (6.5.15)$$

The value of $\tilde{h}$ will depend on the value of $y$. In particular, if $y < 1$, the dominant redshift $z_m$ in (6.5.12) will also be $z_m(y) < 1$, if $1 < y < y_{eq}$, the dominant redshift is $1 < z_m < z_{eq}$ and if $y > y_{eq}$, $z_m(y) > z_{eq}$. We then conclude that the gravitational wave amplitude from equation (6.5.3) becomes

$$\tilde{h}(R, f) = g_1 G\mu\alpha^{2/3} (ft_0)^{-1/3} \varphi_h(y)\Theta(1 - \theta_m(\alpha, R, z)), \qquad (6.5.16)$$

where the redshift is fixed to the value $z_m$, $\Theta$ is the Heaviside step function,

and $\theta_m$ can be obtained if we substite the value $z = z_m(y)$ into equation (6.5.4). However, the cut-off function will not affect the result for the values of frequency $f$ and rate of observation $R$ that we are looking into (see [4]).

## 6.5.2 Model II

We can write the length $l$ of a loop in terms of the amplitude $h$, the frequency $f$ and the redshift $z$

$$l(h, z, f) = \left( \frac{h f^{4/3} (1+z)^{1/3} \varphi_r(z)}{g_1 G \mu H_0} \right)^{3/2}. \tag{6.5.17}$$

The above is obtained by inverting equation (6.4.2). Recall that we have replaced $f_{rec}$ by $f$. We can also express $\theta^{div}$ in terms of $h$, $f$ and $z$. By combining equations (6.4.4) and (6.5.17), we find

$$\theta^{div}(h, f, z) = \left[ g_2 \left( \frac{f^2 h (1+z) \varphi_r(z)}{g_1 G \mu H_0} \right)^{3/2} \right]^{-1/3}. \tag{6.5.18}$$

Finally, we will also define the number of cusps per unit space time volume created by loops of length between $l$ and $l + dl$, which is [3]

$$\nu(l, z, f) dl = \frac{2}{l(h, z, f)} c \, n(l(h, z, f), t(z)) dl. \tag{6.5.19}$$

We can also describe a related quantity, namely the number of cusps per unit space time volume and for GWBs of amplitudes between $h$ and $h + dh$, using the transformation

$$\nu(h, z, f) dh = \nu(l(h, z), z) \frac{dl}{dh} dh = \nu(l(h, z), z) \frac{3}{2h} l dh \tag{6.5.20}$$

By using $\nu(h, z, f)$ defined in equation (5.7.36). The derivative $dl/dh$ has been calculated using equation (6.5.17).

Following the same reasoning as in section 6.5.1, we can define the rate of GWBs in the unit spacetime volume $dV(z)$ and in an interval of amplitudes from $h$ to

$h + dh$, [3]

$$\frac{d^2R}{dV(z)dh} = \left(\frac{\theta^{div}(h,z,f)}{2}\right)^2 (1+z)^{-1}\nu(h,z,f)\Theta(1-\theta^{div}(h,z,f)). \quad (6.5.21)$$

We use equation (6.5.20) to change the coordinates of $\nu$ from $h$ to $l$ in the above expression, and find that

$$\begin{aligned}
\frac{d^2R}{dV(z)dh} &= \left(\frac{\theta^{div}(h,z,f)}{2}\right)^2 (1+z)^{-1})\frac{3}{2h}l(h,z,f)\nu(l,z,f)\times \\
&\quad \times \Theta(1-\theta^{div}(h,z,f)) \\
&= \left(\frac{\theta^{div}(h,z,f)}{2}\right)^2 (1+z)^{-1})\frac{3}{2h}l(h,z,f)\times \\
&\quad \times \frac{2}{l(h,f,z)}c\,n(h,z,f)\Theta(1-\theta^{div}(h,z,f)).
\end{aligned} \quad (6.5.22)$$

Using the fact that

$$dV(z) = \frac{\varphi_V(z)}{H_0^3}dz, \quad (6.5.23)$$

where $\varphi_V(z)$ was defined in equation (5.6.12), and using equation (5.6.25) substituting

$$n(l,z,f) = t^{-4}\mathcal{F}(l,z,f), \quad (6.5.24)$$

we find that

$$\frac{d^2R}{dzdh} = \left(\frac{\theta^{div}(h,z,f)}{2}\right)^2 \frac{3c\varphi_V(z)}{H_0^3(1+z)ht(z)^4}\mathcal{F}(l,z,f)\Theta(1-\theta^{div}(h,z,f)). \quad (6.5.25)$$

Finally, substituting $\theta^{div}$ from equation (6.5.18), we find that

$$\begin{aligned}
\frac{d^2R}{dzdh}(h,f,z) &= \frac{3}{4}\frac{g_1}{g_2^{2/3}}\frac{G\mu c\varphi_V(z)}{H_0^2 f^2(1+z)\varphi_r(z)t^4(z)}\frac{1}{h^2}\mathcal{F}(l,z,f)\times \\
&\quad \times \Theta(1-\theta^{div}(h,f,z)).
\end{aligned} \quad (6.5.26)$$

The above expression is true for any cosmological era, i.e. for all redshifts, and for any cosmic string model.

We will now proceed to calculate the rate of GWBs for the large loop cosmic string model presented in section 5.6.1.1, separating the matter era calculation from the radiation era calculation (i.e. we will no longer use interpolating functions between the two eras as we did in section 6.5.1).

During the radiation era, $z > z_{eq} = 3366$, we substitute equation (5.6.26) into the GWB rate (6.5.26) to find that the rate of GWBs during the radiation era is

$$
\begin{aligned}
\frac{d^2 R_{rad}}{dzdh}(h, f, z) =& 3\frac{g_1}{g_2^{2/3}}\frac{G\mu\pi c}{f^2}\frac{H_0^2 \varphi_r(z)}{(1+z)^5\varphi_t(z)^4\mathcal{H}(z)}\frac{1}{h^2}\times \\
& \times \frac{C_{rad}}{\left[\frac{H_0}{\varphi_t(z)}\left(\frac{hf^{4/3}(1+z)^{1/3}\varphi_r(z)}{g_1 G\mu H_0}\right)^{3/2} + \Gamma G\mu\right]^{5/2}}\times \\
& \times \Theta\left(1 - \theta^{div}(h, f, z)\right)\Theta\left(\alpha - \gamma(h, f, z)\right)
\end{aligned}
\tag{6.5.27}
$$

In the above, we used equation (5.6.12) to substitute $\varphi_V(z)$ with $\varphi_r(z)$ and equation (5.6.8) to substitute $t(z)$ with $\varphi_t(z)$. Also, recall that $\alpha$ is a constant determining the size of the loops formed from the long string network (for the large loop case it takes the value $\alpha = 0.1$), and $\mathcal{H}(z)$ is the Hubble constant at redshift $z$ normalized by $H_0$, defined in equation (5.6.13).

During the matter era, $z < z_{eq} = 3366$, we substitute equation (5.6.30) into the GWB rate (6.5.26) to find that the rate of GWBs during the radiation era is

$$
\frac{d^2 R_{mat}}{dzdh}(h, f, z) = \frac{d^2 R_{mat}^{(1)}}{dzdh}(h, f, z) + \frac{d^2 R_{mat}^{(2)}}{dzdh}(h, f, z),
\tag{6.5.28}
$$

where the first term on the right-hand-side of the equation corresponds to the GWBs originating from loops that formed in the radiation era and survive into the matter era, and the second term on the right-hand-side of the equation corresponds to the GWBs originating from matter era loops. They are given by

$$
\begin{aligned}
\frac{d^2 R_{mat}^{(1)}}{dzdh}(h, f, z) =& 3\frac{g_1}{g_2^{2/3}}\frac{G\mu\pi c}{f^2}\frac{H_0^2 \varphi_r(z)}{(1+z)^5\varphi_t(z)^4\mathcal{H}(z)}\frac{1}{h^2}\left(\frac{\varphi_t(z_{eq})}{\varphi_t(z)}\right)^{1/2}\times \\
& \times \frac{C_{mat}}{\left[\frac{H_0}{\varphi_t(z)}\left(\frac{hf^{4/3}(1+z)^{1/3}\varphi_r(z)}{g_1 G\mu H_0}\right)^{3/2} + \Gamma G\mu\right]^2}\times \\
& \times \Theta\left(1 - \theta^{div}(h, f, z)\right)\Theta\left(-\gamma(h, f, z) + \beta(t)\right),
\end{aligned}
\tag{6.5.29}
$$

and

$$
\begin{aligned}
\frac{d^2 R_{mat}^{(2)}}{dzdh}(h,f,z) =& 3\frac{g_1}{g_2^{2/3}} \frac{G\mu\pi c}{f^2} \frac{H_0^2 \varphi_r(z)}{(1+z)^5 \varphi_t(z)^4 \mathcal{H}(z)} \frac{1}{h^2} \times \\
& \times \frac{C_{mat}}{\left[ \frac{H_0}{\varphi_t(z)} \left( \frac{hf^{4/3}(1+z)^{1/3}\varphi_r(z)}{g_1 G\mu H_0} \right)^{3/2} + \Gamma G\mu \right]^{5/2}} \times \\
& \times \Theta\left(1 - \theta^{div}(h,f,z)\right) \Theta\left(\gamma(h,f,z) - \beta(t)\right) \times \\
& \times \Theta\left(\alpha - \gamma\right),
\end{aligned}
\tag{6.5.30}
$$

respectively. Note that

$$
\gamma(z) = \frac{l(z)}{t(z)} = \frac{H_0}{\varphi_t(z)} \left[ \frac{hf^{4/3}(1+z)^{1/3}\varphi_r(z)}{g_1 G\mu H_0} \right]^{3/2}
\tag{6.5.31}
$$

and

$$
\beta(z) = \alpha \frac{\varphi_t(z_{eq})}{H_0} - \frac{\Gamma G\mu}{H_0} \left( \varphi_t(z) - \varphi_t(z_{eq}) \right),
\tag{6.5.32}
$$

as we can derive from equations (5.6.24) and (5.6.27). The rate of GWBs is obtained by integrating over both the redshift and the amplitude of the GWBs, and it is given by

$$
R(h,z) = \int_0^{z_{max}} \int_{h_{min}(z)}^{h_{max}(z)} \frac{d^2 R(h,f,z)}{dzdh} dz dh
\tag{6.5.33}
$$

The range of integration is limited by the conditions we have imposed. The beaming angle of the cusps satisfies $\theta^{div} < 1$, which implies that

$$
h(z) > h_{min}(z) = \frac{g_1 G\mu H_0}{g_2^{2/3}(1+z)f^2 \varphi_r(z)},
\tag{6.5.34}
$$

and holds for any era.

During the radiation era $\alpha > \gamma$ holds, which provides the upper limit of integration for the amplitude

$$
h(z) < h_{max}(z) = \frac{g_1 G\mu H_0 \alpha^{2/3} t(z)^{2/3}}{f^{4/3}(1+z)^{1/3}\varphi_r(z)}.
\tag{6.5.35}
$$

| Value of $g_2$ | $z_{max}$ |
|---|---|
| Set 0, $g_2 = 1.000$ | $3.81 \times 10^{20}$ |
| Set 1, $g_2 = 0.305$ | $1.16 \times 10^{20}$ |
| Set 2, $g_2 = 0.699$ | $2.66 \times 10^{20}$ |
| Set 3, $g_2 = 0.779$ | $2.97 \times 10^{20}$ |
| Set 4, $g_2 = 0.842$ | $3.21 \times 10^{20}$ |

Table 6.5.1: The value of the limit of integration $z_{max}$ for each value that $g_2$ obtains in the set 0 to set 4 defined in section 6.2.5.

By combining equations (6.5.34) and (6.5.35), we also find the upper limit of integration for the redshift in the radiation era, for $g_2 = 1$,

$$z < z_{max} = 3.81 \times 10^{20}. \tag{6.5.36}$$

To find the above, we first solved equations (6.5.34) and (6.5.35) to find that

$$\varphi_t(z)(1 + z) \geq \frac{H_0}{\alpha g_2 f}. \tag{6.5.37}$$

Since, $z \gg 1$, we can use the large redshift expression for $\varphi_t(z)$, given in equation (5.6.17), and substitute the value of $H_0$ given in section 5.6.1.1. We have also set the value of the frequency to be $f = 100Hz$. We calculate $z_{max}$ for all the set of parameter values defined in section 6.2.5 in Table 6.5.1.

During the matter era, we will examine the limits of integration for the loops surviving from the radiation era, and for the loops formed in the matter era. For the former the inequality

$$-\gamma(z) + \beta(z) \geq 0 \tag{6.5.38}$$

holds. Using equations (6.5.31)-(6.5.32), and substituting $l(h, z, f)$ from equation (6.5.17) the above can be written as

$$h \leq h_{max} = \frac{g_1 G \mu H_0^{1/3}}{f^{4/3}(1 + z)^{1/3} \varphi_r(z)} \left(\varphi_t(z)\beta(z)\right)^{2/3}. \tag{6.5.39}$$

Keep in mind that $\beta(z)$ should be positive, since it is the length of a loop formed at $t(z_{eq})$ evaluated at time $t(z)$. Requiring that $h_{min}$ (given by equation (6.5.34)) is less than $h_{max}$, we find that

$$0 \leq -\frac{H_0}{g_2 f(1+z)} + (\alpha + \Gamma G \mu)\varphi_t(z_{eq}) - \Gamma G \mu \varphi_t(z). \tag{6.5.40}$$

To solve the above inequality, and find $z_{min}$ we will use numerical methods. It is given that we should search for the values of $z$ that satisfy the above inequality in the interval $[0, 3366]$, since we are referring to loops in the matter era. We will pick a large number of points in the above interval, given by

$$z_i = k_i z_1 \tag{6.5.41}$$

where $k_i$ is an integer obtaining values in the interval $[0, N_1]$, and $z_1 = 3366/N_1$. We will calculate the value of $\varphi_t(z)$ at each of these points $z_i$ by splitting the integral, which appears in equation (5.6.8), into two integrals, one with integral limits 0 and 3366 and one with integral limits 3366 and infinity. We calculate the former integral using numerical integration and the latter using the asymptotic behaviour given in equation (5.6.17). By plotting the right-hand-side of equation (6.5.40), we find that the inequality is satisfied for all values of $z$ larger than unity. Therefore, we calculate the value of $\varphi_t(z)$ once more, using the method above, but inside the interval $z \in [0, 1]$, which is the interval of interest for the solution of the inequality. Then, we find that for $g_2 = 1$ and $f = 100Hz$, the solution to (6.5.40) is

$$z \geq z_{min} = 0.288. \tag{6.5.42}$$

We will also calculate it for different values of $g_2$. For $g_2 = 0.305$, we find that

$$z \geq z_{min} = 0.288 \tag{6.5.43}$$

and for $g_2 = 0.699$ we find that

$$z \geq z_{min} = 0.29. \tag{6.5.44}$$

The miniscule change is $z_{min}$ is expected since the first term on the right-hand-size of (6.5.40) is much smaller than the other two terms. Therefore, the value of $g_2$ does not affect the value of redshift $z_{min}$. We find that $z_{min} = 0.288$ for any $g_2$ around 1.

For the loops formed in the matter era, the inequalities $1 - \theta_m \geq 0$, $\gamma(z) - \beta(z) \geq 0$ and $\alpha - \gamma(z) \geq 0$ hold. From these we obtain

$$h \geq h_{min}^{(1)} = h_{min},$$  (6.5.45)

$$h \geq h_{min}^{(2)} = \frac{g_1 G \mu H_0^{1/3} \varphi_t(z)^{2/3} \beta(z)^{2/3}}{f^{4/3}(1+z)^{1/3} \varphi_r(z)},$$  (6.5.46)

$$h \geq h_{max} = \frac{g_1 G \mu H_0^{1/3} \alpha^{2/3} \varphi_t(z)^{2/3}}{f^{4/3}(1+z)^{1/3} \varphi_r(z)},$$  (6.5.47)

respectively. We find that $h_{min}^{(1)} \geq h_{min}^{(2)}$ for all z in the matter era, by plotting both functions. Therefore, $h \geq h_{min}^{(1)} = h_{min}$ should be satisfied.

## 6.6  Results

### 6.6.1  Model I

We are interested in plotting the GW amplitude of cusp bursts for a given rate of observation $R$, following [4]. First of all, note that $\tilde{h}^{cusp}(R, f)$ is a monotonically decreasing function in terms of $R$ and in terms of $\alpha$. We find that the dependence on $R$ is

$$\begin{aligned}
\tilde{h}^{cusp} &\propto R^{-1/3}, \quad \text{for } z_m < 1 \\
\tilde{h}^{cusp} &\propto R^{-8/11}, \quad \text{for } 1 < z_m < z_{eq} \\
\tilde{h}^{cusp} &\propto R^{-5/11}, \quad \text{for } z_{eq} < z_m
\end{aligned}$$  (6.6.1)

and the dependence on $\alpha$ is

$$\begin{aligned}
\tilde{h}^{cusp} &\propto \alpha^{7/9}, \quad \text{for } z_m < 1 \\
\tilde{h}^{cusp} &\propto \alpha^{-3/11}, \quad \text{for } 1 < z_m < z_{eq} \\
\tilde{h}^{cusp} &\propto \alpha^{5/11}, \quad \text{for } z_{eq} < z_m.
\end{aligned}$$  (6.6.2)

We can now plot the GW amplitude of cusp bursts $\tilde{h}^{cusp}$ in terms of $\alpha$ for a given rate of observation $R = 1/year = 10^{-7}/3.15 Hz$, following [4]. We fix the value of the frequency to $f = 150 Hz$, which is the optimal frequency of observation of this signal by LIGO and was used in [4]. We will compare our results with the original plot of [4], which assumed the above mentioned values and the set 0 values ($g_1 = 1, g_2 = 1, c = 1$). Firstly, we will plot the set 1 of values for $g_1$, $g_2$ and $c$, discussed in section 6.2.5. We obtain the Figure 6.6.1, which implies the values of $g_1$ and $g_2$ obtained from the odd-harmonic family of strings suppress the signal by around half an order of magnitude over the range of $\alpha$ values from $10^{-12}$ to $10^{-3}$.

We will also plot the magnitude of the GWB against $\alpha$, applying our results from the toy model presented in section 5.7. The red line in Figure 6.6.2 shows the results from [4], while the green, black and blue lines correspond to the GWB amplitude using set 4, set 3 and set 2 values of $(g_1, g_2, c)$, respectively. We can see that in all cases, the results from section 5.7.8 enhance the signal by around half an order of magnitude compared to the original plot. The uniform distribution assumption enhances the signal the most, having the greatest value of $c$ out of the three, while the geometric distribution enhances the signal the least.

## 6.6.2 Model II

We will plot the quantity

$$R(z) = \int_{z}^{z+\Delta z} \int_{h_{min}(z')}^{h_{max}(z')} \frac{d^2 R}{dz' dh} dh dz' \qquad (6.6.3)$$

for frequency $f = 100 Hz$. Note that we use the value $100 Hz$ for the frequency instead of $150 Hz$, which we used in the previous section, to follow the approach of [3], which is the most recent LIGO publication for gravitational waves from cosmic strings. The range of redshifts appearing in the plot will be from $z_1 = 10^{-12}$ to $z_2 = 10^{32}$. We will divide the redshift axis into 1000 intervals, over which we will apply the integration over $z$ in equation (6.6.3), to reproduce the Figure 7 from [3]. $\Delta z$ is the interval width at redshift z, and the integration has a lower limit

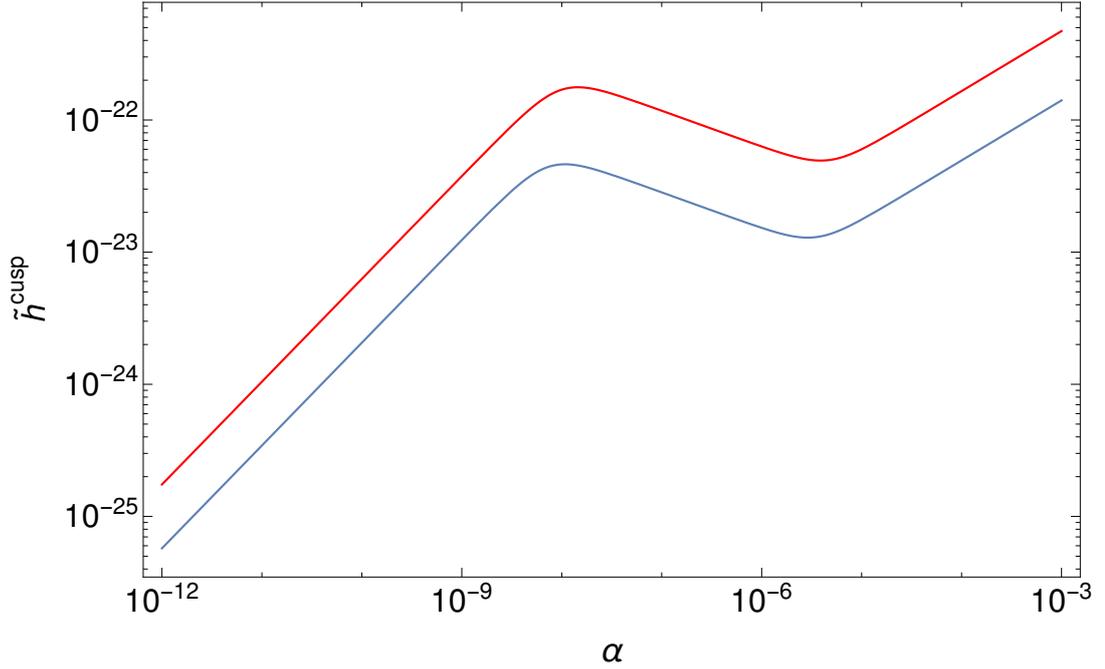

Figure 6.6.1: The amplitude of GWBs from cusps for $c = 1$. The red line assumes the values of set 0, $g_1 = 1 = g_2$ (results from [4]), while the blue line assumes the values of set 1, $g_1 = 0.43$ and $g_2 = 0.305$ (results using the odd-harmonic string from section 6.2.5).

of

$$z = z_1 \left(\frac{z_2}{z_1}\right)^{\frac{b}{N}} \tag{6.6.4}$$

and an upper limit of

$$z + \Delta z = z_1 \left(\frac{z_2}{z_1}\right)^{\frac{b+1}{N}}, \tag{6.6.5}$$

where the counter $b$ takes integer values in the interval $0 \leq b \leq N$ and $N = 1000$. Note that the functional form of $d^2R/dzdh$, as well as the limits of integration $h_{min}(z)$ and $h_{max}(z)$, change with the cosmological era, matter or radiation. Therefore, we will deal with the rate of GWBs from the loops formed in the radiation era, the loops surviving into the matter era and the loops formed in the matter era separately.

For the loops formed in the radiation era, we will integrate the function in equation (6.5.27), with limits of integration $h_{min}(z)$ and $h_{max}(z)$ given by equations (6.5.34) and (6.5.35) respectively, and with the redshift ranging from $z = z_{eq} = 3366$ to $z = z_{max}$, which depends on the value of $g_2$. This corresponds to integer

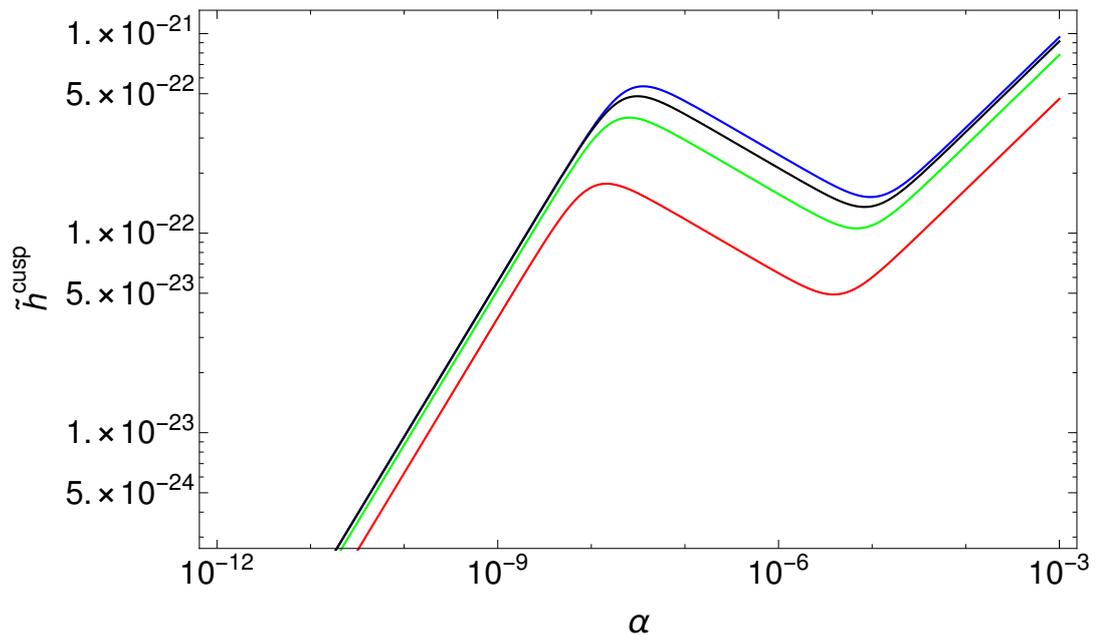

Figure 6.6.2: The amplitude of GWBs from cusps using the results from the toy model to compare with the results from [4] (red line). The green line corresponds to the set 4 values, ($g_1 = 0.839, g_2 = 0.842, c = 4.0$) (the geometric distribution of harmonics), the black line corresponds to set 3 ($g_1 = 0.773, g_2 = 0.779, c = 6.43$) (Benford's law distribution of harmonics) and the blue line corresponds to set 2, ($g_1 = 0.680, g_2 = 0.699, c = 8.82$) (the uniform distribution of harmonics).

values of the counter $b$ from 352 to 476. The integral of $d^2R/dzdh$ over $h$ is calculated analytically. After integrating it we obtain the function

$$\frac{dR_{rad}}{dz}(z, h_{max}(z)) - \frac{dR_{rad}}{dz}(z, h_{min}(z)), \tag{6.6.6}$$

which is a function of redshift only. It is not possible to integrate this function analytically over $z$, and we will use numerical methods to integrate the above over the redshifts in the intervals given by (6.6.3). Since we calculate this for redshifts $z \gg 1$, we will use the asymptotic expressions for $\varphi_t(z)$ and $\varphi_r(z)$, given by equations (5.6.17) and (5.6.20), respectively. For $3366 \leq z \leq 10^9$ (i.e. $b \in [352, 476]$), we set $\mathcal{G}(z) = 1$ (see equation (5.6.15)), and we numerically integrate the above function over $z$, thus obtaining the plot of the rate of GWBs (6.6.3) for this range of redshifts. We apply the same method for $10^9 < z \leq 2 \times 10^{12}$ (i.e. $477 \leq b \leq 551$), setting $\mathcal{G} = 0.83$. Finally, we move to the region of redshifts $2 \times 10^{12} < z \leq 3.8 \times 10^{20}$, which corresponds to the integral bins $552 \leq b \leq 739$. In this region of redshift values, the problem that appears is that the first term and the second term in equation (6.6.3) obtain values that are very close to each other, to at least 12 significant digits whose number increases with $z$. This issue appears for $z > 10^{16}$ in particular. The absolute values of each term in (6.6.6) are of the order $10^{35} - 10^{39}$ in the redshift interval $[10^{18}, 10^{20}]$. If we plot the function (6.6.6) we obtain the blue plot in Figure 6.6.3, which results from numerical error due to the small difference of the left and right term in (6.6.6). To resolve this issue we increase the maximum machine precision in Mathematica, to ensure it calculates all the digits that are significant for the function (6.6.6). Then, we apply a fifth order Taylor series around the point $2 \times 10^{18}$, and we use it to fit the data in the redshift interval $[1.2 \times 10^{18}, 3 \times 10^{18}]$ (which corresponds to the purple plot in Figure 6.6.3). We apply the same method in the redshift interval $[3 \times 10^{18}, 4.5 \times 10^{18}]$, using a Taylor series around the point $4 \times 10^{18}$ (the green plot), in the redshift interval $[4.5 \times 10^{18}, 3 \times 10^{19}]$, using a Taylor series around $2 \times 10^{19}$ (the red plot), in the redshift interval $[3 \times 10^{19}, 10^{20}]$, using a Taylor series around the point $10^{20}$ (the yellow plot), and in the redshift interval $[10^{20}, 3.8 \times 10^{20}]$, using a Taylor series around the point $2 \times 10^{20}$ (the black plot). In this way we resolve the issue that was caused when we plotted the function (6.6.6) directly due to the numerical error. When integrating over $z$, we will use the expressions obtained from the Taylor series to obtain the final result, which is the rate of the GWBs given in equation (6.6.3). In this way, we have finally obtained the plot

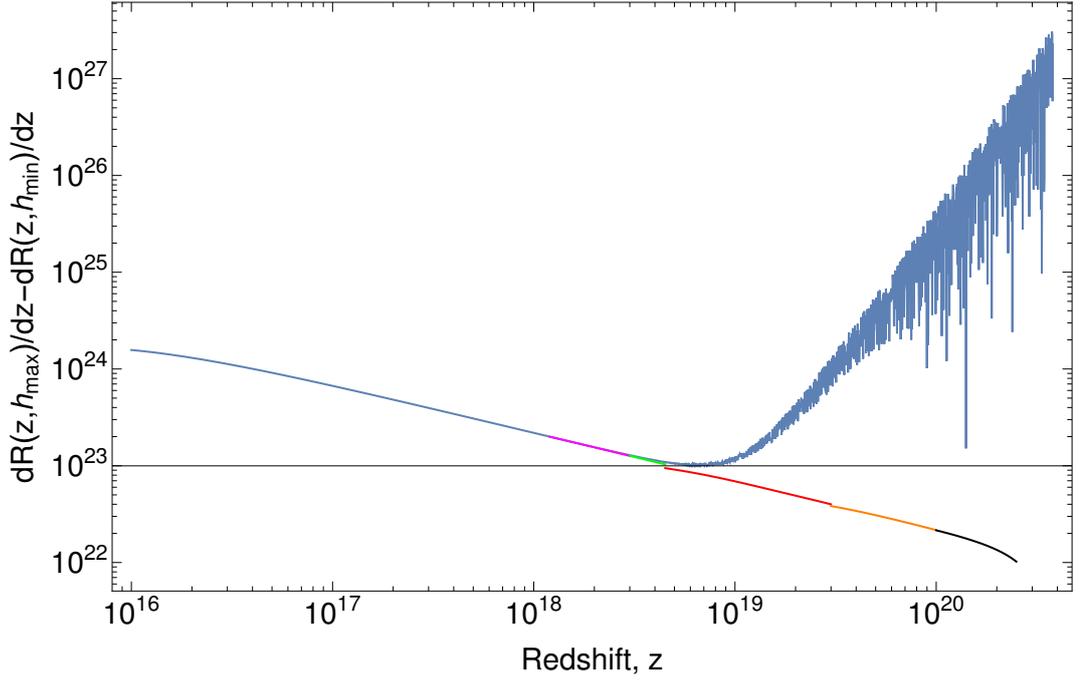

Figure 6.6.3: We show the plot of the function (6.6.6) (the blue plot) and the plot that results if we apply Taylor series to the function (6.6.6) around several points (the multicolour plot). Each color of the multicolour plot indicates a different Taylor series about some redshift $z$.

of $R(z)$ in the radiation era shown in Figure 6.6.4.

During the matter era, we have two types of loops, the ones that formed in the radiation era and survive into the matter era, which are given by equation (6.5.29), and the ones formed in the matter era, which are given by equation (6.5.30). We will plot these in separate plots over the redshifts $[0.288, 3366]$, which correspond to the $b$ values in the interval $[261, 352]$. Note that during the matter era we cannot use the asymptotic expressions for $\varphi_t(z)$ and $\varphi_r(z)$, apart from $z \ll 1$. For this reason we will use the Simpson rule to plot the function (6.6.3) in the matter era. For the calculation, we apply the following procedure. Firstly, we create a list of the redshift values for each interval from $z$ to $z + \Delta z$ in the matter era, which consists of the points

$$z_i = z + i\frac{\Delta z}{n} \tag{6.6.7}$$

where $i$ obtains all the integer values such that $0 \le i \le n$. Using these, we can easily obtain the values of $\varphi_t(z_i)$ and $\varphi_r(z_i)$ at each $z_i$. Then, we have calculated all the unknowns needed to calculate the values of the function

$$\frac{dR_{rad}}{dz}(z_i, h_{max}(z_i)) - \frac{dR_{rad}}{dz}(z_i, h_{min}(z_i)), \qquad (6.6.8)$$

at each of the redshift values $z_i$. By placing these values into a list we can integrate the above quantity using the Simpson rule formula, given in section 2.6.1. This results in the values of the function (6.6.3) which is the desired quantity to plot.

In Figure 6.6.4, we can see the the event rate of GWBs, given in equation (6.6.3), versus redshift. The orange plot corresponds to the event rate during the radiation era, which ranges from $z_{eq} = 3366$, the redshift at matter-radiation equality, and $z = 3.81 \times 10^{20}$. For $z = 10^{-8}$ to $z = z_{eq}$, i.e. during the matter era, we have the light red shaded plot, which corresponds to the event rate arising from loops formed in the radiation era, and the blue shadded plot, which corresponds to the event rate from loops formed in the matter era. Note that this plot has been reproduced using the same assumptions as the ones used in [3], and it corresponds to the upper left-hand-side plot of Figure 7 of Ref [3]. In that plot we can see that a discontinuity appears in the line that gives the event rate of the loops in the matter era. We have removed this discontinuity in the Figure 6.6.4 by using the Simpson's rule we described above.

We will also plot the event rate for different values of $g_1$, $g_2$ and the cusp number, $c$, based on the results we obtained for the odd-harmonic string and the toy model of section 5.7, and presented in section 6.2.5, to compare them with the Set 0 values. Note that the set 0 values, which have been used to plot figure 6.6.4, are the ones used in [3]. Therefore, this figure will serve as a comparison between our results. We find that, out of the three sets of values $(g_1, g_2, c)$ presented in section 6.2.5, it is set 2 that has the greatest impact to the event rate compared to set 0. We can see in figure 6.6.5 that the change is of one order of magnitude. We should mention that in figure 6.6.5, during the era where matter formed loops and radiation formed loops that survive into matter coexist, we have plotted the integral of the quantity on the left-hand-side of equation (6.5.28), unlike in figure (6.6.4) where we have plotted the quantities on the right-hand-side of equation (6.5.28) separately. The results that are provided using the set 1 values are effectively the same as the results of [3], which use the set 0 values. The results using the set 3 and set 4 values have a difference of around half an order of magnitude compared to the set 0 results.

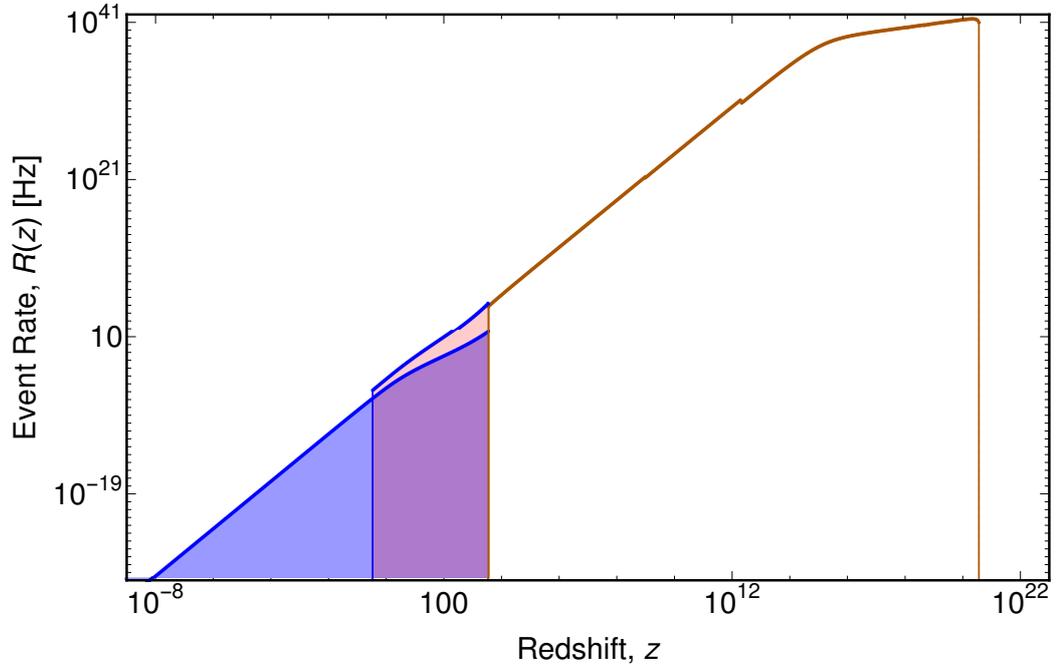

Figure 6.6.4: The plot of the GWB event rate, $R(z)$, versus redshift, $z$, for $G\mu = 10^{-8}$, $g_1 = g_2 = 1$, $c = 1$ and $f = 100 Hz$, for the "one-scale" cosmic string network model. The orange curve corresponds to the event rate of GWBs from cusps during the radiation era. After the matter-radiation equality, the GWB event rate splits into two contributions. The one originates from loops surviving from the radiation era into the matter era, which corresponds to the light red shaded curve. The other originates from loops formed in the matter era, and corresponds to the light blue shaded era. We notice that the two contributions overlap from the matter-radiation equality until the evaporation of the radiation formed loops. During the overlap of the contributions, the radiation ear loops contribution is the one which dominates, since it is by at least two orders of magnitude stronger than the matter era loops contribution. After the evaporation of the last surviving radiation formed loops, only the matter era loops contribute towards the GWB event rate.

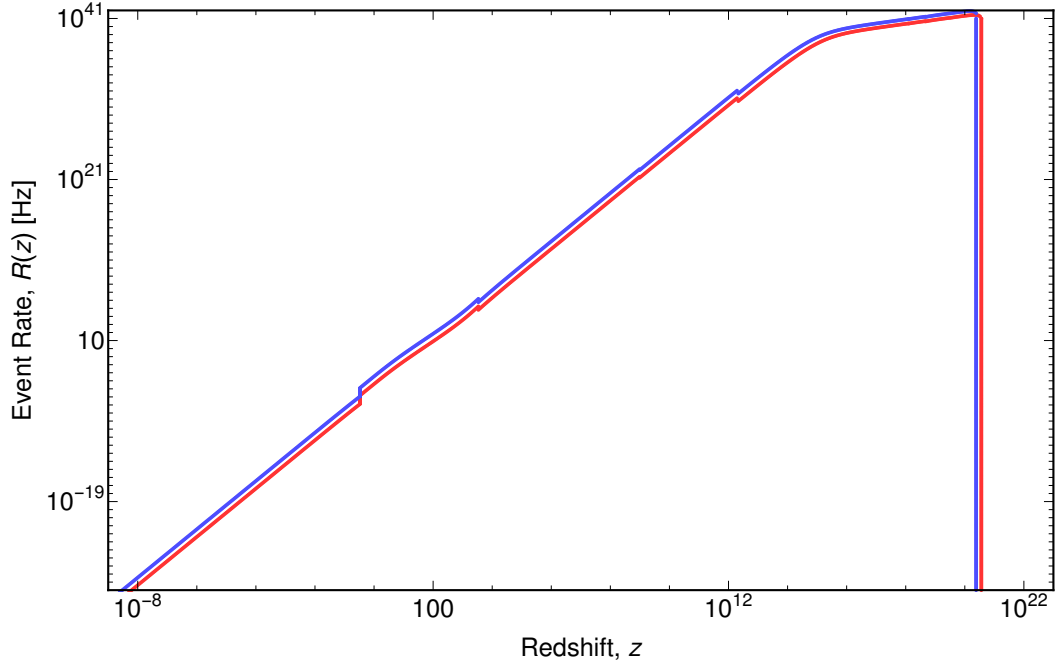

Figure 6.6.5: The plot of the GWB event rate, $R(z)$, versus redshift, $z$, for $G\mu = 10^{-8}$ and $f = 100Hz$. The red line corresponds to the set 0 values $g_1 = g_2 = 1$, $c = 1$ and the blue line corresponds to the set 2 values ($g_1 = 0.680$, $g_2 = 0.699$, $c = 8.82$). We have found that set 1 of values, which leaves the value of $c$ unchanged does not produce a difference in the event rate compared to the set 0 values. Also, the sets 3 and 4 have a difference of around half an order of magnitude compared to the set 0 values. We omit these results from the plot because there would be overlapping of the curves. Note the discontinuity that appears at high redshift is because of the piecewise function $\mathcal{G}$ defined in equation (5.6.15). At the matter-radiation equality another discontinuity appears. This is caused because at redshift greater than $z_{eq}$ we use the asymptotic behaviour of the cosmological functions, while at redshift less that $z_{eq}$ we use Simpson's rule to evaluate the cosmological functions numerically. The discontinuity at lower redshift appears because the last of radiation loops evaporate and they do not contribute towards the GWB event rate. From the expression of the number density of the radiation loops we can see that they evaporate in a discontinuous manner. Finally, in the region of redshift where the contribution of the radiation formed loops that survive into the matter era and of the matter loops overlaps we have summed the two contributions, unlike Figure 6.6.4, to offer a simplified view of the curves.

# CHAPTER 7

---

# Conclusions

---

In this thesis we examined properties of high harmonic cosmic string loops in the wire approximation. Cosmic strings are topological defects that can be produced from grand unification theories during a symmetry breaking. Depending on the energy scale at which they are formed they obtain a unique value of their string tension $\mu$, which is an observable parameter. A potential observation of cosmic strings that will determine the value of $\mu$ can give us an insight about the physics of very high energies. Equivalently, restrictions on the potential values of $\mu$ can restrain parameters in grand unification theory models.

We began by introducing the fundamental theories and mathematical tools we will use, in chapter 2 to chapter 4. We derived the Einstein field equations and the gravitational waves from the linearized Einstein equations. We also introduced the cosmological concepts needed for the following chapters, the FLRW universe and the matter sources in the universe.

We then described the Nambu-Goto cosmic string in spacetime and specialised it to a closed loop in a flat background spacetime in chapter 5. Assuming that these loops lie in the family of the odd-harmonic string and motivated by the sensitivity of the predictions of the GWB from cusps to these parameters, we calculated the cusps per period they support, the values of the second derivates of the left- and right-movers and the distribution of the cusps on the string. We also developed a toy model that estimates the evolution of a single cosmic string loop which self-intersects and chops into smaller lower harmonic daughter loops. By assuming the harmonic distribution of the closed loops chopped off the long string network, we used the above result to estimate the number of cusps produced in a spacetime volume $dV(z)$ from the network of loops per loop period.

In chapter 6, we showed the derivation of the gravitational wave signal from a cosmic string network, following [4]. We described two cosmic string network models, Model I which is a small loop approach that appears in [4], and Model II which is a large loop approach that corresponds to Model 1 in [3]. We then compared our results derived from our high harmonic cosmic string loop approach with the results in the aforementioned publications, where the harmonic order is not taken into account. We found that the rate of gravitational wave signals arriving to Earth should be increased when assuming a high harmonic content in the cosmic string loop network, taking into account their enhanced cusp production, while the amplitude of the gravitational wave remains relatively unchanged.

Our aim in this thesis has been to evaluate the effect of the harmonic distribution in the loop network to the observed gravitational wave signal from the network on Earth. Such a calculation and associated search for a gravitational wave signal could give us an insight in or constrain early universe phase transitions. It is in particular the value of the string tension $\Gamma G\mu$ that can be found observationally that is correlated with the energy scale at which the phase transition occured. In the case of cosmic superstrings, a potential observation of a gravitational wave event emanating from such an object could serve as a direct observation of an M-theory predicted phenomenon. In the recent publication by the LIGO Collaboration [3], it was found that $\Gamma G\mu$ is constrained to be less than $5 \times 10^{-8}$ for Model 1, which is the model that we use to compare our results. The above was found for the case of (topological) cosmic strings and with the stochastic background analysis. When using the gravitational wave burst analysis, the results are only relevant for cosmic superstrings, where the intercommutation probability can be less than one. The restrictions that are applied can be seen in Figure 6 of their publication.

Until now the effect high harmonic loops could have on the signal had not been regarded. It has been our effort in this thesis to discuss the potential consequences that high harmonics could impose on the GWB signal. Further steps taken into this direction is to improve the toy model as discussed in section 5.7.9. This model has a big impact on the results for the gravitational wave signal, and it is currently limiting us to take into account loops of harmonic order only up to $N = 7$. Improvement of the model could help to use a wider range of harmonics. A step towards the direction of superstrings could be to implement our results for type $F$ superstrings, as in [55], which is a continuation of [4]. When it comes to calculations of gravitational waves from superstrings with junctions, as in [125, 103], we believe that implementation is not that direct since the odd-harmonic string does not support junctions. Despite the fact the cusps is a local event, a study of the effects of junctions on $|\vec{a}''|$ and $|\vec{b}''|$ should be applied. When it comes to the calculation of cusps per period, a generalization is not obvious at all. Finally, a study of the effects of pseudocusps could be a further step for our model. Following [57], we could implement their approach for the odd-harmonic string, calculate the frequency of the pseudocusp appearance, and how fast the cosmic string velocity drops around the cusp. As a further note, our high harmonic loop results can be applied to the calculation of the stochastic background of gravitational wave radiation from cosmic string loops, as performed in [4], a calculation that is not included in this thesis.

Additional calculations

## A.1 Transformation of cusp coordinates

The positions of cusps for some particular cosmic string loop solutions are mentioned in [14], where they are solved analytically. To use these cusp solutions we should find the transformation between our formalism and the one used in [14].

The left- and right-movers for the one harmonic loop in [14] are

$$\vec{a}'_A(\sigma_A + \tau_A) = -\sin(\sigma_A + \tau_A)\vec{i} + \cos(\sigma_A + \tau_A)\vec{j} \qquad \text{(A.1.1)}$$

$$\vec{b}'_A(-\sigma_A + \tau_A) = -\sin(-\sigma_A + \tau_A)\vec{i} + \cos\phi\cos(-\sigma_A + \tau_A)\vec{j} + \\ \sin\phi\cos(-\sigma_A + \tau_A)\vec{k} \qquad \text{(A.1.2)}$$

Solving for the cusps $\vec{a}'_A = \vec{b}'_A{}^1$, we reach the following system of equations

$$-\sin(\sigma_A + \tau_A) = -\sin(-\sigma_A + \tau_A) \qquad (A.1.3)$$

$$\cos(\sigma_A + \tau_A) = \cos\phi\cos(-\sigma_A + \tau_A) \qquad (A.1.4)$$

$$0 = \sin\phi\cos(-\sigma_A + \tau_A). \qquad (A.1.5)$$

If we apply the transformations

$$\tau_A = \tau'_A + \pi, \; \sigma_A = \sigma'_A - \frac{\pi}{2} \qquad (A.1.6)$$

the system (A.1.3)-(A.1.5) transforms to

$$\cos(\sigma'_A + \tau'_A) = -\cos(\sigma'_A - \tau'_A), \qquad (A.1.7)$$

$$\sin(\sigma'_A + \tau'_A) = -\cos\phi\sin(\sigma'_A - \tau'_A), \qquad (A.1.8)$$

$$0 = \sin\phi\sin(\sigma'_A - \tau'_A). \qquad (A.1.9)$$

If a point $(\sigma_A^{(c)}, \tau_A^{(c)})$ is a solution of (A.1.3)-(A.1.5), then the point $(\sigma_A'^{(c)}, \tau_A'^{(c)}) = (\sigma_A^{(c)} + \pi/2, \tau_A'^{(c)} - \pi)$ is a solution of (A.1.7)-(A.1.9), and vice versa. Also, the two systems will have the same number of solutions. Since the system (A.1.7)-(A.1.9) corresponds to the cusp condition $\vec{a}' = \vec{b}'$ for the left- and right-movers defined in section 5.4.2, equations (5.4.9)-(5.4.10), this gives us the coordinate transformation between our cusp points and the cusp points in [14]. The one harmonic string forms two simultaneously cusps at $\tau_A = \pi/2$, and therefore, this transforms to $\tau = \pi/2$ (recall that the period of the string motion is $\pi$).

For the case of the Kibble-Turok string, the left- and right-movers in [14] are,

$$\begin{aligned}\vec{a}'_A(\sigma_A + \tau_A) = &\left[\alpha\cos\left(3(\sigma_A + \tau_A)\right) + (1-\alpha)\cos\left(\sigma_A + \tau_A\right)\right]\vec{i} + \\ &\left[\alpha\sin\left(3(\sigma_A + \tau_A)\right) + (1-\alpha)\sin\left(\sigma_A + \tau_A\right)\right]\vec{j} + \\ &2\sqrt{\alpha - \alpha^2}\sin\left(\sigma_A + \tau_A\right)\vec{k}\end{aligned} \qquad (A.1.10)$$

$$\vec{b}'_A(-\sigma_A + \tau_A) = \cos\left(-\sigma_A + \tau_A\right)\vec{i} + \sin\left(-\sigma_A + \tau_A\right)\vec{j} \qquad (A.1.11)$$

---

$^1$Note that the cusp condition in [14] does not exhibit a minus sign in this case, compared to the condition in section 5.3.3, because the left-mover is a function of $(\tau - \sigma)$, instead of $(\sigma - \tau)$.

The system of equations that determine the cusp positions is

$$\alpha \cos\left(3(\sigma_A + \tau_A)\right) + (1-\alpha)\cos\left(\sigma_A + \tau_A\right) = \cos\left(-\sigma_A + \tau_A\right), \quad \text{(A.1.12)}$$

$$\alpha \sin\left(3(\sigma_A + \tau_A)\right) + (1-\alpha)\sin\left(\sigma_A + \tau_A\right) = \sin\left(-\sigma_A + \tau_A\right), \quad \text{(A.1.13)}$$

$$2\sqrt{\alpha - \alpha^2}\sin\left(\sigma_A + \tau_A\right) = 0. \quad \text{(A.1.14)}$$

It has two solutions at the cusp points $(\tau_A = 0, \sigma_A = 0)$ and $(\tau_A = 0, \sigma_A = \pi)$. If we apply the transformations

$$\tau_A = -\tau_A' + \frac{\pi}{2},\ \sigma_A = \sigma_A' + \frac{\pi}{2} \quad \text{(A.1.15)}$$

then the above system transforms to

$$-\alpha\cos\left(3(\sigma_A' - \tau_A')\right) - (1-\alpha)\cos\left(\sigma_A' - \tau_A'\right) = \cos\sigma_A' + \tau_A' \quad \text{(A.1.16)}$$

$$-\alpha\sin\left(3(\sigma_A' - \tau_A')\right) - (1-\alpha)\sin(\sigma_A' - \tau_A') = \sin(\sigma_A' + \tau_A') \quad \text{(A.1.17)}$$

$$2\sqrt{\alpha - \alpha^2}\sin\left(\sigma_A' - \tau_A'\right) = 0. \quad \text{(A.1.18)}$$

These are equivalent to the system of cusp solutions for the Kibble-Turok loop (5.4.20)-(5.4.21). This implies that for our choice of notation, the cusps of the Kibble-Turok string appear at $(\pi/2, \pi/2)$ and $(\pi/2, 3\pi/2)$.

## A.2 Coordinate transformation of the GW waveform

We can decompose $I_+^\lambda$ and $I_-^\lambda$, defined in equations (6.2.2)-(6.2.3), into

$$I_+^\lambda = a_+ l^\lambda + d_+^\lambda \quad \text{(A.2.1)}$$

and

$$I_-^\lambda = a_- l^\lambda + d_-^\lambda, \quad \text{(A.2.2)}$$

where $a_\pm$ are constants. In the above, $a_+ l^\lambda$ ($a_-^\lambda$) corresponds to the first integrated term of equation (6.2.2) ((6.2.3)) and $d_+^\lambda$ ($d_-^\lambda$) corresponds to the remaining integrated terms of equation (6.2.2) ((6.2.3)). We can prove that the term $a_\pm l^\lambda$ can be gauged away, and therefore it does not correspond to a physical effect, following [4].

Under a linear coordinate transformation in the Fourier domain $k^\lambda \to k^{\lambda'}$, the GW waveform $\kappa_{\lambda\nu}$ transforms to

$$\kappa^{\lambda\nu'} = \kappa^{\lambda\nu} + k^\lambda \xi^\nu + k^\nu \xi^\lambda, \tag{A.2.3}$$

with $k^\nu \propto l^\nu$, since we are calculating the GW waveform along the direction of a cusp. We also know that $\kappa^{\lambda\nu} \propto I^{(\lambda}I^{\nu)}$ from equation (6.1.17). Combining the above, we find that

$$\kappa^{\lambda\nu'} = G\mu \left( 2a_+a_- l^\lambda l^\nu + a_+ d^\nu_- l^\lambda + a_- d^\lambda_+ l^\nu + d^\lambda_+ d^\nu_- + a_+ d^\lambda_- l^\nu + a_- d^\nu_+ l^\lambda + d^\nu_+ d^\lambda_- \right) + k^\lambda \xi^\nu + k^\nu \xi^\lambda. \tag{A.2.4}$$

If we choose

$$\xi^\nu = a_+a_- l^\nu + a_+ d^\nu_- + a_- d^\nu_+ \tag{A.2.5}$$

and $k^\nu = G\mu l^\nu$, and substitute in equation (A.2.4) we find

$$\kappa^{\lambda\nu'} = G\mu \left( d^\lambda_+ d^\nu_- + d^\lambda_- d^\nu_+ \right), \tag{A.2.6}$$

which proves that the terms $a_+a_- l^\lambda l^\nu$, $a_+ l^\lambda d^\nu_-$ and $a_- d^\lambda_+ l^\nu$ can be ignored in the expression of the GW waveform, since they can be gauged away. Therefore, the expression of the waveform which exhibits only terms of physical significance is

$$\kappa^{\lambda\nu} = G\mu d^{(\lambda}_+ d^{\nu)}_-. \tag{A.2.7}$$

The above, explains why the GW waveform introduced in [4] differs from the one in [126], where it was not taken into account that the terms in $I^\lambda_\pm$ proportional to $l^\lambda$ are pure gauge. In terms of the logarithmic Fourier transform defined in (6.1.16), we find that $\tilde{\kappa}^{\lambda\nu} = G\mu |f| d^{(\lambda}_+ d^{\nu)}_-$.

## A.3  Obtaining the interpolating functions

In sections 5.6.1.2 and 6.5.1 we substituted quantities that have different expressions in the matter and in the radiation era with one smooth function that interpolated between the two eras. In this section we will show how to obtain the interpolating functions. In doing this, it is important to note that some terms that are important in the matter era are very small in the radiation era, and vice versa, due to the very large difference in redshifts. In particular, during the matter era it holds that $z < z_{eq}$ and during the radiation era $z > z_{eq}$, where $z_{eq}$ is the redshift at matter and radiation equality, i.e. when the matter density of the universe was equal to its radiation density.

In section 5.6.1.2, we defined the function

$$\varphi_l(z) = (1+z)^{-3/2} \left(1 + \frac{z}{z_{eq}}\right)^{-1/2}. \qquad (A.3.1)$$

which allowed us to obtain a single expression for the cosmic time at any redshift $t \simeq t_0 \varphi_l(z)$. In equation (A.3.1), during the matter era ($z < z_{eq}$), the term $(1 + z/z_{eq})$ will be roughly unity. Therefore, $\varphi_l(z) \simeq (1+z)^{-3/2}$ during matter domination. In the radiation era ($z > z_{eq}$), $\varphi_l$ can be written as

$$\varphi_l(z) \simeq z^{-3/2} \frac{z^{-1/2}}{(1+z_{eq})^{-1/2}} = z^{-2}(1+z_{eq})^{1/2}. \qquad (A.3.2)$$

In the above we have approximated

$$\frac{(1+z)^{1/2}}{\left(1 + \frac{z}{z_{eq}}\right)^{1/2}} \simeq z_{eq}^{1/2} \simeq (1+z_{eq})^{1/2}. \qquad (A.3.3)$$

We see that the expressions of $\varphi_l$ in both the matter and in the radiation era correspond to the cosmic time obtained in equation (5.6.38) and to the cosmic time in equation (5.6.39), respectively. Therefore, it indeed acts as an interpolating function between the two eras.

In section 6.5.1, we defined the function

$$\varphi_n(z) = z^3(1+z)^{-7/6}\left(1+\frac{z}{z_{eq}}\right)^{11/6} \qquad (A.3.4)$$

which allows us to use a single expression for the logarithmic density $R(f,z)$ in equation (6.5.10). To obtain the above, we will first need to calculate the logarithmic density $R(f,z)$ for three redshift intervals, for $z < 1$, for $1 < z < z_{eq}$ and for $z > z_{eq}$. In particular, for $1 < z < z_{eq}$, we can combine equations (5.6.38) and (5.6.44) with the expression of $R(f,z)$ from equation (6.5.10) to find that

$$R(f,z) = \frac{54\pi}{2}c\alpha^{-8/3}g_2^{-2/3}f^{-2/3}t_0^{-5/3}z^{11/6}. \qquad (A.3.5)$$

Here, we have approximated $1 + z \simeq z$, $1 + z/z_{eq} \simeq 1$ and $(1+z)^{1/2} - 1 \simeq z$. We can follow the same procedure for $z < 1$. Combining the above equations and fully simplifying the terms, we find that

$$R(f,z) = \frac{54\pi}{8}c\alpha^{-8/3}g_2^{-2/3}f^{-2/3}t_0^{-5/3}z^3. \qquad (A.3.6)$$

Here, we have approximated $1 + z \simeq 1$, $1 + z/z_{eq} \simeq 1$ and $(1+z)^{1/2} - 1 \simeq z/2$. For the latter expression, we have used a first order Taylor expansion of $(1+z)^{1/2} \simeq 1 + z/2$, expanded around $z = 0$. For the expression of $\varphi_n$ during the radiation era ($z < z_{eq}$), we substitute equations (5.6.39) and (5.6.45) into equation (6.5.10) to find that

$$R(f,z) = 36\pi c\alpha^{-8/3}g_2^{-2/3}f^{-2/3}t_0^{-5/3}z_{eq}^{-11/6}z^{11/3}. \qquad (A.3.7)$$

We now want to construct an interpolating function which will consist of the terms, $z$, $(1 + z)$ and $(1 + z/z_{eq})$ to some power. The term $z$ is significant for the redshift range $z < 1$. The terms $z$ and $(1 + z)$ are significant for $1 < z < z_{eq}$ and all of the above mentioned terms are significant in the radiation era. Given the above, we can easily obtain the interpolating function (A.3.4). Note that the factor of $10^2$ in equation (6.5.10) is used to approximate the values of $72\pi/2$, $54\pi/8$ and $54\pi/2$.